\documentclass[a4paper, 11pt, twoside]{Thesis}  
\graphicspath{Figures/}  

\usepackage{verbatim}  
\usepackage{vector}  
\usepackage{braket}
\usepackage{dsfont}
\usepackage{pbox}
\usepackage{slashed}
\usepackage[dvipsnames]{xcolor}
\usepackage{tikz}
\usepackage[sorting=none, backend=biber, style=numeric,citestyle=numeric-comp]{biblatex} 
\usepackage{bibentry}
\nobibliography{Bibliography}
\usetikzlibrary{decorations.pathmorphing}

\tikzset{snake it/.style={decorate, decoration=snake}}

\hypersetup{urlcolor=blue, colorlinks=true}  

\begin{document}


\title  {One Formula To Match Them All: \\ The Bispinor Universal One-Loop Effective Action}
\authors  {\texorpdfstring
            {\href{your web site or email address}{Benjamin Summ}}
            {Benjamin Summ}
            }
\addresses  {\groupname\\\deptname\\\univname}  
\date       {\today}
\subject    {}
\keywords   {}

\maketitle
\frontmatter  

\setstretch{1.3}  

\fancyhead{}  
\rhead{\thepage}  
\lhead{}  

\pagestyle{fancy}  




 
 
 
 
 

 

\pagestyle{plain}  

\null\vfill
\textit{``I want to live my life taking the risk, all the time, that I don't know anything like enough yet, that I haven't understood enough, that I can't know enough, that I am always hungrily operating on the margins of a potentially great harvest of future knowledge and wisdom. I wouldn't have it any other way. ''}

\begin{flushright}
Christopher Hitchens
\end{flushright}

\vfill\vfill\vfill\vfill\vfill\vfill\null
\cleardoublepage  
\pagestyle{fancy} 
\kurzfassung{
Seit der Entdeckung des Higgs Bosons ist das Standardmodell der Teilchenphysik formal vollst{\"a}ndig. Trotz dieser formalen Vollst\"andigkeit existieren beobachtete Ph\"anomene, wie etwa Neutrinooszillationen und Dunkle Materie, die nicht im Kontext des Standardmodells erkl\"art werden k\"onnen. Erweiterungen des Standardmodells, die diese Unzul\"anglichkeiten in Angriff nehmen, enthalten h\"aufig neue Teilchen mit Massen weit \"uber der elektroschwachen Skala. Aufgrund der vorliegenden Massenhierarchie sind effektive Feldtheorien ein geeignetes Werkzeug f\"ur das Studium solcher Erweiterungen. Jedoch muss man, um ein gegebenes Modell der neuen Physik auf eine passende effektive Feldtheorie abzubilden, eine Matchingberechnung durchf\"uhren, was schon auf dem Einschleifen-Niveau eine beachtliche Aufgabe sein kann, wenn das ultraviolette (UV) Modell ausreichend kompliziert ist. In dieser Dissertation pr\"asentieren wir eine L\"osung des Matchingproblems auf Einschleifen-Niveau in der Form der Bispinor Universal One-Loop Effective Action (BSUOLEA), die einen vorberechneten Ausdruck der effektiven Wirkung darstellt. Dieser Ausdruck ist durch Ableitungen der Lagrangedichte des UV Modells parametrisiert. Die BSUOLEA wird durch die Anwendung von Funktionalmethoden hergeleitet und ist ausreichend allgemein um ein renormierbares, Lorentzinvariantes UV Modell, das Vektorfelder, Spin-$1/2$ Fermionen und skalare Felder enth\"alt, an eine effektive Feldtheorie mit Operatoren bis zu und einschlie{\ss}lich Massendimension sechs auf dem Einschleifen-Niveau zu matchen. Dies beinhaltet supersymmetrische Modelle, die durch dimensionale Reduktion regularisiert sind. Da der vorberechnete Ausdruck viele Operatoren enth\"alt, wurde er in ein momentan privates Mathematica-Paket implementiert, welches die Matchingberechnung automatisch ausf\"uhrt und auf der BSUOLEA basiert. Die Anwendung dieses Pakets, das in naher Zukunft \"offentlich zug\"anglich gemacht wird, wird anhand des Matchings des um ein Singlet erweiterten Standardmodells an die Standard Model Effective Field Theory (SMEFT) veranschaulicht. Der Anwendungsrahmen wird anschlie{\ss}end auf die Behandlung einer nicht-renormierbaren Erweiterung des Standardmodells ausgeweitet, um eine Vektorresonanz zu beschreiben, die ihren Ursprung im Confinement einer Eichgruppe an hohen Energien hat. Dieses Modell wirft einige interessante theoretische Fragen auf, da eine solche Vektorresonanz ein System mit Zwangsbedingungen zweiter Klasse beschreibt. Wir f\"uhren die Analyse der Zwangsbedingungen f\"ur dieses Modell durch und leiten das erzeugende Funktional her, um Funktionalmethoden f\"ur das Matching zu nutzen. Abschlie{\ss}end wird das Modell an die SMEFT gematcht und wir pr\"asentieren das komplette Einschleifenergebnis, einschlie{\ss}lich Operatoren bis zu Massendimension sechs.    
}
\cleardoublepage

\abstract{

Since the discovery of the Higgs boson, the Standard Model of particle physics is formally complete. Despite this formal completeness, there are observed phenomena, such as neutrino oscillations and dark matter, that cannot be explained in the context of the Standard Model alone. Extensions of the Standard Model that address these shortcomings often contain new particles with masses far above the electroweak scale. Due to the presence of a mass hierarchy, effective field theory is a suitable tool for the study of such extensions. However, in order to map a given model of new physics onto a suitable effective field theory one has to perform a matching calculation, which can be a formidable task already at the one-loop level, if the ultraviolet (UV) model is sufficiently complicated. In this thesis, we provide a solution to the matching problem at the one-loop level in the form of the Bispinor Universal One-Loop Effective Action (BSUOLEA), which is a pre-computed expression for the effective action, parameterised in terms of derivatives of the Lagrangian of the UV model. The BSUOLEA is derived using functional methods and is sufficiently general to match any renormalisable, Lorentz invariant UV model containing vector fields, spin-$1/2$ fermions and scalar fields to an effective field theory, containing operators of mass dimension up to and including six at the one-loop level. This includes supersymmetric models regularised in dimensional reduction. Since the pre-computed expression contains many operators, it is implemented into a currently private Mathematica package, which automatically performs the matching based on the BSUOLEA. We illustrate the use of this package, which will be made publicly available in the near future, by applying it to the matching of the Singlet Extended Standard Model to the Standard Model Effective Field Theory (SMEFT). We then expand the framework in order to apply it to a non-renormalisable extension of the Standard Model containing a vector resonance, which is assumed to arise from the confinement of a gauge group at high energies. This model raises some interesting theoretical questions since such a vector resonance comprises a system with second-class constraints. We perform the constraint analysis of the model and derive the generating functional in order to apply functional methods to the matching. We finally match the model to the SMEFT and present the full one-loop result including operators up to mass dimension six.  

}

\cleardoublepage  


\lhead{\emph{List of Publications}}  
\lop{
The research presented in this thesis was conducted at the Institute for Theoretical Particle Physics and Cosmology at RWTH Aachen University from October 2017 to September 2020. Parts of this thesis are based on the publications listed below.
\begin{itemize}
    \item[{[BS1]}] \fullcite{Summ:2018oko}. 
    \item[{[BS2]}] \fullcite{Kramer:2019fwz}.
\end{itemize}
Section~\ref{sec: DRED_to_DREG} of this thesis is partially based on my own contributions to Ref.\ [BS1]. Furthermore, Sections~\ref{sec: mixed_statistics_and_open_cds} and \ref{sec: results_ops_coeffs} are based on my contributions to Ref.\ [BS2]. The results and discussions of these sections are entirely based on my own work. Finally, the applications of Sections~\ref{sec: top_quark_out}--\ref{sec: gluinoOut} are based on the presentation of these applications in Ref.\ [BS2]. The scalar contribution, $\Delta \lambda^{1\Loop,\phi}$, to the threshold correction computed in Section~\ref{sec:lambdacalc} was computed by Dr.\ Alexander Voigt. The remaining computations are completely based on my contributions to Ref.\ [BS2].   
}

\cleardoublepage
\lhead{\emph{Contents}}  
\tableofcontents  


\setstretch{1.3}  



\mainmatter	  
\pagestyle{fancy}  


\clearpage
\begin{refsection}
\lhead{\emph{Introduction}}
\chapter{Introduction}
\label{ch: intro}
%
With the discovery of the Higgs boson at the Large Hadron Collider (LHC), see Refs.\  \cite{Aad:2012tfa,Chatrchyan:2012xdj}, the Standard Model of
Particle Physics (SM) is formally complete. However, the SM does not provide a complete description of Nature as it, for instance, fails to provide a dark matter candidate and cannot explain the existence of non-vanishing neutrino masses. Furthermore, in the SM the spontaneous breaking of the electroweak symmetry is achieved through the ad hoc addition to the SM Lagrangian of a classical potential for the Higgs boson without any explanation for the origins of such a potential. 
%

%
Although these shortcomings of the SM are well-known and many possible extensions of the SM have been proposed to address these problems, direct searches for physics beyond the SM have not been successful thus far. Exclusion limits for masses of new particles introduced in SM extensions often exceed the TeV scale. The possibility of a large mass hierarchy between the SM and new physics (NP) suggests the need for an extensive effective field theory (EFT) program\footnote{EFTs have become a widely used tool in particle phenomenology. Some of the foundational works are Refs.\ \cite{decoupling,Ovrut:1979pk,Ovrut:1980dg,Ovrut:1980uv,Collins:1978wz,Weinberg:1978kz,Coleman:1969sm,Callan:1969sn,Witten:1976kx,GILMAN1980129,WEINBERG198051}. The ideas surrounding EFTs are reviewed in Refs.\ \cite{Georgi:1994qn,Manohar:1996cq,Polchinski:1992ed,Pich:1998xt,Rothstein:2003mp,Skiba:2010xn,Burgess:2007pt,Kaplan:1995uv}.} for two reasons.  
%
First of all, the observables predicted in models with large mass hierarchies usually suffer from large logarithmic quantum corrections
which should be resummed in order to obtain precise predictions. Such a resummation of large logarithms is achieved in an EFT through the use of the renormalisation group.
Secondly, in an EFT the effects of heavy particles, which cannot be produced directly at the energies of current experiments, are captured by operators of mass dimension larger than four. These higher dimensional operators are suppressed by increasing powers of the scale of NP and therefore decrease in importance with increasing mass dimension, allowing for the systematic improvement of predictions. In the context of the SM one of the most relevant EFTs is the Standard Model Effective Field Theory (SMEFT)\footnote{For a review of the SMEFT and related topics see Ref.\ \cite{Brivio:2017vri}. There is also the Higgs Effective Field Theory (HEFT), see Refs. \cite{Feruglio:1992wf,Buchalla:2013mpa,Brivio:2013pma,Brivio:2014pfa,Gavela:2014uta,Gavela:2014vra,Alonso:2014wta,Buchalla:2015wfa,Hierro:2015nna,Buchalla:2015qju,Brivio:2016fzo,Gavela:2016vte,Merlo:2016prs,Hernandez-Leon:2017kea,Buchalla:2012qq}. The SMEFT and the HEFT are distinguished by the fact that in the SMEFT a $SU(2)_L$ scalar doublet containing the physical Higgs is present, as is the case in the SM. In the HEFT, a generic real singlet plays the role of the physical Higgs.}, which driven by the previously mentioned exclusion limits, builds on the assumption that NP only appears in the guise of heavy new particles. One may then parameterise the effects of NP in terms of higher dimensional operators, built from the SM field content and respecting the symmetries of the SM, which are added to the SM Lagrangian. Such a parameterisation of NP is model independent except for the underlying assumption that the SMEFT is valid. This parameterisation can be used as a general framework for the interpretation of LHC measurements, see e.g.\ Refs.\  \cite{Butter:2016cvz,Biekotter:2018rhp,Brivio:2019ius,Dawson:2020oco}. However, even if the SMEFT assumptions hold it is important to investigate which conclusions can be drawn about explicit extensions of the SM based on fits of Wilson coefficients, since the ultimate goal is to understand the particular form of NP realised in Nature. First steps in this direction were taken in Ref.\ \cite{Ellis:2018gqa} and in Ref.\ \cite{Dawson:2020oco} the drastic impact of the renormalisation group evolution (RGE) on SMEFT fits based on explicit NP models was demonstrated. This impact results from operator mixing, which implies that even if certain operators are not induced at the matching scale in the tree-level matching, they may be induced by the RGE. If the operators induced by the RGE are strongly constrained, their presence can have large effects on the fits and the constraints derived from these fits. As is clear from the analysis presented in Ref.\ \cite{Dawson:2020oco}, and as was pointed out by the authors of that study, the inclusion of one-loop contributions in the matching is expected to give rise to similar effects, since in general one-loop matching induces new operators not present at tree level. Furthermore, whereas the SMEFT is a useful framework it is bound by its assumption that no new light degrees of freedom are present. This is not guaranteed to be the case and interesting scenarios, such as split supersymmetry introduced in Refs.\ \cite{Wells:2003tf,Giudice:2004tc,ArkaniHamed:2004yi}, exist in which there are both new heavy degrees of freedom as well as light ones. It is therefore desirable to remain flexible in the construction of EFTs at the one-loop level.
Conventional matching procedures using
Feynman diagrams, however, are often cumbersome, in particular if the
NP model contains many new heavy particles and/or complicated
interactions. In particular, in the diagrammatic matching approach one has to first construct the most general Lagrangian compatible with the symmetries and field content of the EFT before imposing a matching condition that is sufficiently general to capture all of the Wilson coefficients present in said effective Lagrangian. This procedure is prone to errors as it is easy to miss operators in the construction of the Lagrangian or to use an insufficient set of matching conditions.\footnote{That this is not a purely academic concern is exemplified by the fact that Refs.\ \cite{Haisch:2020ahr,Jiang:2018pbd} missed the contribution of the operator $Q_{Hud}$ to the matching presented in those references. This oversight was discovered in the context of this thesis and confirmed by the authors of Ref.\ \cite{Haisch:2020ahr}. It has since been corrected.}

An alternative approach to the problem of one-loop matching is through the use of functional methods. First introduced almost 40 years ago in Refs.\ \cite{Gaillard:1985uh,Cheyette:1987qz}, interest in these methods was recently reignited in Ref.\ \cite{Henning:2014wua} where they were used to derive a master formula for the effective Lagrangian of an EFT obtained by integrating out a heavy, mass degenerate scalar multiplet at the one-loop level. In Ref.\ \cite{Drozd:2015rsp} this result was extended to the case of a non-degenerate scalar multiplet and the resulting master formula was dubbed the Universal One-Loop Effective Action (UOLEA). In both references contributions to the matching stemming from loops with both light and heavy fields were not included. The inclusion of such loops was addressed by several authors in Refs.\  \cite{Fuentes-Martin:2016uol,Henning:2016lyp,Ellis:2017jns} with Ref.\ \cite{Ellis:2017jns} providing the UOLEA including the effects of light scalar fields as well.\footnote{Some contributions from fermion loops can be calculated using these
results by squaring the fermionic trace. This treatment is incomplete
when the interactions depend on gamma matrices as is the case in a chiral theory such as the SMEFT. Furthermore,
contributions from loops containing both scalars and fermions as well
as terms with so-called open covariant derivatives were not provided.}
This master formula is expressed in terms of pre-computed, universal coefficients, which include all of the one-loop integrals appearing in the matching condition, and second functional derivatives of the action of the ultraviolet (UV) theory\footnote{In what follows, we will refer to any theory we match from as the UV theory, regardless of its realm of validity.}, which carry the theory dependence. In order to apply this master formula to the matching of any particular UV theory one has to compute these functional derivatives and insert them into the pre-computed formula. In addition traces have to be evaluated. This approach reduces the amount of redundant work present in matching computations by avoiding the repeated computation of one-loop integrals when a new model is studied. Furthermore, in this approach, the effective Lagrangian is generated from the matching condition and the action of the UV theory. The matching condition itself is the requirement that the generating functional of one-light-particle irreducible (1LPI) correlation functions in the UV theory equals the generating functional of one-particle irreducible (1PI) correlation functions in the EFT at the matching scale. It thus effectively matches all of the correlation functions and hence provides a matching condition that is sufficient to fix all Wilson coefficients by construction. Due to the generative nature of the approach it is also impossible to miss operators in the construction of the EFT Lagrangian. A downside of the functional approach is that the result of the matching usually contains redundant operators and the translation into a basis has to be performed separately. In any case, the fact that the matching automatically generates the effective Lagrangian and that the matching condition, by construction, is sufficient to match all possible correlation functions, makes this approach an excellent choice when trying to match complete models to the SMEFT (or to more general low-energy EFTs). Furthermore, since the matching is parameterised in terms of a master formula it is straightforward to automate once this master formula is known. 
%
%
It is then necessary to compute such a master formula that is sufficiently general for phenomenological applications. For this reason, in this thesis the results provided in Refs.\ \cite{Drozd:2015rsp,Ellis:2017jns} are extended such that the resulting UOLEA can be used to match any renormalisable, Lorentz invariant UV theory, containing scalar fields, vector fields and spin-$1/2$ fermions, to an EFT at the one-loop level, including operators up to mass dimension six in the EFT. The precise nature of these extensions is discussed in Chapter~\ref{chap: fcuntional matching} after the necessary theoretical background has been introduced. The extensions are also summarised in Table \ref{table: UOLEA_table}. Note that this table shows the status of the UOLEA before the results of this thesis, which have been partially published in Ref.\ \cite{Kramer:2019fwz}, became available. Since then, further developments have occurred which will be discussed in Chapter~\ref{chap: fcuntional matching}.
\begin{table}[]
\centering
\begin{tabular}{|c|ccc|c}
\cline{1-4}
 & Heavy-only & heavy-light & derivative couplings  &  \\ \cline{1-4}
Bosonic & \cite{Drozd:2015rsp} & \cite{Ellis:2017jns} & \color{red}BS &  \\
Fermionic (pure) & \color{red}BS & \color{red}BS & NR &  \\
Fermionic (mixed) & \color{red}BS & \color{red}BS & NR &  \\
Mixed statistics & \color{red}BS & \color{red}BS & NR &  \\ \cline{1-4}
\end{tabular}
\caption{Overview of the different contributions to the UOLEA divided into contributions arising from heavy-only and heavy-light loops as well as loops depending on derivative couplings. Fermionic (pure) denotes loops with only one kind of fermion, i.e.\ either Dirac or Majorana. Fermionic (mixed) denotes loops with at least one Dirac and one Majorana fermion. Mixed statistics denotes loops containing at least one boson and one fermion. Contributions that were computed as a part of this thesis are marked by BS. Contributions marked by NR only arise when the UV theory is non-renormalisable.}
\label{table: UOLEA_table}
\end{table}

Beyond the extensions shown in Table \ref{table: UOLEA_table}, we also supplement the operators provided in Refs.\ \cite{Drozd:2015rsp,Ellis:2017jns} by a set of operators that are needed when matching a UV model regularised using dimensional reduction (\DRED), as introduced in Refs.\ \cite{Siegel:1979wq,Stockinger:2005gx}, to an EFT regularised using dimensional regularisation (\DREG), introduced in Ref.\ \cite{tHooft:1972tcz}. These latter operators are useful when supersymmetric UV theories are matched to non-supersymmetric low-energy EFTs.

The remainder of this thesis is structured as follows: In Chapter~\ref{chap: fcuntional matching} we introduce the functional matching techniques upon which the UOLEA computations are based. We comment on the applicability of the results presented in Refs.\ \cite{Drozd:2015rsp,Ellis:2017jns} and point to necessary extensions of these results in order to be able to apply the UOLEA to general renormalisable, Lorentz invariant UV theories containing scalar fields, vector fields and spin-$1/2$ fermions. In Chapter~\ref{ch: UOLEA} we show in detail how the missing contributions can be calculated and discuss their implementation into a currently private Mathematica code. In Chapter~\ref{ch: applications} we present several different applications of the master formula. The first application is very simple and included to show explicitly how the master formula is employed in matching computations. The remaining applications present checks of various parts of the master formula. The final application is a full matching computation reproducing the known result of the matching of the singlet extended SM (SSM) to the SMEFT and should be considered as a proof of principal. In Chapter~\ref{ch: resonances} we discuss the application of functional matching techniques to a simplified model with vector resonances originating from a strongly coupled UV completion of the SM. This model is matched to the SMEFT and we present the full matching result. We summarise and conclude in Chapter~\ref{ch: conclusions and outlook}.  

\clearpage
\lhead{\emph{Functional matching in a scalar theory}}
\chapter{Functional matching in a scalar theory}
\label{chap: fcuntional matching}

In this chapter we discuss the functional matching approach up to the one-loop level in a scalar theory. This has been discussed extensively in the literature, see especially Refs.\  \cite{Henning:2014wua,Henning:2016lyp,Fuentes-Martin:2016uol,Zhang:2016pja}, however, it serves to illustrate the main ideas and techniques of the functional approach. In addition it allows for the introduction of notation that will be made use of when the framework is extended to the case of mixed statistics and open covariant derivatives in the following chapter. Finally, there has been some confusion in the literature with regards to the limitations of functional methods, see the discussions in Refs.\ \cite{delAguila:2016zcb,Henning:2016lyp}. It seems that some of this confusion arose due to the use of imprecise language and, whereas this confusion has been resolved, it is in the best interest of the uninitiated reader to spell out all the details of the computation. 

We consider a generic UV theory that contains heavy real scalar fields,
collected in a multiplet denoted by $\Phi$, with masses of the order of $M$ and light real scalar fields, collected in a multiplet denoted by $\phi$, with masses of the order of $m$.
We assume that $m/M < 1$ such that an EFT expansion in the mass
ratio $m/M$ is valid. The objective is then, to find a local action containing only the light fields $\phi$, which allows for the reproduction of correlation functions of the UV theory up to a fixed order $N$ in an expansion in $m/M$. In order to fix the parameters of this effective action one has to impose matching conditions. A sufficient matching condition, which can be imposed order by order in the loop expansion, is given by
\begin{align}
  \hat{\Gamma}_\text{L,UV}[\bg{\phi}]=\Gamma_\EFT[\bg{\phi}],
  \label{eq: scalar_matching_condition}
\end{align}
where $\Gamma_{\text{L,UV}}[\bg{\phi}]$ is the generating functional of 1LPI correlation functions in the UV theory\footnote{A brief discussion of this generating functional can be found in Appendix~\ref{app: 1LPI_gen}. For a general discussion of generating functionals see e.g.\ Ref.\ \cite{Peskin:1995ev}.} and $\Gamma_{\EFT}[\bg{\phi}]$ is the generating functional of 1PI correlation functions in the EFT. The hat over $\Gamma_{\text{L,UV}}[\bg{\phi}]$ signifies the fact that the functional has to be expanded in an EFT expansion up to order $N$ to obtain a local functional in $\bg{\phi}$. This is necessary since we are looking for a local effective action.\footnote{For a very insightful and thorough discussion of the role of non-locality in matching computations see Ref.\ \cite{Henning:2016lyp}.} For the determination of these generating functionals beyond
tree level a regulator must be specified, which is chosen
to be \DREG.\footnote{In principle we may also use \DRED as a regulator, see Section~\ref{sec: DRED_to_DREG}.} This
introduces a dependence on the unphysical renormalisation scale $\mu$
in both generating functionals, so that Eq.\ \eqref{eq: scalar_matching_condition} has to be imposed at some scale $\mu= Q_\text{match}$, which is referred to as the matching scale. A natural choice for the matching scale is $Q_\text{match} = M$ since this choice allows for the avoidance of large logarithms and hence reduced theoretical uncertainties in the perturbative determination of EFT parameters from the UV theory.

It should be noted that at the level of the matching condition, Eq.\ \eqref{eq: scalar_matching_condition}, one still has the choice of performing the matching using Feynman diagrams. This matching would proceed by expanding the correlation functions generated by the functionals in terms of Feynman diagrams and equating the two diagrammatic expansions. This leads to a number of different equations for the unknown parameters of the EFT. It also requires knowledge of the effective Lagrangian prior to the matching. As will be shown below, functional techniques allow for the direct computation of the effective Lagrangian from Eq.\ \eqref{eq: scalar_matching_condition}, without considering individual correlation functions, and the effective Lagrangian is generated entirely from the Lagrangian of the UV theory. 

\section{Computation of $\Gamma_\text{L,UV}$}
\label{sec: Gamma_LUV_comp}
In order to calculate $\Gamma_\text{L,UV}[\bg{\phi}]$ one starts from the generating functional of all correlation functions
\begin{align}
Z_\UV[J_\Phi,J_\phi]=\int \measure \Phi \measure \phi \,\exp\left \{i \int \rd^d x \, \big[\Lag{UV}[\Phi,\phi]+J_{\Phi}(x) \Phi(x)+J_{\phi}(x) \phi(x) \big]\right\}
\end{align}
with sources $J_\Phi$ and $J_\phi$. Since we are working in \DREG the spacetime integration is performed in $d$ dimensions, where $d$ is left arbitrary. Both the heavy and the
light fields are split into background parts $\bg{\Phi}$ and $\bg{\phi}$, respectively,
and fluctuations $\delta \Phi$ and $\delta \phi$, respectively, as
\begin{align}
\Phi&=\bg{\Phi}+\delta \Phi, \\
\phi&=\bg{\phi}+\delta \phi.
\end{align}
The background fields are defined to satisfy the classical equations
of motion in the presence of the sources,
\begin{align}
\label{eq: heavy_classical_eom}
  \frac{\delta S_\text{UV}}{\delta \Phi}[\bg{\Phi},\bg{\phi}]+J_\Phi &= 0, \\
   \label{eq: light_classical_eom}
  \frac{\delta S_\text{UV}}{\delta \phi}[\bg{\Phi},\bg{\phi}]+J_\phi &= 0,
\end{align}
where
\begin{align*}
S_\text{UV}[\Phi,\phi] = \int \rd^d x \, \Lag{UV}[\Phi,\phi],
\end{align*} 
is the action associated with the Lagrangian $\Lag{UV}$. The background fields depend on the source currents, i.e.\ $\bg{\Phi}=\bg{\Phi}[J_\Phi,J_\phi]$ and $\bg{\phi}=\bg{\phi}[J_\Phi,J_\phi]$. This dependence will usually be suppressed. We further introduce $\classical{\Phi} \equiv \bg{\Phi}[J_\Phi = 0, J_\phi]$, which through Eq.\ \eqref{eq: heavy_classical_eom} can be expressed in terms of $\bg{\phi}$, that is $\classical{\Phi}=\classical{\Phi}[\bg{\phi}]$.
The generating functional of 1LPI correlation functions of the UV theory, $\Gamma_\text{L,UV}[\bg{\phi}]$, is then given by
\begin{align}
\Gamma_\text{L,UV}[\bg{\phi}]=-i \log Z_\UV[J_\Phi=0,J_\phi]-\int \rd^d x \, J_\phi(x) \bg{\phi}(x).
\label{eq: Gamma_LUV_formula}
\end{align}
Expanding the action together with the source terms around the background fields yields
\begin{align}
  S_\text{UV}[\Phi,\phi]+\left(J_{\Phi}\right)_x\Phi_x+\left(J_{\phi}\right)_x\phi_x ={}&
  S_\text{UV}[\bg{\Phi},\bg{\phi}] + \left(J_{\Phi}\right)_x\left(\bg{\Phi}\right)_x + \left(J_{\phi}\right)_x\left(\bg{\phi}\right)_x \nonumber \\
  &-\frac{1}{2}\begin{pmatrix} \delta \Phi_x ^T  & \delta \phi_x ^T  \end{pmatrix} 
  \left(\fluct_\text{UV}\right) _{xy}
  \begin{pmatrix}
    \delta \Phi_y \\  \delta \phi_y
  \end{pmatrix} + \cdots ,
  \label{eq:action_expansion}
  \end{align}
where the matrix
\begin{align}
  \left(\fluct_\text{UV}\right)_{xy} &\equiv
    \begin{pmatrix}
      -\frac{\delta ^2 S_\text{UV}}{\delta \Phi_x \delta \Phi_y}[\bg{\Phi},\bg{\phi}] && -\frac{\delta ^2 S_\text{UV}}{\delta \Phi_x \delta \phi_y}[\bg{\Phi},\bg{\phi}] \\
      -\frac{\delta ^2 S_\text{UV}}{\delta \phi_x \delta \Phi_y}[\bg{\Phi},\bg{\phi}] && -\frac{\delta ^2 S_\text{UV}}{\delta \phi_x \delta \phi_y}[\bg{\Phi},\bg{\phi}]
    \end{pmatrix} \equiv  
    \begin{pmatrix}
     \left(\Delta_\Phi\right)_{xy} && \left(X_{\Phi \phi}\right)_{xy} \\
     \left(X_{\phi \Phi}\right)_{xy} && \left(\Delta_\phi\right)_{xy}
    \end{pmatrix} 
    \label{eq: fluctuation_operator}
\end{align}
is referred to as the fluctuation operator and the dots indicate
higher order terms in the expansion. In Eqs.\ \eqref{eq:action_expansion} and \eqref{eq: fluctuation_operator} we introduced the convention that spacetime points are written as indices and repeated spacetime indices are integrated over. We will suppress these indices whenever they do not add clarity to the presentation. Inserting the expansion given in Eq.\ \eqref{eq:action_expansion} into Eq.\ \eqref{eq: Gamma_LUV_formula} and neglecting the higher order terms, which only contribute at higher loop orders, one obtains
\begin{align}
\Gamma_\text{L,UV}[\bg{\phi}] ={}& S_\text{UV}[\classical{\Phi}[\bg{\phi}],\bg{\phi}] \nonumber 
\\ & -i \log \int \measure \delta \Phi \measure \delta \phi \, \exp{\left\{-\frac{\mathrm{i}}{2}\begin{pmatrix} \delta \Phi_x ^T  & \delta \phi_x ^T  \end{pmatrix} 
  \left(\fluct_\text{UV}\right) _{xy}
  \begin{pmatrix}
    \delta \Phi_y \\  \delta \phi_y
  \end{pmatrix}\right\}},
  \label{eq: gamma_LUV_expanded}
\end{align}
where the first term corresponds to the tree-level contribution and the second term corresponds to the one-loop contribution. At this point the functional appearing in Eq.\ \eqref{eq: gamma_LUV_expanded} is non-local with the non-locality arising from the fact that $\classical{\Phi}[\bg{\phi}]$ is non-local in general. Indeed for a real scalar field we may rewrite Eq.\ \eqref{eq: heavy_classical_eom} with $J_\Phi$ set to zero as
\begin{align}
    \left(-P^2+M^2+G[\bg{\phi}]\right) \classical{\Phi} = \lambda F[\bg{\phi},\classical{\Phi}]+ \lambda H[\bg{\phi}],
\label{eq:scalar_eom}
\end{align}
where $P^\mu \equiv iD^\mu$, $D^\mu$ is the gauge covariant derivative and the couplings that lead to linear terms in $\Phi$ in the equation of motion have been separated into the functional $G$. The remaining couplings are contained in $F$ and $H$ and are at least quadratic in $\classical{\Phi}$ and independent of $\classical{\Phi}$, respectively. On the r.h.s.\ we factored out the coupling $\lambda < 1$ between $\classical{\Phi}$ and $\bg{\phi}$ explicitly. The assumption of one universal coupling constant is made for simplicity of presentation and the inclusion of several distinct coupling constants does not present any conceptual difficulties as long as all of the couplings are small, such that a perturbative expansion is allowed. Eq.\ \eqref{eq:scalar_eom} can be solved perturbatively as a series in the coupling $\lambda$ by introducing the Green's function $\mathcal{G}(x,y)$ of the operator $-P^2+M^2+G[\bg{\phi}]$. That is the function $\mathcal{G}(x,y)$ satisfies
\begin{align}
    \left(-P^2+M^2+G[\bg{\phi}]\right)_x \mathcal{G}(x,y) = \delta (x-y).
\end{align}
In order to find a perturbative solution of Eq.\ \eqref{eq:scalar_eom} we make the Ansatz 
\begin{align}
 \classical{\Phi} = \sum_{n=0}^\infty \lambda^n \classical{\Phi}^{(n)},  
\end{align}
so that Eq.\ \eqref{eq:scalar_eom} becomes
\begin{align}
    \sum_{n=0}^\infty \lambda^n \left(-P^2+M^2+G[\bg{\phi}]\right) \classical{\Phi}^{(n)} = \sum_{n=0}^{\infty} \lambda^{n+1} p^{(n)}[\bg{\phi},\{\classical{\Phi}^{(n)}\}]+\lambda H[\bg{\phi}],
    \label{eq: scalar_eom_expanded}
\end{align}
where the series on the r.h.s.\ results from expanding the functional $F$ in powers of $\lambda$. The exact form of the coefficients is irrelevant to the discussion, however, it is important to note that the functions $p^{(n)}[\bg{\phi},\{\classical{\Phi}^{(n)}\}]$ are polynomials in $\classical{\Phi}^{(n)}$ only containing $\classical{\Phi}^{(0)},\classical{\Phi}^{(1)}, \dots, \classical{\Phi}^{(n)}$, but no higher orders than $n$. 
Then the solutions of Eq.\ \eqref{eq: scalar_eom_expanded} at each order are given by
\begin{align}
\classical{\Phi}^{(n)}(x) = 
\begin{cases}
     0, & \text{ if } n = 0  \\
     \int \rd^dy \, \mathcal{G}(x,y) H[\bg{\phi}](y), & \text{ if } n = 1  \\
       \int \rd^dy \, \mathcal{G}(x,y) p^{(n-1)}[\bg{\phi},\{\classical{\Phi}^{(n-1)}\}](y) , & \text{ if } n > 1 
    \end{cases}.
\end{align}
It is clear that the $\classical{\Phi}^{(n)}$ can be determined recursively through the use of the Green's function up to arbitrary $n$ thus providing a perturbative solution to Eq.\ \eqref{eq:scalar_eom}. The non-locality of the functional given in Eq.\ \eqref{eq: gamma_LUV_expanded} comes from the fact that this solution is inserted both into the action $S_\text{UV}$ and into the fluctuation operator $\fluct_\text{UV}$, since this introduces terms into the functional containing light background fields evaluated at different spacetime points. To obtain a fully local generating functional we first note that the Green's function $\mathcal{G}(x,y)$ is the position space representation of the operator $\left(-\hat{P}^2+M^2+G[\bg{\phi}]\right)^{-1}$ in the Hilbert space constructed\footnote{In this construction the operator is of the form $-D^2+Y$, where Y is a positive matrix. The sign of the derivative term is accounted for by Wick rotating to Euclidean space. The positivity of $Y=M^2+G[\bg{\phi}]$ can only be assured if the mass dominates this term, i.e.\ if the couplings are sufficiently small.} in Ref.\ \cite{Ball:1988xg}, i.e.\ 
\begin{align}
    \mathcal{G}(x,y) = \Braket{x|\left(-\hat{P}^2+M^2+G[\bg{\phi}]\right)^{-1}|y}.
\end{align}
We use $\hat{P}$ for the representation independent operator whose position space representation is given by $P$. We may expand the operator $\left(-\hat{P}^2+M^2+G[\bg{\phi}]\right)^{-1}$ in a Neumann series to obtain 
\begin{align}
    \left(-\hat{P}^2+M^2+G[\bg{\phi}]\right)^{-1} = \left(M^{-2}\right)\sum_{n=0}^{\infty} \left[(\hat{P}^2-G[\bg{\phi}])M^{-2}\right]^n,
    \label{eq: local_Green's_func}
\end{align}
which has the position space representation
\begin{align}
  \Braket{x|M^{-2}\sum_{n=0}^{\infty} \left[\left(\hat{P}^2-G[\bg{\phi}]\right)M^{-2}\right]^n|y} = M^{-2}\sum_{n=0}^{\infty} \left[\left(P^2-G[\bg{\phi}]\right)M^{-2}\right]_x^n \delta (x-y).
\end{align}
This indeed yields a local expression as indicated by the presence of the delta function. Using this expansion one can expand the perturbative solution of Eq.\ \eqref{eq:scalar_eom} to a given order in $1/M$. The same procedure can be applied in the general case to obtain a solution of Eq.\ \eqref{eq: heavy_classical_eom}. This truncated, local solution of Eq.\ \eqref{eq: heavy_classical_eom} will be denoted by $\classical{\hat{\Phi}}$ and should be substituted into Eq.\ \eqref{eq: gamma_LUV_expanded} to obtain a local functional, truncated at a given order in $1/M$. Having addressed the question of locality we turn to the second term on the r.h.s.\ of Eq.\ \eqref{eq: gamma_LUV_expanded}. The Gaussian path integral gives rise to the usual functional determinant 
\begin{align}
\label{eq: general_functional_integral}
    - i \log \int \measure \delta \Phi \measure \delta \phi \, \exp{\left\{-\frac{\mathrm{i}}{2}\begin{pmatrix} \delta \Phi_x ^T  & \delta \phi_x ^T  \end{pmatrix} 
  \left(\fluct_\text{UV}\right) _{xy}
  \begin{pmatrix}
    \delta \Phi_y \\  \delta \phi_y
  \end{pmatrix}\right\}} = i c_s \log \det Q_{\text{UV}},
\end{align}
where the factor $c_s$ accounts for the statistics and number of degrees of freedom of the fields integrated over. For example, for real scalar fields it takes the value $c_s = 1/2$, whereas for complex scalar fields it takes the value $c_s= 1$. In this section we will only treat real scalar fields. The extension to more general theories is the subject of Chapter~\ref{ch: UOLEA}. The main results of this section are
\begin{align}
\label{eq: gamma_LUV_tree}
\hat{\Gamma}^\text{\tree}_\text{L,UV}[\bg{\phi}] &= S_\text{UV}[\classical{\hat{\Phi}}[\bg{\phi}],\bg{\phi}], \\
\hat{\Gamma}^\text{1\Loop}_\text{L,UV}[\bg{\phi}] &= \frac{i}{2} \log \det \fluct_\text{UV},
\label{eq: gamma_LUV_1L}
\end{align}
where in $\fluct _\text{UV}$ the local expansion $\classical{\hat{\Phi}}$ is substituted for $\classical{\Phi}$. 

Before moving on to the r.h.s.\ of the matching condition given in Eq.\ \eqref{eq: scalar_matching_condition} we show the useful identity (see also Ref.\ \cite{Fuentes-Martin:2016uol})
\begin{align}
\log \det Q_\text{UV} &= 
\log \det \left(\Delta_\Phi - X_{\Phi \phi} \Delta_\phi^{-1}X_{\phi \Phi}\right)
+\log \det \Delta_\phi \nonumber \\
&= \log \det \left(Q_\text{UV}/\Delta_\phi\right) + \log \det \Delta_\phi,
\label{eq: Schured_Q_UV}
\end{align}
where we introduced, $Q_\text{UV}/\Delta_\phi=\Delta_\Phi - X_{\Phi \phi} \Delta_\phi^{-1}X_{\phi \Phi}$, the Schur complement of $\Delta_\phi$ relative to $Q_\text{UV}$, and where $\Delta_\phi^{-1}$ is the inverse of the differential operator $\Delta_\phi$, c.f.\  $\left(-\hat{P}^2+M^2+G[\bg{\phi}]\right)^{-1}$. The standard way to show the validity of Eq.\ \eqref{eq: Schured_Q_UV} is to introduce appropriate matrices of unit determinant on the l.h.s.\ as in Refs.\ \cite{Fuentes-Martin:2016uol, Zhang:2016pja}. For future reference it is more convenient to transform the argument of the exponential in Eq.\ \eqref{eq: gamma_LUV_expanded} by performing shifts in the integral. Up to a prefactor the argument of the exponential can be written as
\begin{align}
\begin{pmatrix} \delta \Phi ^T  & \delta \phi ^T  \end{pmatrix} 
  \fluct_\text{UV}
  \begin{pmatrix}
    \delta \Phi \\  \delta \phi
  \end{pmatrix}={}&
    \left(\delta \phi^T +\delta \Phi^T X_{\Phi \phi}\Delta^{-1}_\phi\right) \Delta_\phi  \left(\delta \phi +  \Delta_\phi ^{-1} X_{\phi \Phi} \delta \Phi\right) + \delta \Phi^T \Delta_\Phi \delta \Phi \nonumber \\
    &-\delta \Phi^T X_{\Phi \phi}\Delta^{-1}_\phi X_{\phi \Phi} \delta \Phi.
\end{align}
Changing variables in the $\delta \phi$-integral to $\delta \phi' = \delta \phi+  \Delta_\phi ^{-1} X_{\phi \Phi} \delta \Phi$, which is a constant shift w.r.t.\ the integral, the argument of the exponential becomes
\begin{align}
\begin{pmatrix} \delta \Phi ^T  & \delta \phi ^T  \end{pmatrix} 
  \fluct_\text{UV}
  \begin{pmatrix}
    \delta \Phi \\  \delta \phi
  \end{pmatrix} = \delta \phi'^T \Delta_\phi \delta \phi' + \delta \Phi^T \left(\Delta_\Phi - X_{\Phi \phi}\Delta_\phi^{-1} X_{\phi \Phi}\right) \delta \Phi.
  \end{align}
At this point the light and heavy fluctuations are completely decoupled and the two integrals can be performed separately, yielding precisely the two determinants appearing on the r.h.s.\ of Eq.\ \eqref{eq: Schured_Q_UV}. The above line of reasoning to decouple different fluctuations will play a pivotal role in the generalisation to a field content of mixed statistics. 

\section{Computation of $\Gamma_\text{EFT}$}
We now consider the r.h.s.\ of Eq.\ \eqref{eq: scalar_matching_condition}. Since we are going to impose the matching condition order by order in perturbation theory such that the effective action can be determined at a given order in the loop expansion, it is sensible to organise the effective action by loop orders from the beginning. We therefore write
\begin{align}
    S_\text{EFT}[\phi] = S^\tree_\text{EFT}[\phi] + S^{1\Loop}_\text{EFT}[\phi],
\end{align}
where $S^\tree_\text{EFT}$ contains operators with coefficients fixed at tree level and $S^{1\Loop}_\text{EFT}$ contains operators with coefficients fixed at one loop. Following the procedure of the previous section, we expand the action around the background fields of the EFT, that is we write
\begin{align}
\phi = \bg{\phi} + \delta \phi, 
\intertext{with $\bg{\phi}$ defined by}
\funcDer{S_\text{EFT}}{\phi}[\bg{\phi}] + J_\phi^\text{EFT} = 0.
\label{eq: EFT_background}
\end{align}
The definition in Eq.\ \eqref{eq: EFT_background} ensures that the term linear in $\delta \phi$ vanishes and up to second order in fluctuations we therefore find 
\begin{align}
    S_\text{EFT}[\phi]+\left(J^\text{EFT}_\phi\right)_x \phi_x = S^\tree_\text{EFT}[\bg{\phi}]+S^{1\Loop}_\text{EFT}[\bg{\phi}]+\left(J^\text{EFT}_\phi\right)_x \left(\bg{\phi}\right)_x -\frac{1}{2}\delta \phi_x^T \left(\Delta^\text{EFT}_\phi\right)_{xy} \delta \phi _y,
\end{align}
with 
\begin{align}
\label{eq: delta_phi_scalar}
    \left(\Delta_\phi^\text{EFT}\right)_{xy} \equiv -\frac{\delta^2 S_\text{EFT}^\tree}{\delta \phi_x \delta \phi _y}[\bg{\phi}].
\end{align}
In Eq.\ \eqref{eq: delta_phi_scalar} only the tree-level action appears since the contribution from the one-loop action would yield results of two-loop order in the functional determinant. We then obtain the generating functional of 1PI correlation functions accurate up to the one-loop level as
\begin{align}
    \Gamma_\text{EFT}[\bg{\phi}] &= -i \log Z_\EFT [J^\text{EFT}_\phi]-\int \rd^d x \, J^\text{EFT}_\phi(x) \bg{\phi}(x) \nonumber \\ &= S^\tree_\EFT [\bg{\phi}]+S^{1\Loop}_\EFT [\bg{\phi}]+\frac{i}{2}\log \det \Delta_\phi^\EFT,
\end{align}
which can be organised in loop orders as 
\begin{align}
\label{eq: Gamma_EFT_tree}
\Gamma^\tree_\EFT &= S^\tree_\EFT [\bg{\phi}], \\
\label{eq: Gamma_EFT_1Loop}
\Gamma^{1\Loop}_\EFT &= S^{1\Loop}_\EFT [\bg{\phi}]+\frac{i}{2}\log \det \Delta_\phi^\EFT.
\end{align}
The one-loop part consists of two contributions, namely the loop-induced Wilson coefficients, contained in $S^{1\Loop}_\EFT [\bg{\phi}]$, and the genuine one-loop contributions coming from the tree-level matched Wilson coefficients, contained in $\log \det \Delta_\phi^\EFT$.
\section{Matching and soft region cancellation}
\label{sec: ir_cancellation_purely_scalar}
We now impose the matching condition and show explicitly that the infrared (IR) physics is correctly reproduced by the EFT. The main ideas of this proof were introduced in Ref.\ \cite{Fuentes-Martin:2016uol} and the proof was first presented in Ref.\ \cite{Zhang:2016pja}. Imposing Eq.\ \eqref{eq: scalar_matching_condition} and using Eqs.\ \eqref{eq: gamma_LUV_tree} and \eqref{eq: gamma_LUV_1L} as well as Eqs.\ \eqref{eq: Gamma_EFT_tree} and \eqref{eq: Gamma_EFT_1Loop} we find
\begin{align}
\label{eq: tree_level_matching}
    S^\tree_\EFT [\bg{\phi}] &= S_\text{UV}[\classical{\hat{\Phi}}[\bg{\phi}],\bg{\phi}], \\
    \label{eq: 1L_matching}
S^{1\Loop}_\EFT [\bg{\phi}] &=  \frac{i}{2} \log \det \fluct_\text{UV}-\frac{i}{2}\log \det \Delta_\phi^\EFT.
\end{align}
Given the first of these equations, we may compute $\Delta_\phi^\EFT$, which according to Eq.\ \eqref{eq: delta_phi_scalar} is given by
\begin{align}
\label{eq: delta_phi_eft_through_UV}
     \left(\Delta_\phi^\text{EFT}\right)_{xy} &= -\frac{\delta^2 S_\text{EFT}^\tree}{\delta \phi_x \delta \phi _y}[\bg{\phi}] = -\left. \funcDer{}{\phi_x} \right \vert  \funcDer{}{\phi _y}  S_\text{UV}[\classical{\hat{\Phi}}[\phi],\phi] \nonumber \\
     &= -\left. \funcDer{}{\phi _x} \right \vert  \left( \funcDer{\classical{\hat{\Phi}}^z}{\phi _y} \funcDer{S_\text{UV}[\classical{\hat{\Phi}}[\phi],\phi]}{\classical{\hat{\Phi}}^z}+  \funcDer{S_\text{UV}[\classical{\hat{\Phi}}[\phi],\phi]}{\phi_y}\right)\nonumber \\ 
     &= - \left(\left. \funcDer{\classical{\hat{\Phi}}^z}{\phi _x}\right \vert \left. \secondfuncDer{S_\text{UV}[\classical{\hat{\Phi}}[\phi],\phi]}{\classical{\hat{\Phi}}^z}{\phi_y}\right \vert+\left. \secondfuncDer{S_\text{UV}[\classical{\hat{\Phi}}[\phi],\phi]}{\phi_x}{\phi_y}\right \vert\right),
\end{align}
where in going from the second to the third line we used that 
\begin{align}
\label{eq: heavy_eom_local_expansion}
    \funcDer{S_\text{UV}[\classical{\hat{\Phi}}[\phi],\phi]}{\classical{\hat{\Phi}}^z} = 0
\end{align}
up to a given order in the $1/M$ expansion and we introduced the notation
\begin{align}
    \left . \funcDer{}{\phi _y}\right \vert = \left . \funcDer{}{\phi _y}\right \vert _{\phi=\bg{\phi}}.
\label{eq: evaluated_der_def}
\end{align}
Using Eq.\ \eqref{eq: heavy_eom_local_expansion} we also find 
\begin{align}
\label{eq: eft_UV_relation_scalar_only}
    0 &= \left. \funcDer{}{\phi _y} \right \vert \funcDer{S_\text{UV}[\classical{\hat{\Phi}}[\phi],\phi]}{\classical{\hat{\Phi}}^z} = \left. \funcDer{\classical{\hat{\Phi}}^x}{\phi _y} \right \vert \secondfuncDer{S_\text{UV}[\classical{\hat{\Phi}}[\bg{\phi}],\bg{\phi}]}{\classical{\hat{\Phi}}^x}{\classical{\hat{\Phi}}^z} + \left. \secondfuncDer{S_\text{UV}[\classical{\hat{\Phi}}[\phi],\phi]}{\phi_y}{\classical{\hat{\Phi}}^z} \right \vert \nonumber \\
    &= -\left. \funcDer{\classical{\hat{\Phi}}^x}{\phi _y} \right \vert \left(\hat{\Delta}_\Phi \right)_{xz} - \left(X_{\phi \Phi}\right)_{yz},
\end{align}
which implies that 
\begin{align}
   \left. \funcDer{\classical{\hat{\Phi}}^x}{\phi _y} \right \vert =  -\left(X_{\phi \Phi}\right)_{yz} \left(\hat{\Delta}_\Phi^{-1} \right)_{zx},
\end{align}
where $\hat{\Delta}_\Phi^{-1}$ is the local expansion of $\Delta_\Phi^{-1}$ that gives the correct result up to order $N$ in the EFT expansion.
Inserting this result into Eq.\ \eqref{eq: delta_phi_eft_through_UV} then yields
\begin{align}
   \left(\Delta_\phi^\text{EFT}\right)_{xy} &= -\left(X_{\phi \Phi}\right)_{xu} \left(\hat{\Delta}_\Phi \right)^{-1}_{uz}\left(X_{\Phi \phi}\right)_{zy} + \left(\Delta_\phi\right)_{xy} = \left(\hat{\fluct}_\UV/\hat{\Delta}_\Phi\right)_{xy},
\end{align}
where
\begin{align}
    \hat{\fluct}_\UV = \begin{pmatrix}
    \hat{\Delta}_\Phi & X_{\Phi \phi} \\
    X_{\phi \Phi} & \Delta_\phi
    \end{pmatrix},
\end{align}
and in the last equality we used the definition of the Schur complement as introduced below Eq.\ \eqref{eq: Schured_Q_UV}. Finally we use Schur's determinant identity, see Ref.\ \cite{berPotenzreihendieimInnerndesEinheitskreisesbeschrnktsind}, to write 
\begin{align}
    \log \det \Delta_\phi^\EFT &= \log \det \left(\hat{\fluct}_\UV/\hat{\Delta}_\Phi\right) = \log \det \hat{\fluct}_\UV- \log \det \hat{\Delta}_\Phi \nonumber \\ &= \log \det \left(\hat{\fluct}_\UV/\Delta_\phi\right) + \log \det \Delta _\phi - \log \det \hat{\Delta} _\Phi.
\end{align}
Inserting this into Eq.\ \eqref{eq: 1L_matching} and using Eq.\ \eqref{eq: Schured_Q_UV} we find
\begin{align}
\label{eq: final_1L_matching}
    S^{1\Loop}_\EFT [\bg{\phi}] ={}&  \frac{i}{2} \left(\log \det \left(Q_\text{UV}/\Delta_\phi\right)-\log \det \left(\hat{\fluct}_\UV/\Delta_\phi\right) + \log \det \hat{\Delta} _\Phi\right) \nonumber \\ 
    ={}& \frac{i}{2} \left(\log \det \left[\Delta_\Phi - X_{\Phi \phi} \Delta_\phi^{-1}X_{\phi \Phi}\right] - \log \det \left[ \hat{\Delta}_\Phi - X_{\Phi \phi} \Delta_\phi^{-1}X_{\phi \Phi}\right]\right) \nonumber \\
    & +\frac{i}{2} \log \det \hat{\Delta}_\Phi.
\end{align}
When evaluating the functional determinants that appear in Eq.\ \eqref{eq: final_1L_matching} one encounters loop-integrals of the form (see Section~\ref{sec: functional_dets_and_cde})
\begin{align}
\label{eq: generic_loop_int}
    \mathcal{I}(l,n,p) = \int \frac{\rd^dq}{(2\pi)^d} \frac{q^{\mu_1} \cdots q^{\mu_n}}{(q^2 - M_1^2)\cdots(q^2-M_p^2) (q^2-m_1^2)\cdots(q^2-m_l^2)},
\end{align}
with $p \geq 1$ and $l,n \geq 0$. Here the masses $M_1,\dots,M_p$ are masses of heavy particles and are all of the order of the heavy scale $M$, whereas $m_1,\dots, m_l$ are masses of light particles, which are all of the order of the light scale $m$. In order to evaluate these integrals, one can utilise the method of regions, discussed in Refs.\ \cite{Beneke:1997zp,Smirnov:2002pj,Jantzen:2011nz}, where in this case two regions are of relevance. These are the hard region in which $|q^2| \sim M^2 \gg m^2$ and the soft region in which $|q^2| \sim m^2 \ll M^2$. Since the overlap integrals arising in the method of regions are scaleless and scaleless integrals vanish in \DREG, the evaluation of Eq.\ \eqref{eq: generic_loop_int} is entirely reduced to the evaluation of the hard region and the soft region, i.e.\
\begin{align}
   \mathcal{I}(l,n,p) =  \mathcal{I}(l,n,p)_\text{hard} +  \mathcal{I}(l,n,p)_\text{soft},
\end{align}
which leads to
\begin{align}
    \log \det \fluct = (\log \det \fluct)_\text{soft} + (\log \det \fluct)_\text{hard},
    \label{eq: general_soft_hard_separation}
\end{align}
where $\fluct$ is any differential operator.
Note that in the soft region we expand the propagators of heavy particles as
\begin{align}
\frac{1}{q^2-M^2} = -\frac{1}{M^2} \sum_{n=0} ^\infty \left(\frac{q^2}{M^2}\right)^n,
\end{align}
which is the momentum-space version of the local operator expansion\footnote{In this discussion we are expanding in partial derivatives $\partial^\mu$ and hence the $\Box$ operator appears. This is sometimes referred to as a partial derivative expansion. In what follows a manifestly gauge covariant formalism will be used and we will expand in covariant derivatives as in Eq.\ \eqref{eq: local_Green's_func}. This is called a covariant derivative expansion and was introduced in Refs.\ \cite{Gaillard:1985uh, Cheyette:1987qz}. As will be discussed in Section~\ref{sec: functional_dets_and_cde}, the basic argument still holds.}
\begin{align}
    \frac{1}{-\Box - M^2} = -\frac{1}{M^2} \sum_{n=0} ^\infty \left(\frac{-\Box}{M^2}\right)^n.
\end{align}
As heavy propagators enter the computation through $\Delta_\Phi$ and since $\hat{\Delta}_\Phi$ is the local operator expansion of $\Delta_ \Phi$, we have
\begin{align}
    \log \det \hat{\Delta}_\Phi &= \left(\log \det \Delta_ \Phi \right)_\text{soft} = 0,  \\
    \log \det \left[\hat{\Delta}_\Phi - X_{\Phi \phi} \Delta_\phi^{-1}X_{\phi \Phi}\right] &= \left(\log \det \left[\Delta_\Phi - X_{\Phi \phi} \Delta_\phi^{-1}X_{\phi \Phi}\right]\right)_\text{soft},
    \label{eq: soft_local_relations}
\end{align}
where we have used that $M$ is the only scale appearing in $\Delta_\Phi$ so that all of the integrals appearing in $\left(\log \det \Delta _ \Phi\right)_\text{soft}$ are scaleless. Combining Eqs.\ \eqref{eq: soft_local_relations}, \eqref{eq: general_soft_hard_separation} and \eqref{eq: final_1L_matching} we arrive at the final expression for the one-loop contribution to the effective action
\begin{align}
     S^{1\Loop}_\EFT [\bg{\phi}] &= \frac{i}{2} \left(\log \det \left[\Delta_\Phi - X_{\Phi \phi} \Delta_\phi^{-1}X_{\phi \Phi}\right]\right)_\text{hard,trunc},
     \label{eq: final_1l_effective_action_scalar_case}
\end{align}
where \emph{trunc} indicates that the computation should be truncated at order $N$ in the $1/M$-expansion. It is interesting to interpret this result. First of all, as will become clear in the next section, the functional determinant contains only loop integrals that contain at least one propagator with a heavy mass. This is due to the fact that diagrammatically all diagrams with exclusively light fields in the loop are present both in the UV theory and in the EFT and therefore cancel in the matching. Two of the diagrams present in Eq.\ \eqref{eq: final_1l_effective_action_scalar_case} and contributing to the two-point function of $\phi$ are depicted in Figure \ref{Fig: matching_cont_illustration} below.  
\begin{figure}[h]
\center
\begin{tikzpicture}
\draw[thick, dashed] (3.5,2.7) -- (4.3,2.7);
\draw[thick, dashed] (5.3,2.7) -- (6.2,2.7);
\draw[thick, dashed] (4.8,2.7) circle [radius=0.5];;
\node[align=left, above] at (3.8,2.7)%
{$\phi$};
\node[align=left, above] at (5.8,2.7)%
{$\phi$};
\node[align=left, above] at (4.8,3.2)%
{$\Phi$};
\node[align=left, above] at (4.8,1.6)%
{$\Phi$};
\draw[thick,  dashed] (7.5,2.7) -- (8.3,2.7);
\draw[thick,  dashed] (9.3,2.7) -- (10.2,2.7);
\draw[thick, dashed] (8.8,2.7) circle [radius=0.5];;
\node[align=left, above] at (7.8,2.7)%
{$\phi$};
\node[align=left, above] at (9.8,2.7)%
{$\phi$};
\node[align=left, above] at (8.8,3.2)%
{$\Phi$};
\node[align=left, above] at (8.8,1.5)%
{$\phi$};
\end{tikzpicture}
\caption{Diagrams contributing to the two-point function of $\phi$ in Eq.\ \eqref{eq: final_1l_effective_action_scalar_case}.}
\label{Fig: matching_cont_illustration}
\end{figure}
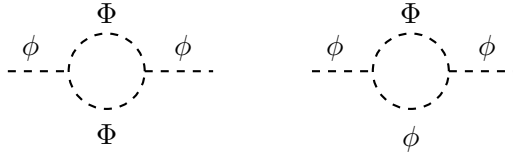
Secondly, only the hard region, that is the high-energy region, of the loop integration contributes to the one-loop Wilson coefficients. The details of the low-energy physics, which come from the soft region of the loop integral, cancel entirely in the matching. Whereas this should not be surprising, as the EFT by construction has to reproduce the IR behaviour of the UV theory, we see this cancellation explicitly without computing a single integral. The origin of the cancellation is also clear, the soft contribution of the UV theory is cancelled by the corresponding local, tree-level matched operators in the EFT (c.f.\ the origin of $\log \det \hat{Q}_\text{UV}$). Furthermore, all of the quantities appearing on the r.h.s.\ of Eq.\ \eqref{eq: final_1l_effective_action_scalar_case} can be determined without any prior knowledge of the effective action.
\section{Evaluation of the functional determinant and the\\UOLEA}
\label{sec: functional_dets_and_cde}
We now discuss the evaluation of Eq.\ \eqref{eq: final_1l_effective_action_scalar_case} using a covaraint derivative expansion (CDE), which is a method that yields manifestly gauge-invariant results by keeping the covariant derivative intact at every step of the computation. This technique was first introduced in Refs.\ \cite{Gaillard:1985uh,Cheyette:1987qz}. For simplicity we first consider the case where $X_{\Phi \phi}=0$. For a multiplet of real scalar fields $\Phi$ we have
\begin{align}
    \Delta_{\Phi} = D^2+M^2+X_{\Phi \Phi} = -P^2+M^2+X_{\Phi \Phi},
    \label{eq: explicit_Delta_Phi}
\end{align}
where as before $P_\mu = i D_\mu$, $D_\mu$ is the covariant derivative and 
\begin{align}
X_{\Phi \Phi} = - \left.\funcDer{^2 S_\text{UV,int.}}{\Phi^2}\right \vert_{\Phi=\classical{\hat{\Phi}}},
\label{eq: example_X_Phi_Phi}
\end{align}
where $S_\text{UV,int.}$ is the interaction part of the action of the UV theory. In Eq.\ \eqref{eq: explicit_Delta_Phi} both $P_\mu$ and $M$ are diagonal matrices with the diagonal of $P_\mu$ being the covariant derivative acting on the respective field in the multiplet $\Phi$ and $M^2$ being the matrix with the squared masses of the heavy particles on its diagonal. The evaluation of $\log \det \Delta _\Phi$ rests on the identity 
\begin{align}
    \log \det \Delta_\Phi = \Tr \log \Delta_\Phi
\end{align}
and on the fact that, after Wick rotation, the operator $\Delta_\Phi$ is elliptic. This latter fact allows for the construction of a Hilbert space in which $\Delta_\Phi$ acts as the position space representation of the Hamiltonian on that Hilbert space.\footnote{This Hilbert space is the same Hilbert space that was discussed in Section~\ref{sec: Gamma_LUV_comp} in the context of the determination of $\classical{\Phi}$.} The functional trace is defined to be the trace in this Hilbert space, see Ref.\ \cite{Ball:1988xg} for details. It shall be noted that the Hilbert space in which the trace is computed has no connection to the Hilbert space of one-particle states of the underlying quantum field theory. Furthermore, it is not necessary to explicitly Wick rotate the expression, since this Wick rotation would be performed on both sides of Eq.\ \eqref{eq: final_1l_effective_action_scalar_case} and the derived effective action would then be rotated back to Minkowski space. This is equivalent to performing the computation directly in Minkowski space. The operator $\Delta_\Phi$ is a function of both the momentum operator and the position operator in Hilbert space $\Delta_\Phi = \Delta_\Phi (\hat{p},\hat{x})$. The position space representation of the momentum operator is given by $i\partial_x$ as usual. The trace can be computed as
\begingroup
\allowdisplaybreaks
\begin{align}
    \Tr \log \Delta_\Phi (\hat{p},\hat{x}) &= \tr \int \frac{\rd^d q}{(2\pi)^d} \bra{q}  \log \Delta_\Phi (\hat{p},\hat{x}) \ket{q} \nonumber \\ 
    &= \tr \int \rd^d x \int \frac{\rd^d q}{(2\pi)^d} \braket{q|x}\bra{x}  \log \Delta_\Phi (\hat{p},\hat{x})\ket{q} \nonumber \\ 
    &= \tr \int \rd^d x \int \frac{\rd^d q}{(2\pi)^d} {\rm e} ^{iqx}  \log \Delta_\Phi (i \partial_x,x){\rm e} ^{-iqx} \nonumber \\ &= \tr \int \rd^d x \int \frac{\rd^d q}{(2\pi)^d} \log \Delta_\Phi (i \partial_x-q,x),
    \label{eq: CDE_equation}
\end{align}
\endgroup
where in the last equality we changed variables from $q$ to $-q$, which is customary in the literature, and $\tr$ is the trace over all indices carried by $\Delta_\Phi$. Since the UV theory is gauge-invariant the partial derivative only ever enters through the covariant derivative. We therefore find
\begin{align}
    \Lag{EFT}^{1\Loop}[\bg{\phi}] &= \frac{i}{2} \tr \left.\int \frac{\rd^d q}{(2\pi)^d}  \left(\log \Delta_\Phi\right) \right|^{P^\mu \rightarrow P^\mu - q^\mu} _\text{hard} \nonumber \\
    &= \frac{i}{2} \tr \left. \int \frac{\rd^d q}{(2\pi)^d} \log \left(-q^2+M^2-P^2+2 q\cdot P +X_{\Phi \Phi} \right)\right|_\text{hard},
    \label{eq: trace_ex}
\end{align}
where $X_{\Phi \Phi}$ has to be evaluated at $P^\mu \rightarrow P^\mu - q^\mu$ as it potentially depends on $P^\mu$. As noted in Ref.\ \cite{Drozd:2015rsp}, it is useful to write 
\begin{align}
    X_{\Phi \Phi} = U_{\Phi \Phi} + P^\mu Z_{\Phi \Phi,\mu} + Z^\dagger_{\Phi \Phi, \mu} P^\mu+\dots,
\end{align}
where $U_{\Phi \Phi}, Z_{\Phi \Phi,\mu}$ and $Z^\dagger_{\Phi \Phi,\mu}$ only depend on $P^\mu$ through commutators and the ellipses denote terms with more covariant derivatives explicitly factored out. These explicit covariant derivatives give rise to further dependence on the loop momentum $q$ after the shift $P^\mu \rightarrow P^\mu - q^\mu$. They are usually referred to as open covariant derivatives as they act on everything to their right as opposed to covariant derivatives which appear in commutators. It can be shown that the result of the trace evaluation can always be written with all covariant derivatives appearing in commutators, see Refs.\ \cite{Gaillard:1985uh,Cheyette:1987qz} for the proof. For the time being we neglect all contributions from open covariant derivatives and set $X_{\Phi \Phi} = U_{\Phi \Phi}$.
We may then expand the logarithm as 
\begin{align}
    \log \left(M^2-q^2
    -P^2+2 q\cdot P +U_{\Phi \Phi} \right) ={}& \log \left(-q^2+M^2 \right) \nonumber \\ & + \log \left(\mathds{1} - (q^2-M^2)^{-1} \left[2 q \cdot P -P^2+ U_{\Phi \Phi}\right]\right) \nonumber \\ 
    ={}& \log \left(-q^2+M^2 \right) \nonumber \\ &- \sum_{n=1}^\infty \frac{1}{n}  \left((q^2-M^2)^{-1} \left[2 q \cdot P -P^2+ U_{\Phi \Phi}\right]\right)^n
\end{align}
to obtain, up to an irrelevant constant,
\begin{align}
    \Lag{EFT}^{1\Loop}[\bg{\phi}] &= -\frac{i}{2}\sum_{n=1}^\infty \frac{1}{n} \tr \int \frac{\rd^d q}{(2\pi)^d} \left((q^2-M^2)^{-1} \left[2 q \cdot P -P^2+ U_{\Phi \Phi}\right]\right)^n.
    \label{eq: effective_action_expansion_ready}
\end{align}
Note that each term in the square bracket in Eq.\ \eqref{eq: effective_action_expansion_ready} contains either at least one covariant derivative or at least one background field. Thus, by including terms up to $n=N$ in the sum, we generate all terms up to mass dimension at least $N$ in the effective action. Hence Eq.\ \eqref{eq: effective_action_expansion_ready} can be used to generate the one-loop effective action up to operators of a given mass dimension. This allows for the determination of a generic effective action, parameterised in terms of the covariant derivative of the UV theory, and functional derivatives of the action of the UV theory. For example, for $n=2$ we have a term 
\begin{align}
    \Lag{EFT}^{1\Loop}[\bg{\phi}] &\supset -\frac{i}{4} \sum_{ij} \int  \frac{\rd^d q}{(2\pi)^d} \frac{1}{q^2-M_i^2}\left(U_{\Phi \Phi}\right)_{ij}\frac{1}{q^2-M_j^2}\left(U_{\Phi \Phi}\right)_{ji} \nonumber \\ & = \sum_{ij}F(M_i,M_j,\mu) \left(U_{\Phi \Phi}\right)_{ij}\left(U_{\Phi \Phi}\right)_{ji}, 
\end{align}
where
\begin{align}
     F(M_i,M_j,\mu) &= -\frac{i}{4} \int  \frac{\rd^d q}{(2\pi)^d} \frac{1}{(q^2-M_i^2)(q^2-M_j^2)} \nonumber \\ &= \frac{1}{4(4\pi)^2}\frac{1}{M^2_i-M^2_j}\left(\Delta M^2_{ij}+M^2_j \log\frac{M_j^2}{\mu^2}-M^2_i \log\frac{M_i^2}{\mu^2}+2\frac{\Delta M^2_{ij}}{\bar{\epsilon}}\right), 
\end{align}
with $\Delta M^2_{ij} = M_i^2 - M_j^2$, is a coefficient that can be calculated once and for all, independent of the details of the UV theory. The indices $i$ and $j$ are to be interpreted as multi-indices, containing all indices of the field $\Phi$. This includes for example flavour and gauge indices. Including all contributions needed to generate all operators up to mass dimension $N$ one generically finds a result of the form
\begin{align}
    \Lag{EFT}^{1\Loop}[\bg{\phi}] = \sum_{\alpha}\sum_{ijk\dots}F^\alpha(M_i,M_j,M_k,\dots)\mathcal{O}^{\alpha}_{ijk\dots}
    \label{eq: generic_UOLEA},
\end{align}
with universal coefficients $F^\alpha$. Once such an expression has been derived, the determination of the effective action, arising by integrating out heavy degrees of freedom from any given UV theory, reduces to computing functional derivatives and performing algebra. No integrals have to be calculated in the matching as they can be computed in a model-independent way. This is the idea behind the UOLEA and performing the computation using the expression in Eq.\ \eqref{eq: effective_action_expansion_ready} including all operators up to mass dimension six yields the heavy-only UOLEA of Ref.\ \cite{Drozd:2015rsp}. It is heavy-only because every loop only contains propagators of heavy particles as can be seen directly from Eq.\ \eqref{eq: effective_action_expansion_ready}. Loops which contain both heavy and light particles are not included due to the assumption that $X_{\Phi \phi} = 0$. This assumption can easily be lifted by shifting
\begin{align}
    X_{\Phi \Phi} \rightarrow X_{\Phi \Phi}- X_{\Phi \phi} \Delta_\phi^{-1}X_{\phi \Phi}
    \label{eq: X_Phi_Phi_shift}
\end{align}
in the discussion above. The general argument is not affected by this shift. However, it is necessary to obtain an expression for $\left.\Delta_\phi^{-1}\right|^{P_\mu \rightarrow P_\mu-q_\mu}_\text{hard}$. We have 
\begin{align}
   \left.\Delta_\phi^{-1}\right|^{P_\mu \rightarrow P_\mu-q_\mu}_\text{hard} &= \left(-q^2+2q \cdot P-P^2+m^2+X_{\phi \phi}\right)^{-1} \nonumber \\ 
   &= -\left(\mathds{1}-q^{-2}\left[2q\cdot P-P^2+m^2+X_{\phi \phi}\right]\right)^{-1} q^{-2} \nonumber \\
   &= -\sum_{k=0}^\infty \left(q^{-2}\left[2q\cdot P-P^2+m^2+X_{\phi \phi}\right]\right)^k q^{-2},
   \label{eq: Delta_phi_inv_expanded}
\end{align}
where the fact that everything is evaluated in the hard region was used in order to expand in $q^{-2}$. Performing the shift given in Eq.\ \eqref{eq: X_Phi_Phi_shift} in Eq.\ \eqref{eq: effective_action_expansion_ready} with
\begin{align}
    X_{\Phi \Phi} &= U_{\Phi \Phi}, \\
    X_{\Phi \phi} &= U_{\Phi \phi}, \\
    X_{\phi \Phi} &= U_{\phi \Phi}, \\
    X_{\phi \phi} &= U_{\phi \phi}, 
\end{align}
and using Eq.\ \eqref{eq: Delta_phi_inv_expanded} one obtains an expression that allows for the computation of a generic effective action of the same form as Eq.\ \eqref{eq: generic_UOLEA} including also heavy-light loops. Such an expression was derived in Ref.\ \cite{Ellis:2017jns} up to and including operators of mass dimension six and comprises the heavy-light UOLEA. There are several contributions missing in the versions of the UOLEA derived in Refs.\ \cite{Drozd:2015rsp} and \cite{Ellis:2017jns}, the inclusion of which would allow for a generic matching from a Lorentz invariant, renormalisable UV theory containing vector fields, scalars and spin-$1/2$ fermions, to an EFT containing operators up to mass dimension six:
\begin{enumerate}
    \item Contributions from fermions
    \item Contributions from open covariant derivatives
    \item Contributions from the gauge sector, including gauge bosons and ghosts
    \item Contributions from massive vector fields.
\end{enumerate}
The first category of missing contributions comes from the presence of fermions, both heavy and light ones, in the UV theory. In Refs.\ \cite{Drozd:2015rsp,Ellis:2017jns} it was assumed that the kinetic parts of $\Delta_\Phi$ and $\Delta_\phi$ are both of the scalar form $-P^2$. Whereas it is possible to use the results derived in this manner to compute the effective action in a purely fermionic theory by squaring the trace as long as the interaction terms are independent of gamma matrices, as was pointed out in Ref.\ \cite{Drozd:2015rsp}, this is no longer viable in a theory where both bosons and fermions are present and couple to each other or in a theory where interactions depend on gamma matrices. This is the case in virtually all models of interest to particle physics. It is therefore important to add these contributions to the existing results. Such contributions were first computed in the context of this thesis in Ref.\ \cite{Kramer:2019fwz}. Since then, results have become available in Refs.\ \cite{Angelescu:2020yzf,Ellis:2020ivx} providing similar contributions to the UOLEA, but organising the computation and final result differently. These latter results only allow for heavy fermions in the UV theory and do not provide operators resulting from mixed bosonic and fermionic loops. The results of Ref.\ \cite{Kramer:2019fwz} include both heavy and light fermions as well as mixed bosonic and fermionic loops. We will briefly discuss the differences between the results of Refs.\ \cite{Kramer:2019fwz,Angelescu:2020yzf,Ellis:2020ivx} in Chapter~\ref{ch: UOLEA}. 

Beyond the possibility of fermions being present in the UV theory, some contributions from open covariant derivatives are necessary for the matching of a general Lorentz invariant, renormalisable theory to an EFT. To see this, consider a vector field $V^\mu$, which, at the renormalisable level, can couple to scalar currents of the form $\Phi P^\mu \Phi$, in which case $X_{V \Phi}$ and $X_{\Phi V}$ have contributions from open covariant derivatives. Note that this is the case for any gauge theory containing scalar fields which are charged under the gauge group of the theory. Under the restriction that the UV theory be renormalisable and Lorentz invariant a counting of mass dimensions shows that indeed only terms containing one open covariant derivative are relevant and that these always are associated with purely bosonic couplings. Hence, no open covariant derivatives appear in fermionic contributions when restricting to this case. 

Finally, one has to address the presence of massive vector fields and massless gauge bosons in the UV theory. This topic has been touched upon in Ref.\ \cite{Zhang:2016pja}, where it was shown how to treat vector fields in the covariant diagram approach, which was developed in that reference. Based on the discussion of said reference we show how vector fields can be included into the UOLEA computation presented in Ref.\ \cite{Kramer:2019fwz} in the following chapter. Thus, in the next chapter we will extend the UOLEA by the terms necessary to perform the full matching from a Lorentz invariant, renormalisable UV theory, containing vector fields, scalar fields and spin-$1/2$ fermions, to an EFT with higher dimensional operators of mass dimension up to and including six. In order to distinguish the results presented here from those of Refs.\ \cite{Drozd:2015rsp,Ellis:2017jns,Ellis:2020ivx} we will refer to them as the Bispinor Universal One-Loop Effective Action, or BSUOLEA for short. The bispinor\footnote{It is, of course, a mere coincidence that the word bispinor has a natural abbreviation that coincides with the initials of the author of this thesis.} in the name refers to the specific parameterisation of the fermionic sector used in this thesis as introduced in Section~\ref{sec: Mixed_statistics}. We will also present operators that are needed to account for the change in regulator when matching a theory regularised in \DRED to one regularised in \DREG. These operators are especially relevant when the UV theory is supersymmetric and were first derived in Ref.\ \cite{Summ:2018oko} as a part of this thesis.


\clearpage
\lhead{\emph{The Bispinor Universal One-Loop Effective Action}}
\chapter{The Bispinor Universal One-Loop Effective Action}
\label{ch: UOLEA}

\section{Vector bosons in the UOLEA}
\label{sec: background_field_method}
So far we have not addressed the presence of vector fields in the UV theory explicitly. Moreover, in the previous section the CDE formalism was described, which is explicitly constructed to maintain gauge covariance at every step of the calculation, but the contributions to the trace originating from the gauge bosons associated with the covariant derivative were ignored. In this section we close this gap, by allowing for two kinds of vector fields in the theory, namely gauge bosons that are associated with an unbroken gauge symmetry, which we collectively denote by $A_\mu^a$, and vector bosons that acquire a mass through spontaneous symmetry breaking (SSB), which will be denoted by $V_\mu^A$ if they are heavy and by $v_\mu^A$ if they are light. We do not consider the existence of vector resonances that arise from some confining gauge subgroup in this section. However, this possibility will be entertained in the context of a specific model in Chapter~\ref{ch: resonances}.  

We first consider the case of an unbroken gauge group $G$, which might be semi-simple.\footnote{This case is well-studied and a textbook treatment can be found in Ref.\ \cite{Peskin:1995ev}.} We denote the covariant derivative by 
\begin{align}
    \cov_\mu = \partial_\mu-ig A_\mu^a T^a,
\end{align}
where $A_\mu^a$ denotes all of the gauge fields, $T^a$ are the relevant generators and the product $g A_\mu^a T^a$ has to be understood to include a sum over the different factors of the gauge group if it is not simple. The background field method proceeds by splitting the gauge fields into a background part $A_{\text{B},\mu}^a$ and a fluctuation $\delta A_\mu^a$ as for the scalar case
\begin{align}
\label{eq: gauge_field_split}
    A_\mu^a = A_{\text{B},\mu}^a + \delta A_\mu^a,
\end{align}
where only the fluctuation is integrated over in the path integral. Due to the gauge redundancy the gauge of these fluctuations has to be fixed. The strength of the background field method is that this can be accomplished without fixing the gauge of the background field itself, thus leaving the result gauge invariant with respect to background field gauge transformations, see Refs.\ \cite{Abbott:1981ke,PhysRev.162.1195,tHooft:1975uxh,DeWitt:1980jv,Boulware:1980av,Abbott:1980hw}. To perform this gauge fixing we first define the background covariant derivative
\begin{align}
    D_\mu = \partial_\mu - ig A_{\text{B},\mu}^a T^a,
\end{align}
such that 
\begin{align}
\label{eq: covariant_der_split}
    \cov_\mu = D_\mu -ig \delta A_\mu^a T^a.
\end{align}
Note that Eq.\ \eqref{eq: covariant_der_split} has exactly the form of the usual covariant derivative with the background covariant derivative taking the role of the partial derivative and the gauge field fluctuation taking the role of the full gauge field. It is clear that if the original Lagrangian is gauge invariant, then, after splitting the gauge fields as in Eq.\ \eqref{eq: gauge_field_split}, the Lagrangian has a gauge symmetry associated with the following transformation of the gauge field fluctuation
\begin{align}
    \delta A_\mu^a \rightarrow \delta A_\mu^a + \frac{1}{g} \cov_\mu \alpha^a =  \delta A_\mu ^a +  \frac{1}{g} D_\mu \alpha^a + f^{abc} \delta A_\mu^b \alpha^c,
\end{align}
keeping the background field fixed. This gauge symmetry can be fixed using the Faddeev-Popov method introduced in Ref.\ \cite{Faddeev:1967fc}. It is necessary to modify the conventional gauge-fixing functional by using the background covariant derivative instead of the partial derivative in order to retain gauge invariance at the level of the background fields. Thus, the gauge-fixing functional $G^a(\delta A)$ reads
\begin{align}
\label{eq: gauge_fixing_func_unbroken}
    G^a(\delta A) = D^\mu \delta A^a_\mu. 
\end{align}
This modifies the associated ghost Lagrangian to
\begin{align}
\label{eq: ghost_Lag_unbroken}
\Lag{gh.} = -\bar{c}^a (D^2)^{ac} c^c - \bar{c}^a D^\mu(f^{abc} \delta A_\mu^b c^c),    
\end{align}
where $c$ and $\bar{c}$ are the Fadeev-Popov ghosts. Considering the full Lagrangian, including the ghosts and the gauge-fixing term
\begin{align}
\label{eq: gauge_fixing_Lag}
    \Lag{g.f.} = -\frac{1}{2\xi} (D^\mu \delta A_\mu^a)^2,
\end{align}
where $\xi$ is the gauge parameter, it is clear that this Lagrangian is gauge-invariant under background gauge transformations as long as the gauge field fluctuation transforms in the adjoint representation of the gauge group. That is, under such background gauge transformations $A_{\text{B},\mu}^a$ acts as the gauge field and $\delta A_\mu^a$ acts as a matter field in the adjoint representation. For all other fields, both the background part and the fluctuation belong to the same representation as the original field, before the split was performed. Using this construction we may evaluate the Gaussian path integral for the gauge field fluctuations keeping the background gauge symmetry intact. 

In order to incorporate the gauge fields in the UOLEA computation we need to consider the contribution to the Lagrangian that is quadratic in gauge field fluctuations. To this end we make the following definitions 
\begin{align}
    -igF_{\mu \nu} &= -igF^a_{\mu \nu}T^a = [\cov_\mu,\cov_\nu], \\
    -igG_{\mu \nu} &= -igG^a_{\mu \nu}T^a = [D_\mu,D_\nu],
\end{align}
so that 
\begin{align}
\label{eq: F_mu_nu}
    -ig F_{\mu \nu} &= [\cov_\mu,\cov_\nu]=[D_\mu,D_\nu] - ig [D_\mu,\delta A_\mu] + ig [D_\nu, \delta A_\nu] - g^2 [\delta A_\mu, \delta A_\nu] \nonumber \\
    &= -ig G_{\mu \nu} - ig \left(D_\mu \delta A^a_\nu\right)T^a + ig \left(D_\nu \delta A^a_\mu\right)T^a -i g^2 f^{abc} \delta A_\mu ^a \delta A_\nu ^b T^c.   
\end{align}
Using Eq.\ \eqref{eq: F_mu_nu} the kinetic term of the gauge field can be expanded to second order in gauge field fluctuations. Keeping only terms with no fluctuations and terms quadratic in fluctuations\footnote{Terms linear in fluctuations vanish since the background fields satisfy the classical equations of motion as in the scalar case, c.f.\ Eqs.\ \eqref{eq: heavy_classical_eom} and \eqref{eq: light_classical_eom}. Higher order terms only contribute at higher loop orders.} the result reads 
\begin{align}
    \Lag{A} &= -\frac{1}{4}F_{\mu \nu}^a F^{\mu \nu a} \nonumber \\
    &= -\frac{1}{4}G_{\mu \nu}^a G^{\mu \nu a} - \frac{g}{2} f^{abc} G^a_{\mu \nu} \delta A^{\mu b} \delta A^{\nu c} - \frac{1}{2} \left(D_\mu \delta A_\nu^a\right)\left(D^\mu \delta A^{\nu a}\right)+\frac{1}{2} \left(D_\mu \delta A_\nu^a\right)\left(D^\nu \delta A^{\mu a}\right).
\end{align}
The part quadratic in fluctuations, including the gauge-fixing term, can be written as
\begin{align}
    \mathcal{L}_{\delta A} = -\frac{1}{2} \delta A_\mu^a \left[-D^2 g^{\mu \nu}-\left(\frac{1}{\xi}-1\right) D^\mu D^\nu + 2ig G^{\mu \nu} \right]^{ab} \delta A_\nu^b.
\end{align}
Using $P_\mu=iD_\mu$ we may then read off the contribution to the fluctuation operator coming from terms quadratic in gauge field fluctuations, which reads
\begin{align}
\label{eq: Delta_A_general_gauge}
    (\Delta_A)_{\mu \nu} = P^2 g_{\mu \nu} +\left(\frac{1}{\xi}-1\right) P_\mu P_\nu + 2 [P_\mu,P_\nu].
\end{align}
Contributions to the fluctuation operator coming from one gauge field fluctuation and one fluctuation of any other field can be computed by taking functional derivatives of the action w.r.t.\ the two fields. Thus, this treatment allows for the construction of the fluctuation operator in a completely background gauge-invariant way. The evaluation of the full fluctuation operator, including gauge bosons, will be performed in Section~\ref{sec: Mixed_statistics}.

We now discuss the inclusion of massive vector fields, which obtain their masses through SSB. The contributions of massive vector fields to the fluctuation operator have previously been discussed in Ref.\ \cite{Henning:2014wua}, where many details can be found. However, we do find an extra term in the ghost Lagrangian coupling the background fields of the vector bosons to the ghosts. This term is consistent with the appearance of such terms in the case of the SM as treated in Ref.\ \cite{Denner:1994xt}. Furthermore, in Ref.\ \cite{Henning:2014wua} the heavy vector boson fluctuations were gauge fixed in isolation. However, in a full matching computation one needs to gauge fix both the fluctuations of the massive vector fields $V^A_\mu$ and those of the remaining gauge bosons $A^a_\mu$ leading to further terms in the ghost Lagrangian. For our discussion we assume that the gauge group $G$ is spontaneously broken to a subgroup $H$ with generators $T^a$. After suitable field rotations and redefinitions of gauge group generators\footnote{see e.g.\ Appendix B of Ref.\ \cite{Henning:2014wua} for a detailed discussion.} one is left with gauge fields $A_\mu^a$ corresponding to the unbroken subgroup $H$ and gauge fields which acquire masses corresponding to the broken generators. These might in principle be heavy or light, but we here restrict our attention to the case where only heavy massive vector fields, denoted by $V^A_\mu$, are present. We choose to restrict to this case as there are significant complications when working in the broken phase of the EFT. These complications come from the fact that higher dimensional operators may contribute to the masses and mixings of the light vector bosons in the EFT, implying that the corresponding fields $v_\mu^A$ present in the UV theory differ from those in the EFT, denoted by $\tilde{v}_\mu^A$, already at tree level. In that case Wilson coefficients are introduced into the ghost interactions once they are expressed in the broken phase, a fact that is well-known from the gauge fixing of the SMEFT as discussed in Ref.\ \cite{Hartmann:2015oia}. The difference between the fields of the UV theory and those of the EFT, as well as the presence of the extra terms in the ghost Lagrangian, affect the soft region cancellation, which was discussed for the scalar case in Section~\ref{sec: ir_cancellation_purely_scalar} and will be extended to a more general case in Section~\ref{sec: soft_region_cancellation_proof}. Indeed, to the best of our knowledge a full proof of the soft region cancellation taking this into account is not available and therefore, in this particular case the soft region cancellation has to be assumed. If this assumption is made, incorporating light, massive vector fields in the computations presented in this thesis is straightforward. Keeping all this in mind, we note that assuming that light massive vector fields are not present in the UV theory is not very restrictive, especially for the case of the EFT being the SMEFT where the most relevant scenario is that in which $H = SU(3)_c \times SU(2)_L \times U(1)_Y$ and one can comfortably work in the unbroken phase.

As shown in Ref.\ \cite{Henning:2014wua} the fields $V_\mu^A$ furnish a representation of $H$ and it is clear that the gauge fields associated with $H$ can be treated in exactly the same way as the gauge bosons in the absence of SSB. Care has to be taken in the definitions of $\cov_\mu$ and $D_\mu$, which now only contain the gauge fields and generators associated with $H$. We next discuss the heavy vector fields $V_\mu^A$, which also need to be gauge-fixed in order to perform the Gaussian path integral. After performing the necessary field rotations the gauge-kinetic term for the heavy vector fields can be written as
\begin{align}
\label{eq: full_vector_gauge_kinetic_term}
  \Lag{V} = \frac{1}{2} V_\mu^A \left(\cov^2 g^{\mu \nu}-\cov^\nu \cov^\mu + [\cov^\mu,\cov^\nu] \right)^{AB} V_\nu^B,
\end{align}
as proved in Refs.\ \cite{osti_4045871,PhysRevD.46.3529,Henning:2014wua}. Next we split both the vector fields and the gauge fields into a background field and a fluctuation. However, we are only interested in the kinetic term of the vector field fluctuations, which comes from simply replacing the vector fields by their fluctuations and the gauge fields by their background fields. The remaining terms that are quadratic in fluctuations will contain mixed fluctuations and will contribute to $X_{VA}$ and $X_{AV}$, or they contain two gauge field fluctuations contributing to $X_{AA}$. Hence the contribution quadratic in fluctuations that we are interested in is given by 
\begin{align}
\label{eq: vector_fluct_gauge_kinetic_term}
    \mathcal{L}_{\delta V} = -\frac{1}{2} \delta V_\mu^A \left(-D^2 g^{\mu \nu}+D^\nu D^\mu - [D^\mu,D^\nu] \right)^{AB} \delta V_\nu ^B.
\end{align}
To this we need to add the $R_\xi$ gauge-fixing term for the fluctuations in a way that does not spoil the background gauge invariance associated with $H$. It should be recalled that the $R_\xi$ gauge-fixing condition is constructed in order to remove the mixing between the Goldstone bosons and the massive vector fields. This mixing term contains a partial derivative, which is why a partial derivative appears in the $R_\xi$ gauge-fixing term. The appearance of such a partial derivative would spoil the background gauge invariance in our formalism. We now show that indeed the partial derivative should be replaced by the background covariant derivative in the case under consideration. Let $\phi$ be the scalar multiplet that acquires a vacuum expectation value (vev) $v$ and hence breaks $G$ to $H$. Without loss of generality this multiplet can be assumed to only contain real scalar fields. The full covariant derivative $\bar{\cov}_\mu$ associated with $G$ can be written as $\bar{\cov}_\mu =\cov_\mu - iV_\mu^A t^A$, where the $t^A$ are linear combinations of the generators that are broken by SSB, see Ref.\ \cite{Henning:2014wua}. Prior to SSB the Lagrangian contains the gauge-kinetic term of the scalar multiplet $\phi=v+\chi$, where $\chi$ is the fluctuation around the vev. We thus have
\begin{align}
    \mathcal{L} \supset \frac{1}{2} \left(\bar{\cov}_\mu \phi_i\right) \left(\bar{\cov}^\mu \phi_i \right) ={}& \frac{1}{2} \left(\cov_\mu \phi_i\right) \left(\cov^\mu \phi_i\right) -iV_\mu^A \left(t_{ij}^A \phi_j\right) \left(\cov^\mu \phi_i\right)\nonumber \\ &-\frac{1}{2}\left(V^A_\mu t_{ij}^A \phi_j\right)\left(V^{\mu B} t_{ik}^B \phi_k\right), 
\end{align}
where the first term on the r.h.s.\ is not affected by SSB, since in particular $T^a v = 0 \, \forall a$. The last term on the r.h.s.\ will give rise to the mass of $V_\mu^A$ after SSB and to interactions between $V_\mu^A$ and $\chi$. The remaining term on the r.h.s.\ is the one that is relevant for the gauge fixing. Inserting $\phi=v+\chi$ into this term we find
\begin{align}
\label{eq: vector_Goldstone_mixing}
    \mathcal{L} \supset -i V_\mu^A (t_{ij}^Av_j) \cov^\mu \chi_i \supset \delta V_\mu^A (-it_{ij}^A v_j) D^\mu \delta \chi_i,
\end{align}
where we also display the contribution to the quadratic fluctuation that is relevant to the gauge fixing of $\delta V_\mu^A$. This mixing can be removed by using a background gauge covariant version of the $R_\xi$ gauge-fixing functional
\begin{align}
    G^A_{R_\xi} = \frac{1}{\sqrt{\xi}}\left(D^\mu \delta V_\mu^A - \xi (-it_{ij}^Av_j)\delta \chi_i\right),
    \label{eq: gauge_fixing_func_broken}
\end{align}
which yields the gauge-fixing Lagrangian
\begin{align}
\label{eq: R_xi_gf_Lag}
   \Lag{g.f.} =  -\frac{1}{2\xi} \left(D^\mu \delta V_\mu^A\right) \left(D^\nu \delta V_\nu^A\right) + D^\mu \delta V_\mu^A (-it_{ij}^Av_j)\delta \chi_i-\frac{\xi}{2} (-it^A_{ij}v_j)(-it^A_{kl}v_l)\delta \chi_i \delta \chi_j.
\end{align}
By construction, the second term in Eq.\ \eqref{eq: R_xi_gf_Lag} cancels with Eq.\ \eqref{eq: vector_Goldstone_mixing} and the last term gives rise to the Goldstone boson masses. The first term has to be added to Eq.\ \eqref{eq: vector_fluct_gauge_kinetic_term} after which $\Delta_V^{\mu \nu}$ can be identified as
\begin{align}
\label{eq: Delta_vectors}
    \Delta_V^{\mu \nu} =  P^2 g^{\mu \nu}-P^\nu P^\mu + \frac{1}{\xi} P^\mu P^\nu + [P^\mu,P^\nu] - M^2_V g^{\mu \nu}+ X_{VV} ^ {\mu \nu},
\end{align}
where we also included the mass term and interaction contributions to $\Delta_V^{\mu \nu}$. There is also a ghost Lagrangian associated with the modified $R_\xi$ gauge fixing, which is given by
\begin{align}
\label{eq: ghost_lag_broken}
    \Lag{gh.,V} = \bar{c}^A \left[-\left(D^2\right)^{AB}+ig D_\mu V_\text{B}^{\mu C} t^C_{AB} + \xi (-it^A_{ij}v_j)(-it^B_{ik}v_k)\right]c^B. 
\end{align}
The second term in brackets in Eq.\ \eqref{eq: ghost_lag_broken} is missing in Ref.\ \cite{Henning:2014wua}, but is consistent with Ref.\ \cite{Denner:1994xt}, where the background field method was applied to the SM. It provides the coupling of the massive vector bosons to the ghosts, which would not be present if the term was absent, as the covariant derivative only contains the gauge bosons associated with $H$. Note that it is not trivially true that one can simply add the ghost Lagrangians in Eqs.\ \eqref{eq: ghost_Lag_unbroken} and \eqref{eq: ghost_lag_broken} to incorporate a full fixing of the gauge symmetry associated with $G$, since a general gauge transformation mixes the fields $A^a_\mu$ and $V^A_\mu$, meaning that the gauge variations of the two gauge-fixing functionals include terms mixing the lower case indices with the upper case indices. Indeed, the full ghost Lagrangian incorporating the gauge variation of both Eqs.\ \eqref{eq: gauge_fixing_func_unbroken} and \eqref{eq: gauge_fixing_func_broken} is given by
\begin{align}
\Lag{gh.} &=
    \bar{c}^a \left(-D^2+iD_\mu V_\text{B}^{\mu A}t^A\right)_{ab} c^b + \bar{c}^A \left(iD_\mu V_\text{B}^{\mu C}t^C\right)_{Ab} c^b + \bar{c}^a \left(iD_\mu V_\text{B}^{\mu C}t^C\right)_{aB} c^B \nonumber \\ & \quad +\bar{c}^A \left[-\left(D^2\right)^{AB}+ig D_\mu V_\text{B}^{\mu C} t^C_{AB} + \xi (-it^A_{ij}v_j)(-it^B_{ik}v_k)\right]c^B,
\end{align}
where we explicitly partitioned the ghost fields into massive ones and massless ones, mirroring the partitioning of the gauge fields. Defining 
\begin{align}
\label{eq: Delta_H}
    \Delta_{\text{H}}^{AB} &= -(P^2)^{AB}-g[P_\mu,V^{\mu C}_\text{B} t^C]^{AB}-gV^{\mu C}_\text{B} \left(t^C P_\mu\right)^{AB} + M^{2}_{AB},\\
\label{eq: X_HL}
    X^{Ab}_{\text{HL}} &=  -g[P_\mu,V^{\mu C}_\text{B} t^C]^{Ab}-gV^{\mu C}_\text{B} \left(t^C P_\mu\right)^{Ab}, \\
\label{eq: X_LH}
    X^{aB}_{\text{LH}} &=  -g[P_\mu,V^{\mu C}_\text{B} t^C]^{aB}-gV^{\mu C}_\text{B} \left(t^C P_\mu\right)^{aB}, \\
\label{eq: Delta_L}
    \Delta_{\text{L}}^{ab} &= -(P^2)^{ab}-g[P_\mu,V^{\mu C}_\text{B} t^C]^{ab}-gV^{\mu C}_\text{B} \left(t^C P_\mu\right)^{ab},
\end{align}
we see that the ghost contribution to the generating functional of 1LPI correlation functions is given by
\begin{align}
    \Gamma_\text{L,UV,gh.} & = \log \det \begin{pmatrix}
       \Delta_\text{H} & X_\text{HL} \\
       X_\text{LH} & \Delta_\text{L}
    \end{pmatrix} = \log \det \left(\Delta_\text{H} - X_\text{HL}\Delta_\text{L}^{-1}X_\text{LH}\right) + \log \det \Delta_\text{L},
\end{align}
where Schur's determinant identity was used. We note that there is only one scale in the functional determinants, which is the mass of the ghosts. This mass defines the hard region of the momentum integrals and hence the soft region contributions completely vanish. It is clear that going through the same steps in the EFT, where the full ghost Lagrangian is given by Eq.\ \eqref{eq: ghost_Lag_unbroken} we get a completely scaleless contribution which vanishes in \DREG.\footnote{These arguments are no longer sufficient to show the soft region cancellation in the presence of light massive vector fields.} Therefore the genuine contribution to the effective Lagrangian coming from the ghosts will be given by 
\begin{align}
    \Lag{EFT,gh.} &= -\left. i \Tr \log \left(\Delta_\text{H} - X_\text{HL}\Delta_\text{L}^{-1}X_\text{LH}\right) \right \vert^{P_\mu \to P_\mu - q_\mu}_\text{hard,trunc},
\end{align}
which should be compared to Eq.\ \eqref{eq: final_1l_effective_action_scalar_case}. The only difference between these two expressions is the factor in front of the trace, which can easily be adjusted. Hence, if we compute the scalar trace, including contributions from terms with one open covariant derivative, the ghost terms can be extracted from that result through the replacements 
\begin{align}
    X_{\Phi \Phi} &\to -g[P_\mu,V^{\mu C}_\text{B} t^C]^{AB}-gV^{\mu C}_\text{B} \left(t^C P_\mu\right)^{AB}, \\
    X_{\phi \phi} &\to -g[P_\mu,V^{\mu C}_\text{B} t^C]^{ab}-gV^{\mu C}_\text{B} \left(t^C P_\mu\right)^{ab}, \\ 
    X_{\Phi \phi} &\to -g[P_\mu,V^{\mu C}_\text{B} t^C]^{Ab}-gV^{\mu C}_\text{B} \left(t^C P_\mu\right)^{Ab}, \\
    X_{\phi \Phi} &\to -g[P_\mu,V^{\mu C}_\text{B} t^C]^{aB}-gV^{\mu C}_\text{B} \left(t^C P_\mu\right)^{aB},
\end{align}
and by replacing the heavy scalar masses by the masses of the ghosts. In the next section we will extend the known UOLEA result to the BSUOLEA and include the terms necessary for this extraction to work. We will therefore not consider the ghost contributions any further. 

\section{Including mixed statistics and open covariant derivatives}
\label{sec: mixed_statistics_and_open_cds}
In this section we extend the UOLEA to a form that can be used to match a renormalisable, Lorentz invariant UV theory, containing scalar fields, vector fields and spin-$1/2$ fermions, to an EFT including effective operators up to mass dimension six. The vector fields are assumed to be associated with a gauge symmetry, either spontaneously broken or unbroken. We assume that the gauge group is semi-simple and include both heavy and light massive vector fields, keeping in mind the issues associated with the latter that were raised in the previous section. We start by briefly discussing the appearance of open covariant derivatives and the issues connected to mixed statistics. Then, the computation is presented. The presentation of the computation is based on Ref.\ \cite{Kramer:2019fwz} where fermions were included into the UOLEA using the bispinor parameterisation as a part of this thesis. In the following $X_{AB}$ denotes the matrix 
\begin{align}
    X_{AB} = - \secondfuncDer{S_\text{UV,int.}}{A}{B},
\end{align}
where $A$ and $B$ are any two field multiplets and $S_\text{UV,int.}$ is the interaction part of the UV action. This matrix is always evaluated at the expanded classical field value of heavy fields, c.f.\ Eq.\ \eqref{eq: example_X_Phi_Phi}. 
\subsection{Open covariant derivatives}
\label{sec: open_cds}
To simplify the computation as much as possible it is useful to determine how open covariant derivatives can enter. To this end, we consider the possible terms appearing in the Lagrangian containing at least one covariant derivative. These are restricted by the assumptions of gauge invariance, Lorentz invariance and renormalisability. Indeed, denoting a generic fermion by $\psi$, a generic scalar field by $\phi$, a generic vector field by $V_\mu$ and the covariant derivative by $\mathcal{D}_\mu$ we have the following allowed terms in the Lagrangian
\begin{align}
\label{eq: open_covs_fermion_current}
    &\bar{\psi}\slashed{\mathcal{D}}\psi, \\
\label{eq: open_covs_scalar_kin}    
    &\cov _\mu \phi \cov ^\mu \phi, \\
\label{eq: open_covs_scalar_current}
    &V^\mu \phi \cov_\mu \phi, \\
\label{eq: open_covs_triple_vector}
    &V_\mu V_\nu \cov ^\mu V^\nu,\\    
\label{eq: open_covs_vector_current}
    &[\cov_\mu,\cov_\nu] \cov ^{[\mu} V^{\nu]},\\
\label{eq: open_covs_double_vector}
    &V^\mu [\cov_\mu,\cov_\nu] V^\nu,\\
\label{eq: open_covs_vector_kin}    
    &\cov_{[\mu} V_{\nu]} \cov ^{[\mu} V^{\nu]},
\end{align}
where we do not list the gauge-kinetic term for the field strength tensor associated with $\cov _\mu$, which was treated in Section~\ref{sec: background_field_method}. We also use the notation $T^{[\mu \nu]}=T^{\mu \nu}-T^{\nu \mu}$ for any Lorentz tensor $T^{\mu \nu}$ carrying two indices. 

Eq.\ \eqref{eq: open_covs_fermion_current} is simply the gauge-kinetic term of the fermions. A covariant derivative appears in the fluctuation operator only when considering two fermionic fluctuations. The resulting term will appear as the leading term in $\Delta_\psi$ (c.f.\ Eq.\ \eqref{eq: explicit_Delta_Phi}). Hence, $X_{\bar{\psi} \psi}$ does not contain any open covariant derivatives. It also follows that open covariant derivatives only appear in the bosonic sector. Considering Eq.\ \eqref{eq: open_covs_scalar_kin} there are two ways in which open covariant derivatives could appear in the fluctuation operator. The first alternative is to consider two scalar fluctuations. This gives rise to the gauge-kinetic operator of the scalar field, which does not contribute to $X_{\phi \phi}$. The second possibility is to consider one scalar fluctuation and one gauge boson fluctuation. This gives rise to a genuine open covariant derivative contribution to $X_{\phi A}$ and $X_{A \phi}$ with one open covariant derivative. Similarly, one open covariant derivative appears in $X_{V \phi}$ and $X_{\phi V}$ by considering one scalar fluctuation and one vector field fluctuation in Eq.\ \eqref{eq: open_covs_scalar_current}. From this term we also obtain an open covariant derivative contribution to $X_{\phi \phi}$. Correspondingly, Eq.\ \eqref{eq: open_covs_triple_vector} gives rise to an open covariant derivative contribution to $X_{VV}$. Eq.\ \eqref{eq: open_covs_vector_current} is not present as long as $V_\mu$ obtains its mass from the spontaneous breaking of a semi-simple gauge group and $\cov_\mu$ is the covariant derivative of the residual, unbroken gauge group. An explicit proof of this fact can be found in Ref.\ \cite{Henning:2014wua}. Eq.\ \eqref{eq: open_covs_double_vector} can be non-vanishing depending on the gauge indices and yields an open covariant derivative contribution to $X_{VA}$ and $X_{AV}$ with one open covariant derivative. Note that it does not yield an open covariant derivative contribution to $X_{VV}$ since here the covariant derivatives appear in a commutator. Finally, Eq.\ \eqref{eq: open_covs_vector_kin} gives rise to the gauge-kinetic term of the vector field fluctuations presented in Eq.\ \eqref{eq: full_vector_gauge_kinetic_term} when considering two vector field fluctuations. Combining this with the $R_\xi$ gauge-fixing term one obtains Eq.\ \eqref{eq: Delta_vectors}, which by choosing $\xi = 1$ can be reduced to
\begin{align}
    \Delta_V^{\mu \nu} =  P^2 g^{\mu \nu} + 2 [P^\mu,P^\nu] - M^2_V g^{\mu \nu}+ X_{VV} ^ {\mu \nu},
    \label{eq: massive_vector_quadratic_fluct}
\end{align}
so that no open covariant derivatives appear in this part. One can then only produce one open covariant derivative contributing to $X_{A V}$ and $X_{V A}$ by considering one gauge boson fluctuation and one vector field fluctuation. In the computation presented in this thesis we will make this simplifying gauge choice. The same gauge choice will be made in Eq.\ \eqref{eq: Delta_A_general_gauge} such that it becomes
\begin{align}
    \Delta_A^{\mu \nu} = P^2 g^{\mu \nu} + 2 [P^\mu,P^\nu],
\end{align}
again avoiding terms with open covariant derivatives. Making these gauge choices we conclude that open covariant derivatives only appear in the bosonic sector and that there can be at most one open covariant derivative contributing to the individual second derivatives $X_{A B}$, which appear in the fluctuation operator for any two bosonic fields $A$ and $B$. 

\subsection{Mixed statistics}
\label{sec: Mixed_statistics}
In the previous chapter in Eq.\ \eqref{eq: general_functional_integral} it was pointed out that the result of the Gaussian path integral depends on the statistics and number of degrees of freedom associated with the field to be integrated over through the factor $c_s$. In the results presented in Refs.\ \cite{Drozd:2015rsp,Ellis:2017jns} this problem is solved by leaving $c_s$ undetermined such that it can be chosen in correspondence with the fields treated in any specific application. However, one still has to raise the question of how to treat cases in which the fluctuation operator couples fluctuations of fields with different statistics and/or different numbers of degrees of freedom. The answer is that in such a case the fluctuation operator has to be diagonalised as discussed in Ref.\ \cite{Henning:2016lyp}. This diagonalisation is equivalent to performing constant shifts in the Gaussian path integral to remove mixed fluctuations. This procedure, in the case of Yukawa theory, was described in Ref.\ \cite{Haba:2011vi}. As an illustration of such a shift consider the fields $\sigma$ and $\rho$ and the corresponding second variation
\begin{align}
\label{eq: general_shift}
  \delta^2 \LagNoT_{\sigma \rho} &=  \delta \rho \Delta_\rho \delta \rho + \delta \rho X_{\rho \sigma} \delta \sigma + \delta \sigma X_{\sigma \rho} \delta \rho + \delta \sigma \Delta_\sigma \delta \sigma \nonumber  \\ 
     &= 
     (\delta \rho + \delta \sigma X_{\sigma \rho} \Delta^{-1}_\rho) \Delta_\rho (\delta \rho + \Delta^{-1}_\rho X_{\rho \sigma} \delta \sigma) + \delta \sigma \left(\Delta_{\sigma} - X_{\sigma \rho} \Delta ^{-1}_\rho X_{\rho \sigma}\right) \delta \sigma \nonumber \\ 
     &=  
     \delta \rho' \Delta _\rho \delta \rho' + \delta \sigma  \left(\Delta_{\sigma} - X_{\sigma \rho} \Delta ^{-1}_\rho X_{\rho \sigma}\right) \delta \sigma,
\end{align}
where in the last line a change of variables from $\delta \rho$ to $\delta \rho'$, with 
\begin{align}
    \delta \rho' = \delta \rho + \delta \sigma X_{\sigma \rho} \Delta^{-1}_\rho
\end{align}
was performed. As this is a constant shift of the integration variable $\delta \rho$ it produces a trivial Jacobian. Note that in Eq.\ \eqref{eq: general_shift} we introduced shorthand notation for the second variation of the action
\begin{align}
    \delta^2 S = \delta \rho_x \left(\Delta_\rho\right)_{xy} \delta \rho_y + \delta \rho_x \left(X_{\rho \sigma}\right)_{xy} \delta \sigma_y + \delta \sigma_x \left(X_{\sigma \rho}\right)_{xy} \delta \rho_y + \delta \sigma_x \left(\Delta_\sigma\right)_{xy} \delta \sigma_y,
\end{align}
where 
\begin{align}
    S = \int \rd^d x \, \LagNoT, 
\end{align}
by omitting spacetime indices and writing $\LagNoT$ instead of $S$. This notation is commonly used in the literature and will be employed in the following, unless the explicit appearance of spacetime indices adds clarity to the presentation.

To cover most cases of interest to particle phenomenology it suffices to treat a theory containing Majorana and Dirac fermions together with real and complex scalar fields as well as real and complex vector fields. The different values of $c_s$ for the different kinds of fields are summarised in Table \ref{table: c_s_table}.
\begin{table}[h]
\centering
\begin{tabular}{|c|c|c}
\cline{1-2}
 Particle type & $c_s$ &   \\ \cline{1-2}
 Dirac Fermion & $-1$  &   \\ \cline{1-2}    
 Majorana Fermion & $-1/2$ &   \\ \cline{1-2}
 Complex Scalar & $1$ &   \\ \cline{1-2}
 Real Scalar & $1/2$ &   \\ \cline{1-2}
 Complex Vector & $1$ &   \\ \cline{1-2}
 Real Vector & $1/2$ &   \\ \cline{1-2}
\end{tabular}
\caption{Values of $c_s$ for the different kinds of fields considered in this work.}
\label{table: c_s_table}
\end{table}
As can be seen from the table it is not sufficient to perform shifts to separate the bosonic sector from the fermionic one as the value of $c_s$ also varies within each individual sector. This is due to the fact that $c_s$ also accounts for the number of independent degrees of freedom of the field that is integrated out. Therefore, there are many shifts to be performed in order to first separate the fermionic sector from the bosonic sector and then fields with different numbers of degrees of freedom within each sector. In addition the distinction between light and heavy fields has to be kept since they play completely different roles in the computation. After the separation of the different sectors has been achieved through shifts of the form of Eq.\ \eqref{eq: general_shift}, every sector gives rise to a separate functional determinant, which generates its own contributions to the UOLEA. In order to keep the final expressions compact and simple it is sensible to keep the number of necessary shifts at a minimum. To achieve this, and to treat real and complex scalar fields on the same footing, one could split all complex fields into a real part and an imaginary part and perform the calculation using these as the fundamental fields. However, it is often desirable to maintain the complex fields as they might have some
physical interpretation in the effective theory. We therefore use the
field and its complex conjugate as independent degrees of freedom. The inclusion of vector fields (including gauge bosons) in the bosonic sector is straightforward in Feynman gauge. Each component of the four-vector can be treated as a scalar field itself. The Lorentz indices are then handled in the same way as the internal indices. 

In order to treat Dirac and Majorana fermions simultaneously without diagonalising
the fluctuation operator among these it is convenient to treat any
Dirac fermion and its charge conjugate as independent degrees of
freedom. We collect all light and heavy scalars, including the components of vector fields, into the multiplets
$\phi$ and $\Phi$, respectively, and all light and heavy fermions into
the multiplets $\xi$ and $\Xi$, respectively, see Table~\ref{table: field_content}.
\begin{table}[tb]
\centering
\begin{tabular}{cll}
  \toprule
  Multiplet & Components & Description \\
  \midrule
  $\Xi$ & $\big(\Omega, \ccfield{\Omega}, \Lambda\big)^T$ & \pbox{20cm}{$\Omega$, $\ccfield{\Omega}$: heavy Dirac fermions \\
  $\Lambda$: heavy Majorana fermions}  \\
  \midrule
  $\Phi$ & $\big(\Sigma, W_\mu, \Sigma^*,  W^*_\mu, \Theta, V_\mu\big)^T$ & \pbox{20cm}{$\Sigma$, $\Sigma^*$: heavy complex scalars \\
  $\Theta$: heavy real scalars \\ $W_\mu$, $W^*_\mu$: heavy complex vector fields \\
  $V_\mu$: heavy real vector fields}  \\
  \midrule
  $\xi$ & $\big(\omega, \ccfield{\omega}, \lambda\big)^T$ & \pbox{20cm}{$\omega$, $\ccfield{\omega}$: light Dirac fermions \\
  $\lambda$: light Majorana fermions} \\
  \midrule
  $\phi$ & $\big(\sigma, w_\mu, \sigma^*,  w^*_\mu, \theta, v_\mu\big)^T$ & \pbox{20cm}{$\sigma$, $\sigma^*$: light complex scalars \\
  $\theta$: light real scalars\\ 
  $w_\mu$, $w^*_\mu$: light complex vector fields \\
  $v_\mu$: light real vector fields (incl. gauge fields)} \\
  \bottomrule
\end{tabular}
\caption{Contents of the different multiplets appearing in the calculation.}
\label{table: field_content}
\end{table}
The charge conjugate of the Dirac spinor $\Omega$ is denoted as
$\ccfield{\Omega}=\cc\bar{\Omega}^T$, with $\cc$ being the charge
conjugation matrix.\footnote{We choose to work in the chiral representation of the Dirac algebra.} Similarly, we define for a light Dirac spinor
$\omega$, $\ccfield{\omega}=\cc\bar{\omega}^T$. As will be shown below, with this setup it will be sufficient to perform two shifts of the form of Eq.\ \eqref{eq: general_shift}. Since collecting the fields into these multiplets is somewhat unconventional and since the starting point of the computation is the second variation of the action, we here elaborate on how this second variation is obtained in the parameterisation that was chosen here. These details were not discussed in Ref.\ \cite{Kramer:2019fwz}. Furthermore, vector fields in the bosonic sector were not included and the discussion that follows allows for a demonstration of this inclusion. We start with the bosonic sector where, with the heavy field content given in Table \ref{table: field_content}, we have the following second variation for heavy bosons only
\begin{align}
    \delta^2 \Lag{B,H} &= -\delta \Sigma^\dagger \Delta_{\Sigma} \delta \Sigma -\frac{1}{2} \delta \Theta ^T \Delta_{\Theta} \delta \Theta - \delta W^\dagger _\mu \Delta^{\mu \nu}_{W} \delta W_\nu - \frac{1}{2} \delta V^T_\mu \Delta^{\mu \nu}_{V} \delta V_\nu \nonumber \\
    & \quad + \delta^2 \Lag{B,H,int.},
   \label{eq: heavy_boson_fluct}
\end{align}
where 
\begingroup
\allowdisplaybreaks
\begin{align}
  \Delta_\Sigma &= -P_{\Sigma}^2+M_\Sigma^2+X_{\Sigma ^* \Sigma}, \\
  \Delta_{\Theta} &= -P_\Theta^2+M_\Theta^2+X_{\Theta \Theta}, \\
  \Delta_{W}^{\mu \nu} &= P_{W}^2 g^{\mu \nu} + 2 [P_W^\mu,P_W^\nu] - M^2_W g^{\mu \nu} + X_{W^* W} ^ {\mu \nu}, \\
  \Delta_{V}^{\mu \nu} &= P_{V}^2 g^{\mu \nu} + 2 [P_V^\mu,P_V^\nu] - M^2_V g^{\mu \nu} + X_{V V} ^ {\mu \nu}.
\end{align}
\endgroup
Here a subscript on $P^\mu = iD^\mu$ indicates the representation on which the covariant derivative acts. For example, $P^\mu_\Sigma$ contains the covariant derivative acting on the (possibly reducible) representation of all heavy complex scalar fields. Furthermore, $\delta ^2 \Lag{B,H,int.}$ contains the remaining interactions and can be written as
\begin{align}
\delta^2 \Lag{B,H,int.} &= -\frac{1}{2}
     \begin{pmatrix}
     \delta \Sigma^T & \delta W_\mu & \delta \Sigma^\dagger & \delta W^\dagger_\mu & \delta \Theta^T   & \delta V^T_\mu 
     \end{pmatrix} \nonumber \\ 
     & \quad
     \begin{pmatrix}
     X_{\Sigma \Sigma} & X_{\Sigma W}^\nu & 0 & X_{\Sigma W^*}^\nu & X_{\Sigma \Theta}   & X_{\Sigma V}^\nu \\
     X_{W \Sigma}^\mu & X_{W W}^{\mu \nu} & X_{W \Sigma^*}^\mu & 0 & X_{W \Theta}^\mu  & X_{W V}^{\mu \nu} \\ 
     0 & X_{\Sigma^* W}^\nu & X_{\Sigma^* \Sigma^*} & X_{\Sigma^* W^*}^\nu & X_{\Sigma^* \Theta} & X_{\Sigma* V}^\nu \\
     X_{W^* \Sigma}^\mu & 0 & X_{W^* \Sigma^*}^\mu & X_{W^* W^*}^{\mu \nu} & X_{W^* \Theta}^\mu  & X_{W^* V}^{\mu \nu} \\
     X_{\Theta \Sigma} & X_{\Theta W}^\nu & X_{\Theta \Sigma^*} & X_{\Theta W^*}^\nu & 0 & X_{\Theta V}^\nu \\
     X_{V \Sigma}^\mu & X_{V W}^{\mu \nu} & X_{V \Sigma^*}^\mu & X_{V W^*}^{\mu \nu} & X_{V \Theta}^\mu  & 0 \\
     \end{pmatrix}
     \begin{pmatrix}
     \delta \Sigma \\ \delta W_\nu  \\ \delta \Sigma^* \\ \delta W^*_\nu \\ \delta \Theta  \\ \delta V_\nu 
     \end{pmatrix},
     \label{eq: interaction_only_fluct}
\end{align}
which can be directly expressed through the multiplet $\Phi$. In order to rewrite the first line of Eq.\ \eqref{eq: heavy_boson_fluct} in terms of said multiplet we note that the kinetic contributions of complex fields can be symmetrised through integration by parts. In particular it holds that 
\begin{align}
    \delta \Sigma^\dagger \Delta_{\Sigma} \delta \Sigma &= \delta \Sigma^\dagger \left(-P_{\Sigma}^2+M_\Sigma^2+X_{\Sigma ^* \Sigma}\right) \delta \Sigma \nonumber \\ 
   &=  \delta \Sigma^T \left(-P^2_{\Sigma^*}+M_{\Sigma^*}^2+X_{\Sigma \Sigma^*}\right) \delta \Sigma^* \nonumber \\ 
   & = \delta \Sigma ^T \Delta_{\Sigma^*} \delta \Sigma^*,
\end{align}
where the covariant derivative now acts on the conjugate representation and we used that $X_{\Sigma \Sigma^*} = X^T_{\Sigma^* \Sigma}$. Similarly, we have that 
\begin{align}
    \delta W^\dagger _\mu \Delta_{W}^{\mu \nu} \delta W_\nu = \delta W^T_\mu \Delta_{W^*}^{\mu \nu} \delta W^*_\nu,
\end{align}
with
\begin{align}
   \Delta_{W^*}^{\mu \nu} &= P_{W^*}^2 g^{\mu \nu} + 2 [P_{W^*}^\mu,P_{W^*}^\nu] - M^2_W g^{\mu \nu} + X_{W W^*} ^ {\mu \nu}.
\end{align}
This allows for Eq.\ \eqref{eq: heavy_boson_fluct} to be written as
\begin{align}
     \delta^2 \Lag{B,H} &= 
     -\frac{1}{2}
     \delta \Phi^T \mathbf{\Delta}_\Phi \delta \Phi, 
\end{align}
with 
\begin{align}
    \mathbf{\Delta}_\Phi &=   \begin{pmatrix}
     X_{\Sigma \Sigma} & X_{\Sigma W}^\nu & \Delta_{\Sigma^*} & X_{\Sigma W^*}^\nu & X_{\Sigma \Theta}   & X_{\Sigma V}^\nu \\
     X_{W \Sigma}^\mu & X_{W W}^{\mu \nu} & X_{W \Sigma^*}^\mu & \Delta_{W^*}^{\mu \nu} & X_{W \Theta}^\mu  & X_{W V}^{\mu \nu} \\ 
     \Delta_\Sigma & X_{\Sigma^* W}^\nu & X_{\Sigma^* \Sigma^*} & X_{\Sigma^* W^*}^\nu & X_{\Sigma^* \Theta} & X_{\Sigma* V}^\nu \\
     X_{W^* \Sigma}^\mu & \Delta_{W} & X_{W^* \Sigma^*}^\mu & X_{W^* W^*}^{\mu \nu} & X_{W^* \Theta}^\mu  & X_{W^* V}^{\mu \nu} \\
     X_{\Theta \Sigma} & X_{\Theta W}^\nu & X_{\Theta \Sigma^*} & X_{\Theta W^*}^\nu & \Delta_{\Theta} & X_{\Theta V}^\nu \\
     X_{V \Sigma}^\mu & X_{V W}^{\mu \nu} & X_{V \Sigma^*}^\mu & X_{V W^*}^{\mu \nu} & X_{V \Theta}^\mu  & \Delta_{V}^{\mu \nu} \\
     \end{pmatrix} \nonumber \\
     & = 
     \tilde{\mathds{1}}_B \left(-P^2_\Phi+M^2_\Phi\right)+\mathbf{\tilde{X}}_{\Phi \Phi},
\end{align}
where we define 
\begin{align}
P^2_\Phi-M^2_\Phi &\equiv \diag \left(P_{\Sigma} ^2 - M_\Sigma ^2,P_{W} ^2 - M_W ^2,P_{\Sigma^*} ^2 - M_\Sigma ^2,P_{W^*} ^2 - M_W ^2,P_{\Theta} ^2 - M_\Theta ^2,P_{V} ^2 - M_V ^2\right),
\end{align}
\begin{align}
\label{eq: bosonic_id}
\tilde{\mathds{1}}_B &\equiv \begin{pmatrix}
    0 & 0 & \mathds{1} & 0 & 0 & 0 \\
    0 & 0 & 0 & -\mathds{1} & 0 & 0 \\
    \mathds{1} & 0 & 0 & 0 & 0 & 0 \\
    0 & -\mathds{1} & 0 & 0 & 0 & 0 \\
    0 & 0 & 0 & 0 & \mathds{1} & 0 \\
    0 & 0 & 0 & 0 & 0 & -\mathds{1} \\
    \end{pmatrix},\\ 
    \intertext{as well as}
\mathbf{\tilde{X}}_{\Phi \Phi} & \equiv
\begin{pmatrix}
     X_{\Sigma \Sigma} & X_{\Sigma W}^\nu & X_{\Sigma \Sigma^*} & X_{\Sigma W^*}^\nu & X_{\Sigma \Theta}   & X_{\Sigma V}^\nu \\
     X_{W \Sigma}^\mu & X_{W W}^{\mu \nu} & X_{W \Sigma^*}^\mu & X_{W W^*}^{\mu \nu} & X_{W \Theta}^\mu  & X_{W V}^{\mu \nu} \\ 
     X_{\Sigma^* \Sigma} & X_{\Sigma^* W}^\nu & X_{\Sigma^* \Sigma^*} & X_{\Sigma^* W^*}^\nu & X_{\Sigma^* \Theta} & X_{\Sigma* V}^\nu \\
     X_{W^* \Sigma}^\mu & X_{W^* W} & X_{W^* \Sigma^*}^\mu & X_{W^* W^*}^{\mu \nu} & X_{W^* \Theta}^\mu  & X_{W^* V}^{\mu \nu} \\
     X_{\Theta \Sigma} & X_{\Theta W}^\nu & X_{\Theta \Sigma^*} & X_{\Theta W^*}^\nu & X_{\Theta \Theta} & X_{\Theta V}^\nu \\
     X_{V \Sigma}^\mu & X_{V W}^{\mu \nu} & X_{V \Sigma^*}^\mu & X_{V W^*}^{\mu \nu} & X_{V \Theta}^\mu  & X_{V V}^{\mu \nu} \\
     \end{pmatrix}.
\end{align}
We here absorb the explicit commutators of covariant derivatives appearing in $\Delta_W^{\mu \nu}$, $\Delta_{W^*}^{\mu \nu}$ and $\Delta_V^{\mu \nu}$ in the corresponding self-interaction matrices $X_{W W^*}^{\mu \nu}$, $X_{W^* W}^{\mu \nu}$ and $X_{V V}^{\mu \nu}$. Furthermore, in Eq.\ \eqref{eq: bosonic_id} the identity $\mathds{1}$ denotes an identity with respect to all indices of the respective multiplets. In the case of vector fields this includes the metric $g^{\mu \nu}$. It is clear that the purely light bosonic fluctuations can be written in exactly the same way by simply replacing upper case letters by lower case ones. Hence, this quadratic fluctuation is given by
\begin{align}
     \delta^2 \Lag{B,L} &= 
     -\frac{1}{2}
     \delta \phi^T \mathbf{\Delta}_\phi \delta \phi, 
\end{align}
where
\begin{align}
    \mathbf{\Delta}_\phi =    \tilde{\mathds{1}}_B \left(-P^2_\phi+M^2_\phi\right)+\mathbf{\tilde{X}}_{\phi \phi},
\end{align}
with 
\begin{align}
    -P^2_\phi+M^2_\phi &= -\text{diag} \left(P_{\sigma} ^2 - M_\sigma ^2,P_{w} ^2 - M_w ^2,P_{\sigma^*} ^2 - M_\sigma ^2,P_{w^*} ^2 - M_w ^2,P_{\theta} ^2 - M_\theta ^2,P_{v} ^2 - M_v ^2\right), 
\end{align}
and
\begin{align}
    \mathbf{\tilde{X}}_{\phi \phi} &=
\begin{pmatrix}
     X_{\sigma \sigma} & X_{\sigma w}^\nu & X_{\sigma \sigma^*} & X_{\sigma w^*}^\nu & X_{\sigma \theta}   & X_{\sigma v}^\nu \\
     X_{w \sigma}^\mu & X_{w w}^{\mu \nu} & X_{w \sigma^*}^\mu & X_{w w^*}^{\mu \nu} & X_{w \theta}^\mu  & X_{w v}^{\mu \nu} \\ 
     X_{\sigma^* \sigma} & X_{\sigma^* w}^\nu & X_{\sigma^* \sigma^*} & X_{\sigma^* w^*}^\nu & X_{\sigma^* \theta} & X_{\sigma* v}^\nu \\
     X_{w^* \sigma}^\mu & X_{w^* w} & X_{w^* \sigma^*}^\mu & X_{w^* w^*}^{\mu \nu} & X_{w^* \theta}^\mu  & X_{w^* v}^{\mu \nu} \\
     X_{\theta \sigma} & X_{\theta w}^\nu & X_{\theta \sigma^*} & X_{\theta w^*}^\nu & X_{\theta \theta} & X_{\theta v}^\nu \\
     X_{v \sigma}^\mu & X_{v w}^{\mu \nu} & X_{v \sigma^*}^\mu & X_{v w^*}^{\mu \nu} & X_{v \theta}^\mu  & X_{v v}^{\mu \nu} \\
     \end{pmatrix}.
\end{align}
We again absorb explicit commutators of $P_\mu$ in the self-interaction matrices of the vector bosons. Besides the purely heavy and purely light bosonic variations there are also mixed bosonic variations. These do not contain any kinetic terms and originate entirely from the interaction Lagrangian like the terms in Eq.\ \eqref{eq: interaction_only_fluct}. They can be directly expressed in terms of $\delta \phi$ and $\delta \Phi$ as
\begin{align}
    \delta^2 \Lag{B,HL} = -\frac{1}{2} \delta \Phi^T \mathbf{\tilde{X}}_{\Phi \phi} \delta \phi -\frac{1}{2} \delta \phi^T \mathbf{\tilde{X}}_{\phi \Phi} \delta \Phi,
\end{align}
where 
\begin{align}
    \mathbf{\tilde{X}}_{\Phi \phi} &= 
    \begin{pmatrix}
    X_{\Sigma \sigma} & X_{\Sigma w}^\nu & X_{\Sigma \sigma^*} & X_{\Sigma w^*}^\nu & X_{\Sigma \theta}   & X_{\Sigma v}^\nu \\
     X_{W \sigma}^\mu & X_{W w}^{\mu \nu} & X_{W \sigma^*}^\mu & X_{W w^{*}} & X_{W \theta}^\mu  & X_{W v}^{\mu \nu} \\ 
     X_{\Sigma^* \sigma} & X_{\Sigma^* w}^\nu & X_{\Sigma^* \sigma^*} & X_{\Sigma^* w^*}^\nu & X_{\Sigma^* \theta} & X_{\Sigma^* v}^\nu \\
     X_{W^* \sigma}^\mu & X_{W^* w} & X_{W^* \sigma^*}^\mu & X_{W^* w^*}^{\mu \nu} & X_{W^* \theta}^\mu  & X_{W^* v}^{\mu \nu} \\
     X_{\Theta \sigma} & X_{\Theta w}^\nu & X_{\Theta \sigma^*} & X_{\Theta w^*}^\nu & X_{\Theta \theta} & X_{\Theta v}^\nu \\
     X_{V \sigma}^\mu & X_{V w}^{\mu \nu} & X_{V \sigma^*}^\mu & X_{V w^*}^{\mu \nu} & X_{V \theta}^\mu  & X_{V v} \\ 
     \end{pmatrix}, \\
     \intertext{and}
      \mathbf{\tilde{X}}_{\phi \Phi} &= 
     \begin{pmatrix}
    X_{\sigma \Sigma} & X_{\sigma W}^\nu & X_{\sigma \Sigma^*} & X_{\sigma W^*}^\nu & X_{\sigma \Theta}   & X_{\sigma V}^\nu \\
     X_{w \Sigma}^\mu & X_{w W}^{\mu \nu} & X_{w \Sigma^*}^\mu & X_{w W^{*}} & X_{w \Theta}^\mu  & X_{w V}^{\mu \nu} \\ 
     X_{\sigma^* \Sigma} & X_{\sigma^* W}^\nu & X_{\sigma^* \Sigma^*} & X_{\sigma^* W^*}^\nu & X_{\sigma^* \Theta} & X_{\sigma^* V}^\nu \\
     X_{w^* \Sigma}^\mu & X_{w^* W} & X_{w^* \Sigma^*}^\mu & X_{w^* W^*}^{\mu \nu} & X_{w^* \Theta}^\mu  & X_{w^* V}^{\mu \nu} \\
     X_{\theta \Sigma} & X_{\theta W}^\nu & X_{\theta \Sigma^*} & X_{\theta W^*}^\nu & X_{\theta \Theta} & X_{\theta V}^\nu \\
     X_{v \Sigma}^\mu & X_{v W}^{\mu \nu} & X_{v \Sigma^*}^\mu & X_{v W^*}^{\mu \nu} & X_{v \Theta}^\mu  & X_{v V} \\ 
     \end{pmatrix}.
\end{align}
The purely bosonic part of the second variation is then
\begin{align}
    \delta^2 \Lag{B} = \delta^2 \Lag{B,H} + \delta^2 \Lag{B,L} + \delta^2 \Lag{B,HL}.
\end{align}
The matrices $\tilde{\mathbf{X}}_{AB}$ defined here differ from those defined in Ref.\ \cite{Kramer:2019fwz} by the inclusion of the vector fields. As in Ref.\ \cite{Kramer:2019fwz} these are not the matrices that will appear in the final result since the effects of the presence of $\tilde{\mathds{1}}_B$, which does not equal the identity matrix, will be absorbed in the redefinition of these matrices. Note that the appearance of the matrix $\tilde{\mathds{1}}_B$ is due to the fact that the propagators of the components of the multiplets $\Phi$ and $\phi$ differ in signs between the scalar fields and the vector fields and that for complex fields, the propagators connect the field and its complex conjugate, rather than connecting the field to itself. 

We next turn to the purely fermionic sector, where various signs, stemming from the Grassmann odd\footnote{For a brief discussion of Grassmann numbers and the Berezin algebra see Appendix~\ref{App: berezin_algebra}.} nature of the fermionic fields, and various charge conjugation matrices, stemming from the parameterisation chosen to perform the computation, appear in intermediate steps of the calculation. In what follows, we discuss how these signs and matrices enter the computation. We start with purely heavy fermions, including both Dirac fermions $\Omega$ and Majorana fermions $\Lambda$. It seems most natural to parameterise the second variation in terms of $\delta \Omega$, the Dirac adjoints $\delta \bar{\Omega}$ and the Majorana fermions $\delta \Lambda$. If this parameterisation is chosen, one has to diagonalise terms in the fluctuation operator which couple Dirac and Majorana fermions. Such terms are present for instance in supersymmetric models in the form of fermion-sfermion-gaugino couplings. To avoid this diagonalisation we instead introduce the 8-component spinor
\begin{align}
    \Psi = \begin{pmatrix}
    \Omega \\
    ~~ \ccfield{\Omega}
    \end{pmatrix},
    \label{eq: bispinor}
\end{align}
also introduced in Ref.\ \cite{roepstorff1994path} as one of three equivalent ways of treating Dirac fermions in the path integral approach. The spinor defined in Eq.\ \eqref{eq: bispinor} is referred to as a bispinor and lends its name to the BSUOLEA. For the intents and purposes of the Gaussian path integral this bispinor behaves like a Majorana fermion and eliminates the need for the extra diagonalisation within the fermionic sector. It is then necessary to express the second variation in terms of this bispinor and ultimately in terms of the multiplet $\Xi = \begin{pmatrix}
\Psi & \Lambda 
\end{pmatrix}^T$. Consider first the gauge-kinetic terms, mass terms and self-interactions of the Dirac Fermions. These can be written as
\begin{align}
\delta \bar{\Omega} \Delta_\Omega \delta \Omega &= \delta \bar{\Omega} \left(\slashed{P}_\Omega-M_\Omega+X_{\bar{\Omega} \Omega} \right) \delta \Omega \nonumber \\
&= \delta \bar{\Omega} \left(-\overleftarrow{\slashed{P}}_{\bar{\Omega}}-M_\Omega+X_{\bar{\Omega} \Omega} \right) \delta \Omega \nonumber \\ 
&= \delta \Omega ^T \cc \cc^{-1} \left(\slashed{P}_{\bar{\Omega}}^T+M_\Omega-X_{\bar{\Omega} \Omega}^T\right) \cc^{-1} \cc \delta \bar{\Omega}^T \nonumber \\
&= \delta \Omega ^T \cc \left(\slashed{P}_{\ccfield{\Omega}} - M_\Omega + \cc^{-1} X_{\Omega \bar{\Omega}} \cc^{-1} \right) \delta \ccfield{\Omega}, 
\end{align}
where from the first to the second line we integrated by parts and from the third to the last line we used that $\cc^{-1} = -\cc$, $\delta \ccfield{\Omega}=C \delta \bar{\Omega}^T$ and $X_{\bar{\Omega}\Omega}^T=-X_{\Omega \bar{\Omega}}$, where the sign comes from the fact that derivatives with respect to fermionic fields anticommute. It then follows that
\begin{align}
    \delta \bar{\Omega} \Delta_\Omega \delta \Omega ={}&\frac{1}{2} \left(\delta \ccfield{\Omega}\right)^T  \cc \left(\slashed{P}_\Omega-M_\Omega+X_{\bar{\Omega} \Omega} \right) \delta \Omega \nonumber \\ 
    &+ \frac{1}{2}\delta \Omega ^T \cc \left(\slashed{P}_{\ccfield{\Omega}} - M_\Omega + \cc^{-1} X_{\Omega \bar{\Omega}} \cc^{-1} \right) \delta \ccfield{\Omega} \nonumber \\  ={}& \frac{1}{2} \delta \Psi^T \begin{pmatrix}
    0 & \cc \left(\slashed{P}_{\ccfield{\Omega}}-M_\Omega+\cc^{-1}X_{\Omega \bar{\Omega}}\cc^{-1}\right) \\
    \cc \left(\slashed{P}_\Omega - M_\Omega + X_{\bar{\Omega} \Omega}\right) & 0
    \end{pmatrix}
    \delta \Psi,
\end{align}
which we combine with the term quadratic in Majorana fermion fluctuations
\begin{align}
\frac{1}{2} \delta \Lambda^T \Delta_\Lambda \delta \Lambda = \frac{1}{2} \delta \Lambda^T \cc \left(\slashed{P}_\Lambda-M_\Lambda+X_{\Lambda \Lambda}\right) \delta \Lambda,
\end{align}
and the pure interaction terms\footnote{The requirements of Lorentz invariance and renormalisability imply that there are no $X_{\bar{\Omega} \bar{\Omega}}$ and $X_{\Omega \Omega}$ contributions as long as there is some global symmetry group (a subgroup of which could be gauged) under which the two components of the Dirac spinors transform in conjugate representations. This requirement can be used to define what is meant by a Dirac fermion as opposed to a Majorana fermion, which would transform in a real representation of the same symmetry group, see Ref.\ \cite{Dreiner:2008tw}. In any case, extending the computation to include these contributions is straightforward.}
\begin{align}
\delta^2 \Lag{F,H,int.} &= \delta \bar{\Omega} X_{\bar{\Omega}\Lambda} \delta \Lambda + \delta \Lambda^T X_{\Lambda \Omega} \delta \Omega\nonumber \\ 
&= \frac{1}{2} \left(\delta \ccfield{\Omega}\right)^T \cc X_{\bar{\Omega} \Lambda} \delta \Lambda + \frac{1}{2} \delta \Lambda^T X_{\Lambda \bar{\Omega}}\cc^{-1} \delta \ccfield{\Omega} + \frac{1}{2} \delta \Lambda ^T X_{\Lambda \Omega} \delta \Omega + \frac{1}{2} \delta \Omega^T X_{\Omega \Lambda} \delta \Lambda \nonumber \\
&= \frac{1}{2} \delta \Lambda^T \begin{pmatrix}
X_{\Lambda \Omega} & X_{\Lambda \bar{\Omega}} \cc^{-1}
\end{pmatrix} \delta \Psi +  \frac{1}{2} \delta \Psi^T \begin{pmatrix}
X_{\Omega \Lambda} \\ \cc X_{\bar{\Omega} \Lambda} 
\end{pmatrix} \delta \Lambda,
\end{align}
to obtain the full contribution to the second variation coming only from heavy fermions
\begin{align}
    \delta^2 \Lag{F,H} =\frac{1}{2} \delta \Xi ^T \mathbf{\Delta}_\Xi \delta \Xi,
\end{align}
where 
\begin{align}
    \mathbf{\Delta}_\Xi &= \begin{pmatrix} 0 & \cc(\slashed{P}_{\ccfield{\Omega}}-M_\Omega+\cc^{-1} X_{\Omega \bar{\Omega}}\cc^{-1}) & X_{\Omega \Lambda} \\
\cc (\slashed{P}_\Omega-M_\Omega+X_{\bar{\Omega} \Omega}) & 0 & \cc X_{\bar{\Omega} \Lambda} \\
X_{\Lambda \Omega} & X_{\Lambda \bar{\Omega}}\cc ^{-1} & \cc (\slashed{P}_\Lambda-M_\Lambda+\cc ^{-1} X_{\Lambda \Lambda})
\end{pmatrix} \nonumber \\
&= \cc \tilde{\mathds{1}}_F (\slashed{P}-M_\Xi) +\tilde{\mathbf{X}}_{\Xi \Xi}, \\
\intertext{with}
\tilde{\mathds{1}}_F &\equiv \begin{pmatrix}
0 && \mathds{1} && 0 \\
\mathds{1} && 0 && 0 \\
0 && 0 && \mathds{1}
\end{pmatrix}, \; \slashed{P}-M_\Xi = \text{diag}\left(
\slashed{P}_{\Omega}-M_\Omega,\, \slashed{P}_{\ccfield{\Omega}}-M_\Omega,\, \slashed{P}_\Lambda-M_\Lambda\right),
\label{eq:fermID} \\
\intertext{and}
\tilde{\mathbf{X}}_{\Xi \Xi} &= \begin{pmatrix}
0 && X_{\Omega \bar{\Omega}}\cc^{-1} && X_{\Omega \Lambda} \\
\cc X_{\bar{\Omega} \Omega} && 0 && \cc X_{\bar{\Omega} \Lambda} \\
X_{\Lambda \Omega} && X_{\Lambda \bar{\Omega}}\cc ^{-1} &&  X_{\Lambda \Lambda} 
\end{pmatrix}.
\end{align}
By replacing the upper case letters with the corresponding lower case ones, one analogously obtains the second variation for the purely light fermions. We present this result for the sake of completeness
\begin{align}
    \delta^2 \Lag{F,L} =\frac{1}{2} \delta \xi ^T \mathbf{\Delta}_\xi \delta \xi,
\end{align}
where 
\begin{align}
    \mathbf{\Delta}_\xi &= \cc \tilde{\mathds{1}}_F (\slashed{P}-M_\xi) +\tilde{\mathbf{X}}_{\xi \xi}, \\
\slashed{P}-M_\xi &= \text{diag}(
\slashed{P}_{\omega}-M_\omega,\,\slashed{P}_{\ccfield{\omega}}-M_\omega,\,\slashed{P}_\lambda-M_\lambda),
\label{eq:fermID} \\
\intertext{and}
\tilde{\mathbf{X}}_{\xi \xi} &= \begin{pmatrix}
0 && X_{\omega \bar{\omega}}\cc^{-1} && X_{\omega \lambda} \\
\cc X_{\bar{\omega} \omega} && 0 && \cc X_{\bar{\omega} \lambda} \\
X_{\lambda \omega} && X_{\lambda \bar{\omega}}\cc ^{-1} &&  X_{\lambda \lambda} 
\end{pmatrix}.
\end{align}
As in the bosonic sector there are also couplings that couple heavy fermions to light fermions. These can be treated in complete analogy to the heavy-heavy couplings that have already been discussed and lead to the following contribution to the second variation
\begin{align}
    \delta^2 \Lag{F,HL} &= \frac{1}{2} \delta \xi ^T \tilde{\mathbf{X}}_{\xi \Xi} \delta \Xi +\frac{1}{2}\delta \Xi ^T \tilde{\mathbf{X}}_{\Xi \xi} \delta \xi,
\end{align}
where the matrices $\tilde{\mathbf{X}}_{\Xi \xi}$ and $\tilde{\mathbf{X}}_{\xi \Xi}$ are given by
\begin{align}
 \tilde{\mathbf{X}}_{\Xi \xi} &= \begin{pmatrix}
X_{\Omega \omega} && X_{\Omega \bar{\omega}}\cc ^{-1} && X_{\Omega \lambda} \\
\cc X_{\bar{\Omega} \omega} && \cc X_{\bar{\Omega} \bar{\omega}} \cc^{-1} && \cc X_{\bar{\Omega} \lambda} \\
X_{\Lambda \omega} && X_{\Lambda \bar{\omega}} \cc ^{-1} && X_{\Lambda \lambda}
\end{pmatrix}, \\
\tilde{\mathbf{X}}_{\xi \Xi} &= \begin{pmatrix}
X_{\omega \Omega} && X_{\omega \bar{\Omega}}\cc ^{-1} && X_{\omega \Lambda} \\
\cc X_{\bar{\omega} \Omega} && \cc X_{\bar{\omega} \bar{\Omega}} \cc^{-1} && \cc X_{\bar{\omega} \Lambda} \\
X_{\lambda \Omega} && X_{\lambda \bar{\Omega}} \cc ^{-1} && X_{\lambda \Lambda}
\end{pmatrix}.
\end{align}
Adding all of these contributions we obtain the purely fermionic part of the second variation, which reads
\begin{align}
    \delta^2 \Lag{F} &= \delta^2 \Lag{F,H} + \delta^2 \Lag{F,L} + \delta^2 \Lag{F,HL}.
\end{align}
Finally, there are couplings between the fermions and the bosons leading to contributions of mixed statistics to the second variation. In the case of only heavy fields we have
\begingroup
\allowdisplaybreaks
\begin{align}
    \delta^2 \Lag{BF,H} ={}& - \delta \bar{\Omega} \tilde{\mathbf{X}}_{\bar{\Omega} \Phi} \delta \Phi +  \delta \Phi^T \tilde{\mathbf{X}}_{\Phi \Omega} \delta \Omega - \frac{1}{2} \delta \Lambda^T \tilde{\mathbf{X}}_{\Lambda \Phi} \delta \Phi + \frac{1}{2} \delta \Phi^T \tilde{\mathbf{X}}_{\Phi \Lambda} \delta \Lambda \nonumber \\
    ={}& -\frac{1}{2} \left(\delta \ccfield{\Omega}\right)^T \cc \tilde{\mathbf{X}}_{\bar{\Omega} \Phi} \delta \Phi + \frac{1}{2} \delta \Phi^T \tilde{\mathbf{X}}_{\Phi \bar{\Omega}} \cc^{-1} \delta \ccfield{\Omega} + \frac{1}{2} \delta \Phi^T \tilde{\mathbf{X}}_{\Phi \Omega} \delta \Omega -\frac{1}{2} \delta \Omega^T \tilde{\mathbf{X}}_{\Omega \Phi} \delta \Phi \nonumber \\
    & - \frac{1}{2} \delta \Lambda^T \tilde{\mathbf{X}}_{\Lambda \Phi} \delta \Phi + \frac{1}{2} \delta \Phi^T \tilde{\mathbf{X}}_{\Phi \Lambda} \delta \Lambda \nonumber \\ 
    ={}& -\frac{1}{2} \delta \Xi^T \tilde{\mathbf{X}}_{\Xi \Phi} \delta \Phi + \frac{1}{2} \delta \Phi^T \tilde{\mathbf{X}}_{\Phi \Xi} \delta \Xi,
    \label{eq: fermion_scalar_coupling}
 \end{align}
 \endgroup
where the coupling matrices are given by
\begingroup
\allowdisplaybreaks
\begin{align}
\tilde{\mathbf{X}}_{\Xi \Phi} &= \begin{pmatrix}
X_{\Omega \Sigma} & X_{\Omega W}^\nu & X_{\Omega \Sigma^{*}} & X_{\Omega W^*}^\nu & X_{\Omega \Theta} & X_{\Omega V}^\nu \\ 
\cc X_{\bar{\Omega} \Sigma} & \cc X_{\bar{\Omega} W}^\nu & \cc X_{\bar{\Omega} \Sigma ^*} & \cc X_{\bar{\Omega} W^*}^\nu & \cc X_{\bar{\Omega} \Theta} & \cc X_{\bar{\Omega} V}^\nu \\ 
X_{\Lambda \Sigma} & X_{\Lambda W}^\nu & X_{\Lambda \Sigma ^*} & X_{\Lambda W^*}^\nu & X_{\Lambda \Theta} & X_{\Lambda V}^\nu
\end{pmatrix},\\ 
\tilde{\mathbf{X}}_{\Phi \Xi} &= \begin{pmatrix}
X_{\Sigma \Omega} & X_{\Sigma \bar{\Omega}} \cc ^{-1} & X_{\Sigma \Lambda} \\
X_{W \Omega}^\nu & X_{W \bar{\Omega}}^\nu \cc ^{-1} & X_{W \Lambda}^\nu \\ 
X_{\Sigma ^* \Omega} & X_{\Sigma ^* \bar{\Omega}} \cc ^{-1} & X_{\Sigma ^* \Lambda} \\
X_{W^* \Omega}^\nu & X_{W^* \bar{\Omega}}^\nu \cc ^{-1} & X_{W^* \Lambda}^\nu \\ 
X_{\Theta \Omega} & X_{\Theta \bar{\Omega}} \cc ^{-1} & X_{\Theta \Lambda} \\
X_{V \Omega}^\nu & X_{V \bar{\Omega}}^\nu \cc ^{-1} & X_{V \Lambda}^\nu \\ 
\end{pmatrix}.
\end{align}
\endgroup
In going from the first to the second line of Eq.\ \eqref{eq: fermion_scalar_coupling} we used that $\tilde{\mathbf{X}}_{\bar{\Omega} \Phi}^T = \tilde{\mathbf{X}}_{\Phi \bar{\Omega}}$ and $\tilde{\mathbf{X}}_{\Phi \Omega}^T = \tilde{\mathbf{X}}_{\Omega \Phi}$ as well as the fact that these matrices are Grassmann odd. The remaining boson-fermion couplings can be treated in a similar fashion and the complete mixed contribution to the second variation reads
\begin{align}
   \delta^2 \Lag{BF} &= - \frac{1}{2} \delta \Xi ^T \tilde{\mathbf{X}}_{\Xi \Phi} \delta \Phi + \frac{1}{2} \delta \Phi ^T \tilde{\mathbf{X}}_{\Phi \Xi} \delta \Xi -\frac{1}{2} \delta \Xi ^T \tilde{\mathbf{X}}_{\Xi \phi} \delta \phi+\frac{1}{2} \delta \phi ^T \tilde{\mathbf{X}}_{\phi \Xi} \delta \Xi \nonumber \\
& \quad -\frac{1}{2} \delta \xi ^T \tilde{\mathbf{X}}_{\xi \Phi} \delta \Phi +\frac{1}{2} \delta \Phi ^T \tilde{\mathbf{X}}_{\Phi \xi} \delta \xi -\frac{1}{2} \delta \xi ^T \tilde{\mathbf{X}}_{\xi \phi} \delta \phi+\frac{1}{2} \delta \phi ^T \tilde{\mathbf{X}}_{\phi \xi} \delta \xi,
\label{eq: full_mixed_fluct}
\end{align}
where we introduced the mixed bosonic-fermionic, heavy-light coupling matrices
\begingroup
\allowdisplaybreaks
\begin{align}
\tilde{\mathbf{X}}_{\Xi \phi} &= \begin{pmatrix}
X_{\Omega \sigma} & X_{\Omega w}^\nu & X_{\Omega \sigma^{*}} & X_{\Omega w^*}^\nu & X_{\Omega \theta} & X_{\Omega v}^\nu \\ 
\cc X_{\bar{\Omega} \sigma} & \cc X_{\bar{\Omega} w}^\nu & \cc X_{\bar{\Omega} \sigma ^*} & \cc X_{\bar{\Omega} w^*}^\nu & \cc X_{\bar{\Omega} \theta} & \cc X_{\bar{\Omega} v}^\nu \\ 
X_{\Lambda \sigma} & X_{\Lambda w}^\nu & X_{\Lambda \sigma ^*} & X_{\Lambda w^*}^\nu & X_{\Lambda \theta} & X_{\Lambda v}^\nu
\end{pmatrix}, \\ 
\tilde{\mathbf{X}}_{\phi \Xi} &= \begin{pmatrix}
X_{\sigma \Omega} & X_{\sigma \bar{\Omega}} \cc ^{-1} & X_{\sigma \Lambda} \\
X_{w \Omega}^\nu & X_{w \bar{\Omega}}^\nu \cc ^{-1} & X_{w \Lambda}^\nu \\ 
X_{\sigma ^* \Omega} & X_{\sigma ^* \bar{\Omega}} \cc ^{-1} & X_{\sigma ^* \Lambda} \\
X_{w^* \Omega}^\nu & X_{w^* \bar{\Omega}}^\nu \cc ^{-1} & X_{w^* \Lambda}^\nu \\ 
X_{\theta \Omega} & X_{\theta \bar{\Omega}} \cc ^{-1} & X_{\theta \Lambda} \\
X_{v \Omega}^\nu & X_{v \bar{\Omega}}^\nu \cc ^{-1} & X_{v \Lambda}^\nu \\ 
\end{pmatrix},
\end{align}
\endgroup
with the definitions of $\tilde{\mathbf{X}}_{\Phi \xi}$ and $\tilde{\mathbf{X}}_{\xi \Phi}$ obtained by the replacements $\phi \rightarrow \Phi$ and $\Xi \rightarrow \xi$ and the definitions of $\tilde{\mathbf{X}}_{\phi \xi}$ and $\tilde{\mathbf{X}}_{\xi \phi}$ obtained by the replacement $\Xi \rightarrow \xi$. The full second variation is then given by
\begin{align}
    \delta^2 \Lag{UV} & = \delta^2 \Lag{B} + \delta^2 \Lag{F} + \delta^2 \Lag{BF}.  
\end{align}
Having derived this parameterisation of the second variation, the calculation now proceeds by diagonalising the corresponding fluctuation operator in terms of statistics in order to be able to perform the Gaussian
path integral. As discussed before, the advantage of this parameterisation is that this is the only diagonalisation that has to be performed. We first eliminate terms that mix bosonic fluctuations
and fluctuations of light fermions $\xi$ by rewriting the part of the second variation that contains $\xi$-fluctuations as
\begin{align}
\delta ^2 \mathcal{L}_\xi ={}& \frac{1}{2} \delta \xi ^T \tilde{\mathbf{X}}_{\xi \Xi} \delta \Xi +\frac{1}{2}\delta \Xi ^T \tilde{\mathbf{X}}_{\Xi \xi} \delta \xi+\frac{1}{2}\delta \xi ^T \mathbf{\Delta}_\xi \delta \xi -\frac{1}{2} \delta \xi ^T \tilde{\mathbf{X}}_{\xi \Phi} \delta \Phi +\frac{1}{2} \delta \Phi ^T \tilde{\mathbf{X}}_{\Phi \xi} \delta \xi \nonumber \\ & -\frac{1}{2} \delta \xi ^T \tilde{\mathbf{X}}_{\xi \phi} \delta \phi+\frac{1}{2} \delta \phi ^T \tilde{\mathbf{X}}_{\phi \xi} \delta \xi \\ 
={}& \frac{1}{2} \left(\delta \xi^T+\left[\delta \Xi^T \tilde{\mathbf{X}}_{\Xi \xi}+\delta \Phi^T \tilde{\mathbf{X}}_{\Phi \xi}+\delta \phi^T \tilde{\mathbf{X}}_{\phi \xi}\right]\overleftarrow{\mathbf{\Delta}}_\xi^{-1}\right)\mathbf{\Delta}_\xi \nonumber \\ & \times \left(\delta \xi+\mathbf{\Delta}_\xi^{-1}\left[\tilde{\mathbf{X}}_{\xi \Xi} \delta \Xi-\tilde{\mathbf{X}}_{\xi \Phi} \delta \Phi-\tilde{\mathbf{X}}_{\xi \phi} \delta \phi\right]\right) \nonumber \\
& -\frac{1}{2} \left[\delta \Xi^T \tilde{\mathbf{X}}_{\Xi \xi}+\delta \Phi^T \tilde{\mathbf{X}}_{\Phi \xi}+\delta \phi^T \tilde{\mathbf{X}}_{\phi \xi}\right]\mathbf{\Delta}_\xi^{-1}\left[\tilde{\mathbf{X}}_{\xi \Xi} \delta \Xi-\tilde{\mathbf{X}}_{\xi \Phi} \delta \Phi-\tilde{\mathbf{X}}_{\xi \phi} \delta \phi\right].
\end{align}
In the last step we introduced $\mathbf{\Delta}_\xi ^{-1}$, which
is the matrix-valued Green's function of $\mathbf{\Delta}_\xi$. Note that, restoring spacetime indices, we have
\begin{multline}
  \left(\mathbf{\Delta}^{-1}_\xi\left[\tilde{\mathbf{X}}_{\xi \Xi} \delta \Xi-\tilde{\mathbf{X}}_{\xi \Phi} \delta \Phi-\tilde{\mathbf{X}}_{\xi \phi} \delta \phi\right]\right)_x\equiv 
   \left(\mathbf{\Delta}^{-1}_\xi\right)_{xy} \left[\tilde{\mathbf{X}}_{\xi \Xi} \delta \Xi-\tilde{\mathbf{X}}_{\xi \Phi} \delta \Phi-\tilde{\mathbf{X}}_{\xi \phi} \delta \phi\right]_y.
\end{multline}
Similar to $\mathbf{\Delta}_\xi ^{-1}$ we define
$\overleftarrow{\mathbf{\Delta}}_\xi^{-1}$ in such a way that
\begin{align}
 f_y \left(\overleftarrow{\mathbf{\Delta}}_\xi^{-1}\right) _{yx} \left(\overleftarrow{\mathbf{\Delta}} _\xi\right)_x =f_x,
\end{align}
where
$\left(\overleftarrow{\mathbf{\Delta}} _\xi\right)_x= \left[\cc \tilde{\mathds{1}}_F (-\overleftarrow{\slashed{P}}-M_\xi) +\tilde{\mathbf{X}}_{\xi \xi}\right]_x$ and we only integrate over $y$. Next, we shift the light fermion multiplet as
\begin{align}
  \delta \xi' &= \delta \xi+\mathbf{\Delta}_\xi^{-1}\left[\tilde{\mathbf{X}}_{\xi \Xi} \delta \Xi-\tilde{\mathbf{X}}_{\xi \Phi} \delta \Phi-\tilde{\mathbf{X}}_{\xi \phi} \delta \phi\right],
  \label{eq:xishift} \\
  \delta \xi'^T &= \delta \xi^T+\left[\delta \Xi^T \tilde{\mathbf{X}}_{\Xi \xi}+\delta \Phi^T \tilde{\mathbf{X}}_{\Phi \xi}+\delta \phi^T \tilde{\mathbf{X}}_{\phi \xi}\right]\overleftarrow{\mathbf{\Delta}}_\xi^{-1},
  \label{eq:xishift_T}
\end{align}
under which the path integral measure is invariant.  Since $\xi$ is a
multiplet of Majorana-like spinors, the two shifts in Eqs.\ \eqref{eq:xishift}
and \eqref{eq:xishift_T} are not independent and it is important to show that Eq.\ \eqref{eq:xishift_T} indeed defines the transpose of Eq.\ \eqref{eq:xishift}. The required relation between the two shifts is proved in Appendix~\ref{app: shifts}. After the
shifts have been performed we arrive at
\begin{align}
\delta ^2 \LagNoT_\xi
={}& \frac{1}{2} \delta \xi'^{T} \mathbf{\Delta}_\xi \delta \xi'-\frac{1}{2}\delta \Xi^T \tilde{\mathbf{X}}_{\Xi \xi} \mathbf{\Delta}_\xi^{-1} \tilde{\mathbf{X}}_{\xi \Xi} \delta \Xi+\frac{1}{2}\delta \Xi ^T \tilde{\mathbf{X}}_{\Xi \xi} \mathbf{\Delta}_\xi ^{-1} \tilde{\mathbf{X}}_{\xi \Phi} \delta \Phi
\nonumber \\ &+\frac{1}{2}\delta \Xi ^T \tilde{\mathbf{X}}_{\Xi \xi} \mathbf{\Delta}_\xi ^{-1} \tilde{\mathbf{X}}_{\xi \phi} \delta \phi +\frac{1}{2} \delta \Phi^T \tilde{\mathbf{X}}_{\Phi \xi} \mathbf{\Delta}_\xi ^{-1} \tilde{\mathbf{X}}_{\xi \Xi} \delta \Xi-\frac{1}{2} \delta \Phi^T \tilde{\mathbf{X}}_{\Phi \xi} \mathbf{\Delta}_\xi ^{-1} \tilde{\mathbf{X}}_{\xi \Phi} \delta \Phi \nonumber \\ &-\frac{1}{2} \delta \Phi^T \tilde{\mathbf{X}}_{\Phi \xi} \mathbf{\Delta}_\xi ^{-1} \tilde{\mathbf{X}}_{\xi \phi} \delta \phi +\frac{1}{2} \delta \phi^T \tilde{\mathbf{X}}_{\phi \xi} \mathbf{\Delta}_\xi ^{-1} \tilde{\mathbf{X}}_{\xi \Xi} \delta \Xi-\frac{1}{2} \delta \phi^T \tilde{\mathbf{X}}_{\phi \xi} \mathbf{\Delta}_\xi ^{-1} \tilde{\mathbf{X}}_{\xi \Phi} \delta \Phi \nonumber \\
& -\frac{1}{2} \delta \phi^T \tilde{\mathbf{X}}_{\phi \xi} \mathbf{\Delta}_\xi ^{-1} \tilde{\mathbf{X}}_{\xi \phi} \delta \phi.
\label{eq:original_2nd_variation}
\end{align}
Before proceeding to eliminate terms
that mix bosonic fluctuations and fluctuations of heavy fermions $\Xi$, it is convenient to introduce the matrices
\begin{align}
    M_{A B} = \begin{cases}
\begin{pmatrix}
    \tilde{\mathbf{X}}_{A B} & \tilde{\mathbf{X}}_{A \xi} \\
    \tilde{\mathbf{X}}_{\xi B} & \mathbf{\Delta}_\xi 
    \end{pmatrix}, &\text{if $A \neq B$}\\
\begin{pmatrix}
   \Delta_{A} & \tilde{\mathbf{X}}_{A \xi} \\
    \tilde{\mathbf{X}}_{\xi B} & \mathbf{\Delta}_\xi 
    \end{pmatrix}, &\text{if $A=B$}
\end{cases}
\label{eq:M_def}
\end{align}
as their Schur complements w.r.t.\ $\mathbf{\Delta}_\xi$
\begin{align}
    M_{A B}/\mathbf{\Delta}_\xi = \begin{cases}
    \tilde{\mathbf{X}}_{A B} - \tilde{\mathbf{X}}_{A \xi} \mathbf{\Delta}_\xi ^{-1} \tilde{\mathbf{X}}_{\xi B},  &\text{if $A \neq B$} \\
\Delta_{A} - \tilde{\mathbf{X}}_{A \xi} \mathbf{\Delta}_\xi ^{-1} \tilde{\mathbf{X}}_{\xi A},  &\text{if $A=B$}
\end{cases}
\label{eq: useful_Schur_complements}
\end{align}
appear in the shifted second variation. We also define $\left(M_{AB}\right)_\xi \equiv M_{AB}/\mathbf{\Delta}_\xi $ as a shorthand for the Schur complement. Then, the shifted second variation is given by
\begin{align}
\delta^2 \LagNoT &= \delta^2 \bar{\LagNoT}_\text{B}+\frac{1}{2}\delta \Xi ^T \left(M_{\Xi \Xi}\right)_\xi \delta \Xi- \frac{1}{2} \delta \Xi ^T \left(M_{\Xi \Phi}\right)_\xi \delta \Phi+\frac{1}{2} \delta \Phi ^T \left(M_{\Phi \Xi}\right)_\xi \delta \Xi \nonumber \\
& \quad -\frac{1}{2} \delta \Xi ^T \left(M _{\Xi \phi}\right)_\xi \delta \phi+\frac{1}{2} \delta \phi ^T \left(M_{\phi \Xi}\right)_\xi \delta \Xi+\frac{1}{2} \delta \xi'^{T} \mathbf{\Delta}_\xi \delta \xi'.
\label{eq:d2_Lag_step_1}
\end{align}
In Eq.\ \eqref{eq:d2_Lag_step_1} the first term on the r.h.s.,
$\delta^2 \bar{\LagNoT}_\text{B}$, is obtained by replacing
$\tilde{\mathbf{X}}_{A B}$ and $\mathbf{\Delta}_A$ in
$\delta^2 \Lag{B}$ with $\left(M_{AB}\right)_\xi$ in accordance with Eq.\
\eqref{eq: useful_Schur_complements}.
By shifting $\delta \Xi$ in a similar way to $\delta \xi$,
\begin{align}
\label{eq:Xishift}
  \delta \Xi' &= \delta \Xi-\left(M_{\Xi \Xi}\right)_\xi^{-1}\left[\left(M_{\Xi \Phi}\right)_\xi \delta \Phi+\left(M_{\Xi \phi}\right)_\xi \delta \phi\right], \\
  \label{eq:Xishift_T}
  \delta \Xi'^T &= \delta \Xi ^T+\left[\delta \Phi^T \left(M_{\Phi \Xi}\right)_\xi +\delta \phi ^T \left(M_{\phi \Xi}\right)_\xi \right] \overleftarrow{\left(M_{\Xi \Xi}\right)_\xi}^{-1},
\end{align}
one finds
\begin{align}
  \delta^2 \LagNoT ={}&
  -\frac{1}{2} \delta \Phi ^T \left[\left(M_{\Phi \Phi}\right)_\xi-\left(M_{\Phi \Xi}\right)_\xi\left(M_{\Xi \Xi}\right)_\xi^{-1}\left(M_{\Xi \Phi}\right)_\xi\right] \delta \Phi \nonumber \\
  & -\frac{1}{2} \delta \phi ^T \left[\left(M_{\phi \phi}\right)_\xi-\left(M_{\phi \Xi}\right)_\xi\left(M_{\Xi \Xi}\right)_\xi^{-1}\left(M_{\Xi \phi}\right)_\xi\right] \delta \phi \nonumber \\ 
  & -\frac{1}{2} \delta \Phi ^T \left[\left(M_{\Phi \phi}\right)_\xi-\left(M_{\Phi \Xi}\right)_\xi\left(M_{\Xi \Xi}\right)_\xi^{-1}\left(M_{\Xi \phi}\right)_\xi\right] \delta \phi \nonumber \\ 
  & -\frac{1}{2}\delta \phi ^T \left[\left(M_{\phi \Phi}\right)_\xi-\left(M_{\phi \Xi}\right)_\xi\left(M_{\Xi \Xi}\right)_\xi^{-1}\left(M_{\Xi \Phi}\right)_\xi\right] \delta \Phi \nonumber \\
  & +\frac{1}{2} \delta \xi'^{T} \mathbf{\Delta}_\xi \delta \xi'+\frac{1}{2} \delta \Xi'^{T} \left(M_{\Xi \Xi}\right)_\xi \delta \Xi' \nonumber \\
\equiv{}& -\frac{1}{2}\begin{pmatrix}
\delta \Phi^T & \delta \phi^T
\end{pmatrix} \fluct_\text{B}
 \begin{pmatrix}
\delta \Phi \\
\delta \phi
\end{pmatrix} +\frac{1}{2} \delta \xi'^{T} \mathbf{\Delta}_\xi \delta \xi'+\frac{1}{2} \delta \Xi'^{T} \left(M_{\Xi \Xi}\right)_\xi \delta \Xi' \\
\equiv{}& \delta^2\breve{\LagNoT}_\text{BF} + \delta^2\breve{\LagNoT}_\text{F},
\label{eq:second_var}
\end{align}
with
\begin{align}
  \delta^2\breve{\LagNoT}_\text{BF} &= -\frac{1}{2}
  \begin{pmatrix}\delta \Phi^T & \delta \phi^T\end{pmatrix} \fluct_\text{B}
  \begin{pmatrix}\delta \Phi \\ \delta \phi\end{pmatrix}, \\
  \delta^2\breve{\LagNoT}_\text{F} &= \frac{1}{2} \begin{pmatrix}
   \delta \Xi'^{T} & \delta \xi'^{T}  \end{pmatrix} \fluct_\text{F} 
   \begin{pmatrix}
   \delta \Xi' \\ \delta \xi'
   \end{pmatrix},
\end{align}
where we defined
\begin{align}
\fluct_\text{F} &\equiv 
\begin{pmatrix}
\left(M_{\Xi \Xi}\right)_\xi & 0 \\
0 & \mathbf{\Delta}_\xi
\end{pmatrix}, \\
\fluct_\text{B} &\equiv
    \begin{pmatrix}
\left(\breve{M}_{\Phi \Phi}\right)_{\left(M_{\Xi \Xi}\right)_\xi} & \left(\breve{M}_{\Phi \phi}\right)_{\left(M_{\Xi \Xi}\right)_\xi} \\
\left(\breve{M}_{\phi \Phi}\right)_{\left(M_{\Xi \Xi}\right)_\xi} &  \left(\breve{M}_{\phi \phi}\right)_{\left(M_{\Xi \Xi}\right)_\xi}
\end{pmatrix}, \\
\intertext{and}
\breve{M}_{A B} &= \begin{pmatrix}
\left(M_{AB}\right)_\xi & \left(M_{A \Xi}\right)_\xi \\
\left(M_{\Xi B}\right)_\xi & \left(M_{\Xi \Xi}\right)_\xi
\end{pmatrix}.
\end{align}
At this point there are no terms including both a bosonic and a
fermionic fluctuation and the path integrals over bosons and fermions
can be performed separately. This concludes the treatment of the UV theory with the result
\begin{align}
\hat{\Gamma}^\text{1\Loop}_\text{L,UV} &= \frac{i}{2} \log \det \fluct_\text{B}-\frac{i}{2} \log \det \fluct_\text{F},
\label{eq: gamma_LUV_1L_full}
\end{align}
where the hat on the generating functional indicates that it is local, which is achieved by inserting local expansions of all heavy background fields as discussed in Chapter~\ref{chap: fcuntional matching}. Following the same route as in Chapter~\ref{chap: fcuntional matching} we would now like to compute the EFT side of Eq.\ \eqref{eq: scalar_matching_condition} and impose the matching condition. We expect that this cancels the soft region contributions to Eq.\ \eqref{eq: gamma_LUV_1L_full}. It is not straightforward to show this using the form of $\hat{\Gamma}^\text{1\Loop}_\text{L,UV}$ given here. We instead note that this one-loop contribution to the generating functional, in general, is a superdeterminant as shown in Ref.\ \cite{Neufeld:1998js} and further discussed in Ref.\ \cite{Henning:2016lyp}, which can be rewritten in terms of sub-determinants in different ways. We here chose an expression that turns the superdeterminant into two determinants of Grassmann even matrices, completely separating fields of different statistics. In the final expression, however, there is no manifest separation of heavy and light fields and the soft region cancellation is obscured. In the next section, we will express the same superdeterminant in a different way, which manifestly separates heavy and light fields and allows for a prove of the soft region cancellation. For the remainder of this section we will simply assume that, after imposing the matching condition, all that is left is the hard region contribution to Eq.\ \eqref{eq: gamma_LUV_1L_full}. Using that $\log \det A = \Tr \log A$ and performing steps analogous to Eq. \eqref{eq: CDE_equation} we find
\begin{align}
    \Lag{EFT}^{1\Loop}[\bg{\phi},\bg{\xi}] &= \frac{i}{2} \tr \left.\int \frac{\rd^d q}{(2\pi)^d}  \left(\log \fluct_\text{B}-\log \fluct_\text{F}\right) \right|^{P^\mu \rightarrow P^\mu - q^\mu} _\text{hard,trunc}.
\end{align}

We would now like to absorb the matrices $\tilde{\mathds{1}}_F$, $\cc$ and $\tilde{\mathds{1}}_B$ in redefinitions of the coupling matrices $\tilde{\mathbf{X}}_{AB}$ in order for them not to appear explicitly in the final BSUOLEA operators. To this end, we first note that, with the exception of the explicit appearance of $\mathbf{\Delta}_\xi$ in $\fluct_\text{F}$, $\mathbf{\Delta}_\xi$ only enters the final result through $\left(M_{AB}\right)_\xi$ and hence through its inverse. We may write $\left. \mathbf{\Delta}_\xi ^{-1} \right \vert^{P_\mu\to P_\mu-q_\mu}_\text{hard}\equiv
\mathbf{\Delta}_\xi ^{-1}(q)$ as
\begingroup
\allowdisplaybreaks
\begin{align}
\mathbf{\Delta}_\xi ^{-1}(q)&=\left[\cc \tilde{\mathds{1}}_F(\slashed{P}_\xi-\slashed{q}-M_\xi) +\tilde{\mathbf{X}}_{\xi \xi}\right]^{-1} \nonumber \\
&= \left[\mathds{1}-\left(-\slashed{q}\right)^{-1}\tilde{\mathds{1}}_F\cc^{-1}\left(-\cc \tilde{\mathds{1}}_F\slashed{P}_\xi+\cc \tilde{\mathds{1}}_FM_{\xi}-\tilde{\mathbf{X}}_{\xi \xi}\right)\right]^{-1}\left(-\slashed{q}\right)^{-1}\tilde{\mathds{1}}_F\cc^{-1} \nonumber \\
&= \sum_{m=0} ^\infty \left[\left(-\slashed{q}\right)^{-1}\tilde{\mathds{1}}_F\cc^{-1}\left(-\cc \tilde{\mathds{1}}_F\slashed{P}_\xi+\cc \tilde{\mathds{1}}_FM_{\xi}-\tilde{\mathbf{X}}_{\xi \xi}\right) \right]^m \left(-\slashed{q}\right)^{-1}\tilde{\mathds{1}}_F\cc^{-1} \nonumber \\
&= \sum_{m=0} ^\infty \left[\left(-\slashed{q}\right)^{-1}\left(-\slashed{P}_\xi+M_{\xi}-\mathbf{X}_{\xi \xi}\right) \right]^m \left(-\slashed{q}\right)^{-1}\tilde{\mathds{1}}_F\cc^{-1},
\label{eq:DeltaxiInv}
\end{align}
\endgroup
where we defined 
\begin{align}
\mathbf{X}_{\xi \xi}\equiv \tilde{\mathds{1}}_F\cc ^{-1} \tilde{\mathbf{X}}_{\xi \xi}.
\end{align}
Then Eq.\
\eqref{eq: useful_Schur_complements} becomes
\begin{align}
\left( M_{A B} \right)_\xi = \begin{cases}
    \tilde{\mathbf{X}}_{A B}-\tilde{\mathbf{X}}_{A \xi}\sum _n \left[\left(-\slashed{q}\right) ^{-1} \left(M_{\xi}-\mathbf{X} _{\xi \xi}-\slashed{P}_\xi\right) \right]^n \left(-\slashed{q}\right) ^{-1} \mathbf{X}_{\xi B},\, \text{if } A \neq B \\
 \mathbf{\Delta}_A-\tilde{\mathbf{X}}_{A \xi}\sum _n \left[\left(-\slashed{q}\right) ^{-1} \left(M_{\xi}-\mathbf{X} _{\xi \xi}-\slashed{P}_\xi\right) \right]^n \left(-\slashed{q}\right) ^{-1} \mathbf{X}_{\xi A},~~ \text{if } A=B
\end{cases}
\end{align}
where we introduced
$\mathbf{X}_{\xi B}\equiv \tilde{\mathds{1}}_F \cc ^{-1}
\tilde{\mathbf{X}}_{\xi B}$ and the sum over $n$ runs over all non-negative integers. Thus, in any Schur complement w.r.t.\ $\mathbf{\Delta}_\xi$ the explicit appearance of $\cc$ and $\tilde{\mathds{1}}_F$ can be avoided by replacing the matrices $\tilde{\mathbf{X}}_{\xi B}$ by the matrices $\mathbf{X}_{\xi B}$. Next we note that $\mathbf{\Delta}_\Xi$ only enters the result through $\left(\breve{M}_{AB}\right)_{\left(M_{\Xi \Xi}\right)_\xi}$ except for the explicit appearance of $\left(M_{\Xi \Xi}\right)_\xi$ in $\fluct_\text{F}$. Considering that
\begin{align}
\left(M_{\Xi \Xi}\right)_\xi ^{-1}(q) \equiv{}& \left. \left(M_{\Xi \Xi}\right)_\xi ^{-1} \right \vert^{P^\mu\rightarrow P^\mu-q^\mu}_\text{hard} \nonumber \\ 
={}& \Bigg[\cc \tilde{\mathds{1}}_F\left(-\slashed{q}-M_{\Xi}\right) +\cc \tilde{\mathds{1}}_F \slashed{P}_\Xi+\tilde{\mathbf{X}}_{\Xi \Xi} \nonumber \\
 & -\tilde{\mathbf{X}}_{\Xi \xi}\sum _{n=0} ^{\infty} \left[\left(-\slashed{q}\right) ^{-1} \left(-\mathbf{X} _{\xi \xi}-\slashed{P}_\xi+M_\xi\right) \right]^n \left(-\slashed{q}\right) ^{-1} \mathbf{X}_{\xi \Xi} \Bigg]^{-1} \nonumber \\
={}&\sum _{m=0} ^{\infty} \Big\{\mathcal{K}_\Xi ^{-1}  \left(-\mathbf{X} _{\Xi \Xi}-\slashed{P}_\Xi\right) \nonumber \\
&- \mathcal{K}_\Xi ^{-1}  \mathbf{X}_{\Xi \xi}\sum _{n=0} ^{\infty} \left[\left(-\slashed{q}\right) ^{-1} \left(M_\xi-\mathbf{X} _{\xi \xi}-\slashed{P}_\xi\right) \right]^n \slashed{q}^{-1} \mathbf{X}_{\xi \Xi}  \Big\}^m  \mathcal{K}_\Xi ^{-1} \cc ^{-1} \tilde{\mathds{1}}_F,
\label{eq:DeltaXiTildeInv}
\end{align}
where
\begin{align}
  \mathcal{K}_\Xi &\equiv \left(-\slashed{q}-M_\Xi\right), \\
  \mathbf{X}_{\Xi \xi} &\equiv \cc ^{-1} \tilde{\mathds{1}}_F \tilde{\mathbf{X}}_{\Xi \xi},
\end{align}
we can again absorb $\cc^{-1}$ and $\tilde{\mathds{1}}_F$ since $\left(\breve{M}_{AB}\right)_{\left(M_{\Xi \Xi}\right)_\xi}$ always contains
\begin{align}
\cc ^{-1} \tilde{\mathds{1}}_F \left(M_{\Xi B}\right)_\xi &=  \cc ^{-1} \tilde{\mathds{1}}_F \left(\tilde{\mathbf{X}}_{\Xi B} - \tilde{\mathbf{X}}_{\Xi \xi} \mathbf{\Delta}_\xi ^{-1} \tilde{\mathbf{X}}_{\xi B}\right) = \left(\mathbf{X}_{\Xi B} - \mathbf{X}_{\Xi \xi} \mathbf{\Delta}_\xi ^{-1} \tilde{\mathbf{X}}_{\xi B}\right), \\ 
\intertext{where}
\mathbf{X}_{\Xi B} = \cc ^{-1} \tilde{\mathds{1}}_F \tilde{\mathbf{X}}_{\Xi B}. 
\end{align}
It is clear from this discussion that in the evaluation of $\fluct_\text{B}$ no explicit factors of $\cc$ and $\tilde{\mathds{1}}_F$ appear in the final result. We can further evaluate the contribution from $\fluct_\text{B}$ by using Eq.\ \eqref{eq: Schured_Q_UV} to write
\begin{align}
\log \det \fluct_\text{B} ={}& 
\log \det \left(\left(\breve{M}_{\Phi \Phi}\right)_{\left(M_{\Xi \Xi}\right)_\xi}-\left(\breve{M}_{\Phi \phi}\right)_{\left(M_{\Xi \Xi}\right)_\xi} \left(\breve{M}_{\phi \phi}\right)_{\left(M_{\Xi \Xi}\right)_\xi}^{-1}\left(\breve{M}_{\phi \Phi}\right)_{\left(M_{\Xi \Xi}\right)_\xi}\right) \nonumber \\
& + \log \det \left(\breve{M}_{\phi \phi}\right)_{\left(M_{\Xi \Xi}\right)_\xi}.
\label{eq: Q_B_contribution}
\end{align}
In Chapter~\ref{chap: fcuntional matching} the contribution from the second line vanished completely in the hard region. Here this is not the case since $\left(\breve{M}_{\phi \phi}\right)_{\left(M_{\Xi \Xi}\right)_\xi}$ depends on the heavy fermions $\Xi$. Expanding this Schur complement out in terms of the original matrices we have
\begingroup
\allowdisplaybreaks
\begin{align}
\left. \log \det \left(\breve{M}_{\phi \phi}\right)_{\left(M_{\Xi \Xi}\right)_\xi} \right|_\text{hard} ={}&  \log \det \left[\left(M_{\phi \phi}\right)_\xi - \left. \left(M_{\phi \Xi}\right)_\xi \left(M_{\Xi \Xi}\right)^{-1}_\xi \left(M_{\Xi \phi}\right)_\xi \right] \right|_\text{hard} \nonumber \\
={}& \left. \log \det \left(M_{\phi \phi}\right)_\xi \right|_\text{hard} \nonumber \\
&+ \left. \log \det \left[\mathds{1} - \left(M_{\phi \phi}\right)_\xi^{-1} \left(M_{\phi \Xi}\right)_\xi \left(M_{\Xi \Xi}\right)^{-1}_\xi \left(M_{\Xi \phi}\right)_\xi \right] \right|_\text{hard} \nonumber \\ 
={}& -\sum_{n=1} ^\infty \frac{1}{n} \Tr \left. \left[\left(M_{\phi \phi}\right)_\xi^{-1} \left(M_{\phi \Xi}\right)_\xi \left(M_{\Xi \Xi}\right)^{-1}_\xi \left(M_{\Xi \phi}\right)_\xi \right]^n \right|_\text{hard}^{P_\mu \rightarrow P_\mu-q_\mu},
\label{eq: bosonic_lower}
\end{align}
\endgroup
where the first term in the second line of the right hand side vanishes in the hard
region as it only contains contributions from light fields and the trace is a functional trace which includes the loop integration as discussed in Chapter~\ref{chap: fcuntional matching}. This trace can now be evaluated by inserting $\left(M_{\phi \phi}\right)_\xi^{-1}(q) \equiv \left. \left(M_{\phi \phi}\right)_\xi^{-1} \right \vert^{P_\mu \rightarrow P_\mu-q_\mu}_\text{hard}$, given by
\begin{align}
\left(M_{\phi \phi}\right)_\xi^{-1}(q) &= \left[\tilde{\mathds{1}}_B\left(-q^2+2q \cdot P_\phi-P_\phi^2+M_\phi^2\right) + \tilde{\mathbf{X}}_{\phi \phi}- \tilde{\mathbf{X}}_{\phi \xi} \mathbf{\Delta}_\xi ^{-1}(q) \tilde{\mathbf{X}}_{\xi \phi}\right]^{-1}  \nonumber \\
   &= -\sum_{k=0}^\infty \left(q^{-2}\left[2q\cdot P_\phi-P_\phi^2+M_\phi^2+\tilde{\mathds{1}}_B \tilde{\mathbf{X}}_{\phi \phi}- \tilde{\mathds{1}}_B \tilde{\mathbf{X}}_{\phi \xi} \mathbf{\Delta}_\xi ^{-1}(q) \tilde{\mathbf{X}}_{\xi \phi}\right]\right)^k q^{-2}\tilde{\mathds{1}}_B,
\label{eq: M_phi_phi_inv}
\end{align}
where $\mathbf{\Delta}_\xi^{-1}(q)$ is given in Eq.\ \eqref{eq:DeltaxiInv}, 
as well as Eq.\ \eqref{eq:DeltaXiTildeInv} and the appropriate expressions for $\left(M_{\Xi \phi}\right)_\xi$ and $\left(M_{\phi \Xi}\right)_\xi$. We abstain from explicitly writing out the resulting, somewhat lengthy, expression here. It should be noted that since $\left(M_{\phi \phi}\right)_\xi^{-1}$ always appears in the combination $\left(M_{\phi \phi}\right)_\xi^{-1}\tilde{\mathbf{X}}_{\phi B}$ we can absorb the $\tilde{\mathds{1}}_B$ arising from $\left(M_{\phi \phi}\right)_\xi^{-1}$ if we define $\mathbf{X}_{\phi B} \equiv \tilde{\mathds{1}}_B \tilde{\mathbf{X}}_{\phi B}$, where $B$ may be any field, including $\phi$. This assures that no explicit factors of $\tilde{\mathds{1}}_B$ appear in the final operators that are generated from Eq.\ \eqref{eq: bosonic_lower}. To evaluate the first line of Eq.\ \eqref{eq: Q_B_contribution} we first note that 
\begin{align}
\left(\breve{M}_{\phi \phi}\right)_{\left(M_{\Xi \Xi}\right)_\xi}^{-1} &=\left[\mathbf{\Delta}_\phi - \tilde{\mathbf{X}}_{\phi \xi} \mathbf{\Delta}_\xi^{-1} \tilde{\mathbf{X}}_{\xi \phi} - \left(M_{\phi \Xi}\right)_\xi \left(M_{\Xi \Xi}\right)_\xi^{-1} \left(M_{\Xi \phi}\right)_\xi \right]^{-1} 
\end{align}
can be computed in exactly the same way as $\left(M_{\phi \phi}\right)_\xi^{-1}$ in Eq.\ \eqref{eq: M_phi_phi_inv} and can be obtained from that result by simply replacing $\tilde{\mathbf{X}}_{\phi \phi}- \tilde{\mathbf{X}}_{\phi \xi} \mathbf{\Delta}_\xi ^{-1} \tilde{\mathbf{X}}_{\xi \phi}$ by 
\begin{align}
    \mathbf{F}_{\phi \phi} = \left. \left(\tilde{\mathbf{X}}_{\phi \phi}- \tilde{\mathbf{X}}_{\phi \xi} \mathbf{\Delta}_\xi^{-1} \tilde{\mathbf{X}}_{\xi \phi} - \left(M_{\phi \Xi}\right)_\xi \left(M_{\Xi \Xi}\right)_\xi^{-1} \left(M_{\Xi \phi}\right)_\xi\right) \right|^{P_\mu \rightarrow P_\mu-q_\mu},
\end{align}
which means that
\begin{align}
\left(\breve{M}_{\phi \phi}\right)_{\left(M_{\Xi \Xi}\right)_\xi}^{-1}(q) &= -\sum_{k=0}^\infty \left(q^{-2}\left[2q\cdot P_\phi-P_\phi^2+M_\phi^2+\tilde{\mathds{1}}_B \mathbf{F}_{\phi \phi}\right]\right)^k q^{-2}\tilde{\mathds{1}}_B.
\label{eq: hat_M_phi_phi_inv}
\end{align}
All of the matrices that enter $\mathbf{F}_{\phi \phi}$ have already been written out explicitly and by inserting these into Eq.\ \eqref{eq: hat_M_phi_phi_inv} one obtains an explicit expression for $\left(\breve{M}_{\phi \phi}\right)_{\left(M_{\Xi \Xi}\right)_\xi}^{-1}$. This expression is again rather lengthy and is therefore not given explicitly. Note that it is again possible to absorb all appearances of $\tilde{\mathds{1}}_B$ by defining $\mathbf{X}_{\phi B} \equiv \tilde{\mathds{1}}_B \tilde{\mathbf{X}}_{\phi B}$. With Eq.\ \eqref{eq: hat_M_phi_phi_inv} at hand all the matrices entering
\begin{align}
\mathbf{F}_{\Phi \Phi} \equiv{}& \left( \tilde{\mathbf{X}}_{\Phi \Phi}-\tilde{\mathbf{X}}_{\Phi \xi} \mathbf{\Delta}_\xi^{-1} \tilde{\mathbf{X}}_{\xi \Phi}-\left(M_{\Phi \Xi}\right)_\xi \left(M_{\Xi \Xi}\right)_\xi^{-1} \left(M_{\Xi \Phi}\right)_\xi\nonumber \right. \\
&\left. \left. -\left(\breve{M}_{\Phi \phi}\right)_{\left(M_{\Xi \Xi}\right)_\xi} \left(\breve{M}_{\phi \phi}\right)_{\left(M_{\Xi \Xi}\right)_\xi}^{-1}\left(\breve{M}_{\phi \Phi}\right)_{\left(M_{\Xi \Xi}\right)_\xi}\right) \right|^{P_\mu \rightarrow P_\mu-q_\mu}
\label{eq: F_Phi_Phi}
\end{align}
have been given explicitly and the first line of Eq.\ \eqref{eq: Q_B_contribution} can be written as
\begin{align*}
&\Tr \log \left. \left[\tilde{\mathds{1}}_B\left(-q^2+2q \cdot P_\Phi-P_\Phi^2+M_\Phi^2\right) + \mathbf{F}_{\Phi \Phi}\right] \right|_\text{hard} \nonumber \\
&= \Tr \log \left. \left[\tilde{\mathds{1}}_B\left(-q^2+M_\Phi^2\right)\right] \right|_\text{hard} \nonumber \\
&\quad + \Tr \log \left. \left[\mathds{1}-\left(q^2-M_\Phi^2\right)^{-1} \left(2q\cdot P_\Phi-P_\Phi^2+\tilde{\mathds{1}}_B \mathbf{F}_{\Phi \Phi}\right) \right] \right|_\text{hard}.
\end{align*}
Finally, the first term yields an infinite constant, which can be ignored and in the second term we can expand the log to yield
\begin{align}
    &\Tr \log \left. \left[\mathds{1}-\left(q^2-M_\Phi^2\right)^{-1} \left(2q\cdot P_\Phi-P_\Phi^2+\tilde{\mathds{1}}_B \mathbf{F}_{\Phi \Phi}\right) \right] \right|_\text{hard} \nonumber \\& = -\sum _{n=1}^\infty \frac{1}{n}\Tr \left. \left[\left(q^2-M_\Phi^2\right)^{-1} \left(2q\cdot P_\Phi-P_\Phi^2+\tilde{\mathds{1}}_B \mathbf{F}_{\Phi \Phi}\right) \right]^n \right|_\text{hard}.
    \label{eq: Q_B_upper_contribution}
\end{align}
By defining $\mathbf{X}_{\Phi B} \equiv \tilde{\mathds{1}}_B\tilde{\mathbf{X}}_{\Phi B}$ we can again remove all explicit factors of $\tilde{\mathds{1}}_B$. This concludes the discussion of the contributions to the effective Lagrangian originating from $\fluct_\text{B}$. We briefly summarise the results below. We have shown that the full contribution to the effective action can be expressed as
\begin{align}
\Lag{EFT,B}^{1\Loop} ={}& \left(-\frac{i}{2} \sum _{n=1}^\infty \frac{1}{n}\Tr \left. \left[\left(q^2-M_\Phi^2\right)^{-1} \left(2q\cdot P_\Phi-P_\Phi^2+\tilde{\mathds{1}}_B \mathbf{F}_{\Phi \Phi}\right) \right]^n \right|_\text{hard} \right. \nonumber \\ & \left.-\frac{i}{2}\sum_{n=1} ^\infty \frac{1}{n} \Tr \left. \left[\left(M_{\phi \phi}\right)_\xi^{-1} \left(M_{\phi \Xi}\right)_\xi \left(M_{\Xi \Xi}\right)^{-1}_\xi \left(M_{\Xi \phi}\right)_\xi \right]^n \right|_\text{hard}^{P_\mu \rightarrow P_\mu-q_\mu}\right)_\text{trunc},
\label{eq: final_Q_B_contribution}
\end{align}
with $\mathbf{F}_{\Phi \Phi}$,  $\left(\breve{M}_{\phi \phi}\right)^{-1}_{\left(M_{\Xi \Xi}\right)_\xi}$, $\left(M_{\phi \phi}\right)_\xi^{-1}$, $\left(M_{\Xi \Xi}\right)_\xi^{-1}$ and $\mathbf{\Delta}_\xi^{-1}$ explicitly given in Eqs.\ \eqref{eq: F_Phi_Phi}, \eqref{eq: hat_M_phi_phi_inv}, \eqref{eq: M_phi_phi_inv}, \eqref{eq:DeltaXiTildeInv}  and \eqref{eq:DeltaxiInv}, respectively. To perform the evaluation of Eq.\ \eqref{eq: final_Q_B_contribution} is a tedious task, which is not feasible by hand. Therefore, a Mathematica code was developed that performs these expansions up to a desired mass dimension. In this code, the coupling matrices in the bosonic sector are assumed to contain up to one open covariant derivative and those in the fermionic sector are assumed to not contain any open covariant derivatives as discussed in Section~\ref{sec: open_cds}. The evaluation of the loop integrals is not a problem since they can be reduced to basis integrals for which a general formula is known, see e.g.\ Refs.\ \cite{Drozd:2015rsp,Ellis:2017jns,Zhang:2016pja} and Appendix~\ref{app:loop_functions}.

To complete the calculation we need to compute the contribution from $\fluct_\text{F}$ which reads
\begin{align}
  \log \det \fluct_\text{F} = \log \det \left(M_{\Xi \Xi}\right)_\xi + \log \det \mathbf{\Delta}_\xi.
\end{align}
Again, we are only interested in the contribution from the hard region
where the light only part $\log \det \mathbf{\Delta} _\xi$ vanishes. Hence we only need to consider
$\left(M_{\Xi \Xi}\right)_\xi$. We find
\begin{align}
  \Tr \log \Big( &\mathbf{\Delta}_\Xi(q) - \tilde{\mathbf{X}}_{\Xi \xi}\mathbf{\Delta}_\xi ^{-1}(q) \tilde{\mathbf{X}}_{\xi \Xi}\Big) \nonumber \\ &= \Tr \log \left(\cc \tilde{\mathds{1}}_F \mathcal{K}_\Xi+\cc \tilde{\mathds{1}}_F \slashed{P}_\Xi+\tilde{\mathbf{X}}_{\Xi \Xi}-\tilde{\mathbf{X}}_{\Xi \xi}\mathbf{\Delta}_\xi ^{-1}(q) \tilde{\mathbf{X}}_{\xi \Xi}\right) \nonumber \\
  &= \Tr \log \left( \cc \tilde{\mathds{1}}_F\mathcal{K}_\Xi \right) + \Tr \log \left[ \mathds{1}-\mathcal{K}_\Xi^{-1}  \left(-\slashed{P}_\Xi-\mathbf{X}_{\Xi \Xi}+\mathbf{X}_{\Xi \xi}\mathbf{\Delta}_\xi ^{-1}(q) \tilde{\mathbf{X}}_{\xi \Xi} \right) \right],
\label{eq: trlog_fermionic}
\end{align}
where the first term on the r.h.s.\ of Eq.\ \eqref{eq: trlog_fermionic} is
absorbed in the normalisation of the path integral. Expanding the logarithm yields
\begin{align}
\Lag{EFT,F}^{1\Loop} &= \left.-\frac{i}{2} \sum_{n=1}^{\infty} \frac{1}{n} \Tr \left[\mathcal{K}_\Xi^{-1}\left(-\slashed{P}_\Xi-\mathbf{X}_{\Xi \Xi} +\mathbf{X} _{\Xi \xi} \mathbf{\Delta}_\xi ^{-1}(q) \tilde{\mathbf{X}}_{\xi \Xi} \right) \right]^n\right|_\text{hard,trunc}^{P_\mu \rightarrow P_\mu-q_\mu},
\end{align}
where $\mathbf{\Delta}_\xi ^{-1}(q)$ from Eq.\ \eqref{eq:DeltaxiInv} has to be inserted. This trace is also implemented in a Mathematica code that generates all terms up to a given mass dimension. The full computation for the sum
\begin{align}
  \Lag{EFT}^{1\Loop} = \Lag{EFT,B}^{1\Loop} + \Lag{EFT,F}^{1\Loop}
  \label{eq:UOLEA_final}
\end{align}
has been performed and operators up to mass dimension six have been retained. The full result, which constitutes the BSUOLEA and suffices to match any renormalisable, Lorentz invariant quantum field theory containing scalar fields, vector fields and spin-$1/2$ fermions to an EFT with operators up to mass dimension six, can be found at \cite{code}. As was already mentioned in the previous chapter, it can be shown that in the final expressions every $P_\mu$ can be written in a commutator of the form
$[P_\mu,\bullet]$, as shown in Refs.\ \cite{Gaillard:1985uh,Cheyette:1987qz}. This has the interpretation that $P_\mu$ acts on whatever is to its right in the commutator, since the covariant derivative satisfies the Leibniz rule. For example, $[P_\mu,U_{\Phi \Phi}]=\left(P_\mu U_{\Phi \Phi}\right)+U_{\Phi \Phi} P_\mu - U_{\Phi \Phi} P_\mu =\left(P_\mu U_{\Phi \Phi}\right)$, where in the first term in the intermediate step and in the final result the covariant derivative only acts within the parentheses. If the covariant derivatives are not brought into commutators they have to be interpreted as acting on everything to their right, which makes the use of the operators obscure and difficult. To combine all $P^\mu$ operators into commutators one can either explicitly use
the Baker-Campbell-Hausdorff formula in the calculation, introducing an extra integration as was done in Ref.\ \cite{Drozd:2015rsp}, or construct a basis for these commutators and
then solve a system of equations to fix the coefficients of the basis
elements as was pointed out in Ref.\ \cite{Zhang:2016pja}. The Mathematica code that was used to expand the traces and compute the corresponding universal integral coefficients can also be used to find such bases and solve the system of equations to translate expressions with open covariant derivatives into expressions where every covariant derivative appears in a commutator. This was used to transform the final BSUOLEA result into a form without open covariant derivatives. 

\subsection{Proof of soft region cancellation}
\label{sec: soft_region_cancellation_proof}
We now turn to the proof of the soft region cancellation, which was postponed in the previous section. As was discussed in Section~\ref{sec: background_field_method} we do not address the subtleties that can arise when working in the broken phase of the EFT and we are not aware of a complete proof of the soft region cancellation that extends to this case. The basic idea of the proof presented here follows the steps outlined in Section~\ref{sec: ir_cancellation_purely_scalar}, however, as mentioned before it is not entirely trivial to make the cancellation manifest due to the presence of mixed statistics. To the best of our knowledge, the proof presented here, which is an extension of the proof for the scalar case given in Ref.\ \cite{Zhang:2016pja}, has not been presented elsewhere. In order to proof the cancellation of the soft contributions to $\Gamma^{1\Loop}_\text{L,UV}$ with the genuine one-loop EFT contributions to $\Gamma_\text{EFT}$, it is necessary to relate $Q_\text{EFT}$ to $Q_\text{UV}$. To this end, we consider the equations
\begin{align}
\label{eq: first_connection_eq}
    0 &= \left. \funcDer{}{\xi_y} \right \vert\left(\funcDer{\barLag}{\classical{\hat{\Phi}}}\right)_z = \left.\funcDer{\classical{\hat{\Phi}}^x}{\xi_y} \right \vert \left(\left. \funcDer{^2\barLag}{\classical{\hat{\Phi}}^2} \right \vert \right)_{xz}+\left. \funcDer{\classical{\hat{\Xi}}^x}{\xi_y} \right \vert\left(\left. \secondfuncDer{\barLag}{\classical{\hat{\Xi}}}{\classical{\hat{\Phi}}}\right \vert \right)_{xz}+\left(\left. \secondfuncDer{\barLag}{\xi}{\classical{\hat{\Phi}}}\right \vert\right)_{yz}, \\
\label{eq: second_connection_eq}
     0 &= \left. \funcDer{}{\xi_y}\right \vert \left(\funcDer{\barLag}{\classical{\hat{\Xi}}}\right)_z = \left.  \funcDer{\classical{\hat{\Xi}}^x}{\xi_y}\right \vert \left(\left. \funcDer{^2\barLag}{\classical{\hat{\Xi}}^2}\right \vert\right)_{xz}+\left. \funcDer{\classical{\hat{\Phi}}^x}{\xi_y}\right \vert \left(\left. \secondfuncDer{\barLag}{\classical{\hat{\Phi}}}{\classical{\hat{\Xi}}}\right \vert \right)_{xz}+\left(\left. \secondfuncDer{\barLag}{\xi}{\classical{\hat{\Xi}}}\right \vert\right)_{yz}, \\ 
\label{eq: third_connection_eq}
      0 &= \left. \funcDer{}{\phi_y}\right \vert\left(\funcDer{\barLag}{\classical{\hat{\Phi}}}\right)_z = \left. \funcDer{\classical{\hat{\Phi}}^x}{\phi_y}\right \vert\left(\left. \funcDer{^2\barLag}{\classical{\hat{\Phi}}^2}\right \vert\right)_{xz}+\left. \funcDer{\classical{\hat{\Xi}}^x}{\phi_y}\right \vert\left(\left. \secondfuncDer{\barLag}{\classical{\hat{\Xi}}}{\classical{\hat{\Phi}}} \right \vert\right)_{xz}+\left(\left. \secondfuncDer{\barLag}{\phi}{\classical{\hat{\Phi}}}\right \vert\right)_{yz}, \\ 
\label{eq: fourth_connection_eq}
      0 &= \left. \funcDer{}{\phi_y}\right \vert\left(\funcDer{\barLag}{\classical{\hat{\Xi}}}\right)_z = \left. \funcDer{\classical{\hat{\Xi}}^x}{\phi_y}\right \vert\left(\left. \funcDer{^2\barLag}{\classical{\hat{\Xi}}^2}\right \vert\right)_{xz}+\left. \funcDer{\classical{\hat{\Xi}}^x}{\phi_y}\right \vert\left(\left. \secondfuncDer{\barLag}{\classical{\hat{\Xi}}}{\classical{\hat{\Phi}}} \right \vert\right)_{xz}+\left(\left. \secondfuncDer{\barLag}{\phi}{\classical{\hat{\Xi}}}\right \vert\right)_{yz},
\end{align}
where $\barLag \equiv S_\text{UV}[\classical{\hat{\Phi}}[\phi,\xi],\classical{\hat{\Xi}}[\phi,\xi],\phi,\xi]$ and derivatives with a vertical bar are evaluated at $\phi = \bg{\phi}$ and $\xi = \bg{\xi}$. Eqs.\ \eqref{eq: first_connection_eq}--\eqref{eq: fourth_connection_eq} are analogous to Eq.\ \eqref{eq: eft_UV_relation_scalar_only}.
Defining
\begin{align}
    N_{A B} = \begin{cases}
\begin{pmatrix}
    \tilde{\mathbf{X}}_{A B} & \tilde{\mathbf{X}}_{A \Xi} \\
    \tilde{\mathbf{X}}_{\Xi B} & \mathbf{\Delta}_\Xi 
    \end{pmatrix}, &\text{if $A \neq B$}\\
\begin{pmatrix}
   \mathbf{\Delta}_{A} & \tilde{\mathbf{X}}_{A \Xi} \\
    \tilde{\mathbf{X}}_{\Xi B} & \mathbf{\Delta}_\Xi 
    \end{pmatrix}, &\text{if $A=B$}
\end{cases},
\end{align}
these equations can be rewritten as 
\begingroup
\allowdisplaybreaks
\begin{align}
\label{eq: Phi_xi_der}
  \left. \funcDer{\classical{\hat{\Phi}}^x}{\xi_y} \right \vert &=-\LocalSchur{\xi}{\Phi}_{yz} \LocalSchur{\Phi}{\Phi}_{zx}^{-1}, \\
\label{eq: Xi_xi_der}
   \left. \funcDer{\classical{\hat{\Xi}}^x}{\xi_y}\right \vert&=\LocalSchur{\xi}{\Phi}_{yz} \LocalSchur{\Phi}{\Phi}_{zu}^{-1}\left(\tilde{\mathbf{X}}_{\Phi \Xi}\right)_{uw}\left(\hat{\mathbf{\Delta}}_\Xi^{-1}\right)_{wx}-\left(\tilde{\mathbf{X}}_{\xi \Xi}\right)_{yz}\left(\hat{\mathbf{\Delta}}_\Xi^{-1}\right)_{zx}, \\
\label{eq: Phi_phi_der}
    \left. \funcDer{\classical{\hat{\Phi}}^x}{\phi_y} \right \vert &=-\LocalSchur{\phi}{\Phi}_{yz} \LocalSchur{\Phi}{\Phi}_{zx}^{-1}, \\
\label{eq: Xi_phi_der}    
   \left. \funcDer{\classical{\hat{\Xi}}^x}{\phi_y} \right \vert&=\LocalSchur{\phi}{\Phi}_{yz} \LocalSchur{\Phi}{\Phi}_{zu}^{-1}\left(\tilde{\mathbf{X}}_{\Phi \Xi}\right)_{uw}\left(\hat{\mathbf{\Delta}}_\Xi^{-1}\right)_{wx}-\left(\tilde{\mathbf{X}}_{\phi \Xi}\right)_{yz}\left(\hat{\mathbf{\Delta}}_\Xi^{-1}\right)_{zx},
\end{align}
\endgroup
where we used that 
\begin{align}
\label{eq: X_def}
    \tilde{\mathbf{X}}_{A B} = - \left. \secondfuncDer{\barLag}{A}{B} \right \vert, \\
\label{eq: Delta_def}
    \mathbf{\Delta}_A = - \left. \funcDer{^2 \barLag}{A^2} \right \vert.
\end{align}
Notice that, as introduced in Chapter~\ref{chap: fcuntional matching}, we denote objects that are expanded in inverse masses to obtain local expressions with hats. We prove Eqs.\ \eqref{eq: Phi_xi_der} and \eqref{eq: Xi_xi_der}, the remaining two equations follow analogously. Inserting Eqs.\ \eqref{eq: X_def} and \eqref{eq: Delta_def} into Eqs.\ \eqref{eq: first_connection_eq} and \eqref{eq: second_connection_eq} we obtain
\begin{align}
\label{eq: translated_first_connection_eq}
   \left. \funcDer{\classical{\hat{\Xi}}^x}{\xi_y} \right \vert \left(\hat{\mathbf{\Delta}}_\Xi\right)_{xz} &= - \left. \funcDer{\classical{\hat{\Phi}}^x}{\xi_y} \right \vert \left(\tilde{\mathbf{X}}_{\Phi \Xi}\right)_{xz} - \left(\tilde{\mathbf{X}}_{\xi \Xi}\right)_{yz}, \\
\label{eq: translated_second_connection_eq}
     \left. \funcDer{\classical{\hat{\Phi}}^x}{\xi_y} \right \vert \left(\hat{\mathbf{\Delta}}_\Phi\right)_{xz} &= - \left. \funcDer{\classical{\hat{\Xi}}^x}{\xi_y}\right \vert \left(\tilde{\mathbf{X}}_{\Xi \Phi}\right)_{xz} - \left(\tilde{\mathbf{X}}_{\xi \Phi}\right)_{yz},
\end{align}
from which we get 
\begin{align}
\label{eq: solved_for_Xi_xi_der}
    \left. \funcDer{\classical{\hat{\Xi}}^u}{\xi_y} \right \vert  &= -\left. \funcDer{\classical{\hat{\Phi}}^x}{\xi_y} \right \vert \left(\tilde{\mathbf{X}}_{\Phi \Xi}\right)_{xz} \left(\hat{\mathbf{\Delta}}_\Xi\right)_{zu}^{-1} - \left(\tilde{\mathbf{X}}_{\xi \Xi}\right)_{yz} \left(\hat{\mathbf{\Delta}}_\Xi\right)_{zu}^{-1},\\ \intertext{and}
    \label{eq: solved_for_Phi_xi_der}
    \left. \funcDer{\classical{\hat{\Phi}}^x}{\xi_y}\right \vert &= \left[\left(\tilde{\mathbf{X}}_{\xi \Xi}\right)\left(\hat{\mathbf{\Delta}}_\Xi\right)^{-1}\left(\tilde{\mathbf{X}}_{\Xi \Phi}\right)-\left(\tilde{\mathbf{X}}_{\xi \Phi}\right)\right]_{yz}\left[\hat{\mathbf{\Delta}}_\Phi-\tilde{\mathbf{X}}_{\Phi \Xi}\hat{\mathbf{\Delta}}_\Xi^{-1} \tilde{\mathbf{X}}_{\Xi \Phi}\right]^{-1}_{zx},
\end{align}
where Eq.\ \eqref{eq: solved_for_Phi_xi_der} is obtained by inserting Eq.\ \eqref{eq: solved_for_Xi_xi_der} into Eq.\ \eqref{eq: translated_second_connection_eq}. Expressing Eq.\ \eqref{eq: solved_for_Phi_xi_der} through the appropriate Schur complements we obtain Eq.\ \eqref{eq: Phi_xi_der}. Inserting that result back into Eq.\ \eqref{eq: solved_for_Xi_xi_der} and expressing this through Schur complements we obtain Eq.\ \eqref{eq: Xi_xi_der}.
We then have 
\begin{align}
\label{eq: delta_xi_eft}
    \left(\mathbf{\Delta}_\xi ^\text{EFT}\right)_{xy} &= -\left. \secondfuncDer{S^{\text{tree}}_\text{EFT}}{\xi_x}{\xi_y} \right \vert = -\left. \funcDer{}{\xi_x}\right \vert \funcDer{\barLag}{\xi_y} = \left. \funcDer{\classical{\hat{\Phi}}^z}{\xi_x}\right \vert\left(\tilde{\mathbf{X}}_{\Phi \xi}\right)_{zy}+\left. \funcDer{\classical{\hat{\Xi}}^z}{\xi_x}\right \vert \left(\tilde{\mathbf{X}}_{\Xi \xi}\right)_{zy} + \left(\hat{\mathbf{\Delta}}_\xi\right)_{xy} \nonumber \\
    &= -\LocalSchur{\xi}{\Phi}_{xw} \LocalSchur{\Phi}{\Phi}^{-1}_{wz} \LocalSchur{\Phi}{\xi}_{zy}+\LocalSchur{\xi}{\xi}_{xy}
\end{align}
and similarly 
\begingroup
\allowdisplaybreaks
\begin{align}
\label{eq: delta_phi_eft}
    \left(\mathbf{\Delta}_\phi ^\text{EFT}\right)_{xy} &= -\LocalSchur{\phi}{\Phi}_{xw} \LocalSchur{\Phi}{\Phi}^{-1}_{wz} \LocalSchur{\Phi}{\phi}_{zy}+\LocalSchur{\phi}{\phi}_{xy}, \\
    \label{eq: X_phi_xi_eft}
     \left(\tilde{\mathbf{X}}_{\phi\xi} ^\text{EFT}\right)_{xy} &= -\LocalSchur{\phi}{\Phi}_{xw} \LocalSchur{\Phi}{\Phi}^{-1}_{wz} \LocalSchur{\Phi}{\xi}_{zy}+\LocalSchur{\phi}{\xi}_{xy}, \\
     \label{eq: X_xi_phi_eft}
      \left(\tilde{\mathbf{X}}_{\xi\phi} ^\text{EFT}\right)_{xy} &= -\LocalSchur{\xi}{\Phi}_{xw} \LocalSchur{\Phi}{\Phi}^{-1}_{wz} \LocalSchur{\Phi}{\phi}_{zy}+\LocalSchur{\xi}{\phi}_{xy}.
\end{align}
\endgroup
This completely determines $Q_\text{EFT}$ in terms of the local operator expansions of the quantities appearing in $Q_\text{UV}$ as
\begin{align}
\fluct_\text{EFT} = \begin{pmatrix}
\mathbf{\Delta}_\phi ^\text{EFT} & -\tilde{\mathbf{X}}_{\phi\xi} ^\text{EFT} \\
\tilde{\mathbf{X}}_{\xi\phi} ^\text{EFT} & -\mathbf{\Delta}_\xi ^\text{EFT} 
\end{pmatrix},
\end{align}
where
\begin{align}
    \delta^2 \Lag{EFT} = -\frac{1}{2} \begin{pmatrix} \delta \phi^T & \delta \xi^T \end{pmatrix}\fluct_\text{EFT} \begin{pmatrix}
       \delta \phi \\
       \delta \xi
    \end{pmatrix}.
\end{align}
The contribution to the one-loop generating functional of 1PI correlation functions is given by\footnote{A discussion of Gaussian functional integrals expressed as superdeterminants can be found in Refs.\ \cite{Henning:2016lyp,Neufeld:1998js}. The sign difference introduced between $\fluct_\text{EFT}$ and $\tilde{\fluct}_\text{EFT}$ is due to the parameterisation chosen here. \label{footnote: sdet}}
\begin{align}
   \Gamma^{1\Loop}_\EFT &= S^{1\Loop}_\EFT +\frac{i}{2}\log \Sdet \, \tilde{\fluct}_\text{EFT},
\end{align}
where $\Sdet \, \tilde{\fluct}_\text{EFT}$ denotes the superdeterminant of the supermatrix
\begin{align}
    \tilde{\fluct}_\text{EFT} = \begin{pmatrix}
\mathbf{\Delta}_\phi ^\text{EFT} & \tilde{\mathbf{X}}_{\phi\xi} ^\text{EFT} \\
\tilde{\mathbf{X}}_{\xi\phi} ^\text{EFT} & \mathbf{\Delta}_\xi ^\text{EFT} 
\end{pmatrix}.
\end{align}
For some details relating to supermatrices see Appendix~\ref{App: berezin_algebra}. $\Gamma^{1\Loop}_\EFT$ has to be equal to the corresponding one-loop contribution to the generating functional of 1LPI correlation functions in the UV theory, which we reconsider now. Introducing the vector of all fluctuations
\begin{align}
    \delta \chi = \begin{pmatrix} \delta \Phi \\
    \delta \phi \\
    \delta \Xi \\
    \delta \xi
    \end{pmatrix},
\end{align}
the second variation of the UV Lagrangian can be written as
\begin{align}
    \delta^2 \Lag{UV} = -\frac{1}{2} \delta \chi^T \fluct_\text{UV} \delta \chi,
\end{align}
where $\fluct_\text{UV}$ is the complete fluctuation operator of the UV theory, which is a supermatrix and can be expressed in analogy to $\fluct_\text{EFT}$ by collecting the terms contributing to the second variation of the action into matrix form. The corresponding Gaussian path integral simply yields the superdeterminant of the supermatrix $\tilde{\fluct}_\text{UV}$
\begin{align}
    \int \measure \delta \chi \exp{\left[-\frac{i}{2} \delta \chi^T Q_\text{UV}\delta \chi\right]} = \Sdet \, \tilde{\fluct}_\text{UV},
\end{align}
which can be expressed in many equivalent ways, one of which was explicitly derived and subsequently used in the previous section. $\tilde{\fluct}_\text{UV}$ differs from $\fluct_\text{UV}$ by the appropriate signs to yield the correct superdeterminant as pointed out in footnote \ref{footnote: sdet} and in analogy to $\tilde{\fluct}_\text{EFT}$. We now derive an alternative form of the same quantity using steps analogous to those used in the previous section with the goal to separate fluctuations of heavy fields from fluctuations of light fields. Using the shift invariance of the integral measure we may write
\begin{align}
    \delta^2 \Lag{UV} \supset{}& \frac{1}{2} \left[\delta \Xi^T+\left(\delta \Phi^T\tilde{\mathbf{X}}_{\Phi \Xi}+\delta \phi^T\tilde{\mathbf{X}}_{\phi \Xi}+\delta \xi^T\tilde{\mathbf{X}}_{\xi \Xi}\right)\overleftarrow{\mathbf{\Delta}}_\Xi^{-1}\right]\mathbf{\Delta}_\Xi \nonumber \\
    &\times \left[\delta \Xi-\mathbf{\Delta}_\Xi ^{-1}\left(\tilde{\mathbf{X}}_{\Xi \Phi}\delta \Phi+\tilde{\mathbf{X}}_{\Xi \phi} \delta \phi-\tilde{\mathbf{X}}_{\Xi \xi}\delta \xi \right)\right] \nonumber \\
    &+\frac{1}{2} \left(\delta \Phi^T \tilde{\mathbf{X}}_{\Phi \Xi}+\delta \phi^T\tilde{\mathbf{X}}_{\phi \Xi}+\delta \xi^T\tilde{\mathbf{X}}_{\xi \Xi}\right)\mathbf{\Delta}_\Xi^{-1}\left(\tilde{\mathbf{X}}_{\Xi \Phi}\delta \Phi+\tilde{\mathbf{X}}_{\Xi \phi}\delta \phi-\tilde{\mathbf{X}}_{\Xi \xi} \delta \xi \right) \nonumber \\
    &-\frac{1}{2} \delta \Phi^T \mathbf{\Delta}_\Phi \delta \Phi -\frac{1}{2} \delta \phi^T \mathbf{\Delta}_\phi \delta \phi + \frac{1}{2} \delta \xi^T \mathbf{\Delta}_\xi \delta \xi  -\frac{1}{2} \delta \Phi^T \tilde{\mathbf{X}}_{\Phi \phi} \delta \phi -\frac{1}{2} \delta \phi^T \tilde{\mathbf{X}}_{\phi \Phi} \delta \Phi \nonumber \\
    &-\frac{1}{2} \delta \xi^T \tilde{\mathbf{X}}_{\xi \Phi} \delta \Phi -\frac{1}{2} \delta \xi^T \tilde{\mathbf{X}}_{\xi \phi} \delta \phi +\frac{1}{2} \delta \Phi^T \tilde{\mathbf{X}}_{\Phi \xi} \delta \xi +\frac{1}{2} \delta \phi^T \tilde{\mathbf{X}}_{\phi \xi} \delta \xi \nonumber \\
     \longrightarrow& -\frac{1}{2} \delta \Phi^T \Schur{\Phi}{\Phi}\delta \Phi - \frac{1}{2} \delta \phi^T \Schur{\Phi}{\Phi} \delta \phi + \frac{1}{2} \delta \xi^T \Schur{\xi}{\xi} \delta \xi \nonumber \\
    &- \frac{1}{2} \delta \Phi^T \Schur{\Phi}{\phi} \delta \phi - \frac{1}{2} \delta \phi^T \Schur{\phi}{\Phi} \delta \Phi -\frac{1}{2} \delta \xi^T \Schur{\xi}{\Phi} \delta \Phi \nonumber \\
    &- \frac{1}{2} \delta \xi^T \Schur{\xi}{\phi} \delta \phi + \frac{1}{2} \delta \Phi^T \Schur{\Phi}{\xi} \delta \xi + \frac{1}{2} \delta \phi^T \Schur{\phi}{\xi}\delta \xi \nonumber \\
    &+\frac{1}{2} \delta \Xi^T \mathbf{\Delta}_\Xi \delta \Xi,
\end{align}
where the arrow indicates that we performed a shift of $\delta \Xi$ which is permissible as this is performed under the path integral.
Then, using the shift invariance again to decouple $\delta \Phi$ from light fluctuations we obtain 
\begin{align}
    \delta^2 \Lag{UV} ={}& \frac{1}{2} \delta \Xi^T \mathbf{\Delta}_\Xi \delta \Xi - \frac{1}{2} \delta \Phi^T \Schur{\Phi}{\Phi} \delta \Phi \nonumber \\
    &- \frac{1}{2} \delta \phi^T \mathbf{\Delta}^\text{EFT,nl}_\phi \delta \phi + \frac{1}{2} \delta \xi^T \mathbf{\Delta}^\text{EFT,nl}_\xi \delta \xi -\frac{1}{2} \delta \xi^T \tilde{\mathbf{X}}^\text{EFT,nl}_{\xi \phi} \delta \phi + \frac{1}{2} \delta \phi^T \tilde{\mathbf{X}}^\text{EFT,nl}_{\phi \xi} \delta \xi,  
\end{align}
where $\mathbf{\Delta}^\text{EFT,nl}_\phi$, $\mathbf{\Delta}^\text{EFT,nl}_\xi$, $\tilde{\mathbf{X}}^\text{EFT,nl}_{\xi \phi}$ and $\tilde{\mathbf{X}}^\text{EFT,nl}_{\phi \xi}$ are the non-local counterparts of the corresponding quantities given in Eqs.\ \eqref{eq: delta_xi_eft}--\eqref{eq: X_xi_phi_eft}. The non-locality arises from the dependence on $\mathbf{\Delta}_\Xi^{-1}$ and $\Schur{\Phi}{\Phi}^{-1}$ as opposed to $\hat{\mathbf{\Delta}}_\Xi^{-1}$ and $\LocalSchur{\Phi}{\Phi}^{-1}$. This non-locality persists even once the classical background fields have been inserted due to the gauge-kinetic terms appearing in $\mathbf{\Delta}_\Xi$ and $\mathbf{\Delta}_\Phi$. This difference between the UV theory and the EFT simply accounts for the fact that in the EFT heavy propagators are not present anymore, but have been expanded in inverse powers of heavy masses.

At this point the heavy and light contributions are separated and we have that
\begin{align}
    \log \Sdet\, \tilde{\fluct}_\text{UV}={}&\log \Sdet \begin{pmatrix} \Schur{\Phi}{\Phi} & 0 \\ 0 & \mathbf{\Delta}_\Xi \end{pmatrix} + \log \Sdet \begin{pmatrix} \mathbf{\Delta}^\text{EFT,nl}_\phi & \tilde{\mathbf{X}}^\text{EFT,nl}_{\phi \xi} \\
    \tilde{\mathbf{X}}^\text{EFT,nl}_{\xi \phi} & \mathbf{\Delta}^\text{EFT,nl}_\xi \end{pmatrix} \nonumber \\ 
    ={}& \log \Sdet \begin{pmatrix} \Schur{\Phi}{\Phi} & 0 \\ 0 & \mathbf{\Delta}_\Xi \end{pmatrix}_\text{hard} + \log \Sdet \begin{pmatrix} \mathbf{\Delta}^\text{EFT,nl}_\phi & \tilde{\mathbf{X}}^\text{EFT,nl}_{\phi \xi} \\
    \tilde{\mathbf{X}}^\text{EFT,nl}_{\xi \phi} & \mathbf{\Delta}^\text{EFT,nl}_\xi \end{pmatrix}_\text{hard} \nonumber \\ & + \log \Sdet \begin{pmatrix} \mathbf{\Delta}^\text{EFT,nl}_\phi & \tilde{\mathbf{X}}^\text{EFT,nl}_{\phi \xi} \\
    \tilde{\mathbf{X}}^\text{EFT,nl}_{\xi \phi} & \mathbf{\Delta}^\text{EFT,nl}_\xi \end{pmatrix}_\text{soft}
    \label{eq: final_Q_UV},
\end{align}
where it was used that 
\begin{align}
    \log \Sdet \begin{pmatrix} \Schur{\Phi}{\Phi} & 0 \\ 0 && \mathbf{\Delta}_\Xi \end{pmatrix}_\text{soft} = 0,
\end{align}
since there are no light mass scales present in this contribution, i.e.\ all of the integrals arising from this contribution are scaleless. It is clear that the last line of Eq.\ \eqref{eq: final_Q_UV} corresponds precisely to the local operator expansion appearing in $Q_\text{EFT}$ since in the soft region all integrands will be expanded in inverse powers of heavy masses. Hence, using that
\begin{align}
   \Gamma^{1\Loop}_\text{L,UV} &= \frac{i}{2}\log \Sdet \, \tilde{\fluct}_\text{UV},
\end{align}
and imposing the matching condition $\hat{\Gamma}^{1\Loop}_\text{L,UV} = \Gamma^{1\Loop}_\EFT$ we find 
\begin{align}
    S^{1\Loop}_\EFT = \frac{i}{2}\log \Sdet \left. \tilde{\fluct}_\text{UV} \right|_\text{hard},
\end{align}
that is the superdeterminant is evaluated in the hard region only. This completes the proof.
\section{Discussion of the results}
In this section we discuss the results of the computation of the previous section and compare them to the results of Refs.\ \cite{Ellis:2020ivx,Angelescu:2020yzf}.  
\subsection{Computed operators and coefficients}
\label{sec: results_ops_coeffs}
In the following we describe the full set of BSUOLEA operators, which can be found in the file \texttt{BSUOLEA.m} at \cite{code}. The file contains the
following four lists:
\begin{itemize}
\item \texttt{noOpenNoP}: Operators without
  $P^\mu$, $Z^\mu$ and $Z^{\dagger \mu}$
\item \texttt{noOpenWithP}: Operators with
  $P^\mu$ and without $Z^\mu$ and $Z^{\dagger \mu}$
\item \texttt{withOpenNoP}: Operators without
  $P^\mu$ with $Z^\mu$ and $Z^{\dagger \mu}$.
\item \texttt{withOpenWithP}: Operators with
  $P^\mu$, $Z^\mu$ and $Z^{\dagger \mu}$.
\end{itemize}
The additional list \texttt{bsuolea}, which
is the union of the four lists defined above, is included for convenience. This list contains the full BSUOLEA, which is made up of a total of 58187 operators. In all of these lists the
BSUOLEA is given in the form $\{F^\alpha(M_i,M_j,\dots),\mathcal{O}^\alpha_{ij\cdots}\}$, where $F^\alpha(M_i,M_j,\dots)$
is the coefficient of the operator $\mathcal{O}^\alpha_{ij\cdots}$. This coefficient is
expressed through the integrals
$\ZI [q^{2n_c}]^{n_i n_j \dots n_L} _{i j \dots 0}$ defined in Appendix~\ref{app:loop_functions}, which are the same integrals used to express the universal coefficients in Refs.\ \cite{Ellis:2017jns,Zhang:2016pja}. The operators $\mathcal{O}^\alpha_{ij\cdots}$ are
expressed in terms of the symbols
$U[\text{A},\text{B}][i,j]$, $Z[\text{A},\text{B}][\text{lor}][i,j]$ and $ZT[\text{A},\text{B}][\text{lor}][i,j]$ with
$\text{A}, \text{B}\in \{\text{S},\text{s},\text{F},\text{f}\}$, which
correspond to the matrices defined in Section~\ref{sec: mixed_statistics_and_open_cds} as elaborated in Appendix~\ref{app: UOLEA_ops}.
The indices $i,j\in\mathbb{N}$ label a specific element of the
respective matrix and the index \emph{lor} indicates a Lorentz index. The full BSUOLEA is given by
\begin{align}
\Lag{\EFT}^{1\Loop} = \kappa\sum_\alpha \sum _{ij \cdots} F^\alpha(M_i,M_j,\dots) \mathcal{O}^\alpha_{ij\cdots},
\label{eq: final_UOLEA}
\end{align}
where $\kappa=1/(4\pi)^2$ and the sum over $\alpha$ runs over all operators and their
corresponding coefficients.
A few general comments regarding the use of the BSUOLEA are in order. First of all, no assumptions have
been made regarding the dependence of the second derivatives,
$X_{A B}$, on gamma matrices. The result is valid for any
spin-$1/2$ spinor structure appearing in these derivatives. This includes the presence of $\gamma_5$. Furthermore, no assumptions have been made about the particular $\gamma_5$-prescription to be used, that is  the use of different prescriptions is possible. Any given $\gamma_5$-prescription can be realised by using the trace relations appropriate to that prescription in the evaluation of spinor traces. 
Secondly, one has to carefully retain the poles in $\epsilon$ of the coefficients $F^\alpha(M_i,M_j,\dots)$ until after the spinor trace has been computed, since the gamma algebra has to be performed in $d = 4 - \epsilon$
dimensions. Hence, the trace may generate finite contributions when combined with
the poles of the coefficients.
Lastly, some of the coefficients diverge in the case of degenerate
masses if the degenerate limit is not taken carefully. Degenerate masses are most easily dealt with by first setting the masses equal, which modifies the integrals appearing in the
coefficients $F^\alpha(M_i,M_j,\dots)$, and then calculating these modified
integrals directly using Eqs.\ \eqref{eq: reduction_formula_1}--\eqref{eq: basis_integrals}. This is exemplified in Section~\ref{sec: top_quark_out}.
\subsection{Comparison to other results}
As mentioned in Chapter~\ref{chap: fcuntional matching}, since the publication of the computations in Ref.\ \cite{Kramer:2019fwz}, which were performed as a part of this thesis, Refs.\ \cite{Ellis:2020ivx,Angelescu:2020yzf} appeared, which also consider the existence of fermions in the UV theory and derive a UOLEA for this case. Both of these works differ from the computation presented here in two essential ways:
\begin{itemize}
    \item In Refs.\ \cite{Ellis:2020ivx,Angelescu:2020yzf} operators arising when integrating out heavy fermions without taking into account contributions from mixed statistics are considered, thus accounting for the same operators that arise from Eq.\ \eqref{eq: trlog_fermionic} when setting $\mathbf{X}_{\Xi \xi} = \mathbf{X}_{\xi \Xi} = 0$,
    \item in Refs.\ \cite{Ellis:2020ivx,Angelescu:2020yzf} $\mathbf{X}_{\Xi \Xi}$\footnote{The notation used in Refs.\ \cite{Ellis:2020ivx,Angelescu:2020yzf} differs from the notation used here and in Ref.\ \cite{Kramer:2019fwz}. We translate the notation for ease of comparison. It should also be noted that using the results of Refs.\ \cite{Ellis:2020ivx,Angelescu:2020yzf} terms with one Dirac fermion fluctuation and one Majorana fermion fluctuation cannot be treated as this requires diagonalisation of the fluctuation operator.} is parameterised by explicitly writing out the various possible spin-$1/2$ structures depending on $\gamma$-matrices, allowing for the a priori evaluation of traces in spinor space. 
\end{itemize}
The first point is a restriction on the generality of the result which can only be overcome by further computations. In fact, the result of Ref.\ \cite{Ellis:2020ivx} is more general than the result of Ref.\ \cite{Angelescu:2020yzf} since the latter does not allow couplings to vector bosons. The second point is a clear simplification with regards to the usage of the operators. As pointed out in Ref.\ \cite{Angelescu:2020yzf} the fact that spinor traces can be pre-computed makes it easier to give a physical interpretation to certain operators. This is especially true with regards to CP properties. Furthermore, if one is only interested in a limited set of operators the results of Refs.\ \cite{Ellis:2020ivx,Angelescu:2020yzf} make it easier to determine which UOLEA operators are relevant. Thus, for hand calculations the results presented in Refs.\ \cite{Ellis:2020ivx,Angelescu:2020yzf} seem to be more straightforward to use. However, for the automation of matching computations we deem the generality of the result the most important quality. Whether the a priori evaluation of spinor traces yields a speed-up for a full matching computation is not clear since one can compute the spinor traces independently from, and in parallel with, the remaining traces. Furthermore, not evaluating spinor traces explicitly allows for flexibility in the choice of the $\gamma_5$-prescription, which is fixed in Refs.\ \cite{Ellis:2020ivx, Angelescu:2020yzf}.

\section{Regularisation scheme translating operators}
\label{sec: DRED_to_DREG}
Up to this point we have assumed that both the UV theory and the EFT are regularised in \DREG as introduced in Ref.\ \cite{tHooft:1972tcz}. Although this regularisation is well suited for computations in non-supersymmetric models, its application to
supersymmetric ones is cumbersome, because it explicitly breaks
supersymmetry as shown in Ref.\ \cite{Delbourgo:1974az}.  To nevertheless perform loop calculations
in a SUSY model using \DREG one would have to restore
supersymmetry, for instance by introducing supersymmetry-restoring
counter terms, as discussed for example in Refs.\
\cite{Martin:1993yx,Mihaila:2009bn,Stockinger:2011gp}. As an alternative, regularisation by \DRED was introduced in Ref.\ \cite{Siegel:1979wq}. This regularisation is 
currently known to not break supersymmetry up to the 3-loop level, see Refs.\
\cite{Capper:1979ns,Stockinger:2005gx,Stockinger:2018oxe},
and is therefore widely adopted in SUSY loop calculations. Supplemented with modified minimal subtraction, DRED defines the \DRbar renormalisation scheme.
In order to apply the BSUOLEA to a scenario, where heavy particles of a
supersymmetric model, renormalised in the \DRbar scheme, are
integrated out at a high scale and a non-supersymmetric EFT,
renormalised in the \MSbar scheme results at low energies, the
change in regularisation from \DRED to \DREG must be
accounted for by shifting the running parameters by finite terms.
For general renormalisable softly broken supersymmetric gauge theories these
parameter shifts were calculated at the one-loop level in Ref.\ \cite{Martin:1993yx}. The corresponding generic UOLEA operators accounting for these shifts were computed as a part of this thesis in Ref.\ \cite{Summ:2018oko} on which this section is based. However, the parameterisation used in that reference differs from the one used in Section~\ref{sec: Mixed_statistics}. Based on the computations already presented in this thesis, it is straightforward to re-derive the result of Ref.\ \cite{Summ:2018oko} in the bispinor parameterisation. In this section we therefore present all one-loop effective operators that appear in the effective Lagrangian when changing the regularisation from
DRED to DREG, assuming that the (not necessarily supersymmetric) UV model
is renormalisable and compare the result to the one derived in Ref.\ \cite{Summ:2018oko}. We show that the results agree once the difference in parameterisation is accounted for.

To perform the calculation we employ the formalism of effective field theories by making use of the fact
that the difference between DRED and DREG can be expressed by the
presence/absence of so-called $\epsilon$-scalars as introduced in Ref.\ \cite{Capper:1979ns}.
The $\epsilon$-scalars are integrated out from the DRED-regularised UV model and the
resulting operators are formulated in the language of the
BSUOLEA. This is justified as the difference between \DRED and \DREG at one loop arises from $1/\epsilon$-poles in the loop integration, which are multiplied by $\epsilon$ to yield finite contributions. This means that the sought after terms arise from the region of the loop integration where the loop momentum becomes arbitrarily large. 

In Section~\ref{sec:epsilon} we briefly review the formalism of
$\epsilon$-scalars in DRED and give projection relations and Lagrangian terms
necessary for the calculation of the regularisation scheme translating
operators, which we derive in Section~\ref{sec:results}. It is unfortunately unavoidable to modify the notation of this section in comparison to Ref.\ \cite{Summ:2018oko}.

\subsection{Epsilon scalars in dimensional reduction}
\label{sec:epsilon}
In DRED an infinite dimensional space, $Q4S$, which shares some of the characteristics of a four-dimensional space, is introduced. This quasi-four-dimensional space is decomposed into two subspaces, $QdS$ and $Q\epsilon S$, which are formally $d$-dimensional and $\epsilon$-dimensional, respectively. Due to the decomposition $Q4S=QdS\oplus Q\epsilon S$ it holds that $\epsilon=4-d$. This construction was first presented in Ref.\ \cite{Stockinger:2005gx}. The metrics of the spaces $Q4S$, $QdS$ and
$Q\epsilon S$ are denoted by $g^\mu_\nu$, $\hat{g}^\mu_\nu$ and
$\check{g}^\mu_\nu$, respectively, and satisfy
\begin{align}
\label{contraction1}
g^\mu_\nu&=\hat{g}^\mu_\nu+\check{g}^\mu_\nu\text{,} & g^\mu_\mu=4\text{,} \\
\label{contraction2}
g^{\mu\nu}\check{g}^\rho_\nu&=\check{g}^{\mu\rho}\text{,} & \check{g}^\mu_\mu=\epsilon\text{,} \\
\label{contraction3}
g^{\mu\nu}\hat{g}^\rho_\nu&=\hat{g}^{\mu\rho}\text{,} & \hat{g}^\mu_\mu=d\text{,}\\
\hat{g}^{\mu\nu}\check{g}^\rho_\nu&=0\text{.}
\end{align}
The signature of the metric of $Q\epsilon S$ is $(-,-,\dots)$. In DRED
momenta are taken to be $d$-dimensional as is the case in DREG. However, in DRED, gauge fields and $\gamma$-matrices are taken to be four-dimensional. This construction avoids a mismatch between bosonic and fermionic degrees of freedom and hence the explicit breaking of supersymmetry. Due to the
decomposition of $Q4S$ it is convenient to split any gauge field
$A^a_\mu\in Q4S$ into two parts,
$A^a_\mu = \hat{A} ^a _\mu + \epsilon ^a _\mu$, with
$\hat{A}^a_\mu \in QdS$ and $\epsilon ^a_\mu \in Q\epsilon S$. The
$\epsilon$-dimensional field $\epsilon ^a_\mu$ is a scalar under
$d$-dimensional Lorentz transformations and, following Ref.\ \cite{Capper:1979ns}, is referred to as
$\epsilon$-scalar. This $\epsilon$-scalar transforms in the adjoint representation of the gauge group associated with $A^a_\mu$. After splitting the gauge field in
this way, the Lagrangian may contain the following terms
with $\epsilon$-scalars,
\begin{align}
  \LagNoT &= \LagNoT_\phi + \LagNoT_\psi + \LagNoT_\epsilon,
  \label{eq:Lag_general_epsion} \\
  \LagNoT_\phi &= \epsilon_\mu^a\epsilon^\mu_b F^a_b[\phi_1,\phi_2,\ldots \phi_n],
  \label{epsilonscalarcoupling} \\
  \LagNoT_\psi &= \epsilon_\mu^a \bar{\psi}_i \check{\gamma}^\mu \Gamma T^a_{ij}\psi_j,
  \label{epsilonfermioncoupling} \\
  \LagNoT_\epsilon &= -\frac{1}{2}(\hat{D}^\mu\epsilon_\nu)^a (\hat{D}_\mu \epsilon^\nu)_a+\frac{1}{2}m_\epsilon^2 \epsilon_\mu^a \epsilon^\mu_a-\frac{1}{4}g^2f^{abc}f^{ade}\epsilon^\mu_b \epsilon_\mu^d\epsilon^\nu_c\epsilon_\nu^e,
  \label{epsilonvectorcoupling}
\end{align}
where $\phi_i$ and $\psi_i$ denote scalars and fermions, respectively.
In Eq.\ \eqref{epsilonscalarcoupling} $F^a_b$ is a function of the
scalar fields and may contain linear and quadratic terms. The symbol
$\check{\gamma}^\mu$ denotes a $\gamma$-matrix projected onto
$Q\epsilon S$, $\check{\gamma}^\mu = \check{g}^{\mu}_\nu \gamma^\nu$,
and $\Gamma$ is some $4\times 4$ matrix that contains products of
$\{\mathbf{1}, \gamma^\mu, \gamma^5 \}$ and $\hat{D}_\mu = \partial_\mu -ig \hat{A}_\mu^a T^a$. In the following we denote
any projection of a Lorentz tensor $T^{\sigma \rho \cdots}$ onto
$Q\epsilon S$ by
$\check{T}^{\mu \nu \cdots}=\check{g}^\mu_\sigma
\check{g}^\nu_\rho\cdots T^{\sigma \rho \cdots}$.  Similarly, tensors
projected onto $QdS$ are denoted by $\hat{T}^{\mu\nu\dots}$.
The $m^2_\epsilon$-dependent term in Eq.\ \eqref{epsilonvectorcoupling}
can be removed by shifting the mass terms of the scalar fields
$\phi_i$ as described in Ref.~\cite{Jack:1994rk}, i.e.\ by changing
the renormalisation scheme from \DRbar to \DRbarPrime.
Nevertheless, due to the remaining extra $\epsilon_\mu^a$-dependent
terms in the Lagrangian defined in Eq.\
\eqref{eq:Lag_general_epsion}, the difference between DRED and DREG
manifests in the presence of extra Feynman diagrams with
$\epsilon$-scalars, which contribute additional finite terms to
divergent loop amplitudes due to the contraction relation of Eq.\ 
\eqref{contraction2}.

\subsection{Regularisation scheme translating operators in the BSUOLEA}
\label{sec:results}

To derive the operators that translate between DRED and DREG we
consider a general renormalisable gauge theory with the gauge
group $G$ and the Lagrangian $\LagNoT$, which may contain scalar fields, vector fields and spin-$1/2$ fermions in the form of Dirac and Majorana fermions.
We assume that the theory is regularised in DRED and split the gauge
field $A^a_\mu$ into a $d$- and an $\epsilon$-dimensional
component, as described in Section~\ref{sec:epsilon}. Since the contributions we are after come from the region of loop integrals where the loop momentum becomes arbitrarily large all of the physical fields can be safely assumed to be light in the language of Section~\ref{sec: Mixed_statistics}. We are thus in a scenario in which we have heavy bosons $\Phi$ given by the $\epsilon$-scalars and light bosons $\phi$ as well as light fermions $\xi$ given by the physical fields. By simply removing the heavy fermions from the computation of Section~\ref{sec: Mixed_statistics} and performing the replacement $\Phi \rightarrow \epsilon_\mu$ we can immediately write the second variation of the action as
\begingroup
\allowdisplaybreaks
\begin{align}
\delta^2\mathcal{L}={}& \frac{1}{2} \delta \xi^T \mathbf{\Delta}_\xi \delta \xi - \frac{1}{2} \delta \phi^T \mathbf{\Delta}_\phi \delta \phi - \frac{1}{2} \delta \xi^T \tilde{\mathbf{U}}_{\xi \phi} \delta \phi + \frac{1}{2} \delta \phi^T \tilde{\mathbf{U}}_{\phi \xi} \delta \xi \nonumber \\
&-\frac{1}{2} \delta \epsilon_\mu^T \check{\mathbf{\Delta}}_\epsilon ^ {\mu \nu} \delta \epsilon_\nu - \frac{1}{2} \delta \epsilon_\mu^T \check{\tilde{\mathbf{X}}}_{\epsilon \phi}^\mu \delta \phi - \frac{1}{2} \delta \phi^T \check{\tilde{\mathbf{X}}}_{\phi \epsilon}^\mu \delta \epsilon_\mu - \frac{1}{2} \delta \xi^T \check{\tilde{\mathbf{U}}}_{\xi \epsilon}^\mu \delta \epsilon_\mu +  \frac{1}{2} \delta \epsilon_\mu^T \check{\tilde{\mathbf{U}}}_{\epsilon \xi}^\mu \delta \xi, 
\label{eq: fullsecondvar}
\end{align}
\endgroup
where we have made manifest that open covariant derivatives can only occur in the bosonic sector. Furthermore we introduced the
abbreviation
\begin{align}
  \check{\mathbf{\Delta}}^{\mu \nu}_{\epsilon}&\equiv\check{g}^{\mu\nu} (\hat{P}^2-m_\epsilon^2)+\check{X}^{\mu\nu}_{\epsilon \epsilon},
  \label{eq:epsilon_prop}
\end{align}
and we write the Lorentz indices of $Q\epsilon S$ explicitly. Eq.~\eqref{eq: fullsecondvar} can be simplified further due to
the constraints on the possible couplings of $\epsilon$-scalars to
other fields as given in
Eqs.~\eqref{eq:Lag_general_epsion}--\eqref{epsilonvectorcoupling}. We
can solve the classical equations of motion in a perturbation
expansion in couplings to obtain the classical field $\classical{\epsilon^\mu}$. The leading term is proportional to an
operator of the form $\bar{\psi}\check{\gamma}^\mu\psi$, as this is the only coupling linear in $\epsilon$-scalars, and thus every
term in the series will either vanish or contain this operator. In the limit
$\epsilon \to 0$ this operator vanishes, which means that the
classical fields of the $\epsilon$-scalars can be set to zero from
the start. This is equivalent to the statement that $\epsilon$-scalars only appear in loops and can be used to simplify
Eq.~\eqref{eq: fullsecondvar}, because from
Eqs.~\eqref{epsilonscalarcoupling} and \eqref{epsilonvectorcoupling}
it follows that
$\check{\tilde{\mathbf{X}}}^\mu_{\phi \epsilon}=\check{\tilde{\mathbf{X}}}^\mu_{\epsilon \phi}=0$ for
vanishing $\classical{\epsilon^\mu}$. The computation now closely follows that of Section~\ref{sec: Mixed_statistics}. We first perform the appropriately modified shift corresponding to Eqs.\ \eqref{eq:xishift} and \eqref{eq:xishift_T} leading to the result
\begin{align}
\delta^2 \LagNoT ={}& -\frac{1}{2} \delta \epsilon_\mu^T \left(\check{M}_{\epsilon \phi}\right)^\mu_\xi \delta \phi -\frac{1}{2} \delta \phi^T \left(\check{M}_{\phi \epsilon}\right)^\mu_\xi \delta \epsilon_\mu -\frac{1}{2}\delta \epsilon_\mu ^T \left(\check{M}_{\epsilon \epsilon}\right)^{\mu \nu}_\xi \delta \epsilon_\nu- \frac{1}{2} \delta \phi ^T \left(M_{\phi \phi}\right)_\xi \delta \phi \nonumber \\ 
&+\frac{1}{2} \delta \xi'^{T} \mathbf{\Delta}_\xi \delta \xi',
\label{eq:eps_Lag_step_1}
\end{align}
with the matrices $M_{AB}$ defined as in Eq.\ \eqref{eq:M_def} and the indices in $Q\epsilon S$ again written out explicitly. The bosonic contribution to the second variation can then be written in terms of the fluctuation operator
\begin{align}
    \fluct_\epsilon = \begin{pmatrix}
    \left(\check{M}_{\epsilon \epsilon}\right)^{\mu \nu}_\xi & \left(\check{M}_{\epsilon \phi}\right)^\mu_\xi \\
    \left(\check{M}_{\phi \epsilon}\right)^\mu_\xi & \left(M_{\phi \phi}\right)_\xi
    \end{pmatrix},
\end{align}
as
\begin{align}
    \delta^2 \LagNoT_\epsilon &= -\frac{1}{2} \begin{pmatrix}
    \delta \epsilon^T_\mu & \delta \phi^T
    \end{pmatrix} \fluct_\epsilon 
    \begin{pmatrix}
    \delta \epsilon_\nu \\ \delta \phi
    \end{pmatrix},
\end{align}
which immediately allows us to write the corresponding contribution to the generating functional of 1LPI correlation functions as
\begin{align}
    \Gamma^{1\Loop}_\epsilon = \frac{i}{2} \log \det \left[\left(\check{M}_{\epsilon \epsilon}\right)^{\mu \nu}_\xi - \left(\check{M}_{\epsilon \phi}\right)^\mu_\xi \left(M_{\phi \phi}\right)_\xi^{-1}  \left(\check{M}_{\phi \epsilon}\right)^\nu_\xi \right] + \frac{i}{2} \log \det  \left(M_{\phi \phi}\right)_\xi.
    \label{eq: epsilon_Gamma}
\end{align}
Since we already proved the soft region cancellation for this scenario in Section~\ref{sec: soft_region_cancellation_proof} we may immediately evaluate the result in the hard region, which in this case is the region in which the loop momenta are much larger than all of the masses of physical fields. This means that we can ignore the second term in Eq.\ \eqref{eq: epsilon_Gamma} and that a potential contribution from the fermionic sector also cancels in the matching. Furthermore it is sufficient to only compute the divergent contributions to the functional trace, which can be identified by a power counting. To this end, we write
\begin{align}
\Gamma^{1\Loop}_\epsilon &\supset   \frac{i}{2} \Tr \log \left[\check{g}^{\mu \nu} \left(q^2-m^2_\epsilon+P^2-2q\cdot P\right)+\check{\mathbf{F}}^{\mu \nu}\right] \nonumber \\ 
 &= \frac{i}{2} \Tr \log  \left[\check{g}^{\mu \sigma}\left(q^2-m_\epsilon^2\right) \right]  +  \frac{i}{2} \Tr \log  \left[\check{\delta}_\sigma^\nu-\left(q^2-m_\epsilon^2\right)^{-1}\left(\check{\delta}_\sigma^\nu 2q\cdot P-P^2\check{\delta}_\sigma^\nu - \check{\mathbf{F}}_\sigma^{\;\nu}\right)\right] \nonumber \\
 &\supset-\frac{i}{2} \sum_{n=1}^\infty \frac{1}{n} \Tr  \left[\left(q^2-m_\epsilon^2\right)^{-1}\left(\check{\delta}_\sigma^\nu 2q\cdot P-P^2\check{\delta}_\sigma^\nu - \check{\mathbf{F}}_\sigma^{\;\nu}\right)\right]^n, 
 \label{eq:eps_trace}
\end{align}
where we omitted the infinite constant in the last line and where we defined
\begin{align}
    \check{\mathbf{F}}^{\mu \nu} = \check{\tilde{\mathbf{U}}}_{\epsilon \epsilon}^{\mu \nu} - \check{\tilde{\mathbf{U}}}_{\epsilon \xi}^\mu \mathbf{\Delta}_\xi^{-1} \check{\tilde{\mathbf{U}}}_{\xi \epsilon}^\nu - \left(\check{M}_{\epsilon \phi}\right)^\mu_\xi \left(M_{\phi \phi}\right)_\xi^{-1}  \left(\check{M}_{\phi \epsilon}\right)^\nu_\xi.
    \label{eq: eps_F}
\end{align}
Note that since $\check{\tilde{\mathbf{X}}}_{\epsilon \phi}^\mu=\check{\tilde{\mathbf{X}}}_{\phi \epsilon}^\mu=0$ both $\left(\check{M}_{\epsilon \phi}\right)^\mu_\xi$ and $\left(\check{M}_{\phi \epsilon}\right)^\nu_\xi$ are directly proportional to $\mathbf{\Delta}_\xi^{-1}$ and hence come with at least one inverse power of the loop momentum as can be seen from Eq.\ \eqref{eq:DeltaxiInv}. Furthermore, inspecting Eq.\ \eqref{eq: M_phi_phi_inv} we see that $\left(M_{\phi \phi}\right)_\xi^{-1}$ contains at least two inverse powers of the loop momentum, which means that together with the leading propagator $\left(q^2-m_\epsilon^2\right)^{-1}$ in Eq.\ \eqref{eq:eps_trace} any contribution to the functional trace containing the last term in Eq.\ \eqref{eq: eps_F} comes with at least six inverse powers of the loop momentum and is therefore finite. Thus this part of $\check{\mathbf{F}}^{\mu \nu}$ can be dropped and we only need to insert Eq.\ \eqref{eq:DeltaxiInv} into Eq.\ \eqref{eq: eps_F}. Performing a power counting it is clear that only the combinations $(n,m)=(1,0)$, $(n,m)=(1,1)$, $(n,m)=(2,0)$, $(n,m)=(3,0)$ and $(n,m)=(4,0)$, with $n$ being the summation index in Eq.\ \eqref{eq:eps_trace} and $m$ being the summation index in Eq.\ $\eqref{eq:DeltaxiInv}$, yield divergent integrals. The integrals needed for the evaluation can all be computed using the relations of Appendix~\ref{app:loop_functions} and hence the evaluation is straightforward. The final result reads
\begin{align}
  16 \pi^2 \epsilon \Lag{reg} ={}&-\sum _{i} (m^2_{\epsilon})_{i} (\check{\mathbf{U}}^\mu _{\epsilon \epsilon \mu})_{ii}
 + \frac{1}{2} \sum_{ij} (\check{\mathbf{U}}^{\mu}_{\epsilon \epsilon \nu})_{ij} (\check{\mathbf{U}}^{\nu}_{\epsilon \epsilon \mu})_{ji} \nonumber  \\ 
&+\frac{1}{2} \sum_{ij}  \left[2 M_{\xi j} (\check{\mathbf{U}}^\mu_{\epsilon \xi})_{ij} (\check{\mathbf{U}} _{\xi \epsilon \mu})_{ji} + (\check{\mathbf{U}}^\mu_{\epsilon \xi})_{ij} i \hat{D}_\nu \hat{\gamma}^\nu (\check{\mathbf{U}}_{\xi \epsilon \mu})_{ji}\right] \nonumber \\
&-\frac{1}{4}\sum_{i j k} (\check{\mathbf{U}}^\mu_{\epsilon \xi})_{ij} \hat{\gamma} ^\nu (\mathbf{U}_{\xi \xi})_{jk} \hat{\gamma}_{\nu} (\check{\mathbf{U}}_{\xi \epsilon \mu})_{ki} \nonumber \\ 
& + \frac{\epsilon}{12} \tr\left[ \hat{G}'_{\mu \nu} \hat{G}'^{\mu \nu} \right],
\label{eq: masterformula}
\end{align}
where $\hat{G}'_{\mu\nu}=-ig\hat{G}^a_{\mu\nu}T^a$,
$\hat{G}^a_{\mu\nu} = \hat{\partial}_\mu \hat{A}^a_\nu -
\hat{\partial}_\nu \hat{A}^a_\mu + g f^{abc} \hat{A}_\mu^b
\hat{A}_\nu^c$ and all quantities with Lorentz indices appearing in Eq.\
\eqref{eq: masterformula} are still projected onto either $QdS$ or
$Q\epsilon S$. After inserting the respective functional derivatives
into this equation and computing the sums each term on the r.h.s.\ will contain a
factor $\epsilon$. One can then divide the equation by $\epsilon$ and
take the limit $\epsilon \to 0$. After this limit has been taken there
is no difference between $d$-dimensional and four-dimensional
quantities anymore and the hats can be removed.

Eq.\ \eqref{eq: masterformula} should be compared to the result of Ref.\ \cite{Summ:2018oko}, which translated to the notation used here reads
\begin{align}
  16 \pi^2 \epsilon \Lag{reg} ={}&
-\sum _{i} (m^2_{\epsilon})_{i} (\check{\mathcal{U}}^\mu _{\epsilon \epsilon \mu})_{ii}
 + \frac{1}{2} \sum_{ij} (\check{\mathcal{U}}^{\mu}_{\epsilon \epsilon \nu})_{ij} (\check{\mathcal{U}}^{\nu}_{\epsilon \epsilon \mu})_{ji}  \nonumber \\ 
&+\sum_{ij} 2^{c_{F_j}} \left[2 M_{\psi j} (\check{\mathcal{U}}^\mu_{\epsilon \psi})_{ij} (\check{\mathcal{U}} _{\bar{\psi} \epsilon \mu})_{ji} + (\check{\mathcal{U}}^\mu_{\epsilon \psi})_{ij} i \hat{D}_\nu \hat{\gamma}^\nu (\check{\mathcal{U}}_{\bar{\psi} \epsilon \mu})_{ji}\right] \nonumber \\
&-\sum_{i j k} 2^{c_{F_j}+c_{F_k}-1} (\check{\mathcal{U}}^\mu_{\epsilon \psi})_{ij} \hat{\gamma} ^\nu (\mathcal{U}_{\bar{\psi} \psi})_{jk} \hat{\gamma}_{\nu} (\check{\mathcal{U}}_{\bar{\psi} \epsilon \mu})_{ki} \nonumber \\ 
& + \frac{\epsilon}{12} \tr\left[ \hat{G}'_{\mu \nu} \hat{G}'^{\mu \nu} \right],
\label{eq: masterformula_old}
\end{align}
where $\psi$ runs over both Dirac and Majorana fermions and $c_F=0$ for Dirac fermions and $c_F=1$ for Majorana
fermions. We use $\mathcal{U}$ instead of $\mathbf{U}$ to distinguish the matrices in the two parameterisations. It is expected that the two results agree when setting $c_F=1$ everywhere since in the computation presented here all fermions were treated like Majorana fermions. Superficially this is not the case, which depends on the fact that in Ref.\ \cite{Summ:2018oko} $\bar{\lambda}$ and $\lambda$ were treated as independent degrees of freedom without using the Majorana condition $\ccfield{\lambda}=\lambda$ when computing the coupling matrices $\mathcal{U}$. This means specifically that $\mathcal{U}_{\bar{\lambda}\lambda}=\frac{1}{2}\mathbf{U}_{\lambda \lambda}$, $\check{\mathcal{U}}_{\bar{\lambda}\epsilon}^\mu=\frac{1}{2}\check{\mathbf{U}}_{\lambda \epsilon}^\mu$, $\check{\mathcal{U}}_{\epsilon \lambda}^\mu=\frac{1}{2}\check{\mathbf{U}}_{\epsilon \lambda}^\mu$. Using these relations together with $\check{\mathcal{U}}_{\epsilon \epsilon}^{\mu \nu} = \check{\mathbf{U}}_{\epsilon \epsilon}^{\mu \nu}$ we find full agreement between Eqs.\ \eqref{eq: masterformula} and \eqref{eq: masterformula_old}. 

\section{Automation of one-loop matching}
\label{sec: tofu}
As the number of BSUOLEA operators contributing to applications relevant to particle phenomenology tends to be rather large, rendering hand calculation infeasible, it is desirable to have an implementation of the full BSUOLEA. In this section we therefore briefly describe a currently private Mathematica code called \textbf{To}ol \textbf{f}or \textbf{U}niversal Matching at \textbf{On}e-\textbf{Loop}, or \tofu for short, which automates matching computations through the BSUOLEA. This code has been developed as a part of this thesis and we intend to make it publicly available in the near future.

To use the code, the user has to define the full UV Lagrangian together with the particle content and a specification of which fields need to be integrated out. The specification of the particle content includes the definition of the different types of indices, including gauge indices and flavour indices. From this information \tofu automatically computes the relevant matrices for the matching as given in Appendix~\ref{app: UOLEA_ops} and inserts these into the pre-computed operator structures at the same time performing all index contractions. The user may then define custom replacement rules, which are used to simplify the resulting expressions. These replacement rules may include, but are not limited to, identities among gamma matrices (including a prescription for the handling of $\gamma_5$), traces of gauge group generators and unitarity relations for mixing matrices. Since the spinor algebra can be performed independently, all spinor structures are extracted at the beginning and computed in parallel with the remaining traces.

As there is no upper limit on the number of replacement rules the user can implement one might be concerned about the time consumption of this step. In order to avoid testing replacement rules on expressions that do not contain all of the factors in a particular replacement rule, the program internally assigns a prime number to every tensor and every field appearing in the UV Lagrangian. Since a given term in the effective Lagrangian is simply a product of such tensors and fields it can be assigned a unique number given by the product of the prime numbers associated with the tensors and fields constituting the term. In the same way, a number can be assigned to any replacement rule. Due to the uniqueness of the prime factorisation only replacement rules, which evaluate to divisors of the number assigned to a given term in the effective Lagrangian have to be considered, thus reducing the number of replacement rules being applied to any particular term.

Once the simplifications of the previous step have been performed the user is left with an effective action that usually contains redundant higher dimensional operators, i.e.\ operators that can be eliminated through the use of equations of motion, integration by parts and Fierz identities. To reduce the set of operators to a basis, the user can assign symbols to custom combinations of fields and tensors and then provide \tofu with a set of replacement rules, which removes symbols associated with redundant operators. Since the problem of finding a basis and a general algorithm to turn a set of redundant operators into a basis has not been solved in general\footnote{For some of the progress on this topic see Refs.\ \cite{Henning:2015alf,Henning:2017fpj,Henning:2019enq,Trautner:2020qqo,Henning:2015daa,Criado:2019ugp,Gripaios:2018zrz}.}, this last step usually requires some work by the user. However, for the case of the EFT being the SMEFT there is a supplementary Mathematica code called \route, also developed as a part of this thesis, which allows for the reduction of any set of redundant dimension six operators in the SMEFT to the Warsaw basis constructed in Refs.\ \cite{Buchmuller:1985jz,Grzadkowski:2010es}. The algorithm was extracted from the proof of the non-redundancy of the Warsaw basis presented in Ref.\ \cite{Grzadkowski:2010es}.\footnote{We thank Ilaria Brivio for pointing out that such an algorithm can be extracted from this proof.}

The result of Section~\ref{sec: DRED_to_DREG} has also been implemented into \tofu, where the user has the option to specify a Lagrangian for the $\epsilon$-scalars. If such a Lagrangian is specified the regularisation scheme translating contributions to the matching are computed in a fully automated way keeping track of contractions in $\epsilon$ and $d$ dimensions and taking the limit $\epsilon \to 0$ in the end.

In the matching calculation presented in Section~\ref{sec: SSM_to_SM} \tofu was used to perform the matching and custom replacement rules were used to translate the result to the Warsaw basis. The translation to the Warsaw basis was repeated with \route and agreement was found. The matching computation of Section~\ref{sec: resonance_matching_res} was performed with \tofu and the translation to the Warsaw basis was carried out using \route.


\clearpage
\lhead{\emph{Applications of the BSUOLEA}}
\chapter{Applications of the BSUOLEA}
\label{ch: applications}
In this chapter we present several applications of the BSUOLEA. In Section~\ref{sec: top_quark_out}, we start with a very simple example in which the top quark is integrated out from the SM in order to illustrate how the BSUOLEA can be applied. This application also serves as a check for the treatment of Dirac fermions in the BSUOLEA. Next, in Section~\ref{sec:lambdacalc}, we reproduce the threshold correction to the quartic Higgs coupling when matching the Minimal Supersymmetric Standard Model\footnote{Familiarity with the MSSM is not required to follow the applications presented in this chapter. For the reader interested in learning about supersymmetry we recommend Ref.\ \cite{Drees:2004jm}.} (MSSM) to the SM, which serves as a check for purely fermionic operators, including those with Majorana fermions, as well as operators that translate between DRED and DREG. Afterwards, in Section~\ref{sec:matching_MSSM_to_SMEFT}, we consider an application in which the gluino and the stops are integrated out from the MSSM. In this matching, contributions from operators with mixed statistics arise. Since we again reproduce known results this serves as a check of these operators. We then modify the scenario slightly, in Section~\ref{sec: gluinoOut}, keeping the stops and integrating out only the gluino. In this application operators with one Majorana fermion, namely the gluino, and one Dirac fermion, the top quark, contribute to the known mass corrections of the stops. As these corrections are reproduced correctly this indicates that the treatment of such operators is also correct. In the process previously unknown Wilson coefficients are computed. These four applications were already presented in Ref.\ \cite{Kramer:2019fwz} as a part of this thesis and we closely follow that reference here. The scalar contributions to the threshold correction of the quartic Higgs coupling were computed by Dr.\ Alexander Voigt. For all of these applications the effective Lagrangian was computed by hand. For our final application, in Section~\ref{sec: SSM_to_SM}, the matching of the SSM to the SMEFT, this is not possible anymore since 500 BSUOLEA operators contribute. Therefore the code \tofu was used to perform this matching. Given that the result is also known from previous computations, this serves as a check of the code itself.

As it is unavoidable to handle many different kinds of indices when applying the BSUOLEA we introduce the index conventions used throughout this chapter in Table~\ref{table: index_conventions}. Note that all spinor indices carry dots to distinguish these from Lorentz indices. We do not use dotted and undotted spinor indices to distinguish left- and right-handed spinors. We also suppress indices whenever they do not add to the clarity of presentation. Finally, lower case letters from the middle of the alphabet, $i,\, j, \ldots$ are used as both flavour indices and multi-indices. The latter only appear in expressions that show the BSUOLEA operators which contribute to a certain computation. Therefore, this should not cause any confusion. Repeated indices are always summed over.

\renewcommand{\arraystretch}{1.2}
\begin{table}[]
\centering
\begin{tabular}{|l|l|lll}
\cline{1-2}
Index type & Example indices  \\ \cline{1-2}
Flavour & $i$, $j$, $k$, $l$   \\ \cline{1-2}
$SU(2)_L$ fundamental & $a$, $b$, $c$, $d$   \\ \cline{1-2}
$SU(2)_L$ adjoint & $A$, $B$, $C$, $D$   \\ \cline{1-2}
$SU(3)_c$ fundamental & $I$, $J$, $K$, $L$ \\ \cline{1-2}
$SU(3)_c$ adjoint & $\dot{A}$, $\dot{B}$, $\dot{C}$, $\dot{D}$ \\ \cline{1-2}
Lorentz & $\mu$, $\nu$, $\rho$, $\sigma$ \\ \cline{1-2}
Spinor & $\dot{\alpha}$, $\dot{\beta}$, $\dot{\rho}$, $\dot{\sigma}$ \\ \cline{1-2}
\end{tabular}
\caption{Index naming conventions used throughout this chapter.}
\label{table: index_conventions}
\end{table}
\renewcommand{\arraystretch}{1.2}
\section{Integrating out the top quark from the SM}
\label{sec: top_quark_out}
We consider the corrections to the Higgs
tadpole and mass parameter that arise when integrating out the top
quark from the SM. This example is purposely chosen to be very simple in order to illustrate the use of the BSUOLEA. We restrict the interaction Lagrangian
to only one coupling
\begin{align}
  \Lag{SM} \supset -\frac{g_t}{\sqrt{2}}h \bar{t}t,
\end{align}
where $h$ denotes the physical Higgs field, $t$ is the top quark and
$g_t$ is the top Yukawa coupling.  The relevant operators of the BSUOLEA of Eq.\
\eqref{eq: final_UOLEA} are given by
\begingroup
\allowdisplaybreaks
\begin{align}
\frac{1}{\kappa} \Lag{EFT}^{1\Loop} = \tr \Bigg\lbrace & \frac{1}{4} M_{\Xi i} M_{\Xi j}^3 \ZI ^{13} _{ij} [P_\mu,(\mathbf{U}_{\Xi \Xi})_{ij}][P^\mu,(\mathbf{U}_{\Xi \Xi})_{ji}]
\nonumber \\ &  -\frac{1}{2} \ZI[q^4] ^{22} _{ij} \gamma^\nu [P_\mu,(\mathbf{U}_{\Xi \Xi})_{ij}]\gamma_\nu[P^\mu,(\mathbf{U}_{\Xi \Xi})_{ji}]
\nonumber \\ &  - \ZI[q^4] ^{22} _{ij} \gamma^\nu [P_\nu,(\mathbf{U}_{\Xi \Xi})_{ij}]\gamma_\mu[P^\mu,(\mathbf{U}_{\Xi \Xi})_{ji}]
\nonumber \\ & +\frac{1}{2} M_{\Xi i} \ZI ^1 _i (\mathbf{U}_{\Xi \Xi})_{ii}
\nonumber \\ & -\frac{1}{4} M_{\Xi i} M_{\Xi j} \ZI ^{11} _{ij} (\mathbf{U}_{\Xi \Xi})_{ij} (\mathbf{U}_{\Xi \Xi})_{ji}
\nonumber \\ & -\frac{1}{4} \ZI[q^2] ^{11} _{ij} \gamma ^\mu (\mathbf{U}_{\Xi \Xi})_{ij} \gamma_\mu (\mathbf{U}_{\Xi \Xi})_{ji}
\Bigg\rbrace
\label{eq:UOLEALAG-topout},
\end{align}
\endgroup
where $M_{\Xi i}$ denotes the $i$th component of $M_\Xi = \begin{pmatrix}
m_t & m_t
\end{pmatrix}^T$ with $m_t$ being the mass of the top quark and $\Xi = \begin{pmatrix}
t_{\dot{\alpha} I} & \ccfield{t}_{\dot{\alpha} I}
\end{pmatrix}^T$. Note that $M_{\Xi1} = M_{\Xi2}$ and thus we are in the case of a degenerate mass as mentioned in Section~\ref{sec: results_ops_coeffs}. The matrix $\mathbf{U}_{\Xi \Xi}$ is given by
\begin{align}
(\mathbf{U}_{\Xi \Xi})_{\dot{\alpha} \dot{\beta} IJ} = \begin{pmatrix}
(U_{\bar{t}t})_{\dot{\alpha} \dot{\beta} IJ} & 0 \\
0 & \cc^{-1} _{\dot{\alpha} \dot{\rho}} (U_{t\bar{t}})_{\dot{\rho} \dot{\sigma} IJ} \cc^{-1} _{\dot{\sigma} \dot{\beta}} 
\end{pmatrix}
= 
-\frac{g_t}{\sqrt{2}}h \delta_{\dot{\alpha} \dot{\beta}} \delta_{IJ} \mathbf{1}_{2},
\label{eq:top-derivative}
\end{align}
where $\mathbf{1}_{2}$ is the $2\times 2$ identity matrix. In Eq.\ \eqref{eq:UOLEALAG-topout} we
included terms with two covariant derivatives in order to obtain the
field-redefinition of the Higgs field that is necessary to canonically
normalise the corresponding Higgs field, $\hat{h}$, in the effective
theory.  Since this redefinition arises from the correction to the
kinetic term only, we can set $P^\mu = i\partial ^\mu$. In order to illustrate the handling of spinor indices, the appearance of derivatives and mass-degenerate coefficients we compute the contribution of the operator in the third line of Eq.\ \eqref{eq:UOLEALAG-topout} in detail. Since the masses are degenerate we have that
\begin{align}
    \ZI[q^4] ^{22} _{11} =  \ZI[q^4] ^{22} _{12} =  \ZI[q^4] ^{22} _{21} = \ZI[q^4] ^{22} _{22} =  \ZI[q^4] ^{4} _{t},
    \label{eq: baby_example_coeff}
\end{align}
meaning that the integral coefficient is independent of the indices. For the operator structure we find
\begin{align}
-\tr \lbrace \gamma^\nu [P_\nu,(\mathbf{U}_{\Xi \Xi})_{ij}]\gamma_\mu[P^\mu,(\mathbf{U}_{\Xi \Xi})_{ji}]\rbrace &= \frac{g_t^2}{2}\tr \lbrace \gamma^\nu_{\dot{\alpha} \dot{\beta}} \partial_\nu \left(h \delta_{\dot{\beta} \dot{\rho}} \delta_{IJ} \mathbf{1}_{2} \right) \gamma^\mu_{\dot{\rho} \dot{\sigma}} \partial_\mu \left(h \delta_{\dot{\sigma} \dot{\alpha}} \delta_{JI} \mathbf{1}_{2} \right) \rbrace \nonumber \\
&= 12g_t^2 \partial_\mu h \partial^\mu h,
\end{align}
which together with Eq.\ \eqref{eq: baby_example_coeff} yields the contribution
\begin{align}
    12 g_t^2 \ZI[q^4] ^{4} _{t} (\partial_\mu h) (\partial^\mu h) = -\frac{1}{2}g_t^2 \log\frac{m_t^2}{\mu^2} \partial_\mu h \partial^\mu h,
\end{align}
where $\mu$ is the matching scale and we used
\begin{align}
    \ZI[q^4]^4_t = \frac{1}{12\eps}-\frac{1}{24}\log\frac{m_t^2}{\mu^2}.
\end{align}
We are here allowed to drop the $\eps$-dependence of the integral since no $\eps$-dependence arises from the operator structure. This is not the case for the second line of Eq.\ \eqref{eq:UOLEALAG-topout}, where the contraction of the gamma matrices produces a $d=4-\eps$. Proceeding in a similar fashion for the remaining contributions to Eq.\ \eqref{eq:UOLEALAG-topout} we find
\begin{align}
  \frac{1}{\kappa}\Lag{EFT} ^{1\Loop}  ={}& -3g_t^2 \left(m_t^4 \ZI ^4 _t-2d\ZI [q^4]^4 _{t}-4 \ZI [q^4] ^4 _{t} \right) (\partial_\mu h) (\partial^\mu h) \nonumber \\
  & -3g_t^2 \left(\ZI ^2_t m_t^2+d \ZI [q^2]^2_t \right)h^2-\frac{12}{\sqrt{2}} g_t m_t \ZI^1_t h.
\label{eq: HiggsEFT}
\end{align}
Introducing the canonically normalised field $\hat{h}$ which is related to $h$
through
\begin{align}
  \hat{h}=\left(1+\frac{1}{2}\delta Z_h \right) h,
\end{align}
one can read off $\delta Z_h$ to be
\begin{align}
  \delta Z_h = -6g_t^2\left(m_t^4 \ZI ^4 _t-2d\ZI [q^4]^4 _{t}-4 \ZI [q^4] ^4 _{t}\right)
  =  -6g_t^2\left(m_t^4 \ZI ^4_t - 12 \ZI [q^4] ^4_{t} + \frac{1}{6}\right).
  \label{eq:delta_Zh_top}
\end{align}
The loop functions that appear in Eqs.\ \eqref{eq: HiggsEFT} and
\eqref{eq:delta_Zh_top} can be calculated using Eqs.\ \eqref{eq: reduction_formula_1}-\eqref{eq: basis_integrals} and are given by
\begin{align}
  \ZI^1_t &= 2 \ZI[q^2]^2_t = m_t^2 \left(\frac{2}{\eps} + 1 - \log\frac{m_t^2}{\mu^2}\right), \\
  \ZI^2_t &= 24 \ZI[q^4]^4_t = \frac{2}{\eps}-\log\frac{m_t^2}{\mu^2}, \\
  \ZI^4_t &= \frac{1}{6 m_t^4}.
\end{align}

\section{MSSM threshold correction to the quartic Higgs coupling}
\label{sec:lambdacalc}

In this section we reproduce the one-loop threshold
correction of the quartic Higgs coupling $\lambda$ when matching the MSSM to the
SM at the one-loop level in the unbroken phase. This result was first reported in Ref.\ \cite{Bagnaschi:2014rsa} and as
discussed in that reference there are several contributions
of distinct origins. The scalar contribution
$\Delta \lambda^{1\Loop,\phi}$ arises from interactions of the SM-like
Higgs with heavy Higgs bosons, squarks and sleptons,  and the relevant
interaction Lagrangian is given by
\begingroup
\allowdisplaybreaks
\begin{align}
  \LagNoT_{\phi} ={}&  - \frac{g_t^2}{2} h^2 (\st{L}^* \st{L} + \st{R}^*\st{R})-\frac{1}{\sqrt{2}} g_t X_t h (\st{L}^* \st{R} + \st{L}\st{R}^*) \nonumber \\ & -\frac{1}{8} c_{2\beta} h^2 \left[\left(g_2^2 - \frac{g_1^2}{5}\right)  \su^* _{Li} \su _{Li}+ \frac{4}{5} g_1^2 \su_{Ri}^* \su_{Ri}- \left(g_2^2 + \frac{g_1^2}{5}\right) \sd_{Li}^* \sd_{Li}- \frac{2}{5} g_1^2 \sd_{Ri}^* \sd_{Ri}\right]
\nonumber \\ &-\frac{1}{8} c_{2\beta} h^2 \left[\left(g_2^2 + \frac{3}{5}g_1^2\right)  \sneu^* _{Li} \sneu _{Li}- \left(g_2^2 - \frac{3}{5}g_1^2\right) \sel_{Li}^* \sel_{Li}- \frac{6}{5} g_1^2 \sel_{Ri}^* \sel_{Ri}\right]
\nonumber \\ & +\frac{1}{16} c_{2\beta}^2 \left(\frac{3}{5} g_1^2 + g_2^2\right) h^2 A^2- \frac{1}{8} \left((1 + s_{2\beta}^2) g_2^2 - \frac{3}{5} g_1^2 c_{2\beta}^2\right) h^2 H^{-} H^{+} \nonumber \\ & - \frac{1}{16} \left(\frac{3}{5} g_1^2 + g_2^2\right) (3 s_{2\beta}^2 - 1) h^2 H^2- \frac{1}{8} \left(\frac{3}{5} g_1^2 + g_2^2\right) s_{2\beta} c_{2\beta} h^3 H \nonumber \\ & + \frac{1}{8} \left(\frac{3}{5} g_1^2 + g_2^2\right) s_{2\beta} c_{2\beta} h^2 (G^{-} H^{+} + H^{-} G^{+})+ \frac{1}{8} \left(\frac{3}{5} g_1^2 + g_2^2\right) s_{2\beta} c_{2\beta} h^2 G^0 A.
\end{align}
\endgroup
Here the three generations
of left- and right-handed squarks are denoted as
$\su_{Li}$, $\su_{Ri}$, $\sd_{Li}$, $\sd_{Ri}$, where
$\st{L} \equiv \su_{L3}$ and $\st{R} \equiv \su_{R3}$ are the left-
and right-handed stops. Correspondingly, the three generations of left- and right-handed sleptons are denoted as $\sel_{Li}$,
$\sel_{Ri}$, $\sneu_{Li}$. The couplings $g_1$ and $g_2$ are the GUT-normalised electroweak gauge
couplings, $X_t$ is the stop mixing parameter, and $g_t = y_t s_\beta$
with $y_t$ being the MSSM top Yukawa coupling and $s_\beta=\sin (\beta)$. The angle $\beta$ should not be
regarded as a ratio of vacuum expectation values, but as the
fine-tuned mixing angle which rotates the two MSSM Higgs doublets,
$H_u$ and $H_d$, into two new doublets, $\mathcal{H}$ and $\mathcal{A}$, with
\begin{align}
\label{eq:rot_H}
    \mathcal{H} &= - c_\beta \epstensor H^{*}_d + s_\beta H_u,\\
    \label{eq:rot_A}
  \mathcal{A} &= s_\beta \epstensor H^{*}_d + c_\beta H_u,
\end{align}
since we are working in the unbroken phase. See Ref.\ \cite{Bagnaschi:2014rsa} for details. In Eqs.\ \eqref{eq:rot_H} and \eqref{eq:rot_A} $\epstensor$ is the antisymmetric tensor with $\epstensor_{12}=1$  and $c_\beta = \cos(\beta)$, $s_{2\beta} = \sin(2\beta)$ and $c_{2\beta} = \cos(2\beta)$. The SM-like Higgs doublet $\mathcal{H}$ gives rise to the SM-like physical Higgs field $h=\sqrt{2}\, \Re (\mathcal{H}^0)$, where $\mathcal{H}^0$ is the neutral component of $\mathcal{H}$ and $\Re (\mathcal{H}^0)$ its real part. The fields $G^0$ and $G^{\pm}$ are Goldstone bosons arising from the same Higgs doublet. The heavy Higgs
bosons of the MSSM, $H$, $A$ and $H^\pm$, originate from the heavy doublet
$\mathcal{A}$. 

The fermionic contribution $\Delta \lambda ^{1\Loop,\chi}$ to the
threshold correction of $\lambda$ originates from interactions of the
Higgs boson with charginos $\tilde{\chi}^{+}_i$ ($i=1,2$) and
neutralinos $\tilde{\chi}^0_i$ ($i=1,\ldots,4$) described by the
interaction Lagrangian
\begin{align}
\LagNoT_\chi ={}& - \frac{g_2}{\sqrt{2}} h c_\beta (\overline{\tilde{\chi}^{+}_1} P_R \tilde{\chi}^{+} _2  + \overline{\tilde{\chi}^{+}_2}  P_L \tilde{\chi}^{+} _1)- \frac{g_2}{\sqrt{2}} h s_\beta (\overline{\tilde{\chi}^{+}_2} P_R \tilde{\chi}^{+} _1 + \overline{\tilde{\chi}^{+}_1} P_L \tilde{\chi}^{+}_2   )\nonumber 
 \\ &  +i  \frac{g_Y}{2\sqrt{2}} (c_\beta - s_\beta) h \overline{\tilde{\chi}^0_1} \gamma^5  \tilde{\chi}^0_3-\frac{g_Y}{2\sqrt{2}} (c_\beta + s_\beta)h \overline{\tilde{\chi}^0_1} \tilde{\chi}^0_4 \nonumber
 \\ & -i \frac{g_2}{2\sqrt{2}} (c_\beta - s_\beta) h \overline{\tilde{\chi}^0_2} \gamma^5  \tilde{\chi}^0_3+  \frac{g_2}{2\sqrt{2}}  (c_\beta + s_\beta) h \overline{\tilde{\chi}^0_2} \tilde{\chi}^0_4 ,
\end{align}
where $\overline{\tilde{\chi}^0_i} = (\tilde{\chi}^0_i)^T \cc$ and
$g_Y = \sqrt{3/5}\, g_1$.

To calculate the one-loop threshold correction for $\lambda$, the
following contributions with purely scalar and purely fermionic
operators of the BSUOLEA are relevant,
\begin{align}
\frac{1}{\kappa} \Lag{EFT}^{1\Loop} = \tr \Bigg\lbrace & \frac{1}{2} \ZI ^{1} _{i} (\mathbf{U}_{\Phi \Phi})_{ii}+\frac{1}{2} \ZI [q^2]^{22} _{ij} [P_\mu, (\mathbf{U}_{\Phi \Phi})_{ij}] [P^\mu, (\mathbf{U}_{\Phi \Phi})_{ji}]\nonumber \\ & +\frac{1}{4} \ZI ^{11} _{ij} (\mathbf{U}_{\Phi \Phi})_{ij}(\mathbf{U}_{\Phi \Phi})_{ji}+\frac{1}{6} \ZI^{111} _{ijk}(\mathbf{U}_{\Phi \Phi})_{ij}(\mathbf{U}_{\Phi \Phi})_{jk}(\mathbf{U}_{\Phi \Phi})_{ki} 
\nonumber \\ &+\frac{1}{8} \ZI ^{1111} _{ijkl} (\mathbf{U}_{\Phi \Phi})_{ij}(\mathbf{U}_{\Phi \Phi})_{jk} (\mathbf{U}_{\Phi \Phi})_{kl} (\mathbf{U}_{\Phi \Phi})_{li} + \frac{1}{2} \ZI ^{1} _{i} (\mathbf{U}_{\Phi \phi})_{ij}(\mathbf{U}_{\phi \Phi})_{ji} \nonumber \\ &  -\frac{1}{8} M_{\Xi i}M_{\Xi j} M_{\Xi k} M_{\Xi l} \ZI ^{1111} _{ijkl}(\mathbf{U}_{\Xi \Xi})_{ij}(\mathbf{U}_{\Xi \Xi})_{jk} (\mathbf{U}_{\Xi \Xi})_{kl} (\mathbf{U}_{\Xi \Xi})_{li}
\nonumber \\ &  -\frac{1}{2} M_{\Xi i}M_{\Xi j} \ZI [q^2] ^{1111} _{ijkl}(\mathbf{U}_{\Xi \Xi})_{ij}(\mathbf{U}_{\Xi \Xi})_{jk}\gamma^\mu (\mathbf{U}_{\Xi \Xi})_{kl} \gamma_\mu (\mathbf{U}_{\Xi \Xi})_{li}
\nonumber \\ &  -\frac{1}{4} M_{\Xi i}M_{\Xi k} \ZI [q^2] ^{1111} _{ijkl}(\mathbf{U}_{\Xi \Xi})_{ij}\gamma^\mu(\mathbf{U}_{\Xi \Xi})_{jk} (\mathbf{U}_{\Xi \Xi})_{kl} \gamma_\mu (\mathbf{U}_{\Xi \Xi})_{li}
\nonumber \\ &  -\frac{1}{8} g_{\mu \nu \rho \sigma} \ZI [q^4] ^{1111} _{ijkl}\gamma^\mu (\mathbf{U}_{\Xi \Xi})_{ij}\gamma^\nu(\mathbf{U}_{\Xi \Xi})_{jk} \gamma^\rho (\mathbf{U}_{\Xi \Xi})_{kl} \gamma^\sigma (\mathbf{U}_{\Xi \Xi})_{li}
\nonumber \\ &  +\frac{1}{4} M_{\Xi i} M_{\Xi j}^3 \ZI ^{13} _{ij} [P_\mu,(\mathbf{U}_{\Xi \Xi})_{ij}][P^\mu,(\mathbf{U}_{\Xi \Xi})_{ji}]
\nonumber \\ &  -\frac{1}{2} \ZI[q^4] ^{22} _{ij} \gamma^\nu [P_\mu,(\mathbf{U}_{\Xi \Xi})_{ij}]\gamma_\nu[P^\mu,(\mathbf{U}_{\Xi \Xi})_{ji}]
\nonumber \\ &  - \ZI[q^4] ^{22} _{ij} \gamma^\nu [P_\nu,(\mathbf{U}_{\Xi \Xi})_{ij}]\gamma_\mu[P^\mu,(\mathbf{U}_{\Xi \Xi})_{ji}]
\Bigg\rbrace.
\label{eq:UOLEAToLambda}
\end{align}
The operators containing covariant
derivatives are again included to obtain the field-redefinition of the
Higgs field, which propagates into every Higgs coupling that has a
non-vanishing tree-level contribution once the Higgs field is cannonically normalised in the EFT. Therefore this field-redefinition contributes to the threshold correction of the quartic
coupling.  

Next, we compute the derivative matrices as defined in Appendix~\ref{app: UOLEA_ops}.  We start with
\begin{align}
\mathbf{U}_{\Phi \Phi}=\begin{pmatrix} 
U_{\Sigma ^* \Sigma} && U_{\Sigma ^{*} \Sigma ^{*}} &&  U_{\Sigma^* \Theta} \\
U_{\Sigma \Sigma} && U_{\Sigma \Sigma ^{*}} &&  U_{\Sigma \Theta} \\
U_{\Theta \Sigma} && U_{\Theta \Sigma ^*} && U_{\Theta \Theta}\end{pmatrix}
\label{eq: heavy-scalar-heavy-scalar}
\end{align}
and define
\begin{align}
\Sigma =
\begin{pmatrix}
\su_{Li} & \su_{Ri} & \sd_{Li} & \sd_{Ri} & \sel_{Li} & \sel_{Ri} & \sneu_{Li} & H^{+} 
\end{pmatrix}^T,\text{ }
\Theta = \begin{pmatrix}
A & H
\end{pmatrix}^T ,
\end{align}
with the corresponding masses 
\begin{align}
    M_\Sigma = \begin{pmatrix}
m_{\sq_{i}} & m_{\su_{i}} & m_{\sq_{i}} & m_{\sd_{i}} & m_{\slep_{i}} & m_{\sel_{i}} & m_{\sneu_{i}} & m_A 
\end{pmatrix}^T,\text{ }
M_\Theta = \begin{pmatrix}
m_A & m_A
\end{pmatrix}^T ,
\end{align}
where we did not include the derivatives w.r.t.\ vector bosons since there are no heavy vector bosons in the theory. The non-vanishing
derivatives with respect to two heavy scalar fields read
\begingroup
\allowdisplaybreaks
\begin{align}
U_{\su_{Li}^* \su_{Lj}}&=U_{\su_{Li} \su_{Lj}^*}=\frac{1}{8}c_{2\beta}h^2 \delta_{ij}\left(g_2^2-\frac{1}{5}g_1^2\right)+\delta_{3i}\delta_{3j}\frac{g_t^2}{2}h^2, \\ 
U_{\su_{Ri}^* \su_{Rj}}&=U_{\su_{Ri} \su_{Rj}^*}=\frac{1}{10}c_{2\beta}h^2 \delta_{ij}g_1^2+\delta_{3i}\delta_{3j}\frac{g_t^2}{2}h^2, \\ 
U_{\sd_{Li}^* \sd_{Lj}}&=U_{\sd_{Li} \sd_{Lj}^*}=-\frac{1}{8}c_{2\beta}h^2 \delta_{ij}\left(g_2^2+\frac{1}{5}g_1^2\right), \\ 
U_{\sd_{Ri}^* \sd_{Rj}}&=U_{\sd_{Ri} \sd_{Rj}^*}=\frac{1}{20}c_{2\beta}h^2 \delta_{ij}g_1^2, \\
U_{\sel_{Li}^* \sel_{Lj}}&=U_{\sel_{Li} \sel_{Lj}^*}=\frac{1}{8}c_{2\beta}h^2 \delta_{ij}\left(g_2^2-\frac{3}{5}g_1^2\right), \\  
U_{\sel_{Ri}^* \sel_{Rj}}&=U_{\sel_{Ri} \sel_{Rj}^*}=-\frac{1}{20}c_{2\beta}h^2 \delta_{ij}g_1^2, \\
U_{\sneu_{Li}^* \sneu_{Lj}}&=U_{\sneu_{Li} \sneu_{Lj}^*}=\frac{1}{8}c_{2\beta}h^2 \delta_{ij}\left(g_2^2+\frac{3}{5}g_1^2\right), \\  
U_{H^+ H^-}&=U_{H^- H^+}=\frac{1}{8}h^2 \left[(1+s_{2\beta}^2)g_2^2-\frac{3}{5}g_1^2 c_{2\beta}^2\right], \\
U_{AA}&=-\frac{1}{16}c_{2\beta}^2\left(\frac{3}{5}g_1^2+g_2^2\right)h^2, \\
U_{HH}&=\frac{1}{16}(2s_{2\beta}^2-1)\left(\frac{3}{5}g_1^2+g_2^2\right)h^2, \\
U_{\su_{Li}^* \su_{Rj}}&=U_{\su_{Li} \su_{Rj}^*}=\frac{1}{\sqrt{2}}\delta_{3i}\delta_{3j}g_t X_t h.
\end{align}
\endgroup
Given these derivatives we find that $\mathbf{U}_{\Phi \Phi}$ is
block-diagonal, where the blocks are given by
\begin{align}
U_{\Sigma^* \Sigma}&=\begin{pmatrix}
U_{\su_{Li}^* \su_{Lj}} & U_{\su_{Li}^* \su_{Rj}} & \mathbf{0}_{1\times 6} \\
U_{\su_{Ri}^* \su_{Lj}} & U_{\su_{Ri}^* \su_{Rj}} & \mathbf{0}_{1\times 6} \\
\mathbf{0}_{6\times 1} & \mathbf{0}_{6\times 1} & U_{\Pi^* \Pi} 
\end{pmatrix}, \\
U_{\Pi^* \Pi}&=\diag(U_{\sd_{Li}^* \sd_{Lj}},U_{\sd_{Ri}^* \sd_{Rj}},U_{\sel_{Li}^* \sel_{Lj}},U_{\sel_{Ri}^* \sel_{Rj}},U_{\sneu_{Li}^* \sneu_{Lj}},U_{H^+ H^-})\, , \\
U_{\Sigma \Sigma^*}&=\begin{pmatrix}
U_{\su_{Li} \su_{Lj}^*} & U_{\su_{Li} \su_{Rj}^*} & \mathbf{0}_{1\times 6} \\
U_{\su_{Ri} \su_{Lj}^*} & U_{\su_{Ri} \su_{Rj}^*} & \mathbf{0}_{1\times 6} \\
\mathbf{0}_{6\times 1} & \mathbf{0}_{6\times 1} & U_{\Pi \Pi^*} 
\end{pmatrix}, \\
U_{\Pi \Pi^*}&=\diag(U_{\sd_{Li} \sd_{Lj}^*},U_{\sd_{Ri} \sd_{Rj}^*},U_{\sel_{Li} \sel_{Lj}^*},U_{\sel_{Ri} \sel_{Rj}^*},U_{\sneu_{Li} \sneu_{Lj}^*},U_{H^- H^+})\,, \\
U_{\Theta \Theta}&=\diag(U_{AA},U_{HH})\,,
\end{align}
with $\mathbf{0}_{m\times n}$ being the $m \times n$ matrix of only
zeros. We next calculate $\mathbf{U}_{\phi \Phi}$ and
$\mathbf{U}_{\Phi\phi}$, which contain derivatives with respect to one
heavy and one light scalar field. As discussed in Chapter~\ref{chap: fcuntional matching} the derivatives w.r.t.\ the fields
are evaluated at the background field configurations, and the heavy
background fields are expressed in terms of the light ones using a
local operator expansion. In the case under consideration here the heavy background fields are
suppressed by at least $1/M^2$ and since we are not interested in these
suppressed contributions, we set the heavy background fields to zero. Defining the light scalar field
multiplets as
\begin{align}
  \sigma = (G^+)\,,\text{ }
  \theta =
           \begin{pmatrix}
             h & G^0
           \end{pmatrix}^T,
\end{align}
the non-vanishing derivatives are found to be
\begin{align}
U_{Hh} &= U_{hH}=\frac{3}{8}\left(\frac{3}{5}g_1^2+g_2^2\right)s_{2\beta}c_{2\beta}h^2 ,\\
U_{AG^0} &= U_{G^0A}=-\frac{1}{8}\left(\frac{3}{5}g_1^2+g_2^2\right)s_{2\beta}c_{2\beta}h^2, \\
U_{H^{+}G^{-}} &= U_{H^{-}G^{+}}=-\frac{1}{8}\left(\frac{3}{5}g_1^2+g_2^2\right)s_{2\beta}c_{2\beta}h^2.
\end{align}
We then find that $\mathbf{U}_{\Phi \phi}$ is block-diagonal with the blocks being
\begin{align}
U_{\Sigma^* \sigma}&=\begin{pmatrix} \mathbf{0}_{7 \times 1} \\ U_{H^{-} G^{+}}
\end{pmatrix}, \\
U_{\Sigma \sigma^*}&=\begin{pmatrix} \mathbf{0}_{7 \times 1} \\ U_{H^{+} G^{-}}
\end{pmatrix}, \\
U_{\Theta \theta}&=\begin{pmatrix}  0 & U_{A G^0}\\
U_{Hh} & 0
\end{pmatrix}.
\end{align}
Similarly, $\mathbf{U}_{\phi \Phi}$ is block-diagonal where the blocks are given by
\begin{align}
U_{\sigma^* \Sigma}&=\begin{pmatrix} \mathbf{0}_{1 \times 7} & U_{G^{-} H^{+}}
\end{pmatrix}, \\
U_{\sigma \Sigma^*}&=\begin{pmatrix} \mathbf{0}_{1 \times 7} & U_{G^{+} H^{-}}
\end{pmatrix}, \\
U_{\theta \Theta}&=\begin{pmatrix}   0 & U_{h H}\\
 U_{G^0 A} & 0
\end{pmatrix}.
\end{align}
Finally, we need the derivatives with respect to two heavy fermions to
construct the matrix $\mathbf{U}_{\Xi \Xi}$. Both Dirac and Majorana fermions are present and hence we define
\begin{align}
\Omega = \begin{pmatrix}
\tilde{\chi}^+ _1 & \tilde{\chi}^+ _2
\end{pmatrix}^T,\text{ } \Lambda = \begin{pmatrix}
\tilde{\chi}^0 _1 & \tilde{\chi}^0 _2 & \tilde{\chi}^0 _3 & \tilde{\chi}^0 _4
\end{pmatrix}^T,
\end{align}
with the corresponding masses $M_\Xi = (M_2\; \mu\; M_1\; M_2 \; \mu \; \mu)^T$,
and find that the matrix $\mathbf{U}_{\Xi \Xi}$ is also block-diagonal with the
non-vanishing entries
\begingroup
\allowdisplaybreaks
\begin{align}
U_{\bar{\Omega} \Omega}&=\cc ^{-1} U^T_{\Omega \bar{\Omega}} \cc^{-1}=-\frac{g_2}{\sqrt{2}}h\begin{pmatrix}
0 & c_\beta P_R+ s_\beta P_L \\
c_\beta P_L+s_\beta P_R & 0
\end{pmatrix}, \\
\cc^{-1} U_{\Lambda \Lambda}&=\frac{h}{2\sqrt{2}}\begin{pmatrix}
0 & 0 & i g_Y(c_\beta-s_\beta)\gamma^5 & -g_Y(c_\beta+s_\beta) \\
0 & 0 & -ig_2(c_\beta-s_\beta)\gamma^5 & g_2(c_\beta+s_\beta) \\
i g_Y(c_\beta-s_\beta)\gamma^5 & -ig_2(c_\beta-s_\beta)\gamma^5 & 0 & 0 \\
-g_Y(c_\beta+s_\beta) & g_2(c_\beta+s_\beta) & 0 & 0
\end{pmatrix}. 
\end{align}
\endgroup
Inserting all of these derivatives into Eq.\ \eqref{eq:UOLEAToLambda} and summing over all indices we find the following field-redefinition of the SM-like Higgs boson
\begin{align}
  h ={}& \left(1 - \frac{1}{2} \delta Z_h\right) \hat{h}, \\
  \delta Z_h ={}& -6 g_t^2 X_t^2 \ZI[q^2]_{\sq\su}^{22}+\frac{s_{2 \beta}}{2} \mu \left(g^2_Y M_1 \mu^2 \ZI ^{13}_{1\mu}+g^2_Y M_1^3  \ZI ^{31}_{1\mu}-3g^2_2 M_2 \mu^2 \ZI ^{13}_{2\mu}-3g^2_2 M_2^3  \ZI ^{31}_{2\mu}\right) 
  \nonumber \\& +2(2+d)\left(-g_Y^2 \ZI[q^4]^{22}_{1\mu}+3g_2^2 \ZI[q^4]^{22}_{2\mu}\right),
\end{align}
where the subscripts $1$ and $2$ of the loop functions are shorthand for $M_1$ and $M_2$, respectively. After canonically normalising the Higgs boson in the EFT the effective Lagrangian reads
\begin{align}
  \Lag{EFT}^{1\Loop} = \frac{1}{2}(\partial \hat{h})^2
  - \frac{\lambda}{8} \hat{h}^4 + \cdots
\end{align}
with
%
%
%
%
%
%
%
\begin{align}
  \lambda &= \frac{1}{4} \left( \frac{3}{5} g_1^2 + g_2^2 \right) c_{2\beta}^2
           + \kappa \Delta\lambda^{1\Loop} , \\
   \Delta\lambda^{1\Loop} &= \Delta\lambda^{1\Loop,\text{reg}}
           + \Delta\lambda^{1\Loop,\phi} + \Delta\lambda^{1\Loop,\chi},
\end{align}
and
\begingroup
\allowdisplaybreaks
\begin{align}
  \Delta \lambda^{1\Loop,\phi} ={}&
  g_t^4 \left[
     -3 X_t^4 \ZI_{\sq\sq\su\su}^{1111}
     -6 X_t^2 \left(\ZI_{\sq\sq\su}^{111} + \ZI_{\sq\su\su}^{111}\right)
     -3 \left(\ZI_{\sq\sq}^{11} + \ZI_{\su\su}^{11}\right)
   \right] \nonumber \\
   &+\frac{3}{10} g_t^2 c_{2\beta} \Big\{ X_t^2 \left[
   2 c_{2\beta} \left(3 g_1^2+5g_2^2\right) \ZI[q^2]_{\sq\su}^{22}
   +\left(g_1^2-5 g_2^2\right) \ZI_{\sq\sq\su}^{111}
   -4 g_1^2 \ZI_{\sq\su\su}^{111}\right] \nonumber \\
   &~~~~~~~~~~~~~~~~ + \left(g_1^2-5 g_2^2\right) \ZI_{\sq\sq}^{11}
   -4 g_1^2 \ZI_{\su\su}^{11}\Big\} \nonumber \\
   & -\frac{c_{2\beta}^2}{200} \sum_{i=1}^3 \Big[
   3 \left(g_1^4+25 g_2^4\right) \ZI_{\sq_i\sq_i}^{11}
   +24 g_1^4 \ZI_{\su_i\su_i}^{11}
   +6 g_1^4 \ZI_{\sd_i\sd_i}^{11} \nonumber \\
   &~~~~~~~~~~~~~~~ +\left(9 g_1^4+25 g_2^4\right) \ZI_{\slep_i\slep_i}^{11}
   +18 g_1^4 \ZI_{\sel_i\sel_i}^{11}
   \Big] \nonumber \\
   &+ \frac{1}{200} \Big\{6 c_{2\beta}^2 \left(c_{2\beta}^2-1\right) \left(3
   g_1^2+5 g_2^2\right)^2 \ZI_{A0}^{11} - \Big[9 \left(3
   c_{2\beta}^4-3 c_{2\beta}^2+1\right) g_1^4 \nonumber \\
   &~~~~~~~~~~ +30 \left(3 c_{2\beta}^4-4
   c_{2\beta}^2+1\right) g_1^2 g_2^2+25 \left(3 c_{2\beta}^4-5
   c_{2\beta}^2+3\right) g_2^4\Big]
   \ZI_{AA}^{11}\Big\},\\
  \Delta\lambda^{1\Loop,\chi} ={}& -\frac{1}{4}
  \Big\{-d \big(2 g_Y^4 M_1^2 \ZI[q^2] ^{22} _{1 \mu} + 
      2 g_2^4 M_2^2 \ZI[q^2] ^{22} _{2 \mu} + 
      g_Y^4 \mu^2 \ZI[q^2] ^{22} _{1 \mu} \nonumber \\ &\qquad ~~~~~~~~ - 
      g_Y^4 \mu^2 c_{4 \beta} \ZI[q^2] ^{22} _{1 \mu} +
      g_2^4 \mu^2 \ZI[q^2] ^{22} _{2 \mu} 
      - g_2^4 \mu^2 c_{4 \beta} \ZI[q^2] ^{22} _{2 \mu} \nonumber \\ &\qquad ~~~~~~~~ + 
      4 g_Y^2 g_2^2 M_1 M_2 \ZI[q^2] ^{112} _{1 2 \mu} + 
      2 g_Y^2 g_2^2 \mu^2 \ZI[q^2] ^{112} _{1 2 \mu} - 
      2 g_Y^2 g_2^2 \mu^2 c_{4 \beta} \ZI[q^2] ^{112} _{1 2 \mu}\big) \nonumber \\ &\qquad~~   - 
   d (2 + d) \big(2 g_Y^4 \ZI[q^4] ^{22} _{1 \mu} + 
      2 g_2^4 \ZI[q^4] ^{22} _{2 \mu} + 
      4 g_Y^2 g_2^2 \ZI[q^4] ^{112} _{1 2 \mu}\big) \nonumber \\ &\qquad~~ 
       - g_2^4 \big[2d (2 + d) (3 + c_{4 \beta}) \ZI[q^4] ^{22} _{2 \mu} 
   + 16 c_{\beta} s_{\beta} (d M_2 \ZI[q^2] ^{22} _{2 \mu} (\mu + M_2 c_{\beta} s_{\beta}) 
   \nonumber \\ &\qquad ~~~~~~~~~ +
          \mu \{M_2^2 \mu c_{\beta} \ZI ^{22} _{2 \mu} s_{\beta} + 
            d \ZI[q^2] ^{22} _{2 \mu} (M_2 + \mu c_{\beta} s_{\beta})\})\big]
            \nonumber \\ &\qquad~~  - 
   4 d \mu \big(2 g_Y^4 M_1 \ZI[q^2] ^{22} _{1 \mu} + 
      2 g_2^4 M_2 \ZI[q^2] ^{22} _{2 \mu} + 
      2 g_Y^2 g_2^2 M_1 \ZI[q^2] ^{112} _{1 2 \mu}  \nonumber \\ &\qquad ~~~~~~~~~~~ + 
      2 g_Y^2 g_2^2 M_2 \ZI[q^2] ^{112} _{1 2 \mu}\big) s_{2 \beta} 
      \nonumber \\ &\qquad~~ - 
   2 \mu^2 \big(g_Y^4 M_1^2 \ZI ^{22} _{1 \mu} + 
      g_2^2 M_2 (g_2^2 M_2 \ZI ^{22} _{2 \mu} + 
         g_Y^2 M_1 2\ZI ^{112} _{12 \mu} + 
            )\big) s_{2 \beta}^2 \nonumber \\ &\qquad~~  - 
   2 g_2^2 (g_Y^2 + g_2^2) c_{
     2 \beta}^2 \big(-4 (2 + d) \ZI[q^4] ^{22} _{2 \mu} + 
      M_2 \mu (\mu^2 \ZI ^{13} _{2 \mu} + 
         M_2^2 \ZI ^{31} _{2 \mu}) s_{2 \beta}\big)\nonumber \\ &\qquad~~  - (g_Y^2 + 
      g_2^2) c_{
     2 \beta}^2 \big(-4 (2 + d) g_Y^2 \ZI[q^4] ^{22} _{1 \mu} - 
      4 (2 + d) g_2^2 \ZI[q^4] ^{22} _{2 \mu} \nonumber \\ &\qquad ~~~~~~ + 
      \mu \{g_Y^2 M_1 \mu^2 \ZI ^{13} _{1 \mu} + 
         g_Y^2 M_1^3 \ZI ^{31} _{1 \mu} + 
         g_2^2 M_2 (\mu^2 \ZI ^{31} _{2 \mu} + 
            M_2^2 \ZI ^{31} _{2 \mu})\} s_{2 \beta}\big)\Big\}.
\end{align}
\endgroup
Here $\lambda$ is expressed entirely in terms of the MSSM gauge couplings,
in contrast to Ref.\ \cite{Bagnaschi:2014rsa} and we use the notation $m_{\sq} \equiv m_{\sq_3}$ as well as $m_{\su} \equiv m_{\su_3}$.

It is sensible to regularise the MSSM using
DRED, whereas the SM is more naturally
regularised in DREG. The change in regularisation leads to further contributions to
the threshold correction denoted by $\Delta\lambda^{1\Loop,\text{reg}}$, which can be obtained using the
DRED--DREG translating operators presented in Section~\ref{sec: DRED_to_DREG}. In this case the contribution originates from the operator
\begin{align}
\frac{1}{\kappa} \eps \LagNoT_{\EFT,\eps}^{1\Loop} = \frac{1}{2} \sum_{ij} (\check{\mathbf{U}}^{\mu}_{\epsilon \epsilon \nu})_{ij} (\check{\mathbf{U}}^{\nu}_{\epsilon \epsilon \mu})_{ji},
\label{eq: epsilonconttolambda}
\end{align}
of Eq.\ \eqref{eq: masterformula}, where in the MSSM we have the following couplings of epsilon scalars to the
SM-like doublet $\mathcal{H}$,
\begin{align}
\LagNoT_{\eps \mathcal{H}}=\mathcal{H}^*_a \epsdim{g}_{\mu \nu}\left(g_2 ^2 T^A_{ab} T^B_{bc} \epsdim{a}^{A\mu} \epsdim{a}^{B\nu}+\sqrt{\frac{3}{5}} g_1 g_2 T^A_{ac} \epsdim{a}^{A\mu} \epsdim{b}^\nu +\frac{3}{20}g_1^2 \epsdim{b}^\mu \epsdim{b}^\nu \delta_{ac}\right)\mathcal{H}_c.
\end{align}
The fields
$\epsdim{a}^{A \mu}$ and $\epsdim{b}^\mu$ denote the epsilon scalars
corresponding to $SU(2)_L$ and $U(1)_Y$, respectively, and the matrices $T^A$ are the generators of the fundamental representation of $SU(2)_L$. From this Lagrangian the matrix $\epsdim{\mathbf{U}}^{\mu \nu}_{\eps \eps}$ is calculated to be
\begin{align}
\epsdim{\mathbf{U}}^{\mu \nu}_{\eps \eps}=-\epsdim{g}^{\mu \nu} \begin{pmatrix}
\mathcal{H}^{*}_a g_2^2 \{T^A,T^B\}_{ac} \mathcal{H}_c & \sqrt{\frac{3}{5}} g_1 g_2 \mathcal{H}^{*}_a T^A _{ac} \mathcal{H}_c \\
\sqrt{\frac{3}{5}} g_1 g_2 \mathcal{H}^{*}_a T^A _{ac} \mathcal{H}_c & \frac{3}{10} g_1^2 \mathcal{H}^{*}_a \mathcal{H}_a  
\end{pmatrix},
\end{align}
and inserting this into Eq.\ \eqref{eq: epsilonconttolambda} we obtain
\begin{align}
\Delta\lambda^{1\Loop,\text{reg}} &=
      - \frac{9}{100}g_1^4 - \frac{3}{10} g_1^2 g_2^2 
      -\frac{3}{4} g_2^4.
\end{align}
We do not find the term proportional to $c_{2\beta}^2$ given in Ref.\
\cite{Bagnaschi:2014rsa} since this term only arises once the
tree-level expression for $\lambda$ is expressed in terms of SM gauge
couplings. Up to terms
arising from this conversion the one-loop threshold corrections presented here agree
with the results of Ref.\ \cite{Bagnaschi:2014rsa}.

\section{Integrating out stops and the gluino from the MSSM}
\label{sec:matching_MSSM_to_SMEFT}

In this section we reproduce known threshold corrections from
the MSSM to the SMEFT from
heavy stops and the gluino in the gaugeless limit ($g_1 = g_2 = 0$). We work in
the unbroken phase and set all Yukawa couplings to zero, except for the
one of the top quark. In this application contributions from BSUOLEA operators with mixed statistics arise and the BSUOLEA must be carefully applied.

We consider the following part of the MSSM Lagrangian
\begin{align}
  \begin{split}
  \Lag{MSSM} \supset{}&
  |\partial\st{L}|^2 - \mstL |\st{L}|^2
  + |\partial\st{R}|^2 - \mstR |\st{R}|^2
  + \frac{1}{2}(\gluino{\dot{A}})^T \cc (i\slashed{\partial} - m_{\gluino{}}) \gluino{\dot{A}}\\
  &
  - \frac{y_t s_\beta}{\sqrt{2}} h \bar{t} t
  - \frac{y_t^2 s_\beta^2}{2} h^2 \left(|\st{L}|^2 + |\st{R}|^2\right)
  - \frac{y_t s_\beta X_t}{\sqrt{2}} h \left(\st{L}^* \st{R} + \hc\right) \\
  &
  - \sqrt{2} g_3 \left[
    \bar{t} P_R \gluino{\dot{A}} T^{\dot{A}} \st{L} - \bar{t} P_L \gluino{\dot{A}} T^{\dot{A}} \st{R}
    + \st{L}^* (\gluino{\dot{A}})^T T^{\dot{A}} \cc P_L t - \st{R}^* (\gluino{\dot{A}})^T T^{\dot{A}} \cc P_R t
  \right] ,
  \end{split}
  \label{eq:LMSSM_stop}
\end{align}
where we use the same notation as in the previous section, $g_3$
is the strong gauge coupling and $T^{\dot{A}}$ are the generators of the fundamental representation of $SU(3)_c$. The top
quark, $t$, is defined as a Dirac fermion built from
the upper component of the left-handed quark-doublet $q_L$ and the
right-handed top $t_R$. In Eq.\ \eqref{eq:LMSSM_stop} it was used that the gluino, $\gluino{\dot{A}}$, is a Majorana spinor which satisfies $\overline{\gluino{\dot{A}}} = (\ccfield{(\gluino{\dot{A}})})^T \cc =
(\gluino{\dot{A}})^T \cc$.

After the heavy stops and the gluino have been integrated out the Lagrangian of
the effective theory reads
\begin{align}
  \Lag{SMEFT} \supset
  - \frac{y_t s_\beta}{\sqrt{2}} h \bar{t} t
  + \Lag{SMEFT}^\text{1\Loop}.
\end{align}
In our limit the one-loop term $\Lag{SMEFT}^\text{1\Loop}$ receives
contributions from the following BSUOLEA operators
\begingroup
\allowdisplaybreaks
\begin{align}
  \frac{1}{\kappa} \Lag{EFT}^{1\Loop} \supset{}& \frac{1}{2} \ZI ^1 _i  (\mathbf{U}_{\Phi \Phi})_{ii}+\frac{1}{4} \ZI ^{11} _{ik} (\mathbf{U}_{\Phi \Phi})_{ik} (\mathbf{U}_{\Phi \Phi})_{ki}+\frac{1}{6} \ZI ^{111} _{lik} (\mathbf{U}_{\Phi \Phi})_{ik} (\mathbf{U}_{\Phi \Phi})_{kl} (\mathbf{U}_{\Phi \Phi})_{li}\nonumber \\
  & +\frac{1}{8} \ZI ^{1111} _{likn} (\mathbf{U}_{\Phi \Phi})_{ik} (\mathbf{U}_{\Phi \Phi})_{kl} (\mathbf{U}_{\Phi \Phi})_{ln} (\mathbf{U}_{\Phi \Phi})_{ni}\nonumber \\ 
  & +\frac{1}{10} \ZI ^{11111} _{iklnp} (\mathbf{U}_{\Phi \Phi})_{ik} (\mathbf{U}_{\Phi \Phi})_{kl} (\mathbf{U}_{\Phi \Phi})_{ln} (\mathbf{U}_{\Phi \Phi})_{np} (\mathbf{U}_{\Phi \Phi})_{pi}\nonumber \\
  & +\frac{1}{12} \ZI ^{111111} _{iklnpr} (\mathbf{U}_{\Phi \Phi})_{ik} (\mathbf{U}_{\Phi \Phi})_{kl} (\mathbf{U}_{\Phi \Phi})_{ln} (\mathbf{U}_{\Phi \Phi})_{np} (\mathbf{U}_{\Phi \Phi})_{pr} (\mathbf{U}_{\Phi \Phi})_{ri}\nonumber \\
  & +\frac{1}{2} \ZI [q^2]^{22} _{ki} [P_\mu, (\mathbf{U}_{\Phi \Phi})_{ik}] [P^\mu, (\mathbf{U}_{\Phi \Phi})_{ki}]\nonumber \\
  & -\ZI [q^2] ^{21}_{il} (\mathbf{U}_{\Phi \Xi})_{il} \gamma^\mu [P_\mu, (\mathbf{U}_{\Xi \Phi})_{li}]\nonumber \\
  & - \frac{1}{2} M_{\Xi_{k}} \ZI ^{111} _{ikl} (\mathbf{U}_{\Phi \Xi})_{ik} (\mathbf{U}_{\Xi \Phi})_{kl} (\mathbf{U}_{\Phi \Phi})_{li},
\label{eq:L_SMEFT_operators}
\end{align}
\endgroup
where we set $P_\mu \equiv i \partial_\mu$ and thus omit contributions from gauge
bosons.  We identify $\Sigma = (\st{L}, \st{R})$ as
the heavy scalar multiplet and $\Lambda = (\gluino{\dot{A}})$ as the multiplet of heavy Majorana fermions. From Eq.\ \eqref{eq:LMSSM_stop} we then obtain the following
non-vanishing derivatives
\begin{align}
  (U_{\st{L}^{*} \st{L}})_{IJ} &= (U_{\st{L} \st{L}^{*}})_{IJ} =
  (U_{\st{R}^{*} \st{R}})_{IJ} = (U_{\st{R} \st{R}^{*}})_{IJ} = \frac{1}{2}(y_t s_{\beta} h)^2 \delta_{IJ},\\
  (U_{\st{L}^{*} \st{R}})_{IJ} &= (U_{\st{L} \st{R}^{*}})_{IJ} =
  (U_{\st{R}^{*} \st{L}})_{IJ} = (U_{\st{R} \st{L}^{*}})_{IJ} = \frac{1}{\sqrt{2}} y_t s_{\beta} h X_t \delta_{IJ},\\
  (U_{\st{L} \gluino{\dot{A}}})_{I \dot{\alpha}}^{\dot{A}} &= (U_{\gluino{\dot{A}} \st{L}})_{I \dot{\alpha}}^{\dot{A}} = -\sqrt{2} g_3 (\bar{t}_J P_R)_{\dot{\alpha}} T^{\dot{A}}_{JI}, \label{eq:sign_1}\\
  (U_{\st{R} \gluino{\dot{A}}})_{I \dot{\alpha}}^{\dot{A}} &= (U_{\gluino{\dot{A}} \st{R}})_{I\dot{\alpha}}^{\dot{A}} = \sqrt{2} g_3 (\bar{t}_J P_L)_{\dot{\alpha}} T^{\dot{A}}_{JI}, \label{eq:sign_2}\\
  (U_{\gluino{\dot{A}} \st{L}^{*}})_{I\dot{\alpha}}^{\dot{A}} &= (U_{\st{L}^{*} \gluino{\dot{A}}})_{I \dot{\alpha}}^{\dot{A}} = \sqrt{2} g_3 T^{\dot{A}}_{IJ} (\cc P_L t_J)_{\dot{\alpha}},\\
  (U_{\gluino{\dot{A}} \st{R}^{*}})_{I \dot{\alpha}}^{\dot{A}} &= (U_{\st{R}^{*} \gluino{A}})_{I \dot{\alpha}}^{\dot{A}} =  -\sqrt{2} g_3 T^{\dot{A}}_{IJ} (\cc P_R t_J)_{\dot{\alpha}},
\end{align}
where the flipped
sign in Eqs.\ \eqref{eq:sign_1}--\eqref{eq:sign_2} is due to one
anti-commutation of the spinor $\bar{t}$ with the derivative w.r.t.\
the spinor $\gluino{\dot{A}}$. The derivative matrices are given by
\begingroup
\allowdisplaybreaks
\begin{align}
  \mathbf{U}_{\Phi \Phi} &=
  \begin{pmatrix} 
    U_{\Sigma ^* \Sigma} & U_{\Sigma ^* \Sigma ^*} \\
    U_{\Sigma \Sigma} & U_{\Sigma \Sigma ^{*}}
  \end{pmatrix} =
  \begin{pmatrix}
    (U_{\st{L}^* \st{L}})_{IJ} & (U_{\st{L}^* \st{R}})_{IJ} & 0 & 0 \\
    (U_{\st{R}^* \st{L}})_{IJ} & (U_{\st{R}^* \st{R}})_{IJ} & 0 & 0 \\
    0 & 0 & (U_{\st{L} \st{L}^*})_{IJ} & (U_{\st{L} \st{R}^*})_{IJ} \\
    0 & 0 & (U_{\st{R} \st{L}^*})_{IJ} & (U_{\st{R} \st{R}^*})_{IJ}
  \end{pmatrix} \nonumber \\
  &= \delta_{IJ} \; \mathbf{1}_{2} \otimes
  \begin{pmatrix}
    \frac{1}{2}(y_t s_\beta h)^2 & \frac{1}{\sqrt{2}} y_t s_\beta h X_t \\
    \frac{1}{\sqrt{2}} y_t s_\beta h X_t & \frac{1}{2}(y_t s_\beta h)^2
  \end{pmatrix}, \\
  \mathbf{U}_{\Phi \Xi} &=
  \begin{pmatrix}
    U_{\Sigma ^* \Lambda} \\
    U_{\Sigma \Lambda}
  \end{pmatrix}
  =
  \begin{pmatrix}
    (U_{\st{L}^* \gluino{\dot{A}}})_{I\dot{\alpha}}^{\dot{A}} \\
    (U_{\st{R}^* \gluino{\dot{A}}})_{I \dot{\alpha}}^{\dot{A}} \\
    (U_{\st{L} \gluino{\dot{A}}})_{I \dot{\alpha}}^{\dot{A}} \\
    (U_{\st{R} \gluino{\dot{A}}})_{i\dot{\alpha}}^{\dot{A}}
  \end{pmatrix}
  = \sqrt{2} g_3
  \begin{pmatrix}
    T^{\dot{A}}_{IJ} (\cc P_L t_J)_{\dot{\alpha}} \\
     -T^{\dot{A}}_{IJ} (\cc P_R t_J)_{\dot{\alpha}} \\
     -(\bar{t}_J P_R)_{\dot{\alpha}} T^{\dot{A}}_{JI} \\
    (\bar{t}_J P_L)_{\dot{\alpha}} T^{\dot{A}}_{JI}
  \end{pmatrix},
  \\
  \mathbf{U}_{\Xi \Phi} &=
  \begin{pmatrix}
    \cc^{-1} U_{\Lambda \Sigma} & \cc^{-1} U_{\Lambda \Sigma ^*}
  \end{pmatrix} \nonumber \\
  &= (\cc^{-1})_{\dot{\alpha} \dot{\beta}}
  \begin{pmatrix}
    (U_{\gluino{\dot{A}} \st{L}})_{I\dot{\beta}}^{\dot{A}} &
    (U_{\gluino{\dot{A}} \st{R}})_{I \dot{\beta}}^{\dot{A}} &
    (U_{\gluino{\dot{A}} \st{L}^*})_{I \dot{\beta}}^{\dot{A}} &
    (U_{\gluino{\dot{A}} \st{R}^*})_{I \dot{\beta}}^{\dot{A}}
  \end{pmatrix} \nonumber \\
  &= \sqrt{2} g_3 (\cc^{-1})_{\dot{\alpha}\dot{\beta}}
  \begin{pmatrix}
    -(\bar{t}_J P_R)_{\dot{\beta}} T^{\dot{A}}_{JI} &
   (\bar{t}_J P_L)_{\dot{\beta}} T^{\dot{A}}_{JI} &
   T^{\dot{A}}_{IJ} (\cc P_L t_J)_{\dot{\beta}} &
    -T^{\dot{A}}_{IJ} (\cc P_R t_J)_{\dot{\beta}}
  \end{pmatrix} \nonumber \\
  &= \sqrt{2} g_3
  \begin{pmatrix}
    -(\bar{t}_J P_R (\cc^{-1})^T)_{\dot{\alpha}} T^{\dot{A}}_{JI} &
   (\bar{t}_J P_L (\cc^{-1})^T)_{\dot{\alpha}} T^{\dot{A}}_{JI} &
   T^{\dot{A}}_{IJ} (P_L t_J)_{\dot{\alpha}} &
    -T^{\dot{A}}_{IJ} (P_R t_J)_{\dot{\alpha}}
  \end{pmatrix} .
\end{align}
\endgroup
Inserting these matrices into Eq.\
\eqref{eq:L_SMEFT_operators} and summing over all fields and colors we
obtain
\begin{align}
  \Lag{EFT}^\text{1\Loop} &=
     c_t h\bar{t}t + c_L\bar{t}i\slashed{\partial}P_Lt + c_R \bar{t}i\slashed{\partial}P_Rt
     + c_2' (\partial h)^2 + c_2 h^2 + c_4 h^4 + c_6 h^6 + \cdots,
\end{align}
where
\begin{align}
  c_t &= -\frac{4 \sqrt{2}}{3}\kappa g_3^2 y_t s_\beta m_{\gluino{}} X_t \ZI ^{111} _{\gluino{}\sq\su},\\
\begin{split}
  c_L &= \frac{16}{3}\kappa g_3^2 \ZI [q^2] ^{21} _{\su \gluino{}},
\end{split}\\
  c_R &= c_L|_{\sq \to \su},\\
  c_2' &= -3 \kappa (y_t s_\beta)^2 X_t^2 \ZI [q^2] ^{22} _{\sq \su},\\
  c_2 &= \frac{3}{2}\kappa (y_t s_\beta)^2 \left[\ZI ^{1} _{\sq} + \ZI ^{1} _{\su} + X_t^2 \ZI ^{11} _{\sq\su}\right], \\
  c_4 &= \frac{3}{8}\kappa (y_t s_\beta)^4 \left[
         \ZI ^{11} _{\sq\sq} + \ZI ^{11} _{\su\su} + 2 X_t^2 (\ZI ^{111} _{\sq\sq\su} + \ZI ^{111} _{\sq\su\su}) + X_t^4 \ZI ^{1111} _{\sq\sq\su\su}\right],\\
\begin{split}
  c_6 &= \frac{1}{8}\kappa (y_t s_\beta)^6 \big[
         \ZI ^{111} _{\sq\sq\sq} + \ZI ^{111} _{\su\su\su} + 3 X_t^2 ( \ZI ^{1111} _{\sq\sq\sq\su} + \ZI ^{1111} _{\sq\sq\su\su} + \ZI ^{1111} _{\sq\su\su\su} ) \\
    & ~~~~~~~~~~~~~~~~~~
    + 3 X_t^4 ( \ZI ^{11111} _{\sq\sq\sq\su\su} + \ZI ^{11111} _{\sq\sq\su\su\su} )
    + X_t^6 \ZI ^{111111} _{\sq\sq\sq\su\su\su} \big] .
\end{split}
\end{align}
In order to canonically normalise both the top quark field and the Higgs field present in $\Lag{SMEFT}$ we introduce $\hat{t}_L,\, \hat{t}_R$ and $\hat{h}$ defined through
\begingroup
\allowdisplaybreaks
\begin{align}
  h  &= \left(1 - \frac{1}{2} \delta Z_h\right) \hat{h} , \\
  t_L &= \left(1 - \frac{1}{2} \delta Z_L\right) \hat{t}_L , \\
  t_R &= \left(1 - \frac{1}{2} \delta Z_R\right) \hat{t}_R ,
\end{align}
\endgroup
where the field-redefinitions $\delta Z_{h/L/R}$ are given by
\begin{align}
  \delta Z_h &= 2c_2', \\
  \delta Z_L &= c_L, \\
  \delta Z_R &= c_R.
\end{align}
Expressing the SMEFT Lagrangian through these fields as
\begin{align}
  \Lag{SMEFT} \supset
  - \frac{g_t}{\sqrt{2}} \hat{h} \bar{\hat{t}} \hat{t}
  + \frac{m^2}{2} \hat{h}^2 - \frac{\lambda}{8} \hat{h}^4
  - \frac{\tilde{c}_6}{8} \hat{h}^6,
\end{align}
we find that the SMEFT parameters are given by
\begin{align}
  g_t &= y_t s_\beta \left[1 - \frac{1}{2}(c_L + c_R) - c_2' - \frac{\sqrt{2}c_t}{y_t s_\beta} \right], \\
  m^2 &= 2 c_2,\\
  \lambda &= -8 c_4 ,\\
  \tilde{c}_6 &= -8 c_6,
\end{align}
which agrees with the results calculated in Refs.\ 
\cite{Bagnaschi:2014rsa,Bagnaschi:2017xid,Huo:2015nka,Drozd:2015rsp}.\footnote{It
  was noted in Ref.\ \cite{Bagnaschi:2017xid} that the logarithmic term in
  the last line of Eq.\ (D.4) in Ref.\ \cite{Drozd:2015rsp} should come with a
  minus sign.}

\section{Integrating out the gluino from the MSSM with light stops}
\label{sec: gluinoOut}
In this section we calculate some of the terms that arise when
integrating out the gluino from the MSSM, which is a relevant scenario when there is a large hierarchy between the gluino mass and
the stop masses in the MSSM. This example is also a direct
application of operators where Majorana and Dirac fermions appear in loops at the same time.

We consider the following part of the MSSM Lagrangian
\begin{align}
 \Lag{MSSM} \supset{}&
  |\partial\st{L}|^2 - \mstL |\st{L}|^2
  + |\partial\st{R}|^2 - \mstR |\st{R}|^2
  + \frac{1}{2}(\gluino{\dot{A}})^T \cc (i\slashed{\partial} - m_{\gluino{}}) \gluino{\dot{A}} \nonumber \\
  & -\sqrt{2} g_3 \left(
    \bar{t} P_R \gluino{\dot{A}} T^{\dot{A}} \st{L} - \bar{t} P_L \gluino{\dot{A}} T^{\dot{A}} \st{R}
    + \st{L}^* (\gluino{\dot{A}})^T T^{\dot{A}} \cc P_L t - \st{R}^* (\gluino{\dot{A}})^T T^{\dot{A}} \cc P_R t
  \right)\nonumber \\
  & +\left(-y_t^2+\frac{g_3^2}{2}\right)(\st{L}^*\st{R})(\st{L}\st{R}^*)-\frac{g_3^2}{6}|\st{L}|^2|\st{R}|^2,
  \label{eq: gluinoOut_MSSM_Lag}
\end{align}
where we use the same notation as in the previous section.
We determine the one-loop Wilson coefficients of the operators
\begingroup
\allowdisplaybreaks
\begin{align}
\Lag{EFT}^{1\Loop}  \supset{}& c_{t_L} \bar{t}_Li\slashed{\partial} t_L+c_{t_R} \bar{t}_Ri\slashed{\partial} t_R+c_{\st{L}} \partial_\mu\st{L}^*\partial^\mu \st{L}-\delta m_{\sq} ^2 |\st{L}|^2+c_{\st{R}} \partial_\mu\st{R}^*\partial^\mu \st{R}-\delta m_{\su} ^2 |\st{R}|^2 
\nonumber \\ & +c^L_{41} \left(\st{LI}^*  \st{LI} \right)^2+c^L_{42} \left(\st{LI}^*  \st{LJ} \right) \left(\st{LJ}^*  \st{LI}\right)+c^R_{4} \left(\st{R}^* \st{R}\right)^2\nonumber 
\\ &  +c^{LR}_{41} \left(\st{LI}^* \st{LI}\right)\left(\st{RJ}^* \st{RJ}\right)+c^{LR}_{42} \left(\st{LI}^* \st{LJ}\right)\left(\st{RJ}^* \st{RI}\right)+ c_G G^{\dot{A}}_{\mu \nu} G_{\dot{A}} ^{\mu \nu}\nonumber 
\\ &  +[c^{LL} _{51} (\bar{t}_{LI} T^{\dot{A}} _{IJ} \st{LJ})(\ccfield{t}_{RK} T^{\dot{A}} _{KM} \st{LM})+c^{LL} _{52} (\st{LI}^* T^{\dot{A}} _{IJ} \overline{\ccfield{t_{RJ}}})(\st{LK}^* T^{\dot{A}} _{KM} t_{LM})+(L \leftrightarrow R)]\nonumber 
\\ &  +[c^{LR} _{51} (\bar{t}_{LI} T^{\dot{A}} _{IJ} \st{LJ})(\st{RK}^* T^{\dot{A}}_{KM} t_{RM})+c^{LR} _{52} (\st{LI} \st{RI}^*) (\bar{t}_{LJ} t_{RJ})+(L \leftrightarrow R)]
\nonumber \\ &  +c_{61} ^L (\tilde{t}_{LI}^* \tilde{t}_{LI})^3+c_{62} ^L (\tilde{t}_{LI}^* \tilde{t}_{LI})(\tilde{t}_{LJ}^* \tilde{t}_{LK})(\tilde{t}_{LK}^* \tilde{t}_{LJ})+c_{63} ^L (\tilde{t}_{LI}^* \tilde{t}_{LJ})(\tilde{t}_{LJ}^* \tilde{t}_{LK})(\tilde{t}_{LK}^* \tilde{t}_{LI}) \nonumber \\ 
&+c_6^R (\tilde{t}_{RI}^* \tilde{t}_{RI})^3 +[c_{61} ^{LR} (\tilde{t}_{LI}^* \tilde{t}_{LI})^2(\tilde{t}_{RI}^* \tilde{t}_{RI}) +c_{62} ^{LR} (\tilde{t}_{LI}^* \tilde{t}_{LI})(\tilde{t}_{LJ}^* \tilde{t}_{LK})(\tilde{t}_{RK}^* \tilde{t}_{RJ}) \nonumber \\ 
&+c_{63} ^{LR} (\tilde{t}_{LI}^* \tilde{t}_{LJ})(\tilde{t}_{LJ}^* \tilde{t}_{LI})(\tilde{t}_{RK}^* \tilde{t}_{RK}) \nonumber + c_{64} ^{LR} (\tilde{t}_{LI}^* \tilde{t}_{LJ})(\tilde{t}_{LJ}^* \tilde{t}_{LK})(\tilde{t}_{RK}^* \tilde{t}_{RI}) \nonumber \\ & +c_{61} ^{RL} (\tilde{t}_{RI}^* \tilde{t}_{RI})^2(\tilde{t}_{LI}^* \tilde{t}_{LI})  +c_{62} ^{RL} (\tilde{t}_{RI}^* \tilde{t}_{RI})(\tilde{t}_{RJ}^* \tilde{t}_{RK})(\tilde{t}_{LK}^* \tilde{t}_{LJ})]
\nonumber \\ &  +[c_{61} ^{L^\mu L_\mu}\left(\bar{t}_{LI} \gamma^\mu t_{LI}\right)\left(\bar{t}_{LJ} \gamma_\mu t_{LJ}\right)+c_{62} ^{L^\mu L_\mu}\left(\bar{t}_{LI} \gamma^\mu t_{LJ}\right)\left(\bar{t}_{LJ} \gamma_\mu t_{LI}\right)+(L \leftrightarrow R)]\nonumber 
 \\ &  +c_{61}^{(LR)^\mu (RL)_\mu} \left(\overline{\ccfield{t_{RI}}} \gamma^\mu t_{RJ}\right)\left(\bar{t}_{RJ} \gamma_\mu \ccfield{t}_{RI}\right)+ c_{62}^{(LR)^\mu (RL)_\mu} \left(\overline{\ccfield{t_{RJ}}} \gamma^\mu t_{RI}\right)\left(\bar{t}_{RJ} \gamma_\mu \ccfield{t}_{RI}\right)\nonumber \\ 
 &  + [c_{61} ^{LL}\left(\overline{\ccfield{t_{RI}}} t_{LI}\right)\left(\bar{t}_{LJ} \ccfield{t}_{RJ}\right)+c_{62}^{LL}\left(\overline{\ccfield{t_{RI}}} t_{LJ}\right)\left(\bar{t}_{LJ} \ccfield{t}_{RI}\right)+(L\leftrightarrow R)]\nonumber 
 \\ &  +c_{61} ^{(LR)(RL)} \left(\bar{t}_{RI} t_{LJ}\right)\left(\bar{t}_{LJ} t_{RI}\right) +c_{62} ^{(LR)(RL)} \left(\bar{t}_{RJ}t_{LI}\right)\left(\bar{t}_{LJ} t_{RI}\right),
 \label{eq: gluonOutFirstEFTLag}
\end{align}
\endgroup
which represent all one-loop stop interactions in
the gaugeless limit and in the unbroken phase, excluding operators with covariant derivatives. Note that the terms with the couplings $c_{41}^L$ and $c_{42}^L$ have the same structure as presented in Eq.\ \eqref{eq: gluonOutFirstEFTLag}. However, these terms arise from the two distinct $SU(2)_L$ and $SU(3)_c$ invariant combinations $(\tilde{q}^\dagger_{LI}\tilde{q}_{LI})(\tilde{q}^\dagger_{LJ}\tilde{q}_{LJ})$ and
$(\tilde{q}^\dagger_{LI}\tilde{q}_{LJ})(\tilde{q}^\dagger_{LJ}\tilde{q}_{LI})$,
where $\tilde{q}_{L}$ is the $SU(2)_L$  squark doublet and the $SU(2)_L$ indices are contracted within parentheses. Therefore, these couplings can, in principle, take different values. 

The dimension five operators have contributions already at tree level,
which stem from the insertion of the classical gluino field
$\classical{\hat{\gluino{}}}$ into the Lagrangian of the MSSM. As discussed in Chapter~\ref{chap: fcuntional matching}, the classical gluino field is determined from the equation of motion
\begin{align}
[\cc (i\slashed{\partial}-m_{\gluino{}})]_{\dot{\alpha} \dot{\beta}} (\gluinocl)_{\dot{\beta}}^{\dot{A}}=\sqrt{2} g_3 \left(-
    \bar{t}_{L \dot{\alpha}} T^{\dot{A}} \st{L} + \bar{t}_{R \dot{\alpha}}  T^{\dot{A}} \st{R}
    + \st{L}^* T^{\dot{A}} (\cc t_L)_{\dot{\alpha}} - \st{R}^* T^{\dot{A}} (\cc t_R)_{\dot{\alpha}}
  \right),
\end{align}
which yields
\begin{align}
(\gluinocl)_{\dot{\beta}}^{\dot{A}} &= \sqrt{2} g_3 (i\slashed{\partial}-m_{\gluino{}})_{\dot{\beta} \dot{\alpha}}^{-1} \left[-
    (\bar{t}_L \cc) _{\dot{\alpha}} T^{\dot{A}} \st{L} + (\bar{t}_R \cc)_{\dot{\alpha}}  T^{\dot{A}} \st{R}
    + \st{L}^* T^{\dot{A}}  t_{L\dot{\alpha}} - \st{R}^* T^{\dot{A}} t_{R\dot{\alpha}} \right]
\nonumber \\ &= \frac{\sqrt{2} g_3}{m_{\gluino{}}} \left[
    (\bar{t}_L \cc) _{\dot{\beta}} T^{\dot{A}} \st{L} - (\bar{t}_R \cc)_{\dot{\beta}}  T^{\dot{A}} \st{R}
    - \st{L}^* T^{\dot{A}}  t_{L\dot{\beta}} + \st{R}^* T^{\dot{A}} t_{R\dot{\beta}} + \cdots \right] ,
    \label{eq: ClassicalGluinoField}
\end{align}
where the ellipsis designate higher order terms of
$\mathcal{O}(\partial/m_{\gluino{}})$ with at least one
derivative. Inserting Eq.\ \eqref{eq: ClassicalGluinoField} into Eq.\ \eqref{eq: gluinoOut_MSSM_Lag} one finds
the tree-level values of $c_{5i} ^{AB}$ ($A,B\in \{L,R\}$) to be
\begin{align}
c_{51} ^{LL,\tree}&=c_{52} ^{LL,\tree}=c_{51} ^{RR,\tree}=c_{52} ^{RR,\tree}=\frac{g^2_3}{m_{\gluino{}}}, \\
c_{51} ^{LR,\tree}&=c_{51} ^{RL,\tree}=-\frac{2g^2_3}{m_{\gluino{}}}, \\
c_{52} ^{LR,\tree}&=c_{52} ^{RL,\tree}=0.
\end{align}
At the one-loop level the relevant contributions from the BSUOLEA are
\begingroup
\allowdisplaybreaks
\begin{align}
\frac{1}{\kappa}\Lag{EFT}^{1\Loop} = \tr  \Big\{&(-\ZI [q^4]^{31} _{\gluino{} 0} +\frac{m^2_{\gluino{}}}{12} \ZI [q^2] ^{22} _{\gluino{} 0}) \gamma_\mu [P^\nu,\mathbf{U}_{\Xi \xi}] \gamma ^\mu [P_\nu,\mathbf{U}_{\xi \Xi}] \nonumber \\
&  +(-2\ZI [q^4]^{31} _{\gluino{} 0} +\frac{m^2_{\gluino{}}}{6} \ZI [q^2] ^{22} _{\gluino{} 0}) \gamma_\mu [P^\mu,\mathbf{U}_{\Xi \xi}] \gamma ^\nu [P_\nu,\mathbf{U}_{\xi \Xi}] \nonumber \\ &  + (-\ZI [q^{2}]^{12} _{\gluino{}0}-2 m^2_{\phi_i} \ZI[q^2] ^{13} _{\gluino{}0}) (\mathbf{U}_{\phi \Xi})_i \gamma ^\mu [P_\mu,(\mathbf{U}_{\Xi \phi})_i]\nonumber
\\
&  +\frac{1}{4} \ZI [q^2] ^{22}_{\gluino{}0} (\mathbf{U}_{\phi \Xi})_i \gamma^\mu (\mathbf{U}_{\Xi \phi})_j (\mathbf{U}_{\phi \Xi})_j \gamma_\mu (\mathbf{U}_{\Xi \phi})_i
\nonumber \\ &  -\frac{1}{2}m_{\gluino{}} \ZI ^{12}_{\gluino{}0}(\mathbf{U}_{\phi \phi})_{ij} (\mathbf{U}_{\phi \Xi})_j  (\mathbf{U}_{\Xi \phi})_i 
\nonumber \\ &  +\frac{1}{4} m^2_{\gluino{}} \ZI  ^{22}_{\gluino{}0} (\mathbf{U}_{\phi \Xi})_i (\mathbf{U}_{\Xi \phi})_j (\mathbf{U}_{\phi \Xi})_j (\mathbf{U}_{\Xi \phi})_i-\frac{1}{2} \ZI [q^2]^{11} _{{\gluino{}} 0} \gamma^\mu \mathbf{U}_{\Xi \xi} \gamma_\mu \mathbf{U}_{\xi \Xi} \nonumber 
\\ &  -\frac{1}{4}m^2_{\gluino{}} \ZI [q^2] ^{22} _{\gluino{}0} \mathbf{U}_{\Xi \xi} \gamma^\mu \mathbf{U}_{\xi \Xi} \mathbf{U}_{\Xi \xi} \gamma_\mu \mathbf{U}_{\xi \Xi} 
\nonumber 
\\
&  -\frac{1}{4} \ZI [q^4] ^{22} _{\gluino{}0} g_{\mu \nu \rho \sigma}  \mathbf{U}_{\Xi \xi} \gamma^\mu \mathbf{U}_{\xi \Xi} \gamma^\nu \mathbf{U}_{\Xi \xi} \gamma^\rho \mathbf{U}_{\xi \Xi} \gamma^\sigma
\nonumber
\\
&  -\frac{1}{2}m^2 _{\gluino{}} \ZI [q^4] ^{33} _{\gluino{}0}  g_{\mu \nu \rho \sigma}  \mathbf{U}_{\Xi \xi} \gamma^\mu \mathbf{U}_{\xi \Xi}  \mathbf{U}_{\Xi \xi} \gamma^\nu \mathbf{U}_{\xi \Xi}  \gamma^\rho \mathbf{U}_{\Xi \xi} \gamma^\sigma \mathbf{U}_{\xi \Xi}
\nonumber
\\
&  -\frac{1}{6} \ZI [q^6] ^{33} _{\gluino{}0} g_{\mu \nu \rho \sigma \kappa \lambda} \mathbf{U}_{\Xi \xi} \gamma^\mu \mathbf{U}_{\xi \Xi} \gamma^\nu \mathbf{U}_{\Xi \xi} \gamma^\rho \mathbf{U}_{\xi \Xi}  \gamma^\sigma \mathbf{U}_{\Xi \xi} \gamma^\kappa \mathbf{U}_{\xi \Xi} \gamma^\lambda
\nonumber \\ &  +\frac{1}{6}\ZI ^{2} _{\gluino{}}[P_\mu,P_\nu][P^\mu,P^\nu]\Big\},
\label{eq: UOLEAContrGluino}
\end{align}
\endgroup
where $g_{\mu \nu \cdots}$ is the combination of metric tensors which
is totally symmetric in all indices, see
Appendix~\ref{app:loop_functions}. In Eq.\ \eqref{eq: UOLEAContrGluino} we omitted the multi-indices of the fermionic fields as both the light and heavy fermionic multiplets only contain one field. The derivatives with respect to the stops
and the gluino were already calculated in the previous section and are given by
\begingroup
\allowdisplaybreaks
\begin{align}
  \mathbf{U}_{\phi \Xi} &=
  \begin{pmatrix}
    U_{\sigma ^* \Lambda} \\
    U_{\sigma \Lambda}
  \end{pmatrix}
  =
  \begin{pmatrix}
    (U_{\st{L}^* \gluino{\dot{A}}})_{I \dot{\alpha}}^{\dot{A}} \\
    (U_{\st{R}^* \gluino{\dot{A}}})_{I\dot{\alpha}}^{\dot{A}} \\
    (U_{\st{L} \gluino{\dot{A}}})_{I\dot{\alpha}}^{\dot{A}} \\
    (U_{\st{R} \gluino{\dot{A}}})_{I \dot{\alpha}}^{\dot{A}}
  \end{pmatrix}
  = \sqrt{2} g_3
  \begin{pmatrix}
    T^{\dot{A}}_{IJ} (\cc P_L t_J)_{\dot{\alpha}} \\
     -T^{\dot{A}}_{IJ} (\cc P_R t_J)_{\dot{\alpha}} \\
     -(\bar{t}_J P_R)_{\dot{\alpha}} T^{\dot{A}}_{JI} \\
    (\bar{t}_J P_L)_{\dot{\alpha}} T^{\dot{A}}_{JI}
  \end{pmatrix},
  \\
  \mathbf{U}_{\Xi \phi} &=
  \begin{pmatrix}
    \cc^{-1} U_{\Lambda \sigma} & \cc^{-1} U_{\Lambda \sigma ^*}
  \end{pmatrix} \nonumber \\
  &= (\cc^{-1})_{\dot{\alpha}\dot{\beta}}
  \begin{pmatrix}
    (U_{\gluino{\dot{A}} \st{L}})_{I\dot{\beta}}^{\dot{A}} &
    (U_{\gluino{\dot{A}} \st{R}})_{I\dot{\beta}}^{\dot{A}} &
    (U_{\gluino{\dot{A}} \st{L}^*})_{I\dot{\beta}}^{\dot{A}} &
    (U_{\gluino{\dot{A}} \st{R}^*})_{I\dot{\beta}}^{\dot{A}}
  \end{pmatrix} \nonumber \\
  &= \sqrt{2} g_3
  \begin{pmatrix}
    -(\bar{t}_J P_R \cc)_{\dot{\alpha}} T^{\dot{A}}_{JI} &
   (\bar{t}_J P_L \cc)_{\dot{\alpha}} T^{\dot{A}}_{JI} &
   T^{\dot{A}}_{IJ} (P_L t_J)_{{\dot{\alpha}}} &
    -T^{\dot{A}}_{IJ} (P_R t_J)_{\dot{\alpha}}
  \end{pmatrix},
\end{align}
\endgroup
with the difference that the stops are now considered to be light
fields.
We further need the derivatives with
respect to a top and a gluino, which read
\begin{align}
(U_{\bar{t} \gluino{\dot{A}}})_{I \dot{\alpha} \dot{\beta}}^{\dot{A}}&=-\sqrt{2}g_3T^{\dot{A}}_{IJ}\left[(P_R)_{\dot{\alpha}\dot{\beta}}\st{LJ}-(P_L)_{\dot{\alpha}\dot{\beta}}\st{RJ}\right],\\
(U_{t \gluino{\dot{A}}})_{I \dot{\alpha} \dot{\beta}}^{\dot{A}} &=-\sqrt{2}g_3T^{\dot{A}}_{JI}\left[-\st{LJ}^*(\cc P_L)_{\dot{\beta} \dot{\alpha}}+\st{RJ}^*(\cc P_R)_{\dot{\beta} \dot{\alpha}}\right],\\
(U_{\gluino{\dot{A}}\bar{t}})_{I \dot{\alpha} \dot{\beta}}^{\dot{A}}&=\sqrt{2}g_3T^{\dot{A}}_{IJ}\left[(P_R)_{\dot{\beta} \dot{\alpha}}\st{LJ}-(P_L)_{\dot{\beta} \dot{\alpha}}\st{RJ}\right],\\
(U_{ \gluino{\dot{A}} t})_{I \dot{\alpha} \dot{\beta}}^{\dot{A}}&=\sqrt{2}g_3T^{\dot{A}}_{JI}\left[-\st{LJ}^*(\cc P_L)_{ \dot{\alpha} \dot{\beta}}+\st{RJ}^*(\cc P_R)_{\dot{\alpha} \dot{\beta}}\right],
\end{align}
and are collected into
\begin{align}
\mathbf{U}_{\Xi \xi}&=\begin{pmatrix}
\cc ^{-1} U_{\Lambda \omega} & \cc ^{-1} U_{\Lambda \bar{\omega}} \cc ^{-1}
\end{pmatrix} \nonumber \\
&=\begin{pmatrix}
(\cc ^{-1}U_{ \gluino{\dot{A}} t})_{I \dot{\alpha} \dot{\beta}}^{\dot{A}}  & (\cc ^{-1} U_{\gluino{\dot{A}}\bar{t}} \cc ^{-1})_{I \dot{\alpha} \dot{\beta}}^{\dot{A}} 
\end{pmatrix} \nonumber \\
&= \begin{pmatrix}
-\sqrt{2}g_3T^{\dot{A}}_{JI}\left[\st{LJ}^*(P_L)_{\dot{\alpha} \dot{\beta}}-\st{RJ}^*(P_R)_{\dot{\alpha} \dot{\beta}}\right]  & -\sqrt{2}g_3T^{\dot{A}}_{IJ}\left[(P_R)_{\dot{\alpha} \dot{\beta}}\st{LJ}-(P_L)_{\dot{\alpha} \dot{\beta}}\st{RJ}\right]
\end{pmatrix}, \\ 
\mathbf{U}_{\xi \Xi}&=\begin{pmatrix}
 U_{\bar{\omega} \Lambda} \\
  \cc^{-1} U_{\omega \Lambda} 
\end{pmatrix}=\begin{pmatrix}
 (U_{\bar{t} \gluino{\dot{A}}})_{I \dot{\alpha} \dot{\beta}}^{\dot{A}} \\
  (\cc^{-1}U_{t \gluino{\dot{A}}})_{I \dot{\alpha} \dot{\beta}}^{\dot{A}}
\end{pmatrix}=\begin{pmatrix}
 -\sqrt{2}g_3T^{\dot{A}}_{IJ}\left[(P_R)_{\dot{\alpha} \dot{\beta}}\st{LJ}-(P_L)_{\dot{\alpha}\dot{\beta}}\st{RJ}\right]\\
  -\sqrt{2}g_3T^{\dot{A}}_{JI}\left[\st{LJ}^*(P_L)_{\dot{\alpha} \dot{\beta}}-\st{RJ}^*(P_R)_{\dot{\alpha} \dot{\beta}}\right]
\end{pmatrix}.
\end{align}
Finally we give the derivatives with respect to two stops
\begin{align}
\mathbf{U}_{\phi \phi} &= \begin{pmatrix}
\mathbf{Y}_{\phi \phi} & \mathbf{0}_{2\times2} \\
\mathbf{0}_{2\times2} & (\mathbf{Y}_{\phi \phi})^*
\end{pmatrix},\\
\mathbf{Y}_{\phi \phi} &=
\begin{pmatrix}
x_t \st{RJ}^* \st{RI}-\frac{g_3 ^2}{6 }\st{R}^* \st{R} \delta_{IJ} && x_t  \delta_{IJ} \st{L} \st{R}^*-\frac{g_3^2}{6}\st{LI} \st{RJ}^* \\
x_t  \delta_{IJ} \st{L}^* \st{R}-\frac{g_3^2}{6}\st{RI} \st{LJ}^* && x_t \st{LJ}^* \st{LI}-\frac{g_3 ^2}{6 }\st{L}^* \st{L} \delta_{IJ}
\end{pmatrix},
\end{align}
where we introduced the abbreviation $x_t \equiv y_t^2-g_3 ^2/2$.
Substituting these derivatives into Eq.\ \eqref{eq: UOLEAContrGluino} and
summing over all indices one finds
\begingroup
\allowdisplaybreaks
\begin{align}
\frac{1}{\kappa}c_{t_L}&=\frac{16}{3} g^2_3\left(\ZI [q^{2}]^{12} _{\gluino{}0}+2 m^2_{\tilde{q}} \ZI[q^2] ^{13} _{\gluino{}0}\right), \\
\frac{1}{\kappa}c_{t_R}&=\frac{16}{3} g^2_3\left(\ZI [q^{2}]^{12} _{\gluino{}0}+2 m^2_{\tilde{u}} \ZI[q^2] ^{13} _{\gluino{}0}\right), \\
\frac{1}{\kappa}c_{\st{L}}&=\frac{1}{\kappa}c_{\st{R}}=\frac{32}{3} g^2_3(d+2)\left(-\ZI [q^{4}]^{31} _{\gluino{}0}+ \frac{m^2_{\tilde{q}}}{2} \ZI[q^2] ^{22} _{\gluino{}0}\right),\\
\frac{1}{\kappa}c_{61}^{L^\mu L_\mu}&=\frac{1}{\kappa}c_{61} ^{R^\mu R_\mu}=\frac{7}{6} g^4_3 \ZI [q^2] ^{22}_{\gluino{}0}, \\
\frac{1}{\kappa}c_{62} ^{L^\mu L_\mu}&=\frac{1}{\kappa}c_{62} ^{ R^\mu R_\mu}=\frac{1}{18} g^4_3 \ZI [q^2] ^{22}_{\gluino{}0}, \\
\frac{1}{\kappa}c_{61} ^{(LR)^\mu (RL)_\mu}&=\frac{10}{9} g_3^4 \ZI [q^2] ^{22}_{\gluino{}0}, \\
\frac{1}{\kappa}c_{62}^{(LR)^\mu (RL)_\mu}&=-\frac{2}{9} g_3^4 \ZI [q^2] ^{22}_{\gluino{}0}, \\
\frac{1}{\kappa}c_{61} ^{LL}&=\frac{1}{\kappa}c_{61} ^{ R R}=\frac{5}{18} g^4_3 m^2_{\gluino{}}\ZI [q^2] ^{22}_{\gluino{}0}, \\
\frac{1}{\kappa}c_{62} ^{LL}&=\frac{1}{\kappa}c_{62} ^{ R R}=-\frac{1}{6} g^4_3 m^2_{\gluino{}}\ZI [q^2] ^{22}_{\gluino{}0}, \\
\frac{1}{\kappa}c_{61} ^{(LR)(RL)}&=\frac{7}{6} g_3^4 m^2_{\gluino{}}\ZI [q^2] ^{22}_{\gluino{}0}, \\
\frac{1}{\kappa}c_{62} ^{(LR) (RL)}&=\frac{1}{18} g_3^4 m^2_{\gluino{}}\ZI [q^2] ^{22}_{\gluino{}0}, \\
\frac{1}{\kappa}\delta m_{\sq} ^2 &= \frac{1}{\kappa}\delta m_{\su} ^2 =\frac{16}{3} dg^2 _3 \ZI[q^2] ^{11} _{\gluino{}0}, \\
\frac{1}{\kappa}c^L _{41} &= -\frac{40}{9}m^2_{\gluino{}} g_3^4 \ZI[q^2] ^{22} _{\gluino{}0}-\frac{1}{9}d(d+2)g_3^4 \ZI[q^4] ^{22} _{\gluino{}0}, \\
\frac{1}{\kappa}c^R _{4} &= -\frac{16}{3}m^2_{\gluino{}} g_3^4 \ZI[q^2] ^{22} _{\gluino{}0}-\frac{22}{9}d(d+2)g_3^4 \ZI[q^4] ^{22} _{\gluino{}0}, \\
\frac{1}{\kappa}c^L _{42} &= \frac{8}{3}m^2_{\gluino{}} g_3^4 \ZI[q^2] ^{22} _{\gluino{}0}-\frac{7}{3}d(d+2)g_3^4 \ZI[q^4] ^{22} _{\gluino{}0}, \\
\frac{1}{\kappa}c^{LR} _{41} &= -\frac{8}{9}m^2_{\gluino{}} g_3^4 \ZI[q^2] ^{22} _{\gluino{}0}-\frac{20}{9}d(d+2)g_3^4 \ZI[q^4] ^{22} _{\gluino{}0}, \\
\frac{1}{\kappa}c^{LR} _{42} &= -\frac{56}{3}m^2_{\gluino{}} g_3^4 \ZI[q^2] ^{22} _{\gluino{}0}+\frac{4}{9}d(d+2)g_3^4 \ZI[q^4] ^{22} _{\gluino{}0}, \\
\frac{1}{\kappa}c^{L} _{61} &= \frac{1}{54}d(d+2) g_3^6 m^2_{\gluino{}} \ZI[q^4] ^{33} _{\gluino{}0}+\frac{2}{81}d(d^2+6d+8)g_3^6  \ZI[q^6] ^{33} _{\gluino{}0},	 \\
\frac{1}{\kappa}c^{L} _{62} &=- \frac{2}{3}d(d+2) g_3^6 m^2_{\gluino{}} \ZI[q^4] ^{33} _{\gluino{}0}-\frac{2}{9}d(d^2+6d+8)g_3^6  \ZI[q^6] ^{33} _{\gluino{}0}, \\
\frac{1}{\kappa}c^{L} _{63} &= \frac{1}{2}d(d+2) g_3^6 m^2_{\gluino{}} \ZI[q^4] ^{33} _{\gluino{}0}-\frac{4}{3}d(d^2+6d+8)g_3^6  \ZI[q^6] ^{33} _{\gluino{}0}, \\
\frac{1}{\kappa}c^{R} _{6} &=-\frac{4}{27}d(d+2) g_3^6 m^2_{\gluino{}} \ZI[q^4] ^{33} _{\gluino{}0}-\frac{124}{81}d(d^2+6d+8)g_3^6  \ZI[q^6] ^{33} _{\gluino{}0}, \\
\frac{1}{\kappa}c^{LR} _{61} &= \frac{1}{18}d(d+2) g_3^6 m^2_{\gluino{}} \ZI[q^4] ^{33} _{\gluino{}0} +\frac{2}{27}d(d^2+6d+8)g_3^6  \ZI[q^6] ^{33} _{\gluino{}0},\\
\frac{1}{\kappa}c^{LR} _{62} &=- \frac{12}{9}d(d+2) g_3^6 m^2_{\gluino{}} \ZI[q^4] ^{33} _{\gluino{}0}-\frac{10}{9}d(d^2+6d+8)g_3^6  \ZI[q^6] ^{33} _{\gluino{}0}, \\
\frac{1}{\kappa}c^{LR} _{63} &= -\frac{1}{6}d(d+2) g_3^6 m^2_{\gluino{}} \ZI[q^4] ^{33} _{\gluino{}0}-\frac{14}{9}d(d^2+6d+8)g_3^6  \ZI[q^6] ^{33} _{\gluino{}0}, \\
\frac{1}{\kappa}c^{LR} _{64} &= \frac{2}{9}d(d^2+6d+8) g_3^6  \ZI[q^6] ^{33} _{\gluino{}0}, \\
\frac{1}{\kappa}c^{RL} _{61} &= -\frac{1}{9}d(d+2) g_3^6 m^2_{\gluino{}} \ZI[q^4] ^{33} _{\gluino{}0}-\frac{40}{27}d(d^2+6d+8)g_3^6  \ZI[q^6] ^{33} _{\gluino{}0},\\
\frac{1}{\kappa}c^{RL} _{62} &=- \frac{12}{9}d(d+2) g_3^6 m^2_{\gluino{}} \ZI[q^4] ^{33} _{\gluino{}0}+\frac{8}{9}d(d^2+6d+8)g_3^6  \ZI[q^6] ^{33} _{\gluino{}0}, \\
\frac{1}{\kappa}c_{51} ^{LR\text{,1\Loop}}&=\frac{1}{\kappa}c_{51} ^{RL\text{,1\Loop}}=-\frac{g_3^4}{3} m_{\gluino{}} \ZI ^{12}_{\gluino{}0}, \\
\frac{1}{\kappa}c_{52} ^{LR\text{,1\Loop}}&=\frac{1}{\kappa}c_{52} ^{RL\text{,1\Loop}}=-\frac{8}{3} g_3^4 x_t m_{\gluino{}} \ZI ^{12}_{\gluino{}0}, \\
\frac{1}{\kappa}c_G&=-\frac{g_3^2}{2} \ZI ^{2} _{\gluino{}}.
\end{align}
\endgroup
The threshold corrections for the two stop mass parameters were previously computed in Ref.\ \cite{Aebischer:2017aqa} and the results agree when the effect of the sbottom quarks is neglected.

As in Section~\ref{sec:lambdacalc} it is again convenient
to use DRED as a regulator in the MSSM. Since supersymmetry is explicitly broken once the gluino is
integrated out from the theory, it is natural to regularise the EFT in
DREG. This switch in regularisation again introduces further
contributions to the couplings of the EFT coming from the epsilon
scalars, which can be computed using Eq.\ \eqref{eq: masterformula}.
We obtain the following
additional contributions
\begingroup
\allowdisplaybreaks
\begin{align}
\frac{1}{\kappa}(\delta m^2 _{\sq})_\eps &=\frac{1}{\kappa}(\delta m ^2 _{\su})_\eps=-\frac{4}{3} g_3^2 m_\eps ^2, \label{eq:delta_m2_eps} \\
\frac{1}{\kappa}(c_{t_L})_\eps &= \frac{1}{\kappa}(c_{t_R})_\eps=\frac{4}{3}g_3^2, \\
\frac{1}{\kappa}(c^L_{41})_\eps &= \frac{1}{72}g_3^4, \\
\frac{1}{\kappa}(c^L_{42})_\eps&=\frac{7}{24}g_3^4, \\
\frac{1}{\kappa}(c^R_{4})_\eps &= \frac{11}{36}g_3^4, \\
\frac{1}{\kappa}(c^{LR}_{41})_\eps &= \frac{1}{36}g_3^4, \\
\frac{1}{\kappa}(c^{LR}_{42})_\eps &= \frac{7}{12}g_3^4, \\
\frac{1}{\kappa}(c^{LL}_{51})_\eps &= \frac{1}{\kappa}(c^{LL}_{52})_\eps = \frac{1}{\kappa}(c^{RR}_{51})_\eps=\frac{1}{\kappa}(c^{RR}_{52})_\eps = \frac{3g_3^4 }{2 m_{\gluino{}}}d, \\
\frac{1}{\kappa}(c^{LR}_{51})_\eps &= \frac{1}{\kappa}(c^{RL}_{52})_\eps = -\frac{3g_3^4 }{m_{\gluino{}}}d, \\
\frac{1}{\kappa}(c_G)_\eps &= -\frac{g_3^2}{4}.
\end{align}
\endgroup
%
Notice that the one-loop DRED--DREG conversion corrections to the
coefficients of the dimension five operators arise from the term
\begin{align}
  \left(\epsdim{U}_{\eps t}\right)^\mu \gamma ^\nu U_{\bar{t} \gluino{\dot{A}}} \gamma_{\nu} \left(\epsdim{U}_{\gluino{\dot{A}} \eps}\right)_\mu.
\end{align}
Here $\left(\epsdim{U}_{\gluino{\dot{A}} \eps}\right)^\mu$ has an explicit
dependence on the gluino spinor $\gluino{\dot{A}}$,
\begin{align}
\left(\epsdim{U}_{\gluino{\dot{A}} \eps}\right)^\mu=ig_3\epsdim{\gamma}^\mu f^{\dot{A}\dot{B}\dot{C}}\gluino{\dot{C}},
\end{align}
which must be eliminated by inserting the classical field from Eq.\
\eqref{eq: ClassicalGluinoField}. 
\section{Matching of the SSM to the SMEFT}
\label{sec: SSM_to_SM}
In this section we consider the SSM, which is the SM with an additional real scalar field $S$, transforming as a singlet under the SM gauge group. We assume the singlet to be heavy and integrate it out of the theory using the BSUOLEA to obtain SMEFT operators up to mass dimension six. The final result is then expressed in the Warsaw basis. This particular matching was first computed in Ref.\ \cite{Jiang:2018pbd} using the UOLEA presented in Ref.\ \cite{Ellis:2017jns} supplemented by Feynman diagrammatic techniques for loops of mixed statistics and loops including gauge bosons. The matching was also re-computed using Feynman diagrammatic techniques in Ref.\ \cite{Haisch:2020ahr} for the purpose of studying the effect of SMEFT operators on one-loop probes of pseudo Nambu-Goldstone boson dark matter. We here, for the first time, perform the full matching using the UOLEA approach, illustrating that the extensions that were presented in this thesis are needed even for one of the simplest SM extensions of interest to particle phenomenology. Furthermore, this section can be viewed as both a check of the BSUOLEA and its implementation into \tofu as well as an illustration of how to consistently treat a UV theory that contains the SM field content and gauge group in the BSUOLEA framework.

As the SSM is a chiral gauge theory it is appropriately described using Weyl fermions when working in the unbroken phase. Strictly speaking, we only considered Dirac and Majorana fermions in this thesis. However, for the purpose of a perturbative calculation Weyl fermions may be implemented as either Dirac fermions or Majorana fermions, as the chiral nature of the interactions will always project out the physically relevant components of these four-component spinors.\footnote{A discussion of this and related issues can be found in Ref.\ \cite{Bonora:2020upc}. Interestingly, in non-perturbative considerations Weyl fermions cannot be treated as either Dirac fermions or Majorana fermions.} We here choose to treat them as Majorana fermions. To this end we introduce the left-handed two-component spinors $q,\bar{u},\bar{d},l$ and $\bar{e}$, where the bar on $u,d,e$ is part of the name and does not signify any kind of conjugation. These spinors all carry flavour indices. The charges under the SM gauge group $SU(3)_c \times SU(2)_L \times U(1)_Y$ carried by these fields are summarised in Table~\ref{table: gauge_charges} below, where we also include the Higgs doublet $H$ and the singlet $S$.
\begin{table}[h]
\centering
\begin{tabular}{c|c|c|c|c}
\cline{2-4}
                       & $SU(2)_L$ & $U(1)_Y$ & $SU(3)_c$ &  \\ \cline{1-4}
\multicolumn{1}{|c|}{$q$}  & $2$ & $1/6$ & $3$ &  \\ \cline{1-4}
\multicolumn{1}{|c|}{$\bar{u}$} & $1$ & $-2/3$ & $\bar{3}$ &  \\ \cline{1-4}
\multicolumn{1}{|c|}{$\bar{d}$} & $1$ & $1/3$ & $\bar{3}$ &  \\ \cline{1-4}
\multicolumn{1}{|c|}{$l$}  & $2$ & $-1/2$ & $1$ &  \\ \cline{1-4}
\multicolumn{1}{|c|}{$\bar{e}$} & $1$ & $1$ & $1$ &  \\ \cline{1-4}
\multicolumn{1}{|c|}{$H$} & $2$ & $1/2$ & $1$ &  \\ \cline{1-4}
\multicolumn{1}{|c|}{$S$} & $1$ & $0$ & $1$ &  \\ \cline{1-4}
\end{tabular}
\caption{Gauge charges of the fields in the SSM.}
\label{table: gauge_charges}
\end{table}
Note that the $U(1)_Y$ charges of the barred fields are opposite to those of the corresponding SM fields since the barred fields are left-handed and the corresponding SM fields are related to these by Hermitian conjugation. We then introduce the Majorana spinors 
\begin{align}
\label{eq: singletExtension: Majorana_spinor_definitions}
    \Psi_q = \begin{pmatrix}
    q \, \\ q^\dagger
    \end{pmatrix}, \Psi_u = \begin{pmatrix}
    \bar{u} \; \\ \bar{u}^\dagger \end{pmatrix}, 
    \Psi_d = \begin{pmatrix}
    \bar{d} \; \\ \bar{d}^\dagger
    \end{pmatrix},
    \Psi_l = \begin{pmatrix}
    l \, \\ l^\dagger
    \end{pmatrix},
    \Psi_e = \begin{pmatrix}
    \bar{e} \; \\ \bar{e}^\dagger \end{pmatrix},
\end{align}
which transform in chiral representations of the SM gauge group. Observe that the fields $e = \bar{e}^\dagger$, $d = \bar{d}^\dagger$ and $u = \bar{u}^\dagger$ are the usual right-handed spinors introduced in the SM. The generators for the $U(1)_Y$- and $SU(2)_L$-transformation of $\Psi_f$ can be written as
\begin{align}
    Y_{\Psi_f} = Y_f P_L-Y_f P_R \text{\; and \;} & T^A_{\Psi_f} = T^A_f P_L + \bar{T}^A_f P_R,
\end{align}
respectively, where $T^A_f$ is the usual generator of $SU(2)_L$ in the representation under which the fermion $f\in \{q,d,u,l,e\}$ transforms and $\bar{T}^A_f$ is the generator of the corresponding conjugate representation. Furthermore, $Y_f$ is the hypercharge of the left-handed component of $\Psi_f$. As already mentioned, the details of this construction are inconsequential to the computation since the SSM interactions are chirality preserving and therefore always project out the physical components of these spinors. 

With these preparations we may write the Lagrangian of the SSM as
\begin{align}
\label{eq: SSM_Lag}
\LagNoT &= \LagNoT_S + (D_\mu H)^\dagger (D^\mu H) + \mu_h^2 |H|^2 - \frac{1}{2} \lambda_h |H|^4 - \frac{1}{4} B^{\mu \nu} B_{\mu \nu} - \frac{1}{4} W^{A \mu \nu} W^A_{\mu \nu} \nonumber \\ & \quad  +\frac{i}{2} \sum \limits_{i \in \{q,u,d,l,e\}}\bar{\Psi}_i \slashed{D} \Psi_i - \left(\bar{\Psi}_q y_u \tilde{H} P_R \Psi_u + \bar{\Psi}_q y_d H P_R \Psi_d + \bar{\Psi}_l y_e H P_R \Psi_e + \text{h.c.} \right), 
\end{align}
with 
\begin{align}
    \LagNoT_S = \frac{1}{2}\left(\partial_\mu S\right)\left(\partial^\mu S\right)-\frac{1}{2} M^2 S^2-A |H|^2 S-\frac{\kappa}{2}|H|^2 S^2 -\frac{\mu}{3!}S^3-\frac{\lambda_S}{4!}S^4. 
\end{align}
In Eq.\ \eqref{eq: SSM_Lag} we introduced $\tilde{H}_a = \epstensor_{ab} H^*_b$ with $\epstensor_{ab}$ the totally antisymmetric symbol with two indices and $\epstensor_{12}=1$. The field strength tensors $W^A_{\mu \nu}$ and $B_{\mu \nu}$ correspond to the gauge-kinetic terms of the $SU(2)_L$ and $U(1)_Y$ gauge bosons, respectively. The corresponding gauge couplings will be denoted by $g_1$ and $g_2$. We ignore QCD as it is not relevant for the matching presented here.

We start by briefly describing the tree-level matching. To perform this matching we need to compute the classical field $\classical{S}$, which is determined from the equation of motion of the singlet
\begin{align}
    \label{eq: singlet_eom}
    \left(-\Box - M^2\right) S = A |H|^2 + \kappa |H|^2 S + \frac{\mu}{2} S^2 + \frac{\lambda_S}{3!} S^3.  
\end{align}
This can be solved iteratively in an expansion in $1/M$. Care has to be taken as the parameters $A$ and $\mu$ are dimensionful and are taken to be of $\mathcal{O}(M)$. The lowest order solution is 
\begin{align}
\classical{S}^{\left(1\right)} = -\frac{A}{M^2} |H|^2,
\end{align}
which inserted into the r.h.s.\ of Eq.\ \eqref{eq: singlet_eom} allows for the determination of $\classical{S}^{\left(2\right)}$
\begin{align}
\left(-\Box-M^2\right) \classical{S}^{\left(2\right)} &= A |H^2| -\frac{A \kappa}{M^2} |H|^4 + \mathcal{O}\left(M^{-3}\right) \nonumber \\ 
\Rightarrow \classical{S}^{\left(2\right)} &= -\frac{A}{M^2} |H|^2 + \frac{A}{M^4} \Box |H|^2+ \frac{A \kappa}{M^4} |H|^4.
\end{align}
By repeatedly inserting the obtained solution into the r.h.s.\ of Eq.\ \eqref{eq: singlet_eom} one can determine the classical solution to the desired order in $M^{-1}$. The final result is given by 
\begin{align}
\label{eq: full_classical_singlet}
    \classical{\hat{S}} &= -\frac{A}{M^2}|H|^2+\left(\frac{A\kappa}{M^4}-\frac{A^2\mu}{2M^6}\right)|H|^4 + \frac{A}{M^4} \Box |H|^2 - \left(\frac{A\kappa}{M^6}-\frac{A^2\mu}{M^8} \right)|H|^2\Box |H|^2 \nonumber \\
    & \quad - \left(\frac{A\kappa}{M^6}-\frac{A^3\lambda_\phi+9A^2\kappa \mu}{6M^8}+\frac{A^3 \mu^2}{2M^10}\right)|H|^6,
\end{align}
which was already reported in Ref.\ \cite{Haisch:2020ahr}. Inserting this into the UV Lagrangian and truncating at mass dimension six one finds the tree-level results reported in that reference. The computation of the classical heavy fields and the tree-level matching is also fully automated in \tofu.

We now move to the one-loop matching. The multiplets defined in Table~\ref{table: field_content} containing the different kinds of fields are in the case at hand
\begin{align}
\label{eq: singletApp_Phi_def}
    \Phi &= \begin{pmatrix}
    S
    \end{pmatrix}, \\
    \label{eq: singletApp_xi_def}
    \xi &= \begin{pmatrix}
    \Psi_{qi}^{\dot{\alpha} a} & \Psi_{ui}^{\dot{\alpha}} & \Psi_{di}^{\dot{\alpha}} & \Psi_{li}^{\dot{\alpha} a} & \Psi_{ei}^{\dot{\alpha}}
    \end{pmatrix}^T,\\
    \label{eq: singletApp_phi_def}
    \phi &= \begin{pmatrix}
    H_a & H^*_a & B_\mu & W^A_\mu
    \end{pmatrix}^T,
\end{align}
  where there is no multiplet $\Xi$ since there are no heavy fermions present. Also, since we are treating all of the fermions as Majorana fermions we have $\xi=\lambda$. Next, the relevant derivative matrices have to be computed. The matrices associated with at least one derivative w.r.t.\ a heavy field originate from $\LagNoT_S$ and are given by
\begin{align}
\label{eq: singletExtension_X_Phi_Phi}
    \mathbf{U}_{\Phi \Phi} &= \begin{pmatrix}
    \kappa |H|^2 + \mu \classical{\hat{S}} + \frac{\lambda_S}{2} \classical{\hat{S}}
    \end{pmatrix}, \\
    \label{eq: singletExtension_X_phi_Phi}
    \mathbf{U}_{\phi \Phi} &= \begin{pmatrix}
    A H_a + \kappa H_a \classical{\hat{S}} \\ 
    A H^*_a + \kappa H^*_a \classical{\hat{S}} \\ 
    0 \\
    0
    \end{pmatrix}, \\
    \label{eq: singletExtension_X_Phi_phi}
    \mathbf{U}_{\Phi \phi} &= \begin{pmatrix}
    A H^*_a + \kappa H^*_a \classical{\hat{S}} & 
    A H_a + \kappa H_a \classical{\hat{S}} & 
    0 &
    0
    \end{pmatrix}.
    \end{align}
In Eqs.\ \eqref{eq: singletExtension_X_Phi_Phi}--\eqref{eq: singletExtension_X_Phi_phi} the expression of Eq.\ \eqref{eq: full_classical_singlet} has to be inserted. Many of the contributions to the remaining matrices originate entirely from the SM part of the Lagrangian and will be present in any application in which the UV theory contains the SM. We therefore present these matrices in their entirety starting with $\mathbf{U}_{\phi \phi}$, which is given by
    \begin{align}
    \mathbf{U}_{\phi \phi} &= \begin{pmatrix}
    \left(U_{H^* H}\right)_{ab} & \left(U_{H^* H^*}\right)_{ab} & \left(U_{H^* B}\right)_{a \nu} & \left(U_{H^* W}\right)_{a \nu B} \\
    \left(U_{H H}\right)_{ab} & \left(U_{H H^*}\right)_{ab} & \left(U_{H B}\right)_{a \nu} & \left(U_{H W}\right)_{a \nu B} \\
    -\left(U_{B H}\right)_{\mu b} & -\left(U_{B H^*}\right)_{\mu b} & -\left(U_{B B}\right)_{\mu \nu} & -\left(U_{B W}\right)_{\mu \nu B} \\
    -\left(U_{W H}\right)_{\mu A b} & -\left(U_{W H^*}\right)_{\mu A b} & -\left(U_{W B}\right)_{\mu A \nu} & -\left(U_{W W}\right)_{\mu A \nu B}
    \end{pmatrix}, 
    \end{align} 
    where
    \begingroup
    \allowdisplaybreaks
    \begin{align}
    \left(U_{H^* H}\right)_{ab} &= \lambda_h |H|^2 \delta_{ab}+\lambda_h H_a H^*_b+A \delta_{ab} \classical{\hat{S}} +\frac{\kappa}{2}\delta_{ab} \classical{\hat{S}}^2,\\
    \left(U_{H^* H^*}\right)_{ab} &=  \lambda_h H_a H_b,\\
    \left(U_{H^* B}\right)_{a\mu} &= \left(U_{B H^*}\right)_{\mu a} = \frac{g_1}{2} [P_\mu,H_a], \\ 
    \left(U_{H^* W}\right)_{a\mu A} &= \left(U_{W H^*}\right)_{\mu A a} = g_2 T^A_{ab} [P_\mu, H_b], \\
    U_{H H} &= \lambda_h H^*_a H^*_b, \\
    \left(U_{H H^*}\right)_{a b} &= \lambda_h |H|^2 \delta_{ab}+\lambda_h H_b H^*_a+A \delta_{ab} \classical{\hat{S}} +\frac{\kappa}{2}\delta_{ab} \classical{\hat{S}}^2,\\ 
    \left(U_{H B}\right)_{a \mu} &= \left(U_{B H}\right)_{\mu a} = -\frac{g_1}{2} [P_\mu,H^*_a], \\
    \left(U_{H W}\right)_{a \mu A} &= \left(U_{W H}\right)_{\mu A a} =  -g_2 T^A_{ca} [P_\mu,H^*_c], \\
    \left(U_{B B}\right)_{\mu \nu} &= -\frac{g_1^2}{2} |H|^2 g_{\mu \nu}, \\
    \left(U_{B W}\right)_{\mu \nu A} & = \left(U_{W B}\right)_{\mu A \nu} = -g_1 g_2 H^*_a T^A_{ab} H_b g_{\mu \nu}, \\
    \left(U_{W W}\right)_{\mu A \nu B} & = -g_2^2 H^*_a\{T^A,T^B\}_{ab} H_b g_{\mu \nu}.
    \end{align}
    \endgroup
The matrices with derivatives of mixed statistics are
    \begin{align}
    \mathbf{U}_{\xi \phi} &= \cc^{-1}_{\dot{\alpha}\dot{\beta}} 
    \begin{pmatrix}
    \left(U_{\Psi_q H}\right)_{\dot{\beta}aib} & \left(U_{\Psi_q H^*}\right)_{\dot{\beta}aib} & \left(U_{\Psi_q B}\right)_{\dot{\beta}ai\mu} & \left(U_{\Psi_q W}\right)_{\dot{\beta}ai\mu A} \\
    \left(U_{\Psi_u H}\right)_{\dot{\beta}ib} & \left(U_{\Psi_u H^*}\right)_{\dot{\beta}ib} & \left(U_{\Psi_u B}\right)_{\dot{\beta}i\mu} & \left(U_{\Psi_u W}\right)_{\dot{\beta}ai\mu A} \\
    \left(U_{\Psi_d H}\right)_{\dot{\beta}ib} & \left(U_{\Psi_d H^*}\right)_{\dot{\beta}ib} & \left(U_{\Psi_d B}\right)_{\dot{\beta}i\mu} & \left(U_{\Psi_d W}\right)_{\dot{\beta}ai\mu A} \\
    \left(U_{\Psi_l H}\right)_{\dot{\beta}aib} & \left(U_{\Psi_l H^*}\right)_{\dot{\beta}aib} & \left(U_{\Psi_l B}\right)_{\dot{\beta}ai\mu} & \left(U_{\Psi_l W}\right)_{\dot{\beta}ai\mu A} \\
    \left(U_{\Psi_e H}\right)_{\dot{\beta}ib} & \left(U_{\Psi_e H^*}\right)_{\dot{\beta}ib} & \left(U_{\Psi_e B}\right)_{\dot{\beta}i\mu} & \left(U_{\Psi_e W}\right)_{\dot{\beta}ai\mu A}
    \end{pmatrix},
    \end{align}
    and
    \begin{align}
    \mathbf{U}_{\phi \xi} &=  
    \begin{pmatrix}
    \left(U_{H^* \Psi_q}\right)_{a\dot{\beta}bi} & \left(U_{H^* \Psi_u}\right)_{a\dot{\beta}i} & \left(U_{H^* \Psi_d}\right)_{a\dot{\beta}i} & \left(U_{H^* \Psi_l}\right)_{a\dot{\beta}bi} & \left(U_{H^* \Psi_e}\right)_{a\dot{\beta}i} \\
   \left(U_{H \Psi_q}\right)_{a\dot{\beta}bi} & \left(U_{H \Psi_u}\right)_{a\dot{\beta}i} & \left(U_{H \Psi_d}\right)_{a\dot{\beta}i} & \left(U_{H \Psi_l}\right)_{a\dot{\beta}bi} & \left(U_{H \Psi_e}\right)_{a\dot{\beta}i} \\
    -\left(U_{B \Psi_q}\right)^{\mu}_{ \dot{\beta}bi} & -\left(U_{B \Psi_u}\right)^{\mu}_{\dot{\beta}i} & -\left(U_{B \Psi_d}\right)^\mu_{\dot{\beta}i} & -\left(U_{B \Psi_l}\right)^\mu_{\dot{\beta}bi} & -\left(U_{B \Psi_e}\right)^\mu_{\dot{\beta}i} \\
    -\left(U_{W \Psi_q}\right)^{\mu A}_{\dot{\beta}bi} & -\left(U_{W \Psi_u}\right)^{\mu A}_{\dot{\beta}i} & -\left(U_{W \Psi_d}\right)^{\mu A}_{\dot{\beta}i} & -\left(U_{W \Psi_l}\right)^{\mu A}_{\dot{\beta}bi} & -\left(U_{W \Psi_e}\right)^{\mu A}_{\dot{\beta}i}
    \end{pmatrix},
    \end{align}
with 
\begingroup
\allowdisplaybreaks
\begin{align}
    \left(U_{\Psi_q H}\right)_{\dot{\beta}akb} &= \left(U_{H \Psi_q }\right)_{b\dot{\beta}ak} = -\left(y^*_u\right)^{ki} \Psi_{ui}^{\dot{\alpha}} \cc_{\dot{\alpha}\dot{\rho}} \epsilon_{ab}P_L^{\dot{\rho}\dot{\beta}}+y_d^{ki} \cc^{\dot{\beta}\dot{\rho}}P_R^{\dot{\rho}\dot{\sigma}} \Psi_{di}^{\dot{\sigma}}\delta_{ab}, \\
    \left(U_{\Psi_q H^*}\right)_{\dot{\beta}akb} &= \left(U_{H^* \Psi_q}\right)_{b\dot{\beta}ak} = y_u^{kj} \cc_{\dot{\beta}\dot{\rho}} \epsilon_{ab}P_R^{\dot{\rho}\dot{\sigma}}\Psi_{uj}^{\dot{\sigma}}-\left(y_d^*\right)^{ki} \Psi_{di}^{\dot{\alpha}} \cc^{\dot{\alpha}\dot{\rho}}P_L^{\dot{\rho}\dot{\beta}} \delta_{ab}, \\
      \left(U_{\Psi_u H}\right)_{\dot{\beta}kb} &= \left(U_{H \Psi_u}\right)_{b\dot{\beta}k} =  \left(y^*_u\right)^{ik}  \cc_{\dot{\beta}\dot{\rho}}\epsilon_{ab}P_L^{\dot{\rho}\dot{\sigma}}\Psi_{qi}^{\dot{\sigma}a},\\
      \left(U_{\Psi_u H^*}\right)_{\dot{\beta}kb} &= \left(U_{H^* \Psi_u}\right)_{b\dot{\beta}k} = -y_u^{jk} \Psi_{qj}^{\dot{\alpha}a}\cc_{\dot{\alpha}\dot{\rho}} \epsilon_{ab}P_R^{\dot{\rho}\dot{\beta}},\\
      \left(U_{\Psi_d H}\right)_{\dot{\beta}kb} &= \left(U_{H \Psi_d}\right)_{b\dot{\beta}k} = -y_d^{jk} \Psi_{qj}^{\dot{\alpha}b}\cc_{\dot{\alpha}\dot{\rho}}P_R^{\dot{\rho}\dot{\beta}},\\
     \left(U_{\Psi_d H^*}\right)_{\dot{\beta}kb} &= \left(U_{H^*\Psi_d}\right)_{b\dot{\beta}k} = \left(y^*_d\right)^{ik}  \cc_{\dot{\beta}\dot{\rho}}P_L^{\dot{\rho}\dot{\sigma}}\Psi_{qi}^{\dot{\sigma}b},\\
       \left(U_{\Psi_l H}\right)_{\dot{\beta}akb} &= \left(U_{H \Psi_l}\right)_{b\dot{\beta}ak} = y_e^{ki} \cc^{\dot{\beta}\dot{\rho}}P_R^{\dot{\rho}\dot{\sigma}} \Psi_{ei}^{\dot{\sigma}}\delta_{ab}, \\
    \left(U_{\Psi_l H^*}\right)_{\dot{\beta}akb} &= \left(U_{H^* \Psi_l }\right)_{b \dot{\beta}ak} = -\left(y_e^*\right)^{ki} \Psi_{ei}^{\dot{\alpha}} \cc^{\dot{\alpha}\dot{\rho}}P_L^{\dot{\rho}\dot{\beta}} \delta_{ab}, \\
    \left(U_{\Psi_f B}\right)_{\dot{\beta}ai\mu} &= \left(U_{B \Psi_f}\right)_{\mu \dot{\beta}ai} = -\frac{g_1}{2}\cc_{\dot{\beta}\dot{\rho}}\gamma^\mu_{\dot{\rho}\dot{\sigma}}Y_{\Psi_f}^{\dot{\sigma}\dot{\lambda}}\Psi_{fi}^{\dot{\lambda}a}+\frac{g_1}{2}\Psi_{fi}^{\dot{\lambda}a}\cc_{\dot{\lambda}\dot{\rho}}\gamma^\mu_{\dot{\rho}\dot{\sigma}}Y_{\Psi_f}^{\dot{\sigma}\dot{\beta}}, \\
     \left(U_{\Psi_f W}\right)_{\dot{\beta}ai\mu A} &= \left(U_{W \Psi_f}\right)_{\mu A\dot{\beta}ai} =  -\frac{g_2}{2}\cc_{\dot{\beta}\dot{\rho}}\gamma^\mu_{\dot{\rho}\dot{\sigma}}\left(T^A_{\Psi_f}\right)^{ab}_{\dot{\sigma}\dot{\lambda}}\Psi_{fi}^{\dot{\lambda}b}+\frac{g_2}{2}\Psi_{fi}^{\dot{\lambda}b}\cc_{\dot{\lambda}\dot{\rho}}\gamma^\mu_{\dot{\rho}\dot{\sigma}}\left(T^A_{\Psi_f}\right)_{\dot{\sigma}\dot{\beta}}^{ba}.
\end{align}
\endgroup
Finally we have the matrix of purely fermionic derivatives
\begin{align}
    \mathbf{U}_{\xi \xi} & =
    \cc^{-1}_{\dot{\alpha}\dot{\beta}}
    \begin{pmatrix}
    0 & \left(U_{\Psi_q \Psi_u}\right)_{\dot{\beta}ai \dot{\rho}j} & 
    \left(U_{\Psi_q \Psi_d}\right)_{\dot{\beta}ai \dot{\rho}j} & 
    0 & 
   0 \\
    \left(U_{\Psi_u \Psi_q}\right)_{\dot{\beta}i \dot{\rho}bj} & 0 & 
    0 & 
    0 & 
    0 \\
    \left(U_{\Psi_d \Psi_q}\right)_{\dot{\beta}i \dot{\rho}bj} & 0 & 
    0 & 
    0 & 
    0 \\
    0 & 0 & 
    0 & 
   0 & 
    \left(U_{\Psi_l \Psi_e}\right)_{\dot{\beta}ai \dot{\rho}j} \\
    0 & 0 & 0 & 
    \left(U_{\Psi_e \Psi_l}\right)_{\dot{\beta}i \dot{\rho}bj} & 
    0
    \end{pmatrix},
\end{align}    
where the nonzero entries are given by
\begin{align}
    %
    \left(U_{\Psi_q \Psi_u}\right)_{\dot{\beta}ai \dot{\rho}j} &= -\left(U_{\Psi_u \Psi_q}\right)_{\dot{\rho}j \dot{\beta}ai} = -y_u^{ij} \cc_{\dot{\beta}\dot{\sigma}} \epsilon_{ab} H^*_b P_R^{\dot{\sigma}\dot{\rho}}+(y^*_u)^{ij} \cc_{\dot{\rho}\dot{\sigma}}\epsilon_{ab}H_b P_L^{\dot{\sigma}\dot{\beta}},\\
     \left(U_{\Psi_q \Psi_d}\right)_{\dot{\beta}ai \dot{\rho}j} &= -\left(U_{\Psi_d \Psi_q}\right)_{\dot{\rho}j \dot{\beta}ai} = -y_d^{ij} \cc_{\dot{\beta}\dot{\sigma}} H_a P_R^{\dot{\sigma}\dot{\rho}}+(y^*_d)^{ij} \cc_{\dot{\rho}\dot{\sigma}}H^*_a P_L^{\dot{\sigma}\dot{\beta}},\\
     %
     %
     %
     %
     %
     %
     %
     %
     %
     %
     \left(U_{\Psi_l \Psi_e}\right)_{\dot{\beta}ai \dot{\rho}j} &= -\left(U_{\Psi_e \Psi_l}\right)_{\dot{\rho}j \dot{\beta}ai} = -y_e^{ij} \cc_{\dot{\beta}\dot{\sigma}} H_a P_R^{\dot{\sigma}\dot{\rho}}+(y^*_e)^{ij} \cc_{\dot{\rho}\dot{\sigma}}H^*_a P_L^{\dot{\sigma}\dot{\beta}}.
\end{align}
In the bosonic sector there are also contributions from open covariant derivatives. The nonzero matrix entries of $\mathbf{Z}_{\phi \phi}^\mu$ are 
\begin{align}
    \left(Z^\mu_{H^* B}\right)_{a \nu} &= -\frac{g_1}{2} H_a g^\mu _\nu, \\
     \left(Z^\mu_{H^* W}\right)_{a \nu A} &= -g_2 T^A_{ab} H_b g^\mu _\nu, \\
     \left(Z^\mu_{H B}\right)_{a \nu} &= \frac{g_1}{2} H^*_a g^\mu _\nu, \\
     \left(Z^\mu_{H W}\right)_{a \nu A} &= g_2 T^A_{ba} H^*_b g^\mu _\nu.
\end{align}
As discussed in Section~\ref{sec: tofu} \tofu computes these matrices automatically and in the matching step inserts the classical fields where appropriate. For this particular model the matrices were also computed by hand as a check. There are 500 BSUOLEA operators contributing to the matching, highlighting that hand calculation is not feasible. The reduction of the redundant effective Lagrangian to a set of non-redundant dimension six operators was performed in two different ways.

In the first approach the feature of \tofu was used which allows for the identification of custom operators and the implementation of replacement rules among these. This is most easily realised by writing the SM Lagrangian in terms of the Majorana spinors $\Psi_f$ as in the SM part on the r.h.s.\ of Eq.\ \eqref{eq: SSM_Lag} and then applying the equations of motion compatible with this Lagrangian to the matching result. These equations of motion read
\begin{align}
&-D^2 H + \mu_h^2 H - \lambda_h |H|^2 H - \Psi_q y_u \epsilon P_R \Psi_u - \bar{\Psi}_d y_d^\dagger P_L \Psi_q - \bar{\Psi}_e y_e^\dagger P_L \Psi_l = 0, \\
& \left(D^\mu W_{\mu \nu}\right)^A + g_2 \left(\bar{\Psi}_q T^A_{\Psi_q} \Psi_q + \bar{\Psi}_l T^A_{\Psi_l} \Psi_l +\frac{i}{2}H^\dagger \overleftrightarrow{D}^AH \right) = 0, \\
& D^\mu B_{\mu \nu} + g_1 \sum \limits_{f \in \{u,d,q,e,l\}} \bar{\Psi}_f Y_{\Psi_f} \gamma_\nu \Psi_f + \frac{i}{2}H^\dagger \overleftrightarrow{D}_\nu   H = 0,\\
 & i \cc \slashed{D}\Psi_{q} - H y_d \cc P_R \Psi_d + \bar{\Psi}_d  y_d^\dagger  P_L H^\dagger - \tilde{H} y_u \cc P_R \Psi_u + \bar{\Psi}_u  y_u^\dagger  P_L \tilde{H}^\dagger = 0, \\ 
& i \cc \slashed{D}\Psi_{u} + \bar{\Psi}_q y_u \tilde{H} P_R - \tilde{H}^\dagger y_u^\dagger \cc P_L \Psi_q = 0, \\
& i \cc \slashed{D}\Psi_{d} + \bar{\Psi}_qy_d H P_R - H^\dagger y_d^\dagger \cc P_L \Psi_q = 0,\\
& i \cc \slashed{D}\Psi_{l} - H y_e \cc P_R \Psi_e + \bar{\Psi}_e  y_e^\dagger  P_L H^\dagger  = 0, \\ 
& i \cc \slashed{D}\Psi_{e} + \bar{\Psi}_ly_e H P_R - H^\dagger y_e^\dagger \cc P_L \Psi_l = 0,
\end{align}
and the dimension six operators that are generated in the matching are 
\begingroup
\allowdisplaybreaks
\begin{align}
    Q_{H\Box} &= |H|^2\Box |H|^2, \\
    Q_H &= |H|^6, \\ 
    Q_{HD} &= \left(H^\dagger D_\mu H\right)^\dagger \left(H^\dagger D^\mu H\right), \\
    Q_{HB} &= |H|^2 B_{\mu \nu}B^{\mu \nu}, \\
    Q_{HW} &= |H|^2 W^A_{\mu \nu}W^{\mu \nu A}, \\
    Q_{HWB} &= \left(H^\dagger \tau^A H \right) W^A_{\mu \nu}B^{\mu \nu}, \\
    Q_{uH} &= |H|^2 \left(\bar{q}\tilde{H}u\right) = |H|^2 \left(\bar{\Psi}_q \tilde{H} P_R \Psi_u\right),\\
    Q_{dH} &= |H|^2 \left(\bar{q} H d\right) = |H|^2 \left(\bar{\Psi}_q \tilde{H} P_R \Psi_d\right),\\
    Q_{eH} &= |H|^2 \left(\bar{l} H e\right) = |H|^2 \left(\bar{\Psi}_l \tilde{H} P_R \Psi_e\right),\\
    Q_{Hu} & = \left(H^\dagger i \overleftrightarrow{D}_\mu H\right)\left(\bar{u}\gamma^\mu u\right) = \left(H^\dagger i \overleftrightarrow{D}_\mu H\right)\left(\bar{\Psi}_u\gamma^\mu P_R \Psi_u\right), \\
    Q_{Hd} & = \left(H^\dagger i \overleftrightarrow{D}_\mu H\right)\left(\bar{d}\gamma^\mu d\right) = \left(H^\dagger i \overleftrightarrow{D}_\mu H\right)\left(\bar{\Psi}_d\gamma^\mu P_R \Psi_d\right), \\
    Q_{Hud} &= \left(\tilde{H}^\dagger i D_\mu H\right)\left(\bar{u}\gamma^\mu d\right) = \left(\tilde{H}^\dagger i D_\mu H \right)\left(\bar{\Psi}_u\gamma^\mu P_R \Psi_d\right),\\
    Q_{He} & = \left(H^\dagger i \overleftrightarrow{D}_\mu H\right)\left(\bar{e}\gamma^\mu e\right) = \left(H^\dagger i \overleftrightarrow{D}_\mu H\right)\left(\bar{\Psi}_e\gamma^\mu P_R \Psi_e\right), \\
    Q_{Hq}^{\left(1\right)} & = \left(H^\dagger i \overleftrightarrow{D}_\mu H\right)\left(\bar{q}\gamma^\mu q\right) = \left(H^\dagger i \overleftrightarrow{D}_\mu H\right)\left(\bar{\Psi}_q\gamma^\mu P_L \Psi_q
    \right), \\
    Q_{Hq}^{\left(3\right)} & = \left(H^\dagger i \overleftrightarrow{D}^A_\mu H\right)\left(\bar{q} \sigma^A \gamma^\mu q\right) = \left(H^\dagger i \overleftrightarrow{D}^A_\mu H\right)\left(\bar{\Psi}_q \sigma^A \gamma^\mu \Psi_q\right),\\
    Q_{Hl}^{\left(1\right)} & = \left(H^\dagger i \overleftrightarrow{D}_\mu H\right)\left(\bar{l}\gamma^\mu l\right) = \left(H^\dagger i \overleftrightarrow{D}_\mu H\right)\left(\bar{\Psi}_l\gamma^\mu P_L \Psi_l\right), \\
    Q_{Hl}^{\left(3\right)} & = \left(H^\dagger i \overleftrightarrow{D}^A_\mu H\right)\left(\bar{l} \sigma^A \gamma^\mu l\right) = \left(H^\dagger i \overleftrightarrow{D}^A_\mu H\right)\left(\bar{\Psi}_l \sigma^A \gamma^\mu \Psi_l\right), \\
    Q_{2y} &= |\bar{q}^b y_u u \epsilon_{ba} + \bar{d}y_d^\dagger q^a + \bar{e} y_e^\dagger l^a|^2 = |\bar{\Psi}^b_q P_R y_u \Psi_u \epsilon_{ba} + \bar{\Psi}_d y_d^\dagger P_L \Psi_q^a + \bar{\Psi}_e y_e^\dagger \Psi_l^a|^2,
\end{align}
\endgroup
where the last operator $Q_{2y}$ is a combination of several four-fermion operators in the Warsaw basis, see Ref.\ \cite{Wells:2015uba}. We here introduced the Pauli matrices $\tau^A$ and the Hermitian derivatives
\begin{align}
    H^\dagger i\overleftrightarrow{D}^\mu H &=  H^\dagger iD^\mu H - \left(D^\mu H\right)^\dagger iH,\\
    H^\dagger i\overleftrightarrow{D}^A_\mu H &=  H^\dagger i\tau^A D_\mu H - \left(D_\mu H\right)^\dagger i\tau^A H.
\end{align}
The Wilson coefficients corresponding to these dimension six operators are found to be in agreement with the ones cited in Ref.\ \cite{Haisch:2020ahr} and are therefore not repeated here, with one exception, namely the presence of the additional operator $Q_{Hud}$ and its Hermitian conjugate. This operator was originally missed both in Ref.\ \cite{Haisch:2020ahr} and in Ref.\ \cite{Jiang:2018pbd} since it is obtained from correlation functions contributing to $HH\rightarrow\bar{f}^{'}f$, which were not calculated in the matching of those references. The Wilson coefficient at $Q_\text{match} = M$ is given by
\begin{align}
    C_{Hud} =  -\frac{5A^2}{4M^4} y_u^\dagger y_d, 
\end{align}
which has been confirmed by the authors of Ref.\ \cite{Haisch:2020ahr}. The fact that this operator was missed twice using diagrammatic techniques highlights a strength of the UOLEA approach: It is impossible to miss a contribution to the matching once a sufficiently general UOLEA has been computed. This is due to the fact that the effective Lagrangian is fully generated from the UV Lagrangian through a matching condition which, by construction, includes all correlation functions relevant to the matching. It is not necessary to know the effective Lagrangian prior to the matching and to construct a matching condition that captures all possible contributions to the different operators contained in that effective Lagrangian.

In the second approach \tofu was again used to perform the matching and the raw, redundant result was passed to \route, which first translates operators containing the Majorana fermions $\Psi_f$ into the usual SM notation and then reduces the set of structures to the Warsaw basis. The result of this computation matches the result of the previous calculation once $Q_{2y}$ is translated into the Warsaw basis. Note that this illustrates the main difference between the two approaches. The first approach is flexible in the sense that one can use custom operators or operator bases. This also implies more work since an algorithm has to be devised which translates redundant structures into these bases without being circular. The benefit of using \route is that it is guaranteed to work and terminate. The cost is that one always ends up in the Warsaw basis and that this is limited to the SMEFT.  

\clearpage
\lhead{\emph{Matching of a triplet benchmark model}}
\chapter{Matching of a triplet benchmark model}
\label{ch: resonances}
In this chapter we apply the functional matching formalism to a specific simplified model, which is not covered by the BSUOLEA, described by the Lagrangian
\begin{align}
    \LagNoT ={}& -\frac{1}{4}V^{\mu \nu A} V_{\mu \nu}^A + \frac{m_V^2}{2}V^{\mu A}V_{\mu}^A+V^{\mu A}\left(\gamma_H g_{*}J_\mu^{HA}+\gamma_V\frac{g_2}{g_{*}}J_\mu^{A}+\sum_{f \in \{l,q\}}\gamma_f g_{*}J_{\mu}^{fA}\right) \nonumber 
    \\&+ \Lag{SM},
    \label{eq: model_lag}
\end{align}
where $\Lag{SM}$ is the SM Lagrangian, $V^{\mu A}$ is a $SU(2)_L$ triplet vector field and the currents are given by
\begin{align}
    J_\mu^{HA} &= \frac{i}{2} H^\dagger \overleftrightarrow{D}^A_\mu H,\\
    J_\mu^{A} &= \left(D^\nu W_{\nu \mu}\right)^{A},\\
    J_\mu^{fA} &=\frac{1}{2} \bar{f} \gamma_\mu \tau^A f,
\end{align}
with $\tau^A$ $A=1,2,3$ being the Pauli matrices. The field strength tensor $V^{\mu\nu A}$ is defined as
\begin{align}
    V^{\mu\nu A} = D^{[\mu}V^{\nu]A} = D^\mu V^{\nu A} - D^\nu V^{\mu A},
\end{align}
and the SM conventions are the same as in Section~\ref{sec: SSM_to_SM}. This model was introduced in Ref.\ \cite{Biekoetter:2014jwa} and is a special case of the model discussed in Ref.\ \cite{Pappadopulo:2014qza}. The explicit mapping is given in Section~\ref{sec: resonance_matching_res} and the matching presented in that section also applies to the more general model. Phenomenologically, these models are interesting for two reasons. Firstly, the model of Ref.\ \cite{Pappadopulo:2014qza} acts as a simplified model that accurately describes several UV completions of the SM. In Ref.\ \cite{Pappadopulo:2014qza} it was shown explicitly that both the weakly coupled UV completion of the SM discussed in Ref.\ \cite{Barger:1980ix} and the Minimal Composite Higgs model discussed in Refs.\ \cite{Contino:2011np,Contino:2013gna} can be mapped onto this simplified model under the assumption that the remainder of the NP sector is heavy compared to the vector field $V^{\mu A}$. In Ref.\ \cite{Biekoetter:2014jwa} the model defined by Eq.\ \eqref{eq: model_lag} was studied in the context of SMEFT fits. It was argued that the extended energy reach of the LHC can lead to an increased sensitivity to some higher dimensional operators, the effects of which are enhanced by powers of $E/\Lambda$, where $E$ is some energy scale in the process under consideration and $\Lambda$ is the scale of NP. This enhancement is most pronounced in kinematic regions where one might suspect the EFT expansion to have broken down, rendering any derived constraints inconsistent. However, whether or not this is the case cannot be inferred by only considering the general SMEFT Lagrangian since the Wilson coefficients cannot be disentangled from the unknown NP scale $\Lambda$. Nevertheless, in specific UV completions the scale $\Lambda$ is known and has a clear physical meaning. It is the pole mass of the lightest particle in the NP sector. The question then becomes whether there are specific UV completions of the SM for which the apparently inconsistent constraints are meaningful. One way of realising an energy enhancement without breaking the EFT assumptions is to have large Wilson coefficients, which can be seen from the following argument presented in Ref.\ \cite{Biekoetter:2014jwa}. Assume that the dimension six operator $\op$ with Wilson coefficient $C_\op$ is enhanced by $E^2/\Lambda^2$ and let $\Lambda$ be the pole mass of the lightest particle in the NP sector. With the naive expectation of $C_\op \sim 1$ the EFT expansion indeed breaks down as $E/\Lambda \to 1$ and no enhancement arises. However, if $C_\op = g_e^2 \hat{C}_\op$, where $\hat{C}_\op \sim 1$ and the enhancement factor $g_e^2 > 1$, we get an effective suppression scale $f = \Lambda/g_e$ such that $C_\op \op/\Lambda^2 = \hat{C}_\op \op / f^2$ and the energy enhancement becomes $E^2/f^2$ rather than $E^2/\Lambda^2$. Nothing special happens as $E/f \to 1$ and the EFT expansion is perfectly valid in the region where the operator is enhanced (as long as $E < \Lambda$ is satisfied). Clearly, this argument crucially depends on the fact that the scale $\Lambda$ has a well defined physical meaning, which it only carries once a certain UV completion has been specified. It was precisely this effect that was studied at tree level in Ref.\ \cite{Biekoetter:2014jwa} with $g_{*}$ playing the role of the enhancement. Since it is necessary that $g_{*}>1$ Eq.\ \eqref{eq: model_lag} was interpreted as the effective description of a composite sector and it was shown that in this case an enhancement does occur for certain operator combinations that contribute to Higgs physics.

Given the fact that the model defined by Eq.\ \eqref{eq: model_lag} (or rather its generalisation as given in Ref.\ \cite{Pappadopulo:2014qza}) can be interpreted as the effective description of several interesting UV completions of the SM at an intermediate scale it is an interesting benchmark model to study. Taking into account that parametrically large Wilson coefficients can occur when the model is interpreted as the effective description of a composite sector, this latter scenario becomes particularly interesting. Adding that the one-loop matching is expected to induce operators that are not present at tree level, which might have a dramatic impact on the SMEFT fit similar to the RGE induced operators in Ref.\ \cite{Dawson:2020oco}, we here focus on the one-loop matching of this model using the methods described in Chapters~\ref{chap: fcuntional matching} and \ref{ch: UOLEA}. There are two complications that arise when trying to directly apply the results of Chapter~\ref{ch: UOLEA} to the model studied in this section. Both of these issues are connected to the interpretation of the vector field as a resonance arising from a strongly coupled UV completion. Firstly, an explicit vector resonance comprises a system with second-class constraints. It is well-known that the generating functional of correlation functions in such theories might have to be modified as compared to the naive expectation to take these constraints into account.\footnote{See for example Eq.\ (3.4.26) of Ref.\ \cite{guitman1990quantization}.} Since the starting point of the UOLEA derivation is the generating functional, it is imperative to check whether any relevant modifications arise in this particular instance. As will be shown in Section~\ref{sec: constraint_analysis} this is not the case and one can proceed with the naive expectation for the generating functional. However, it turns out that for a self-consistent field theory the Lagrangian of Eq.\ \eqref{eq: model_lag} is insufficient and a coupling of the form $V^{\mu A}V_{\mu}^A |H|^2$ has to be added. Secondly, as the resonance $V^{\mu A}$ is not a gauge boson, it does not have a gauge-fixing term and the gauge choice $\xi = 1$ is not possible. This leads to a different propagator as compared to the one used in Chapter~\ref{ch: UOLEA} and to extra terms with two open covariant derivatives. This means that a dedicated computation of the relevant UOLEA operators has to be performed. We will briefly discuss this computation in Section~\ref{sec: constraints_computation}.  

\section{Constraint analysis and derivation of the generating functional}
\label{sec: constraint_analysis}
It is well-known that a massless vector field has two physical degrees of freedom and a massive vector field has three, yet in four dimensions each vector field $V^{\mu A}$ is described by four components, which are needed to make Lorentz invariance of the theory manifest. This mismatch between the number of physical degrees of freedom and the total number of degrees of freedom means that constraints are present in the theory. In the case of gauge theories\footnote{For a thorough discussion of the quantisation of gauge theories see Ref.\ \cite{Henneaux:1992ig} and for a general discussion of quantisation of constrained systems see Ref.\ \cite{guitman1990quantization}.}, one way of dealing with these constraints is the Fadeev-Popov method, which introduces ghosts into the generating functional of the theory. More generally it is expected that the presence of constraints in a theory introduces unphysical degrees of freedom into the generating functional, which have to be taken into account when computing correlation functions. The systematic construction of the generating functional in these theories is well understood and reviewed in Ref.\ \cite{guitman1990quantization}, which we follow closely here. Moreover, it is known that the quantisation of an interacting field theory is not necessarily equivalent to the free field quantisation of the individual fields comprising the theory if constraints are present, as pointed out in Refs.\ \cite{Johnson:1960vt,Velo:1969bt} in the case of spin-$3/2$ fields. It may very well be the case that a consistent formulation of the interacting field theory, even at the classical level, restricts the couplings of the theory. This fact, together with the requirement of renormalisability in the sense of EFTs, was used in Ref.\ \cite{Djukanovic:2010tb} to show that the self-couplings of the $\rho$-meson are restricted to the Yang-Mills form. Hence, performing a constraint analysis of the model under consideration here is imperative to check its self-consistency.

We next introduce the main algorithm and concepts needed to derive the generating functional and refer the interested reader to Ref.\ \cite{guitman1990quantization} for more details. A collection of useful definitions and relations can also be found in Appendix~\ref{App: berezin_algebra}. In what follows we will make a clear distinction between the Lagrangian function $L$ and the Lagrangian density $\LagNoT$ as well as the Hamiltonian function $H$ and the Hamiltonian density $\mathcal{H}$. Consider a theory whose Lagrangian function depends on the fields $q^i(t,\mathbf{x})$ with $i=1,\dots,n$ and the corresponding time derivatives\footnote{In a Lorentz covariant theory we expect only derivatives of the form $\partial_\mu q^i$ to appear. However, in the Hamiltonian formalism the time derivative plays a special role and we separate it from the spatial derivatives.}, which we will refer to as velocities, $\dot{q}^i(t,\mathbf{x})$. In going from the Lagrangian formulation in terms of $L[q^i,\dot{q}^i]$ to the Hamiltonian formulation one introduces the canonical momenta $\pi_i(t,\mathbf{x})$ defined by
\begin{align}
    \pi_i(t,\mathbf{x}) \equiv \frac{\delta_t L}{\delta \dot{q}^i},
    \label{eq: canonical_momentum_def}
\end{align}
where the functional derivative $\frac{\delta_t}{\delta \dot{q}^i}$ is taken at a fixed time. One then uses Eq.\ \eqref{eq: canonical_momentum_def} to express the velocities in terms of the canonical fields and their corresponding momenta. As shown in Ref.\ \cite{guitman1990quantization}, in general, one might only be able to solve for $N \leq n$ velocities, $\dot{q}^I(t,\mathbf{x})$, expressed through corresponding fields $q^I(t,\mathbf{x})$ and their conjugate momenta $\pi^I(t,\mathbf{x})$, where $I=1,\dots,N$. The remaining $n-N$ equations then yield constraints on the possible values of the fields and conjugate momenta. These constraints can be brought into the form 
\begin{align}
    \Phi^a(q^i,\pi_i) = 0, \,a = 1,\dots,n-N,
\end{align}
where $\Phi^a$ is some function. Such constraints, arising from the inability to determine certain velocities, will be referred to as primary constraints in the following. For a system with such constraints the classical equations of motion for all degrees of freedom $\eta = (q^i,\pi_i)$ can be expressed in terms of a modified Hamiltonian 
\begin{align}
    H_T = H(\eta)+\lambda^a \Phi^a(\eta),
\end{align}
where $H(\eta)$ is given by
\begin{align}
   H(\eta) = \left(\frac{\delta_t L}{\delta \dot{q}^i_\mathbf{x}}\dot{q}_\mathbf{x}^i-L\right)_{\dot{q}^I=\dot{q}^I(q^i,\Pi_I,\dot{q}^a)},
    \label{eq: Hamiltonian_from_Lagrangian}
\end{align}
and the fields $\lambda^a = \lambda^a(t,\mathbf{x})$ can be thought of as Lagrange multiplier fields. In Eq.\ \eqref{eq: Hamiltonian_from_Lagrangian} the repeated boldface indices imply an integration over space. The usual equations of motion have to be supplemented by the constraint equations, i.e.\ by the requirement that the constraints are satisfied on the phase-space trajectory. For a consistent solution, this property has to hold at all times and hence 
\begin{align}
    0 &\overset{!}{\approx} \{\Phi^a,H_T\} = \{\Phi^a,H\}+ \lambda^b \{\Phi^a,\Phi^b\},
    \label{eq: conservation_condition}
\end{align}
where we use $\approx$ to mean weak equality, that is equality holds after the constraints have been imposed.\footnote{However, Poisson brackets have to be computed before imposing the constraints.} The curly brackets denote the Poisson bracket (PB). For any given constraint Eq.\ \eqref{eq: conservation_condition} may lead to one of three outcomes. The first possibility is that the equation is identically satisfied and no further constraints arise due to the conservation of the constraint under consideration. The second possibility is that one of the Lagrange multipliers can be solved for, after which the equation becomes an identity and no further constraints arise. The final possibility is that the equation is neither identically satisfied, nor does it allow for the determination of a Lagrange multiplier. In this case, the conservation of the constraint on the trajectory has to be imposed as a further constraint. We will refer to constraints arising from the requirement that other constraints be conserved as secondary constraints. If secondary constraints are present, these again have to be conserved on the trajectory. One thus proceeds in this manner, imposing the conservation of the new constraints, thereby determining a subset of the Lagrange multipliers and obtaining new constraints, until no new constraints appear. At this point some of the Lagrange multipliers may still be undetermined. Indeed, for a primary constraint $\Phi$ whose PB with any other constraint vanishes weakly, the corresponding Lagrange multiplier remains undetermined and introduces an arbitrariness into the solutions of the equations of motion. In general, a constraint whose PB with any other constraint vanishes weakly is referred to as a first-class constraint and primary first-class constraints introduce a gauge redundancy into the theory, see Refs.\ \cite{Henneaux:1992ig,guitman1990quantization} for thorough discussions. A constraint that is not first-class is referred to as second-class. In the model under consideration in this chapter first-class constraints arise due to the presence of the gauge fields and second-class constraints due to the presence of the vector resonances.\footnote{There are also second-class constraints associated with the fermionic degrees of freedom.} 

In order to systematically construct the generating functional of the corresponding quantum theory one first canonically quantises the theory and then passes from the Hamiltonian path integral to the Lagrangian path integral. This is desirable in the case under consideration as the inclusion of constraints into the canonical formalism and the Hamiltonian path integral is well understood. As is well-known, for a gauge theory one has to break the gauge redundancy of the theory in order to obtain meaningful results from the generating functional. One way to accomplish this in the canonical formalism is to introduce supplementary constraints $\Phi_{G}$, which turn the first-class constraints associated with the gauge freedom into second-class ones, thereby fixing the gauge. Indeed the Hamiltonian generating functional in the presence of constraints $\mathbf{\Phi}$ can be generally written as (c.f.\ Eq.\ (3.4.17) of Ref.\ \cite{guitman1990quantization})
\begin{align}
    Z[J] = \int \measure \eta \; \Sdet^{1/2} \{\mathbf{\Phi}, \mathbf{\Phi}\} \delta(\mathbf{\Phi}) \exp \left[i\left(S+J_x \eta_x\right)\right],
    \label{eq: constrained_Hamiltonian_generating_functional}
\end{align}
where $\mathbf{\Phi}$ is the vector of all constraints, including the supplementary constraints $\Phi_{G}$ and any secondary constraints that may arise from their conservation, $\{\mathbf{\Phi},\mathbf{\Phi}\}$ is the matrix that results from taking the PBs of the different constraints and $\delta(\mathbf{\Phi})$ is a delta functional enforcing the constraints. As before, the symbol $\eta$ in Eq.\ \eqref{eq: constrained_Hamiltonian_generating_functional} runs over all of the canonical variables and their conjugate momenta, including the unphysical ones, and the action $S$ is given by 
\begin{align}
    S = \int dt \; \left[\pi_{i}^{\mathbf{x}} \dot{q}^i_\mathbf{x} - H\right].
\end{align}
Both the superdeterminant and the delta functional in Eq.\ \eqref{eq: constrained_Hamiltonian_generating_functional} can be brought into the exponential by introducing new, unphysical degrees of freedom such as ghosts, which then have to be included in the computation of correlation functions.

We now show that for the model defined by Eq.\ \eqref{eq: model_lag} these new degrees of freedom are irrelevant, except for the usual Fadeev-Popov ghosts associated with the gauge fixing, by explicitly deriving the generating functional of the model in the Hamiltonian formalism. As it will turn out in the analysis, the Lagrangian of Eq.\ \eqref{eq: model_lag} does not define a self-consistent theory. We therefore add an extra coupling term, which is allowed by the symmetries of the theory. Furthermore, it is more convenient to convert the $V^{\mu A}J_\mu^A$ coupling into a kinetic mixing term through integration by parts. It should be mentioned that at this point the kinetic mixing term cannot be removed by a field redefinition since we are working with a classical theory that is yet to be quantised. The field redefinition is not a canonical transformation and therefore not allowed classically. It is only at the level of the path integral that such field redefinitions are valid. We thus consider the Lagrangian 
\begin{align}
    \LagNoT ={}& -\frac{1}{4}V^{\mu \nu A} V_{\mu \nu}^A - \frac{g_M}{2}V^{\mu \nu A} W_{\mu \nu}^A -\frac{1}{4}W^{\mu \nu A} W_{\mu \nu}^A +  \frac{m_V^2}{2}V^{\mu A}V_{\mu}^A+g_{H} V^{\mu A} J_\mu^{HA} \nonumber \\ &+g_{f} V^{\mu A} J_{\mu}^{fA}+ \bar{f}i\slashed{D}f+ \left(D_\mu H\right)^\dagger \left(D^\mu H\right)+ \frac{g_{VH}}{2}|H|^2V^{\mu A}V_{\mu}^A + \Lag{rest},
\end{align}
where we focus on the $SU(2)_L$ part of the SM since the $SU(3)_c$ and $U(1)_Y$ contributions are inconsequential. We also only consider one kind of fermion as the introduction of several fermions does not alter the treatment significantly and we simplified the notation with the identifications
\begin{align}
    g_H = \gamma_H g_{*}, \; g_M=\gamma \frac{g}{g_{*}},\; g_f=\gamma_f g_{*},
\end{align}
where in this section we denote the $SU(2)_L$ gauge coupling by $g$. We further introduced the term 
\begin{align}
    \frac{g_{VH}}{2}|H|^2V^{\mu A}V_{\mu}^A,
\end{align}
which, as will be shown, is needed for the self-consistency of the theory. The Lagrangian density $\Lag{rest}$ contains an $SU(2)_L$ invariant polynomial in $H$, $H^{\dagger}$ and the fermionic fields. We find the following expressions for the conjugate momenta
\begin{align}
\label{eq: conj_mom_1}
    \pi(W)^{\mu A} &= -W^{0 \mu A}-g_M V^{0 \mu A}, \\
\label{eq: conj_mom_2}    
    \pi(V)^{\mu A} &= -V^{0 \mu A}-g_M W^{0 \mu A}, \\
    \pi(\Hdagg) &= D^0H-\frac{i}{2}g_H V^{0A}\tau^A H,\\
    \pi(H) &= (D^0H)^\dagger+\frac{i}{2}g_H V^{0A}\Hdagg \tau^A, \\
    \bar{\pi}(f) &= i\bar{f}\gamma^0, \\
    \label{eq: conj_mom_last}
    \pi(\bar{f}) &= 0,
\end{align}
where we denote the conjugate momentum corresponding to the field $q$ by $\pi(q)$. As there are four conjugate momenta that are entirely independent of velocities, there are four primary constraints in the system, which read
\begingroup
\allowdisplaybreaks
\begin{align}
\label{eq: constr_1}
    \Phi_1^A &= \pi(W)^{0 A}, \\
    \Phi_2^A &= \pi(V)^{0 A}, \\
    \Phi_3 &= \bar{\pi}(f) - i\bar{f}\gamma^0, \\ 
\label{eq: constr_last}
    \Phi_4 &= \pi(\bar{f}).
\end{align}
\endgroup
For the remaining fields the corresponding velocities can be expressed through fields and conjugate momenta using Eqs.\ \eqref{eq: conj_mom_1}--\eqref{eq: conj_mom_last}, provided that $g_M \neq \pm 1$ in which case Eqs. \eqref{eq: conj_mom_1} and \eqref{eq: conj_mom_2} become linearly dependent and lead to three further constraints. This renders the theory inconsistent as the number of independent degrees of freedom will not match the number of physical degrees of freedom. Therefore, the exclusion of these values of $g_M$ is a consistency condition. For the determinable velocities one finds  
\begin{align}
    \partial_0 W^{i A} &= \frac{1}{g_M^2-1}\pi(W)^{iA}-\frac{g_M}{g_M^2-1}\pi(V)^{iA}-g f^{ABC}W_{0}^B W^{iC}+\partial ^i W^{0A},\\
    \partial_0 V^{i A} &= \frac{1}{g_M^2-1}\pi(V)^{iA}-\frac{g_M}{g_M^2-1}\pi(W)^{iA}-g f^{ABC} W^{[0|B|}V^{i]C}+\partial ^i V^{0 A},\\
    \partial_0 H &= \pi(\Hdagg)+\frac{i}{2}g W_0^A \tau^A H +\frac{i}{2} g_H V_0^A \tau^A H, \\
    \partial_0 \Hdagg &= \pi(H)-\frac{i}{2}g W_0^A \Hdagg \tau^A -\frac{i}{2} g_H V_0^A \Hdagg \tau^A,
\end{align}
where lower case Roman indices $i$, $j$, $k$ etc.\ take the values $1,2,3$ and denote spatial components of a four-vector.\footnote{We use a metric with the mostly minus signature so one has to be careful when raising and lowering spatial indices.}
Using these expressions as well as the constraints in Eqs.\ \eqref{eq: constr_1}--\eqref{eq: constr_last} we can construct the modified Hamiltonian density $\mathcal{H}_T$ as 
\begin{align}
    \mathcal{H}_T ={}& \frac{1}{2(g_M^2-1)}\pi(W)^A_i \pi(W)^{iA}+\frac{1}{2(g_M^2-1)}\pi(V)^A_i \pi(V)^{iA}-\frac{g_M}{g_M^2-1}\pi(W)^A_i \pi(V)^{iA}\nonumber \\ &+\pi(W)^A_i D^i W^{0A} + \pi(V)^A_i \left(D^i V^{0A}-g f^{ABC}W^{0B}V^{iC}\right)\nonumber \\
    &-\frac{i}{2}\Hdagg \tau^A \pi(\Hdagg) \left(g W^{0A}+g_H V^{0A}\right)+\frac{i}{2}\pi(H) \tau^A H \left(g W^{0A}+g_H V^{0A}\right)\nonumber \\
    &+\frac{1}{4}g_H^2V_0^{A}V_0^{B} \Hdagg \tau^A \tau^B H - \frac{g_{VH}}{2}|H|^2 V^{\mu A} V^{A}_{\mu} -g_H V^{iA}J^{HA}_i-g_f V^{\mu A}J_\mu^{f A} \nonumber \\ &+\frac{1}{4}W^{ijA} W_{ij}^A +\frac{1}{4}V^{ijA} V_{ij}^A +\frac{g_M}{2}W^{ijA} V_{ij}^A -\frac{m_V^2}{2}V_{\mu}^A V^{\mu A} - \left(D_i H\right)^\dagger \left(D^i H\right) \nonumber \\
    &-\bar{f}i \gamma^i \partial_i f -g \bar{f}\slashed{W}f - \Lag{rest} \nonumber \\ &+\lambda_1^A \pi(W)^{0A} +\lambda_2^A \pi(V)^{0A} +\lambda_3 \left(\bar{\pi}(f)-i\bar{f}\gamma^0\right) +\lambda_4 \pi(\bar{f}), 
    \label{eq: modified_Hamiltonian_density}
\end{align}
from which the modified Hamiltonian is obtained as 
\begin{align}
    H_T = \int d^3x \, \mathcal{H}_T(x).
\end{align}
In Eq.\ \eqref{eq: modified_Hamiltonian_density} we defined $\slashed{W}=W^{A\mu}\frac{\tau^A}{2}\gamma_\mu$. Given the Hamiltonian we may now impose the conservation of constraints according to Eq.\ \eqref{eq: conservation_condition} for each constraint. Starting with $\Phi_4$ we find 
\begin{align}
    \{\Phi_4,H_T\} &= -\frac{g_f}{2} V^{\mu A} \gamma_\mu \tau^{A} f- i\gamma^i \partial_i f - g \slashed{W} f + \frac{\partial_r \Lag{rest}}{\partial \bar{f}} + i \gamma^0 \lambda_3 \overset{!}{\approx} 0,
\end{align}
which determines the Lagrange multiplier $\lambda_3$ for all field values.\footnote{Here the subscript $r$ on the partial derivative indicates that we are taking the right derivative. For a definition see Appendix~\ref{App: berezin_algebra}.} Hence, no secondary constraint arises in this case. Similarly
\begin{align}
    \{\Phi_3,H_T\} &= \frac{g_f}{2} V^{\mu A} \bar{f} \gamma_\mu \tau^{A} + \bar{f} i\gamma^i \overleftarrow{\partial}_i + g \bar{f} \slashed{W} + \frac{\partial_r \Lag{rest}}{\partial f} - i  \lambda_4 \gamma^0 \overset{!}{\approx} 0
\end{align}
determines $\lambda_4$ without the appearance of secondary constraints. This concludes the discussion of the constraints arising in the fermionic sector. For the constraints associated with the gauge field we find 
\begin{align}
\label{eq: secondary_gauge_constraint_old}
    \{\Phi_1^A,H_T\} ={}& D^i \pi(W)_i^A + g f^{DAC} \pi(V)^D_i V^{iC}+\frac{i}{2}g\Hdagg \tau^A \pi(\Hdagg) -\frac{i}{2}g\pi(H) \tau^A H \nonumber \\ &+ \frac{g}{2} \bar{f}\gamma^0 \tau^A f,
\end{align}
which does not depend on any Lagrange multiplier and hence requiring it to vanish weakly constitutes a secondary constraint. We may add to Eq.\ \eqref{eq: secondary_gauge_constraint_old} any function that vanishes weakly since the resulting constraint is equivalent to the original one. It is customary to write the secondary constraint as
\begin{align}
    \Phi_{1(2)}^A &= \{\Phi_1^A,H_T\} - \frac{ig}{2} \Phi_3 \tau^A f - \frac{ig}{2} \bar{f} \tau^A \Phi_4 + g f^{DAC} \Phi_2^{D} V^{0C} \nonumber \\
    &=  D^i \pi(W)_i^A + g f^{DAC} \pi(V)^D_\mu V^{\mu C} +\frac{i}{2}g\Hdagg \tau^A \pi(\Hdagg) -\frac{i}{2}g\pi(H) \tau^A H \nonumber \\ &\quad - \frac{ig}{2} \bar{\pi}(f) \tau^A f -\frac{ig}{2} \bar{f} \tau^A \pi(\bar{f}).
    \label{eq: secondary_gauge_constraint}
\end{align}
Similarly, the conservation of $\Phi_{2}^A$ gives rise to a secondary constraint $\Phi_{2(2)}^A$ with 
\begin{align}
    \Phi_{2(2)}^A =\{\Phi^A_2,H_T\} ={}&D^i \pi(V)_i^A+\frac{i}{2}g_H \Hdagg \tau^A \pi(\Hdagg)-\frac{i}{2}g_H \pi(H) \tau^A H \nonumber \\ &-\frac{1}{4}g_H^2 V^{0B} \Hdagg [\tau^A,\tau^B]_{+} H+g_{VH}|H|^2 V^{0A} + g_f J_0^{fA} + m_V^2 V^{0A}\nonumber \\
    ={}& D^i \pi(V)_i^A+\frac{i}{2}g_H \Hdagg \tau^A \pi(\Hdagg)-\frac{i}{2}g_H \pi(H) \tau^A H \nonumber \\
    &+ V^{0A} |H|^2 \left(g_{VH}-\frac{g_H^2}{2}\right)+ g_f J_0^{fA} + m_V^2 V^{0A},
\end{align}
where $[A,B]_{+}=AB+BA$ is the anticommutator and we used that $[\tau^A,\tau^B]_{+}=2I\delta^{AB}$, where $I$ is the $2\times 2$ identity matrix. Since non-trivial secondary constraints exist we have to consider their conservation in time. The computation of $\{\Phi_{1(2)}^A,H_T\}$ is rather lengthy and we simply quote the result. As we were unable to find the computation in the literature, even for the case of a gauge theory without vector resonances, we present some of the details of the computation in Appendix~\ref{app: constraint_comp}. The result of the calculation is  
\begin{align}
    \{\Phi_{1(2)}^A,H_T\} \approx -g f^{ABC} \left(W^{0B} \Phi^C_{1(2)}+V^{0B} \Phi^C_{2(2)}\right) \approx 0,
    \label{eq: conservation_secondary_gauge_constraint}
\end{align}
which means that no further constraints arise from the gauge bosons. This is expected from the known analysis of Yang-Mills theory with scalar and spin-$1/2$ matter fields, see Ref.\ \cite{guitman1990quantization}. However, as is clear from Eq.\ \eqref{eq: conservation_secondary_gauge_constraint} the presence of the resonances leads to modifications of the computation. We finally turn to the conservation of $\Phi^A_{2(2)}$, which is expected to determine $\lambda_2^A$ in order to yield the correct number of degrees of freedom for a massive vector field. To see that this is indeed the case we note that $\Phi^A_{2(2)}$ contains terms involving $V^{0A}$ which in the PB with $H_T$ will provide terms proportional to $\lambda_2^A$. These are also the only terms that give rise to such contributions. Furthermore, $\Phi^A_{2(2)}$ is independent of $W^{0A}$ and hence $\lambda_1^A$ will indeed remain undetermined. Noting that
\begin{align}
    \{\Phi^A_{2(2)},H_T\} ={}&\{D^i \pi(V)_i^A,H_T\}+\frac{i}{2}g_H\{\Hdagg \tau^A \pi(\Hdagg),H_T\}-\frac{i}{2}g_H\{\pi(H) \tau^A H,H_T\} \nonumber \\
    &+ g_f\{J_0^{fA},H_T\}+V^{0A}\left(g_{VH}-\frac{g_H^2}{2}\right) \{|H|^2,H_T\} \nonumber \\ 
    &+ \left(m_v^2 +g_{VH} |H|^2 -\frac{g_H^2}{2}|H|^2\right) \{V^{0A},H_T\},
\end{align}
we define 
\begin{align}
    \mathcal{F}^A \equiv{}& \{D^i \pi(V)_i^A,H_T\}+\frac{i}{2}g_H\{\Hdagg \tau^A \pi(\Hdagg),H_T\}-\frac{i}{2}g_H\{\pi(H) \tau^A H,H_T\} \nonumber \\
    &+ g_f\{J_0^{fA},H_T\}+V^{0A}\left(g_{VH}-\frac{g_H^2}{2}\right) \{|H|^2,H_T\}, 
\end{align}
which is independent of $\lambda_2^A$ and write 
\begin{align}
    \{\Phi^A_{2(2)},H_T\}\overset{!}{\approx}0
\end{align}
as 
\begin{align}
    \left(m_v^2 +g_{VH} |H|^2 -\frac{g_H^2}{2}|H|^2\right) \lambda_2^A = - \mathcal{F}^A.
\end{align}
As pointed out before, self-consistency requires this equation to determine $\lambda_2^A$ for all field values, which in turn requires that
\begin{align}
    m_V^2+g_{VH}|H|^2-\frac{g_H^2}{2}|H|^2 \neq 0
\end{align}
for all values of $|H|^2$. For this condition to be satisfied it is necessary and sufficient that $g_{VH} \geq \frac{g_H^2}{2}$. We thus see that it is pivotal to introduce the $V^{\mu A}V_{\mu}^A |H|^2$-coupling into the Lagrangian from the point of view of consistency. 

At this point of the analysis no further constraints arise and all Lagrange multipliers except $\lambda_1^A$, which is associated with the gauge freedom, are determined. In order to fix the gauge we impose the constraint 
\begin{align}
    \Phi_G^A = \partial_i W^{iA},
\end{align}
which may be recognised as the Coulomb gauge condition. Like the previous constraints, this constraint has to be conserved in time. Its PB with the modified Hamiltonian is given by 
\begin{align}
    \{\Phi^A_{G},H_T\} = \frac{1}{g_M^2-1}\partial_i \pi(W)^{iA} -\frac{g_M}{g_M^2-1}\partial_i \pi(V)^{iA}+D^i\partial_i W^{0A} \equiv \Phi_{G(2)}^A,  
\end{align}
which yields a secondary constraint. Imposing the conservation of this secondary constraint we obtain
\begin{align}
     \{\Phi^A_{G(2)},H_T\} = -\Delta \lambda_1^A + g f^{ABC} W^{iB}\partial_i \lambda_1^C + \mathcal{G}^A \overset{!}{\approx} 0,
     \label{eq: lambda_1_cond}
\end{align}
where $\mathcal{G}^A$ is independent of $\lambda_1^A$ and $\Delta$ is the Laplace operator. Eq.\ \eqref{eq: lambda_1_cond} determines the final Lagrange multiplier through a differential equation.\footnote{One might be concerned that the differential operator depends on the spatial gauge field components, which means that there might be field values for which the differential operator is not invertible. However, the particular differential operator arising here is elliptic independent of the values of the field $W^{iA}$ and hence always invertible.} We are now in a position to formulate the generating functional in the Hamiltonian formalism using Eq.\ \eqref{eq: constrained_Hamiltonian_generating_functional}. To this end we need to compute the matrix of PBs of all constraints. It is sufficient to evaluate the result on the constraint surface since the constraints are enforced by the delta functional. Defining 
\begin{align*}
    \mathbf{\Phi} = \begin{pmatrix}
    \Phi^A_{G(2)} & 
    \Phi_1^A &
    \Phi_2^A &
    \Phi_{G}^A &
    \Phi_{1(2)}^A &
    \Phi_{2(2)}^A &
    \Phi_3 &
    \Phi_4
    \end{pmatrix}^T
\end{align*}
we find
\begin{align*}
\{\mathbf{\Phi},\mathbf{\Phi}\} &= \begin{pmatrix}
A & B \\
C & D
\end{pmatrix}
\end{align*}
with
\begingroup
\setlength\arraycolsep{2.7pt}
\allowdisplaybreaks
\begin{align}
A &= \begin{pmatrix}
-\frac{gf^{ABD}}{g_M^2-1} \Delta W^{0D} & \left(D_i\partial^i\right)^{AB}&  0& \frac{\delta^{AB}}{g_M^2-1}\Delta & \{\Phi_{G(2)}^A,\Phi_{1(2)}^B\} & \{\Phi_{G(2)}^A,\Phi_{1(2)}^B\}  \\
-\left(D_i\partial^i\right)^{AB}& 0 & 0 & 0 & 0 & 0  \\
0 & 0 & 0 & 0 & 0 & -M_V^2 \delta^{AB}  \\
-\frac{\delta^{AB}}{g_M^2-1}\Delta & 0 & 0 & 0 & \left(D_i\partial^i\right)^{AB}&0  \\   
\{\Phi_{1(2)}^A,\Phi_{G(2)}^B\} & 0 & 0& -\left(D_i\partial^i\right)^{AB}& 0 & 0 \\
\{\Phi_{2(2)}^A,\Phi_{G(2)}^B\} & 0 & M_V^2\delta^{AB} & 0 & 0 & \{\Phi_{2(2)}^A,\Phi_{2(2)}^B\} 
\end{pmatrix},\\
B&= \begin{pmatrix}
0 & 0\\
0 & 0\\
0 & 0\\
0 & 0\\
0 & 0\\
\frac{g_f}{2} \bar{f} \gamma^{0} \tau^{A} & -\frac{g_f}{2} \gamma^{0} \tau^{A} f
\end{pmatrix},\text{ }
C=\begin{pmatrix}
0 & 0 & 0 & 0 & 0 & -\frac{g_f}{2} \bar{f} \gamma^{0} \tau^{A} \\
0 & 0 & 0 & 0 & 0 & \frac{g_f}{2} \gamma^{0} \tau^{A} f
\end{pmatrix},\text{ }
D = \begin{pmatrix}
0 & -i \gamma^0 \\
i\gamma^0 & 0
\end{pmatrix},
\end{align}
\endgroup
where 
\begin{align}
    M_V^2 = m_V^2+g_{VH}|H|^2-\frac{g_H^2}{2}|H|^2.
\end{align}
We only care about field dependent contributions to the superdeterminant since any constant factors can be absorbed in the normalisation of the generating functional. Using the definition of the superdeterminant, see Appendix~\ref{App: berezin_algebra}, we obtain 
\begin{align}
    \Sdet \{\mathbf{\Phi},\mathbf{\Phi}\} &= \det\left(A-BD^{-1}C\right) \det(D)^{-1} \propto \det(-M_V^2)^2 \det(D_i \partial^i)^4.
\end{align}
The second factor is precisely what is expected in a gauge theory without resonances, whereas the first factor is entirely due to the presence of the resonances. We next bring the generating functional into the Lagrangian form following Refs.\ \cite{guitman1990quantization,Djukanovic:2010tb}. In this discussion we omit the source terms since these can always be adjusted by the equivalence theorem of Ref.\ \cite{Kallosh:1972ap}. We thus write
\begin{align}
    Z \equiv Z[0] = \int \measure \mu \exp{\left\{i\int d^4x \left[\mathcal{P}_I \partial^0 \mathcal{Q}^I - \mathcal{H} \right]\right\}},
\end{align}
where
\begingroup
\setlength\arraycolsep{3pt}
\begin{align*}
    \mathcal{P} = \begin{pmatrix}
    \pi(W)^A_\mu & \pi(V)^A_\mu & \bar{\pi}(f) & \pi(\bar{f}) & \pi(H) & \pi(\Hdagg)
    \end{pmatrix}^T, \text{ }
    \mathcal{Q} = \begin{pmatrix} W^{\mu A} & V^{\mu A} & f & \bar{f} & H & \Hdagg
    \end{pmatrix}^T,
\end{align*}
 \endgroup
and the integration measure is given by
\begin{align*}
    \measure \mu ={}& \measure \mathcal{Q} \, \measure \mathcal{P} \, \delta\left(\pi(W)^{0A}\right) \delta\left(\pi(V)^{0A}\right) \delta\left(\bar{\pi}(f)-i\bar{f}\gamma^0\right)\delta\left(\pi(\bar{f})\right)\delta\left(\partial_i W^{iA}\right)\nonumber \\
    &\times \delta\left(\frac{1}{g_M^2-1}\partial_i \pi(W)^{iA}-\frac{g_M}{g_M^2-1}\partial_i \pi(V)^{iA}+D^i\partial_i W^{0A}\right) \nonumber \\
    & \times \delta\left(\Phi^A_{1(2)}\right) \delta\left(\Phi^A_{2(2)}\right) \det\left(-M_V^2\right) \det\left(D_i \partial^i\right)^2.
\end{align*}
Due to the first four delta functionals in the measure the integrals over $\bar{\pi}(f)$, $\pi(\bar{f})$, $\pi(W)^{0A}$ and $\pi(V)^{0A}$ can be trivially performed. For the $W^{0A}$ integration we define $\tilde{W}^{0A}$ through 
\begin{align*}
  W^{0A}   = (\tilde{D}_i \partial^i)^{-1}\left[\tilde{W}^{0A}-\frac{1}{g_M^2-1}\partial_j \tilde{\pi}(W)^{jA}+\frac{g_M}{g_M^2-1}\partial_j \tilde{\pi}(V)^{jA}\right],
\end{align*}
where we also define $\tilde{\pi}(W)^{jA} =\pi(W)^{jA}$, $\tilde{\pi}(V)^{jA} =\pi(V)^{jA}$ and $\tilde{W}^{jA} =W^{jA}$ where the latter appears in $\tilde{D}^i$. The Jacobian of the change of variables from the fields without tilde to the fields with tilde is $\det \left(D_i\partial^i\right)^{-1}$, which removes one of the determinants in the measure. After performing the integration over $\tilde{W}^{0A}$ using the delta functional and renaming the integration variables to ones without tilde, the argument of the exponential contains
\begin{align}
    &{}\left(D_{i} \partial^{i}\right)^{-1} \left[-\frac{1}{g_M^2-1}\partial_{j} \tilde{\pi}(W)^{jA}+\frac{g_M}{g_M^2-1}\partial_{j} \tilde{\pi}(V)^{jA}\right] \nonumber \\  &\times \left[D^{k}\pi(W)^{A}_{k}+g f^{DAC}\pi(V)^D_{k} V^{kC}+\frac{i}{2}g\Hdagg\tau^{A}\pi(\Hdagg)-\frac{i}{2}g\pi(H)\tau^{A} H+\frac{g}{2}\bar{f}\gamma^{0} W^{A}_{0} \tau^{A} f\right].
\end{align}
Comparing this to 
\begin{multline}
\delta\left(\Phi^{A}_{1(2)}\right) = \int \measure W_{0}^{A} \exp \Big\{i\int \rd^{d} x\, W_{0}^{A} \Big[D^{i} \pi(W)_{i}^{A}+g f^{DAC}\pi(V)^{D}_{k} V^{kC} \\ +\frac{i}{2}g\Hdagg\tau^{A}\pi(\Hdagg)-\frac{i}{2}g\pi(H)\tau^{A} H+\frac{g}{2}\bar{f}\gamma^{0} W^{A}_{0} \tau^{A} f\Big]\Big\},
\end{multline}
where we have used that some of the integrations over delta functionals have already been performed, we see that introducing this constraint into the exponential we may shift the field $W_{0}^{A}$ to absorb the previously generated term. Up to this point the treatment presented here is textbook procedure as outlined in Ref.\ \cite{guitman1990quantization} and produces the generating functional
\begin{align}
    Z = \int \measure \mu' \exp{\left\{i\int \rd^d x\, \mathcal{B} \right\}},
\end{align}
where 
\begingroup
\allowdisplaybreaks
\begin{align}
\label{eq: B}
    \mathcal{B} ={}& \pi(W)_i^A \partial_0 W^{iA} + \pi(V)_i^A \partial_0 V^{iA} + \pi(H) \partial_0 H  + \pi(\Hdagg) \partial_0 \Hdagg + \Lag{rest} \nonumber \\
    &-\frac{1}{2(g_M^2-1)}\pi(W)^{iA}\pi(W)_i^{A}-\frac{1}{2(g_M^2-1)}\pi(V)^{iA}\pi(V)_i^{A}+\frac{g_M}{g_M^2-1}\pi(W)^{iA}\pi(V)_i^{A} \nonumber \\
    & - \pi(H)\pi(\Hdagg) + \frac{g_{VH}}{2}V_i^A V^{iA}|H|^2 -\frac{1}{4}W^A_{ij} W^{ijA} -\frac{1}{4}V^A_{ij} V^{ijA} -\frac{g_M}{2}V^A_{ij} W^{ijA} \nonumber \\
    &+ \frac{m_V^2}{2} V_i^A V^{iA} + g_H V^{iA} J_i^{HA} + g_f V^{iA} J_i^{fA} + \bar{f} i \slashed{D} f + \left(D_i H\right)^\dagger \left(D^i H\right) \nonumber \\
    &+ W^{0A} \left[D^i \pi(W)^A_i + \frac{i}{2}g\Hdagg \tau^A \pi(\Hdagg)-\frac{i}{2}g \pi(H) \tau^A H+gf^{BAC}\pi(V)^B_i V^{iC}\right] \nonumber \\
    &+ V^{0A} \left[D^i \pi(V)^A_i + \frac{i}{2}g_H\Hdagg \tau^A \pi(\Hdagg)-\frac{i}{2}g_H \pi(H) \tau^A H+g_f J_0^{fA}+\frac{M_V^2}{2}V^{0A}\right],
\end{align}
\endgroup
and 
\begin{align}
     \measure \mu' ={}& \measure \mathcal{Q} \, \measure \mathcal{P'} \delta\left(\partial_i W^{iA}\right) \delta\left(\Phi^A_{2(2)}\right) \det\left(-M_V^2\right) \det\left(D_i \partial^i\right).
\end{align}
In the measure $\mathcal{Q}$ is unchanged and
\begin{align*}
    \mathcal{P'} &= \begin{pmatrix}
    \pi(W)^A_i & \pi(V)^A_i& \pi(H) & \pi(\Hdagg)
    \end{pmatrix}^T.
\end{align*}
We now bring $\Phi^A_{2(2)}$ into the exponential by writing
\begin{align}
     \delta\left(\Phi^A_{2(2)}\right) = \int \measure \lambda^A \exp\Big\{i\int \rd^dx \, \lambda^A \Big[&D^i \pi(V)_i^A+\frac{i}{2}g_H\Hdagg\tau^A\pi(\Hdagg)\nonumber \\ &-\frac{i}{2}g_H\pi(H)\tau^A H+g_f J_0^{fA}+M_V^2V_0^A\Big]\Big\},
\end{align}
which we may combine with the last line of Eq.\ \eqref{eq: B} to yield the following contribution to the argument of the exponential
\begin{align}
    &\left(\lambda^A+V^{0A}\right) \Big[D^i \pi(V)_i^A+\frac{i}{2}g_H\Hdagg\tau^A\pi(\Hdagg)-\frac{i}{2}g_H\pi(H)\tau^A H+g_f J_0^{fA}\Big] \nonumber \\
    &+ M_V^2 \lambda^A V_0^A + \frac{M_V^2}{2} V^{0A}V^{0A}.
\end{align}
Shifting $V^{0A}\rightarrow V^{0A}+\lambda^A$ we can arrange for only this field to couple to the terms in the brackets giving rise to the remaining terms 
\begin{align}
-\frac{M_V^2}{2}\lambda^A \lambda^A + \frac{M_V^2}{2} V^{0A} V^{0A},
\end{align}
which leaves us with a term quadratic in the field $\lambda^A$. The fact that this is the only $\lambda^A$-dependent term will become important later. We now integrate over the remaining canonical momenta, starting with the scalar ones. The terms involving the scalar momenta can be arranged as
\begin{align}
    &-\left[\pi(H)+\partial_0 \Hdagg+\frac{i}{2}g_H V^{0B}\Hdagg \tau^B +\frac{i}{2}g W^{0B}\Hdagg \tau^B \right] \nonumber \\ 
    &\times\left[\pi(\Hdagg)+\partial_0 H-\frac{i}{2}g_H V^{0A}\tau^A H-\frac{i}{2}g W^{0A}\tau^A H\right] \nonumber \\
    &+ \left[\left(D_0 H\right)^\dagger + \frac{i}{2}g_H \Hdagg \tau^A V^{0A}\right]\left[\left(D_0 H\right) - \frac{i}{2}g_HV^{0B} \tau^B H\right],
\end{align}
where we may perform a constant shift of the conjugate momenta to completely decouple them. The Gaussian path integral over these variables then yields an irrelevant constant. This only leaves the second term which, using that $V^{0A}V^{0B} [\tau^A,\tau^B]_{+} = 2 V^{0A} V^{0A} I$, yields
\begin{align}
    \left(D_0 H\right)^\dagger \left(D_0 H\right) + g_H V^{0A} J_0^{HA}+\frac{g_H^2}{4}V^{0A}V^{0A}|H|^2.
\end{align}
The first and second term combine with $\left(D_i H\right)^\dagger \left(D^i H\right)$ and $g_H V^{iA} J_i^{HA}$, respectively, to yield $\left(D_\mu H\right)^\dagger \left(D^\mu H\right)$ and $g_H V^{\mu A} J_\mu^{HA}$. The last term cancels the corresponding contribution to $\frac{M_V^2}{2}V^{0A}V^{0A}$ leaving only  $\frac{m_V^2}{2}V^{0A}V^{0A}$ and $\frac{g_{VH}}{2}|H|^2 V^{0A}V^{0A}$, which combine with $\frac{m_V^2}{2}V^{iA}V_i^{A}$ and $\frac{g_{VH}}{2}|H|^2 V^{iA}V_i^{A}$, respectively, to yield the corresponding Lorentz invariant expressions. At this point the scalar and fermionic contributions to the argument of the exponential are of the same form as in the original Lagrangian and there remains an integration over $\pi(W)^{iA}$ and $\pi(V)^{iA}$. To perform this final integration we decouple  $\pi(W)^{iA}$ and $\pi(V)^{iA}$ by changing variables to $F^{iA}$ and $G^{iA}$ defined by
\begin{align}
    \begin{pmatrix}
    F^{iA} \\
    G^{iA}
    \end{pmatrix}
    =
    \frac{1}{\sqrt{2}}
    \begin{pmatrix}
     1 & 1 \\
    -1 & 1 
    \end{pmatrix}
    \begin{pmatrix}
    \pi(W)^{iA} \\
    \pi(V)^{iA}
    \end{pmatrix}.
\end{align}
The Jacobian of the transformation is unity and the terms involving $\pi(W)^{iA}$ and $\pi(V)^{iA}$ can be written as
\begin{align}
    &\frac{1}{2(g_M+1)}\left[F_i^A+\frac{g_M+1}{\sqrt{2}}\left(W_{0i}^A+V_{0i}^A\right)\right]^2   -\frac{1}{2(g_M-1)}\left[G_i^A-\frac{g_M-1}{\sqrt{2}}\left(V_{0i}^A-W_{0i}^A\right)\right]^2 \nonumber \\
    &-\frac{g_M+1}{4} \left(W^{0iA}+V^{0iA}\right)^2+ \frac{g_M-1}{4} \left(V^{0iA}-W^{0iA}\right)^2,
\end{align}
where for the spatial components of any Lorentz vector $v^i$ we write $(v^i)^2 \equiv v_i v^i$. After shifting $F^{iA}$ and $G^{iA}$ the integrals over these variables yield irrelevant constants and the remaining terms add 
\begin{align}
    -\frac{1}{4}\left(W^{0iA}W_{0i}^A+W^{i0A}W_{i0}^A+V^{0iA}V_{0i}^A+V^{i0A}V_{i0}^A\right)-\frac{g_M}{2}\left(W^{0iA}V_{0i}^A+W^{i0A}V_{i0}^A\right)
\end{align}
to the argument of the exponential. These terms combine with
\begin{align}
    -\frac{1}{4}\left(W^{ijA}W_{ij}^A+V^{ijA}V_{ij}^A\right)-\frac{g_M}{2}W^{ijA}V_{ij}^A
\end{align}
to yield the corresponding fully Lorentz invariant terms. The generating functional at this point is of the form
\begin{align}
    Z = \int \measure \mathcal{Q}\, \measure \lambda^A \, \delta\left(\partial_i W^{iA}\right) \det\left(-M_V^2\right) \det\left(D_i \partial^i\right) \exp{\left\{i\int \rd^d x \, \left[\LagNoT - \frac{M_V^2}{2} \lambda^A \lambda^A \right]\right\}},
\end{align}
and introducing ghost fields $\bar{v}^A$ and $v^A$ we may bring $M_V^2$ up into the exponential and write
\begin{align}
    Z = \int \measure \eta\, \delta\left(\partial_i W^{iA}\right) \det\left(D_i \partial^i\right) \exp{\left\{i\int \rd^dx \, \left[\LagNoT - \frac{M_V^2}{2} \left(\lambda^A \lambda^A + \bar{v}^A v^A\right) \right]\right\}},
\end{align}
where $\eta = \begin{pmatrix} \mathcal{Q} & \lambda^A & \bar{v}^A & v^A \end{pmatrix}^T$. Comparing this functional to Eq.\ (4.3.30) of Ref.\ \cite{guitman1990quantization} we note that it matches the naive expectation for a theory quantised in Coulomb gauge up to terms involving the unphysical degrees of freedom $\lambda^A$, $\bar{v}^A$ and $v^A$. It should be noted that these degrees of freedom can never be produced and hence only appear as internal lines in any Feynman diagram. Furthermore, they only appear quadratically meaning that whenever any of these fields is present as an internal line it has to be part of a closed loop. That is, whenever there is a $\lambda^A$ present internally it has to be part of a closed $\lambda^A$-loop and similarly for $v^A$. Furthermore, none of these fields has a kinetic term, which implies that any closed loop of such fields yields a scaleless integral. Hence, working in \DREG, these fields never contribute to any correlation function, and therefore the theory described by the generating functional omitting these fields is equivalent to the one described by the generating functional containing the fields. We may therefore drop these extra degrees of freedom altogether. The generating functional is then of the form that is naively expected for the quantisation of the theory in Coulomb gauge, which in Ref.\ \cite{guitman1990quantization} is shown to be equivalent to the generating functional obtained through the Fadeev-Popov procedure quantising the theory in a manifestly Lorentz invariant way. We have thus shown that we may proceed with a naively constructed generating functional, as long as we work in \DREG and obey the consistency conditions
\begin{align}
\label{eq: coupling_constr_1}
g_M&=\gamma \frac{g}{g_{*}} \neq \pm 1, \\
\label{eq: coupling_constr_2}
g_{VH}&\geq \frac{g_H^2}{2} = \frac{\gamma_H^2 g_{*}^2}{2}.
\end{align}

\section{Computation of UOLEA operators}
\label{sec: constraints_computation}
In this section we briefly discuss the computation of the necessary UOLEA operators. The situation is very similar to the one encountered in Section~\ref{sec:results}, but instead of $\epsilon$-scalars we want to integrate out the heavy resonances. It is clear that the appropriate second variation for this model can be directly obtained from Eq.\ \eqref{eq: fullsecondvar} by simply replacing $\epsilon_\mu$ by $V_\mu$ and forgetting about the fact that certain matrices were projected onto $Q\epsilon S$. It should be kept in mind that the term quadratic in fluctuations of the resonance fields is not of the form given in Eq.\ \eqref{eq:epsilon_prop}, but rather takes the form
\begin{align}
    \Delta_V^{\mu \nu} =  P^2 g^{\mu \nu}-P^\nu P^\mu  - m^2_V g^{\mu \nu} + X_{VV} ^ {\mu \nu},
    \label{eq:resonance_prop}
\end{align}
where, as has been mentioned before, there is no $R_\xi$ gauge-fixing contribution. Due to this fact, the quadratic term of the resonances cannot be brought into the scalar form by a choice of gauge and it is sensible to not treat the Lorentz indices of resonances as internal indices as was done in Section~\ref{sec: Mixed_statistics} for gauge bosons.\footnote{A possible way of still treating these indices as internal indices is to introduce a projector onto the subspace of resonances for the terms that deviate from the scalar form. This, however, turns out to be very tedious and the treatment presented here is more straightforward.} Furthermore, there are two slightly different ways of including light scalars and gauge bosons into the computation. The choice concerns the contributions with open covariant derivatives. First consider a scalar coupling to resonances. The only way to obtain an open covariant derivative is through a coupling of the resonance to the scalar current as 
\begin{align*}
   i V^\mu \phi D_\mu \phi,
\end{align*}
where we omitted gauge indices as they do not affect the argument. This gives rise to the open covariant derivative term
\begin{align*}
    P_\mu Z_{\phi V}^{\mu \rho}, 
\end{align*}
with 
\begin{align*}
    Z_{\phi V}^{\mu \rho} = -\phi g^{\mu \rho},
\end{align*}
so that effectively we may write 
\begin{align*}
     P_\mu Z_{\phi V}^{\mu \rho} = -P^{\rho} \phi = P^{\rho} \tilde{Z}_{\phi V}.
\end{align*}
In the case of the gauge bosons the corresponding couplings come from the gauge-kinetic term of the resonances and one finds contributions to $\frac{\delta^2 \mathcal{L}}{\delta W_{\rho} \delta V_{\sigma}}$ which mix Lorentz indices of gauge bosons and resonances such as
\begin{align*}
    -D_\mu g^{\rho \sigma} V^\mu = iP_\mu g^{\rho \sigma} V^\mu.
\end{align*}
It is clear that, in order to capture such contributions in $Z_{\phi V}$, in general one has to attribute three Lorentz indices to the object. Alternatively, one can separate $Z_{W V}$ from $Z_{\phi V}$. In the computation performed here we choose the second path for the following reasons. First of all, the mass dimension of $Z_{\phi V}$ is one and hence this object can appear up to six times in UOLEA operators. If each copy carries three Lorentz indices a vast number of possible contractions is created, which in the end turn out to be equivalent as they are simply contractions of metric tensors. If instead $Z_{\phi V}$ does not carry any Lorentz indices, this issue is avoided. Secondly, the mass dimension of $Z_{W V}$ is three. This is due to the fact that $Z^{\mu \rho \sigma}_{W V} \propto V^{\mu A}$ and, since $V^{\mu A}$ is a heavy field it will be replaced by the corresponding classical field. As discussed in Chapter~\ref{chap: fcuntional matching} this is the inverse mass expansion of the solution of the classical equations of motion, whose leading contribution for the model under consideration comes from the fermionic and scalar currents, which both are of mass dimension three. Hence, separating $Z_{\phi V}$ from $Z^{\mu \rho \sigma}_{W V}$ has the benefit of avoiding a large number of Lorentz contractions, which turns out to simplify the computation significantly. In practice one may perform the derivation of the functional trace keeping the Lorentz indices of the gauge bosons as internal indices and then perform the split of $Z_{\phi V}$ into a scalar-resonance part and a gauge-boson-resonance part at the level of the expansion of the functional trace. This is how the computation has been performed. Whereas this will affect the explicit computation of the functional trace, it is entirely inconsequential for the derivation of $\Gamma^{1 \Loop}_\text{L,UV}$, which means that all of the steps leading to Eq.\ \eqref{eq: epsilon_Gamma} go through without modification. Also, the soft-region cancellation is clearly covered by the proof of Section~\ref{sec: soft_region_cancellation_proof}. It then follows that the one-loop contribution to the effective Lagrangian is given by
\begin{align}
   \Lag{EFT}^{1\Loop} = \frac{i}{2} \left. \Tr \log \left[\left(M_{V V}\right)^{\mu \nu}_\xi - \left(M_{V \phi}\right)^\mu_\xi \left(M_{\phi \phi}\right)_\xi^{-1}  \left(M_{\phi V}\right)^\nu_\xi \right]\right|^{P_\mu \to P_\mu-q_\mu}_\text{hard,trunc},
    \label{eq: resonance_EFT_lag}
\end{align}
with the usual notation, see in particular Eq.\ \eqref{eq: useful_Schur_complements} . Here we are of course interested in all contributions up to and including mass dimension six and we may not set $X_{\phi V}^\mu$ and $X_{V \phi}^\mu$ to zero as was possible for the corresponding matrices in Section~\ref{sec:results}. Writing the argument of the logarithm as
\begin{align}
    (P^2 +q^2 -2q\cdot P - m^2_V) g^{\mu \nu} - P^\nu P^\mu - q^\nu P^\mu - P^\nu q^\mu - q^\mu q^\nu   + F_{VV} ^ {\mu \nu},
\end{align}
where 
\begin{align}
F_{VV} ^ {\mu \nu} = \left. \left( X_{VV}^{\mu \nu}-\mathbf{U}_{V \xi}^\mu \mathbf{\Delta}_\xi^{-1} \mathbf{U}_{\xi V}^\nu - \left(M_{V \phi}\right)^\mu_\xi \left(M_{\phi \phi}\right)_\xi^{-1}  \left(M_{\phi V}\right)^\nu_\xi \right) \right|^{P_\mu \to P_\mu - q_\mu},
\end{align}
one can expand the logarithm as
\begin{align}
   &\log \left[(P^2 +q^2 -2q\cdot P - m^2_V) g^{\mu \nu} - P^\nu P^\mu - q^\nu P^\mu - P^\nu q^\mu - q^\mu q^\nu   + F_{VV} ^ {\mu \nu}\right] \nonumber \\
   &= \log \left[-g^{\mu \sigma} \left(-q^2+m_V^2\right)-q^\mu q^\sigma\right] \nonumber \\
   &~~~+ \log \left[\delta^\nu_\sigma - (q^2-m_V^2)^{-1}T_{\lambda \sigma}\left(g^{\nu \lambda} \left(2q \cdot P-P^2\right)+P^\nu P^\lambda - q^{(\nu} P^{\lambda)} - F^{\lambda \nu}\right)\right] \nonumber \\
    &= \log \left[-g^{\mu \sigma} \left(-q^2+m_V^2\right)-q^\mu q^\sigma\right] \nonumber \\
    &~~~-\sum_{n=1}^{\infty}\frac{1}{n} \left[(q^2-m_V^2)^{-1}T_{\lambda \sigma}\left(g^{\nu \lambda} \left(2q \cdot P-P^2\right)+P^\nu P^\lambda - q^{(\nu} P^{\lambda)} - F^{\lambda \nu}\right)\right]^n,
\end{align}
where 
\begin{align}
    T_{\lambda \sigma} = g_{\lambda \sigma} - q_\sigma q_\lambda m_V^{-2}, \\
    \intertext{and} q^{(\nu}P^{\lambda)} = q^\nu P^\lambda + q^\lambda P^\nu.
\end{align}
Inserting this into Eq.\ \eqref{eq: resonance_EFT_lag} and ignoring the infinite constant as usual, we find the following expression for the effective Lagrangian at one loop
\begin{align}
    \Lag{EFT}^{1\Loop} = -\frac{i}{2} \sum_{n=1}^{\infty} \frac{1}{n} \Tr \left[(q^2-m_V^2)^{-1}T_{\lambda \sigma}\left(g^{\nu \lambda} \left(2q \cdot P-P^2\right)+P^\nu P^\lambda - q^{(\nu} P^{\lambda)} - F^{\lambda \nu}\right)\right]^n,
\end{align}
which has to be evaluated in the hard region and truncated at the desired mass dimension. The differences between this trace and the one computed in Section~\ref{sec: Mixed_statistics} are the extra $P^\mu$-dependent terms and the modified propagator. Both modifications introduce extra dependencies on the loop momentum, thus modifying the universal coefficients as compared to the previous computation. Furthermore, the presence of the extra terms leads to further UOLEA operators which were not present before. Whereas the evaluation of this trace is computationally more intensive as there are more possible Lorentz structures in the final result, the computation can be performed using the same Mathematica code that was used to perform the calculation of Section~\ref{sec: Mixed_statistics}. The evaluation was executed as a part of this thesis and the resulting operators were implemented into \tofu. The result of the matching based on this implementation is presented in the next section.
\section{Matching results}
\label{sec: resonance_matching_res}
It is most convenient to remove the kinetic mixing term in order to perform the matching utilising the pre-computed UOLEA operators. This can be achieved by starting from the Lagrangian 
\begin{align}
\label{eq: og_lag}
\LagNoT ={}& -\frac{1}{4}\tilde{V}^{\mu \nu A} \tilde{V}_{\mu \nu}^A - \frac{\tilde{g}_M}{2}\tilde{V}^{\mu \nu A} \tilde{W}_{\mu \nu}^A  +  \frac{\tilde{m}_V^2}{2}\tilde{V}^{\mu A}\tilde{V}_{\mu}^A+\tilde{g}_{H} \tilde{V}^{\mu A} J_\mu^{HA}\nonumber \\ & +\tilde{V}^{\mu A} \sum_{f} \tilde{g}_{f}  J_{\mu}^{fA} + \frac{\tilde{g}_{VH}}{2}|H|^2\tilde{V}^{\mu A}\tilde{V}_{\mu}^A + \Lag{SM},
\end{align}
and performing the following field redefinitions, see Refs.\ \cite{delAguila:2010mx,Pappadopulo:2014qza}
\begin{align}
\begin{cases}
    \tilde{W}^{\mu A} =  W^{\mu A} -\frac{\tilde{g}_M}{\sqrt{1-\tilde{g}_M^2}} V^{\mu A} \\
    \tilde{V}^{\mu A} = \frac{1}{\sqrt{1-\tilde{g}_M^2}} V^{\mu A}
\end{cases},
    \label{eq: field_redefs}
\end{align}
which removes the kinetic mixing and brings the Lagrangian into the form
\begin{align}
\label{eq: redef_lag}
     \LagNoT ={}& -\frac{1}{4}V^{\mu \nu A} V_{\mu \nu}^A  +  \frac{m_V^2}{2}V^{\mu A}V_{\mu}^A+g_{H} V^{\mu A} J_\mu^{HA}\nonumber \\ & + V^{\mu A} \sum_{f} g_{f} J_{\mu}^{fA}+ \frac{g_{VH}}{2}|H|^2V^{\mu A}V_{\mu}^A + \frac{g_{3V}}{2} f^{ABC} V^{\mu A}  V^{\nu B}  V_{\mu \nu}^C \nonumber \\ 
     &- \frac{g_{2VW}}{2} f^{ABC} V^{\mu B}  V^{\nu C}  W_{\mu \nu}^A+ \Lag{SM}.
\end{align}
The new couplings expressed through the old ones are then given by\footnote{Note that the gauge coupling appears in these relations since the redefinition of the gauge field affects the gauge-kinetic terms.}
\begingroup
\allowdisplaybreaks
\begin{align}
\label{eq: coupling_rel_1}
    m_V^2 &= \frac{1}{1-\tilde{g}_M^2}\tilde{m}_V^2, \\
    g_H  &=\frac{1}{\sqrt{1-\tilde{g}_M^2}}(\tilde{g}_H-g_2\tilde{g}_M), \\
    g_f  &=\frac{1}{\sqrt{1-\tilde{g}_M^2}}(\tilde{g}_f-g_2 \tilde{g}_M), \\
    g_{VH}  &=\frac{1}{1-\tilde{g}_M^2}\left(\tilde{g}_{VH}+g_2^2 \tilde{g}_M^2-g_2 \tilde{g}_H \tilde{g}_M\right), \\
    g_{2VW} &= -\frac{1}{1-\tilde{g}_M^2}\tilde{g}_M^2g_2, \\
    g_{3V}  &=\frac{\tilde{g}_M g_2}{(1-\tilde{g}_M^2)^{3/2}}\left(2+\tilde{g}_M^2-3\tilde{g}_M\right)
    \label{eq: coupling_rel_-1}
\end{align}
\endgroup
and we express the result of the matching through these couplings. Notice that we have assumed a universal coupling to fermions $g_f = g_q = g_l$ for the matching below. If the coupling relations Eqs.\ \eqref{eq: coupling_rel_1}-\eqref{eq: coupling_rel_-1} are not enforced, the result presented below amounts to a general matching of the simplified model proposed in Ref.\ \cite{Pappadopulo:2014qza}. However, one has to be careful when interpreting this result since there is no guarantee that the model is self-consistent. There might very well be constraints on certain couplings as derived in Section~\ref{sec: constraint_analysis} for the more restricted case. We leave the constraint analysis of the more general model for future work. The interpretation in terms of the Lagrangian in Eq.\ \eqref{eq: og_lag} is unproblematic as long as the conditions represented by Eqs.\ \eqref{eq: coupling_constr_1} and \eqref{eq: coupling_constr_2} are taken into account. In what follows the SM conventions are the same as in Section~\ref{sec: SSM_to_SM} with the exception that we drop the subscript $h$ on the quartic Higgs coupling.
The tree-level matching gives rise to the following effective Lagrangian
\begin{align}
\Lag{EFT}^{\text{tree}} ={}& \Lag{SM} + \frac{g_{H}^2 \mu_h^2 |H|^4}{2 m_V^2} 
-\frac{g_{H}^2 \lambda Q_H}{2 m_V^2}
-\frac{3g_{H}^2Q_{H\Box}}{8 m_V^2} \nonumber \\
 &-\frac{g_{H}^2 Q^{ij}_{eH} y_{eij}}{4 m_V^2} 
 -\frac{g_{H}^2 Q^{ij}_{dH} y_{dij}}{4 m_V^2} 
 -\frac{g_{H}^2 Q^{ij}_{uH} y_{uij}}{4 m_V^2} \nonumber \\
 &-\frac{g_{H}^2Q^{\dagger ij}_{eH} y^\dagger_{eij}}{4 m_V^2} 
 -\frac{g_{H}^2 Q^{\dagger ij}_{dH} y^\dagger_{dij}}{4 m_V^2} 
 -\frac{g_{H}^2 Q^{\dagger ij}_{uH} y^\dagger_{uij}}{4 m_V^2} \nonumber \\ 
 &-\frac{g_{f}g_{H} Q^{(3)ii}_{Hl}}{4 m_V^2}
 -\frac{g_{f} g_{H} Q^{(3) ii}_{Hq}}{4 m_V^2} 
 +\frac{g_{f}^2 (Q^{{iijj}}_{ll} - 
    2 Q^{{ij}ji}_{ll})}{8 m_V^2} 
 -\frac{g_{f}^2 Q^{(3) iijj}_{qq}}{8 m_V^2}-\frac{g_f^2 Q_{lq}^{(3) iijj}}{4 m_V^2},
\end{align}
where repeated flavour indices are summed over and we use the naming conventions of Ref.\ \cite{Grzadkowski:2010es} with the modification that the Higgs doublet is denoted by $H$ rather than $\varphi$. We see that, at tree level, ten independent operators are induced and there is a shift in $\lambda$ which is suppressed by $\mu_h^2/m_V^2$. Due to this suppression this shift is inconsequential for the application of equations of motion in the translation to Warsaw basis. This tree-level matching was already performed in Ref.\ \cite{Biekoetter:2014jwa}, where the four fermion operators were omitted. The result presented here agrees with that of Ref.\ \cite{Biekoetter:2014jwa} for the operators present in both results.

Next we consider the renormalisable part of the effective action induced by the one-loop matching. Since corrections to the kinetic terms give rise to one-loop contributions to Wilson coefficients of effective operators induced at tree level once the fields of the EFT are canonically normalised, the gauge-kinetic terms are particularly interesting. These read
\begin{align}
   \frac{1}{\kappa}\LagNoT_{\text{EFT,kin.}}^{1\Loop}  ={}& \frac{g_{H}^2(-60 m_V^2 - \mu_h^2 + 6(12 m_V^2 + \mu_h^2) \reslog)}
  {32 m_V^2} \left(D_\mu H\right)^\dagger \left(D^\mu H\right) \nonumber \\ 
  &-\frac{1}{4} \left(-4 (g_2 - 2g_{2VW}) g_{2VW} + 3 g_2 (g_2 + 4 g_{2VW})
    \reslog\right) W^{A\mu \nu} W_{\mu \nu} ^A \nonumber \\ 
    & -\frac{9}{8}g_f^2  \left(\bar{l}i\slashed{D}l+\bar{q}i\slashed{D}q\right),
\end{align}
where we defined $L\equiv \log \left(m_V^2/Q^2\right)$ with $Q$ being the matching scale and $\kappa = 1/(16\pi^2)$. Introducing the canonically normalised fields in the EFT as
\begin{align}
    \hat{\varphi} &= \left(1+\frac{1}{2}\delta Z_\varphi\right) \varphi,\, \varphi \in \{H,l,q,W^A_\mu\}
\end{align}
we find 
\begingroup
\allowdisplaybreaks
\begin{align}
    \delta Z_H &= \kappa\frac{g_{H}^2(-60 m_V^2 - \mu_h^2 + 6(12 m_V^2 + \mu_h^2) \reslog)}{32 m_V^2},\\
  \delta Z_l &= \delta Z_q = -\frac{9}{8}\kappa g_{f}^2, \\
  \delta Z_{W^A_\mu} &= \kappa\left(-4(g_2 - 2g_{2VW}) g_{2VW} + 3 g_2 (g_2 + 4 g_{2VW})
    \reslog\right),
\end{align}
\endgroup
which gives rise to the following contributions to one-loop Wilson coefficients
\begingroup
\allowdisplaybreaks
\begin{align}
\label{eq: Wilson_from_field_redef_1}
     C^{1\Loop,\delta Z}_{H} &=\lambda\frac{g_{H}^4(-45  + 54 \reslog)}{16 m_V^2}, \\
    C^{1\Loop,\delta Z}_{H\Box} &= \frac{g_{H}^4(-45   + 54   \reslog)}{32m_V^2},\\
 C^{1 \Loop ,\delta Z ij}_{\psi H}& = \left(\frac{3g_{H}^2(-15  + 18
 \reslog)}{16}-\frac{9}{16}g_f^2\right)\frac{g_{H}^2 }{4 m_V^2}y_{\psi ij}, \\
 C^{1\Loop, \delta Z ,ij}_{H\Psi^{(3)}} &= \frac{g_{f}g_{H} }{4 m_V^2}\left(-\frac{g_{H}^2(-15  + 18\reslog)}{8}+\frac{9}{8}g_{f}^2\right)\delta^{ij},\\
C^{1\Loop, \delta Z ijkl}_{ll} &= \frac{9}{32 m_V^2}g_{f}^4 (\delta^{ij} \delta^{kl} - 
    2 \delta^{il} \delta^{jk}),\\
    \label{eq: Wilson_from_field_redef_last}
C^{1\Loop,\delta Z ijkl}_{qq^{(3)}}&= -\frac{9}{32 m_V^2}g_{f}^4 \delta^{ij}\delta^{kl},\\
C^{1\Loop,\delta Z ijkl}_{lq^{(3)}}&= -\frac{9}{16 m_V^2}g_{f}^4 \delta^{ij}\delta^{kl},
\end{align}
\endgroup
where $\psi$ runs over all $SU(2)_L$ singlet fermions and $\Psi$ over all $SU(2)_L$ doublet fermions and we define 
\begin{align}
    C_Q = C^{\text{tree}}_Q +  \kappa C^{1\Loop}_Q,
\end{align}
that is we extract a factor of $1/(16\pi^2)$ from the one-loop part of each Wilson coefficient. The remainder of the renormalisable part of the Lagrangian reads
\begingroup
\allowdisplaybreaks
\begin{align}
     \frac{1}{\kappa} \LagNoT_{\text{EFT},d \leq 4}^{1\Loop}  ={}& 3g_{VH} m_V^2 |H|^2 (1 - 3\reslog)
    \nonumber \\
   &-\big[\frac{-216 g_2^2 g_{H}^2 m_V^2 + 3456 g_{VH}^2 m_{V}^2 + 144 g_2^4 \mu_h^2 + 
     229 g_2^2 g_{H}^2 \mu_h^2 - 90 g_{H}^4 \mu_h^2}{576 m_V^2} \nonumber \\ 
     &+ \frac{2088 g_{H}^2 g_{VH} \mu_h^2 - 
     192 g_2^2 g_{H} g_{3V} \mu_h^2 - 720 g_{H}^3 g_{3V} \mu_h^2 + 216 g_{H}^2 g_{3V}^2 \mu_h^2}{576 m_V^2} 
    \nonumber \\
     &+ 
     \frac{384 g_2^3 g_{2VW} \mu_h^2 - 1152 g_2 g_{H} g_{3V} g_{2VW} \mu_h^2 + 960 g_2^2 g_{2VW}^2 \mu_h^2 + 
     18 g_{H}^2 \lambda \mu_h^2}{576 m_V^2} \nonumber \\ 
     & +\frac{- 1350 g_{H}^2 \lambda \mu_h^2 - 
     9 g_1^2 g_{H}^2 (8 m_V^2 - 15\mu_h^2) + 6(24 g_2^4 \mu_h^2 + 96g_2^3 g_{2VW} \mu_h^2}{576 m_V^2} \nonumber \\ 
     & + \frac{-96 g_2 g_{H} (g_{H} + 6 g_{3V})g_{2VW} \mu_h^2 + 9 g_1^2 g_{H}^2 (4 m_V^2 - 3 \mu_h^2)}{576 m_V^2}\nonumber \\
       &+ 
       \frac{18 (48 g_{VH}^2 m_V^2 - 16 g_{H}^2 g_{VH} \mu_h^2 + 
         g_{H}^2 (g_{H}^2 + 8 g_{H} g_{3V} - 12 g_{3V}^2 +14 \lambda)\mu_h^2)}{576 m_V^2} \nonumber \\
         &+ 
       \frac{g_2^2(192 g_{H} g_{3V} \mu_h^2 - 32 g_{2VW}^2 \mu_h^2 + 
         g_{H}^2 (108 m_V^2 - 17 \mu_h^2)))\reslog}{576 m_V^2}\big] |H|^4 \nonumber
    \\ &+ g_{H} \mu_h^2 \frac{12g_{f} - g_{H} + 6(-4g_{f} + g_{H})\reslog}
   {64 m_V^2} \left(\bar{q}y_d d H + \bar{l}y_e e H + \bar{q}y_u u \tilde{H} + \text{h.c.} \right).
\end{align}
\endgroup
In addition to the ten operators present at tree level, the one-loop matching induces 33 further effective operators. Thus, at the one-loop level there are 43 dimension six operators induced in total. The Wilson coefficients, without the inclusion of Eqs.\ \eqref{eq: Wilson_from_field_redef_1}--\eqref{eq: Wilson_from_field_redef_last}, are given below. They are organised according to the classes defined in Ref.\ \cite{Grzadkowski:2010es}, again replacing $\varphi$ by $H$.
\newpage
\begingroup
\allowdisplaybreaks
\begin{align}
     &~~~~~~~~~~~~~~~~~~~~~~~~~~~~~~~~~\boxed{X^3} \nonumber \\
    C_W^{1\Loop} ={}& \frac{(g_2^2 + 4 g_{2VW}^2)(g_{2VW}(4 - 6\reslog) + g_2(2 + 3\reslog))}{24 m_V^2},\\
     &~~~~~~~~~~~~~~~~~~~~~~~~~~~~~~~~~\boxed{H^6}\nonumber \\
     C^{1\Loop}_{H} ={}& \frac{(-3456 g_{VH}^3 - 72 g_{H}^2 g_{VH} \lambda (-29 + 24\reslog) + 
    \lambda (144 g_2^4 (1 + \reslog)}{576 m_V^2} \nonumber \\
    & + \frac{192 g_2^3 g_{2VW} (2 + 3\reslog) + 
      g_2^2 (g_{H}^2 (229 - 102 \reslog) - 192 g_{2VW}^2 (-5 + \reslog)}{576 m_V^2}
      \nonumber \\ &+ 
        \frac{192 g_{H} g_{3V} (-1 + 6\reslog)) - 576 g_2 g_{H} g_{2VW}
       (g_{H} \reslog + g_{3V} (2 + 6\reslog))}{576 m_V^2}\nonumber  \\ 
       &+ 
      \frac{9 g_{H}^2 (-3 g_1^2(-5 + 6 \reslog) + 2 (g_{3V}^2 (12 - 72\reslog) + 
          g_{H}^2 (-5 + 6\reslog)}{576 m_V^2}\nonumber  \\
          & + \frac{8 g_{H} g_{3V} (-5 + 6 \reslog) + 
          2 \lambda (-37 + 42 \reslog)))))}{576 m_V^2},\\
     &~~~~~~~~~~~~~~~~~~~~~~~~~~~~~~~\boxed{H^4D^2}\nonumber \\
    C^{1\Loop}_{H \Box} ={}& \frac{(2 g_1^2 g_{H}^2 (47 - 42\reslog) + 144 g_2^4 (1 + \reslog) + 
    192 g_2^3 g_{2VW} (2 + 3 \reslog)}{768 m_V^2} \nonumber \\
    &+ \frac{-576 g_2 g_{H} g_{2VW}
     (g_{H} \reslog + g_{3V} (2 + 6 \reslog)) - 
    2 g_2^2 (96 g_{2VW}^2 (-5 + \reslog)}{768 m_V^2} \nonumber \\
    & +\frac{- 96 g_{H} g_{3V} (-1 + 6 \reslog) + 
      g_{H}^2 (-137 + 78 \reslog))}{768 m_V^2} 
     \nonumber  \\ &+ \frac{9 (-384 g_{VH}^2 \reslog + 
      g_{H}^4 (-5 + 6 \reslog) + 16 g_{H}^3 g_{3V} (-5 + 6 \reslog)}{768 m_V^2}\nonumber  \\
      &+\frac{- 
      8 g_{H}^2 (3 \lambda (5 - 6 \reslog) + 4 g_{VH} (-4 + 3 \reslog) + 
        3 g_{3V}^2 (-1 + 6 \reslog))))}{768 m_V^2},\\
        C^{1\Loop}_{HD} ={}& \frac{g_1^2 g_{H}^2 (-7 + 12\reslog)}{24 m_V^2},\\
     &~~~~~~~~~~~~~~~~~~~~~~~~~~~~~~~\boxed{X^2H^2}\nonumber \\
     C^{1\Loop}_{HW} ={}&-\frac{(24 g_{H}^2 g_{2VW}^2 (3 - 2\reslog) + 
     6 g_2 g_{2VW}(96 g_{VH} + g_{H}^2 (-11 + 6 \reslog))}{96 m_V^2}\nonumber  \\
     &- \frac{g_2^2 (72 g_{VH} + g_{H}^2(5 + 9 \reslog)))}{96 m_V^2},\\
C^{1\Loop}_{HB} ={}& \frac{-5 g_1^2 g_{H}^2}{64 m_V^2},\\ 
C^{1\Loop}_{HWB} ={}& g_1 \Big(\frac{(9 g_2^3 (1 + \reslog) + 12 g_2^2 g_{2VW}(2 + 3 \reslog)}{288 m_V^2}\nonumber  \\ 
     &+ 
    \frac{18 g_{H} g_{2VW}(g_{3V} - 3 g_{3V} \reslog + g_{H}(-1 + 2\reslog))}{288 m_V^2}\nonumber  \\ 
    &+ \frac{2 g_2 (3 g_{H} g_{3V} (3 + \reslog) + 3 g_{2VW}^2 (3 + \reslog) + 
      g_{H}^2 (-8 + 6 \reslog))}{288 m_V^2} \Big), \\ 
     &~~~~~~~~~~~~~~~~~~~~~~~~~~~~~~\boxed{\psi^2XH}\nonumber \\
C^{1\Loop,ij}_{e/d W} ={}&   \big(\frac{g_{f}(-72(g_{f} - g_{H})g_{2VW}(-3 + 2 \reslog) + 18 g_2 g_{H} (-9 + 4 \reslog)}{576m_V^2}\nonumber \\ &+ \frac{g_2 g_{f}(-85 + 12 \reslog))}{576m_V^2}\big)y_{e/d}^{ij}, \\
C^{1\Loop,ij}_{u W} ={}&   \big(\frac{g_{f}(-72(g_{f} + g_{H})g_{2VW}(-3 + 2 \reslog) + 18 g_2 g_{H} (-9 + 4 \reslog)}{576m_V^2}\nonumber \\ 
&+ \frac{g_2 g_{f}(-85 + 12 \reslog))}{576m_V^2}\big)y_u^{ij}, \\
C^{1\Loop,ij}_{eB} ={}& \frac{g_1 g_{f} (4 g_{f} - 3 g_{H})}{32 m_V^2} y_e^{ij}, C^{1\Loop,ij}_{dB} = -\frac{g_1 g_{f} (4 g_{f} + 9 g_{H})}{96 m_V^2} y_d^{ij}, \\
C^{1\Loop,ij}_{uB} ={}& -\frac{g_1 g_{f} (4 g_{f} - 9 g_{H})}{96 m_V^2} y_u^{ij}, \\
C_{dG}^{1\Loop,ij} ={}& -\frac{g_3 g_{f}^2}{4m_V^2}y_d^{ij},\hspace{1mm} C_{uG}^{1\Loop,ij} = -\frac{g_3 g_{f}^2}{4m_V^2}y_u^{ij}, \\
&~~~~~~~~~~~~~~~~~~~~~~~~~~~~~~~\boxed{\psi^2 H^3}\nonumber \\
C^{1\Loop,ij}_{eH} ={}&  \frac{(-144 g_{f}^2 (y_e y_e^\dagger y_e)_{ij} + y_{e}^{ij} 
    (144 g_2^4 (1 + \reslog) + 192 g_2^3 g_{2VW} (2 + 3 \reslog)} {1152 m_V^2}\nonumber  \\
    &+ 
      \frac{g_2^2 (g_{H}^2 (229 - 102 \reslog) - 192 g_{2VW}^2 (-5 + \reslog) + 
        192 g_{H} g_{3V}(-1 + 6 \reslog))}{1152 m_V^2}\nonumber  \\
        &+\frac{- 576g_2 g_{H} g_{2VW}
       (g_{H} \reslog + g_{3V} (2 + 6 \reslog))}{1152 m_V^2}\nonumber  \\
       &+ 
      \frac{9 g_{H} (-3 g_1^2 g_{H} (-5 + 6 \reslog) + 2 (12 g_{f} \lambda (-1 + 2 \reslog) + 
          g_{H}^3(-5 + 6 \reslog)}{1152 m_V^2}\nonumber  \\
          & +\frac{8 g_{H}^2 g_{3V} (-5 + 6 \reslog) + 
          g_{H} (g_{VH} (116 - 96 \reslog)}{1152 m_V^2}\nonumber  \\ 
          &+ \frac{g_{3V}^2 (12 - 72 \reslog) + 
            \lambda (-73 + 78 \reslog)))))}{1152 m_V^2}, \\
C_{dH}^{1\Loop,ij} ={}& \frac{-144 g_{f}^2 (y_d y_d^\dagger y_d)_{ij} + 72 g_{f}^2 (-11 + 6 \reslog)
    (y_u y_u^\dagger y_d)_{ij}}{1152 m_V^2}\nonumber  \\ 
     &+ \frac{(144 g_2^4(1 + \reslog) + 
      192 g_2^3 g_{2VW} (2 + 3 \reslog) + g_2^2 (g_{H}^2 (229 - 102 \reslog)}{1152 m_V^2}\nonumber \\
      &+\frac{- 
        192 g_{2VW}^2 (-5 + \reslog) + 192 g_{H} g_{3V} (-1 + 6 \reslog))}{1152 m_V^2}\nonumber \\ 
        &+ \frac{- 
      576 g_2 g_{H} g_{2VW} (g_{H} \reslog + g_{3V} (2 + 6 \reslog))}{1152 m_V^2}\nonumber  \\
      &+\frac{ 
      9 g_{H} (-3 g_1^2 g_{H} (-5 + 6 \reslog) + 2 (12 g_{f} \lambda (-1 + 2 \reslog) + 
          g_{H}^3(-5 + 6 \reslog)}{1152 m_V^2}\nonumber  \\
          &+ \frac{8 g_{H}^2 g_{3V} (-5 + 6 \reslog) + 
          g_{H} (g_{VH} (116 - 96 \reslog) + g_{3V}^2 (12 - 72 \reslog)}{1152 m_V^2}\nonumber  \\
          &+\frac{ 
            \lambda (-73 + 78 \reslog)))))y^{ij}_d)}{1152 m_V^2},\\
C_{uH}^{1\Loop,ij} =&{} \frac{-144 g_{f}^2 (y_u y_u^\dagger y_u)_{ij} + 72 g_{f}^2 (-11 + 6 \reslog)
    (y_d y_d^\dagger y_u)_{ij}}{1152 m_V^2} \nonumber \\ 
     &+ \frac{(144 g_2^4(1 + \reslog) + 
      192 g_2^3 g_{2VW} (2 + 3 \reslog) + g_2^2 (g_{H}^2 (229 - 102 \reslog)}{1152 m_V^2}\nonumber \\
      &+\frac{- 
        192 g_{2VW}^2 (-5 + \reslog) + 192 g_{H} g_{3V} (-1 + 6 \reslog))}{1152 m_V^2}\nonumber \\ 
        &+ \frac{- 
      576 g_2 g_{H} g_{2VW} (g_{H} \reslog + g_{3V} (2 + 6 \reslog))}{1152 m_V^2}\nonumber  \\
      &+\frac{ 
      9 g_{H} (-3 g_1^2 g_{H} (-5 + 6 \reslog) + 2 (12 g_{f} \lambda (-1 + 2 \reslog) + 
          g_{H}^3(-5 + 6 \reslog)}{1152 m_V^2}\nonumber  \\
          &+ \frac{8 g_{H}^2 g_{3V} (-5 + 6 \reslog) + 
          g_{H} (g_{VH} (116 - 96 \reslog) + g_{3V}^2 (12 - 72 \reslog)}{1152 m_V^2}\nonumber  \\
          &+\frac{ 
            \lambda (-73 + 78 \reslog)))))y^{ij}_u)}{1152 m_V^2},\\
    &~~~~~~~~~~~~~~~~~~~~~~~~~~~~~~~\boxed{\psi^2 H^2 D}\nonumber \\
    C_{He}^{1\Loop,ji} ={}& -\frac{g_1^2 g_H^2 (19 + 6 \reslog)}{384 m_V^2} \delta^{ij} +\frac{- 
   12 g_f(9 g_H - 6 g_H \reslog + g_f (-7 + 6\reslog))
    (y_e^\dagger y_e)_{ji}}{384 m_V^2},\\    
C_{Hu}^{1\Loop,ji} ={}&\frac{g_1^2 g_H^2(19 + 6 \reslog)}{576 m_V^2} \delta^{ij} +\frac{ 18 g_f (9 g_H - 6 g_H \reslog + g_f(-7 + 6\reslog))}{576 m_V^2}(y_u^\dagger y_u)_{ji}, \\
C_{Hd}^{1\Loop,ji} ={}&  \frac{-g_1^2 g_H^2(19 + 6 \reslog)}{1152 m_V^2}\delta^{ij} -\frac{36 g_f(9 g_H - 6 g_H \reslog + g_f (-7 + 6 \reslog))}{ 1152 m_V^2}(y_d^\dagger y_d)_{ji}, \\ 
C_{Hud}^{1\Loop,ji} =&{} -\frac{g_f(g_H(9 - 6\reslog) + g_f(-7 + 6 \reslog))}
 {16 m_V^2}(y_u^\dagger y_d)_{ji},\\
C_{Hl^{(1)}}^{1\Loop,ij} ={}& -\frac{g_1^2(g_{H}^2(19 + 6\reslog) + g_{f}^2(58 + 60 \reslog))}{768 m_V^2} \delta^{ij} \nonumber \\
&+ 
     \frac{36 g_{f}(g_{f} + g_{H}(7 - 10 \reslog) + 6 g_{f} \reslog)
      (y_e y_e^\dagger)_{ij}}{768 m_V^2},\\ 
C_{Hl^{(3)}}^{1\Loop,ij}  ={}&
     \Big(\frac{((288 g_2^4 (1 + \reslog) + 384 g_2^3 g_{2VW} (2 + 3 \reslog) 
     + 
    g_2^2(-384 g_{2VW}^2 (-5 + \reslog)}{2304 m_V^2} \nonumber \\
    & + \frac{192 g_{H}g_{3V}(-1 + 6\reslog) + 
      6 g_{f}^2 (-17 + 54 \reslog) + g_{H}^2 (49 + 114\reslog)}{2304 m_V^2}\nonumber \\ 
      &+\frac{ 
      48 g_{f} (g_{H} (45 - 54 \reslog) + 4 g_{3V} (-1 + 6 \reslog)))}{2304 m_V^2}\nonumber \\
      &+ \frac{- 
    288 g_2 g_{2VW} (g_{f}^2 (1 + 4 \reslog) + 
      2 g_{H} (g_{H} \reslog + g_{3V} (2 + 6 \reslog))}{2304 m_V^2}\nonumber \\
      &+\frac{ 
      g_{f} (g_{H} (5 - 6 \reslog) + 4 (g_{3V} + 3 g_{3V} \reslog)))}{2304 m_V^2}\nonumber  \\
    &+\frac{-72 g_{f} g_{H} (3 g_{f}^2 + 12 g_f g_{3V}+ g_f g_H (5 - 6 \reslog) + 
      4 g_{VH} (-5 + 6 \reslog)}{2304 m_V^2}\nonumber  \\
      &+ \frac{2 g_{3V}(g_H (5 - 6\reslog) + 
        3 g_{3V}(-1 + 6 \reslog)))}{2304 m_V^2}\Big) \delta^{ij}
        \nonumber \\&+ 
    \frac{36 g_{f} (g_{f} + 6 g_{f} \reslog + g_{H} (-1 + 6 \reslog))}{2304 m_V^2}(y_e y_e^\dagger)_{ij}, \\
C_{Hq^{(1)}}^{1\Loop,ij} ={}& \frac{g_1^2 (g_H^2 (19 + 6 \reslog) + g_f^2(58 + 60\reslog))}{2304 m_V^2}\delta^{ij}  \nonumber \\
   &+ \frac{108 g_f (g_f + g_H (7 - 10 \reslog) + 6 g_f \reslog)}{2304 m_V^2} (y_u y_u^\dagger - y_d y_d^\dagger)_{ij}, \\ 
C_{Hq^{(3)}}^{1\Loop,ij} ={}& \Big(\frac{288 g_2^4 - 102 g_2^2 g_f^2 + 2160 g_2^2 g_f g_H - 216 g_f^3 g_H + 49 g_2^2 g_H^2 - 360 g_f^2 g_H^2}{2304 m_V^2}\nonumber  \\ 
&+ \frac{1440 g_f g_H g_{VH}- 
    192 g_2^2 g_f g_{3V} - 192 g_2^2 g_H g_{3V} - 864 g_{f}^2 g_H g_{3V}}{2304 m_V^2}\nonumber  \\
    &+\frac{- 
    720 g_f g_H^2 g_{3V} + 432 g_f g_H g_{3V}^2 + 768 g_2^3 g_{2VW}- 
    288 g_2 g_f^2 g_{2VW}}{2304 m_V^2} \nonumber \\ 
    &+\frac{- 1440 g_2 g_f g_H g_{2VW} - 1152 g_2 g_f g_{3V}g_{2VW} - 1152 g_2 g_H g_{3V} g_{2VW}}{2304 m_V^2} \nonumber \\
    &+ \frac{1920 g_2^2 g_{2VW}^2 + (288 g_2^4 + 
    324 g_2^2 g_f^2 - 2592 g_2^2 g_f g_H  + 
    114 g_2^2 g_H^2 + 432 g_f^2 g_H^2}{2304 m_V^2}\nonumber \\
    &+\frac{- 
    1728 g_f g_H g_{VH} + 1152 g_2^2 g_f g_{3V} + 
    1152 g_2^2 g_H g_{3V} + 
    864 g_f g_H^2 g_{3V}}{2304 m_V^2} \nonumber \\ 
    &+\frac{- 2592 g_f g_H g_{3V}^2 + 
    1152 g_2^3 g_{2VW} - 1152 g_2 g_f^2 g_{2VW} + 
    1728 g_2 g_f g_H g_{2VW}}{2304 m_V^2} \nonumber \\
    &+\frac{- 576 g_2 g_H^2 g_{2VW}- 
    3456 g_2 g_f g_{3V} g_{2VW}}{2304 m_V^2} \nonumber \\ 
    &+\frac{- 3456 g_2 g_H g_{3V} g_{2VW} - 
    384 g_2^2 g_{2VW}^2)\reslog}{2304 m_V^2}\Big) \delta^{ij}\nonumber  \\
    &+\frac{36 g_f (g_f (1 + 6\reslog) + 
      g_H (-1 + 6\reslog))}{2304 m_V^2}((y_d y_d^\dagger)_{ij}+ (y_u y_u^\dagger)_{ij}),\\
          &~~~~~~~~~~~~~~~~~~~~~~~~~~~~~~~\boxed{(\bar{L}L)(\bar{L}L)}\nonumber \\
    C^{1\Loop,ijkl}_{ll} ={}& \Big(\frac{-24 g_{2}^4 (1 + \reslog) - 32 g_{2}^3 g_{2VW} (2 + 3\reslog)}{384m_{V}^2} \nonumber \\ & + 
    \frac{g_{f}^2(144g_{3V}g_{f} + g_{1}^2(137 - 42L) + 2g_{f}^2(127 - 36\reslog) + 
      36g_{3V}^2(-1 + 6\reslog))}{384m_{V}^2} \nonumber \\ & +\frac{96g_{2} g_{2VW} g_{f} (-(g_{f} (-3 + \reslog)) + g_{3V} (2 + 6\reslog))}{384m_{V}^2} \nonumber \\ & + \frac{ 
    g_{2}^2 (32 g_{2VW}^2 (-5 + \reslog) + g_{f} (32 g_{3V} - 795 g_{f} - 192 g_{3V} \reslog + 594 g_{f} \reslog))}{384m_{V}^2}\Big) \delta^{ij} \delta^{kl} \nonumber \\
&+\Big(\frac{ 
  2 24 g_{2}^4 (1 + \reslog) + 32 g_{2}^3 g_{2VW} (2 + 3\reslog)}{384m_{V}^2} \nonumber \\ & + 
    \frac{96 g_{2} g_{2VW} g_{f} (g_{f} (-3 + \reslog) - 2 g_{3V} (1 + 3 \reslog))}{384m_{V}^2} \nonumber \\ & + 
    \frac{2 g_{f}^2 (-72 g_{3V} g_{f} + 18 g_{1}^2 (-3 + 2\reslog) - 18 g_{3V}^2 (-1 + 6 \reslog) + 
      g_{f}^2 (-73 + 36 \reslog))}{384m_{V}^2} \nonumber \\ & + \frac{g_{2}^2 (-32 g_{2VW}^2 (-5 + \reslog) + 
      g_{f} (471 g_{f} - 378 g_{f} \reslog + 32 g_{3V} (-1 + 6 \reslog)))}{384m_{V}^2}\Big)
   \delta^{il} \delta^{jk},\\
C^{1\Loop,ijkl}_{qq^{(1)}}={}& \frac{g_f^2(474 g_3^2 + 972 g_f^2 - 612 g_3^2 \reslog + 972 g_2^2 (-3 + 2 \reslog) + 
      g_1^2 (29 + 30 \reslog))}{3456 m_V^2}\delta^{ij} \delta^{kl} \nonumber \\
      & +\frac{9 g_f^2 g_3^2 (-79 + 102 \reslog)}{3456 m_V^2}\delta^{il} \delta^{jk}, \\
C^{1\Loop,ijkl}_{qq^{(3)}}={}&  \Big(\frac{24 g_{2}^4 (1 + \reslog) + 32 g_{2}^3 g_{2VW} (2 + 3 \reslog)}{384m_{V}^2} \nonumber \\ & + 
    \frac{96 g_{2} g_{2VW} g_{f} (g_{f} (-3 + \reslog) - 2 g_{3V} (1 + 3 \reslog))}{384m_{V}^2}\nonumber\\  & + 
    \frac{2 g_{f}^2 (-72 g_{3V} g_{f} + g_{1}^2 (-6 + 4 \reslog) - 18 g_{3V}^2 (-1 + 6\reslog) + 
      g_{f}^2 (-73 + 36 \reslog))}{384m_{V}^2} \nonumber \\ & + \frac{g_{2}^2 (-32 g_{2VW}^2 (-5 +\reslog) + 
      g_{f} (471 g_{f} - 378 g_{f} \reslog + 32 g_{3V}(-1 + 6 \reslog)))}{384m_{V}^2}\Big)
   \delta^{ij}\delta^{kl} \nonumber \\ &+ \frac{g_{3}^2 g_{f}^2 (-79 + 102 \reslog)}{384m_{V}^2}
   \delta^{il} \delta^{jk} \\ 
C^{1\Loop,ijkl}_{lq^{(1)}} ={}& \frac{g_f^2(-(g_1^2(29 + 30 \reslog)) + 324(g_f^2 + g_2^2(-3 + 2 \reslog)))}{576m_V^2} \delta^{ij} \delta^{kl}, \\
C^{1\Loop,ijkl}_{lq^{(3)}}  ={}& \Big(\frac{48 g_{2}^4 (1 + \reslog) + 64 g_{2}^3 g_{2VW} (2 + 3 \reslog) - 
   48 g_{2} g_{2VW} (8 g_{3V} g_{f} (1 + 3 \reslog)}{384 m_{V}^2} \nonumber \\ &
   +\frac{2 g_{f}^2 (1 + 4 \reslog) - 
     2 g_{f}^2 (-5 + 6 \reslog)) - 4 g_{f}^2 (72 g_{3V} g_{f} + 73 g_{f}^2 - 36 g_{f}^2 \reslog}{384 m_{V}^2} \nonumber \\& + 
     \frac{6 g_{1}^2 (-3 + 2 \reslog) + 18 g_{3V}^2 (-1 + 6 \reslog)) + 
   g_{2}^2 (942 g_{f}^2 - 64 g_{2VW}^2 (-5 + \reslog)}\nonumber \\ &+ \frac{- 756 g_{f}^2 \reslog + 
     64 g_{3V} g_{f} (-1 + 6 \reslog))}{384 m_{V}^2}\Big)\delta^{ij} \delta^{kl}, \\
     &~~~~~~~~~~~~~~~~~~~~~~~~~~~~~~~\boxed{(\bar{L}L)(\bar{R}R)}\nonumber \\
     C_{le}^{1\Loop,ijkl} ={}& \frac{g_1^2 g_f^2 (29 + 30 \reslog)\delta^{ij} \delta^{kl}}{192 m_V^2}\nonumber  \\ & + \frac{3 (12 g_f g_H (1 - 2\reslog) + 
      4 g_f^2(-7 + 6\reslog) + g_H^2(-1 + 6\reslog))
     y_e^{il} y_e^{\dagger kj}}{192 m_V^2},\\
     C^{1\Loop,ijkl}_{lu} ={}& -\frac{g_1^2 g_f^2 (29 + 30 \reslog)}{288 m_V^2} \delta^{ij} \delta^{kl},  \\
    C^{1\Loop,ijkl}_{ld} ={}& \frac{g_1^2 g_f^2 (29 + 30 \reslog)}{576 m_V^2} \delta^{ij} \delta^{kl}, \\
    C^{1\Loop,ijkl}_{qe} ={}& -\frac{g_1^2 g_f^2 (29 + 30 \reslog)}{576 m_V^2} \delta^{ij} \delta^{kl}, \\
     C_{qd^{(1)}}^{1\Loop,ijkl} ={}& -\frac{g_1^2 g_f^2 (29 + 30\reslog)  \delta^{ij} \delta^{lk}}{1728 m_V^2} \nonumber \\ &
   + \frac{9(12 g_f g_H(1 - 2\reslog) + 
      4 g_f^2 (-7 + 6 \reslog) + g_H^2 (-1 + 6 \reslog))
     y_d^{il} y_d^{\dagger kj}}{1728 m_V^2},\\   
    C^{1\Loop,ijkl}_{qd^{(8)}} ={}& \frac{g_3^2 g_f^2(29 + 30 \reslog)}{96 m_V^2} \delta^{ij} \delta^{kl} \nonumber \\
      &+ \frac{3 (12 g_f g_H (1 - 2 \reslog) + 
      4 g_f^2 (-7 + 6\reslog) + g_H^2 (-1 + 6\reslog))
     y_d^{il} y_d^{\dagger jk}
   }{96 m_V^2},\\
   C_{qu^{(1)}}^{1\Loop,ijkl} ={}& \frac{2 g_1^2 g_f^2 (29 + 30\reslog)  \delta^{ij} \delta^{lk}}{1728 m_V^2} \nonumber \\ &
   + \frac{9(12 g_f g_H(1 - 2\reslog) + 
      4 g_f^2 (-7 + 6 \reslog) + g_H^2 (-1 + 6 \reslog))
     y_u^{il} y_u^{\dagger kj}}{1728 m_V^2}, \\
      C^{1\Loop,ijkl}_{qu^{(8)}}={}& \frac{g_3^2 g_f^2(29 + 30 \reslog) \delta^{ij} \delta^{lk}}{96 m_V^2}\nonumber  \\ 
 &+\frac{3(12g_f g_H (1 - 2 \reslog) + 4 g_f^2 (-7 + 6 \reslog) + 
      g_H^2 (-1 + 6 \reslog)) y_u^{il} y_u^{\dagger kj}}{96 m_V^2 },\\
     &~~~~~~~~~~~~~~~~~~~~~~\boxed{(\bar{L}R)(\bar{R}L)\text{ and }(\bar{L}R)(\bar{L}R)}\nonumber \\
 C_{ledq}^{1\Loop,ijlk} ={}& \frac{((g_H^2(1 - 6\reslog) + 12g_f g_H (-1 + 2 \reslog) - 
    4 g_f^2 (-7 + 6 \reslog))}{32 m_V^2} y_e^{ij} y_d^{\dagger lk},\\
C_{quqd^{(1)}}^{1\Loop,ijkl} ={}& \frac{-8 g_f^2 (-3 + 2 \reslog) y_d^{il} y_u^{kj}}{32 m_V^2} \nonumber \\
&+ \frac{(g_H^2 (1 - 6 \reslog) + 12 g_f g_H (-1 + 2 \reslog) + 
      4 g_f^2 (-11 + 6 \reslog))y_d^{kl} y_u^{ij}}{32 m_V^2},\\
C_{lequ^{(1)}}^{1\Loop,ijkl} ={}& \frac{8 g_f^2 + 12 g_f g_H(1 - 2 \reslog) + g_H^2(-1 + 6 \reslog)}{32 M_V^2}y_e^{ij} y_u^{kl},\\
C^{1\Loop,ijkl}_{lequ^{(3)}} ={}& \frac{3 g_f^2 (-3 + 2 \reslog)}{32 m_V^2} y_e^{ij} y_u^{kl},\\
C^{1\Loop,ijkl}_{quqd^{(8)}} ={}& \frac{-3g_f^2 (-3 + 2 \reslog)}{2 m_V^2} y_d^{il} y_u^{kj}. 
\end{align}
\endgroup
As is clear from the results presented here, the matching of the model defined by the Lagrangian of Eq.\ \eqref{eq: redef_lag} is a non-trivial task. Yet, the model is still very simple as it only extends the SM by one new field. Comparing this to SM extensions such as supersymmetry it becomes clear that for the study of such models in the context of the SMEFT or other low-energy EFTs a tool such as \tofu is unavoidable.

In the near future the Wilson coefficients computed here will be used in a phenomenological study aimed at investigating SMEFT fits in the context of explicit extensions of the SM. Such a study, however, is beyond the scope of this thesis.

\section{A remark on renormalisation}
As discussed in Chapter~\ref{chap: fcuntional matching}, the matching condition used to derive all of the results presented in this thesis is the equality of renormalised correlation functions. However, in UOLEA applications the renormalisation is never performed explicitly. Indeed, the final result contains divergences represented by $1/\epsilon$-poles, which in principle can be subtracted systematically by adding counterterms to the tree-level Lagrangians and carrying them over to the final result. In practice, however, the $1/\epsilon$-poles are simply dropped corresponding to $\overline{\text{MS}}$-renormalisation. In cases where one starts from a renormalisable theory and matches it to an EFT this is clearly justified. The model considered in this chapter, however, does not belong to this class of theories as it has to be interpreted as an EFT itself. Indeed, in similar models restrictions on the couplings arise from the condition of renormalisability in the sense of EFTs, as pointed out for the case of the $\rho$-meson in Ref.\ \cite{Djukanovic:2004mm}. In order to assure that no further constraints on the couplings arise from this condition one has to show that all terms appearing in the renormalisation of the model can be absorbed in terms corresponding to higher dimensional operators constructed from the fields, and constrained by the symmetries, of the model. Such an analysis is beyond the scope of this thesis. 

The non-renormalisability of the UV model also has practical implications for the matching, as it affects the structure of logarithmic contributions to one-loop Wilson coefficients and in principle requires the addition of higher dimensional operators in the UV theory for a complete matching. This can be seen as follows. For a renormalisable UV theory, one would expect logarithmic one-loop matching contributions only in Wilson coefficients which have tree-level contributions or mix with tree-level induced Wilson coefficients under the renormalisation group, for the following reason. Assume that a certain correlation function in the UV theory, which yields a contribution to a given Wilson coefficient in the matching, is divergent at one loop. Then, this correlation function must have a non-zero tree-level contribution as this is needed to provide the counterterms that cancel the divergence. Therefore, there is a tree-level contribution to the Wilson coefficient. A divergence in the corresponding correlation function of the EFT can only arise due to a tree-level induced operator that mixes with the operator whose Wilson coefficient is fixed by the divergent correlation function. Since divergences and logarithms are in one-to-one correspondence this justifies the statement. For the Wilson coefficients presented here, the statement is not true. For example, the Wilson coefficient $C_W$ contains logarithmic terms although $Q_W$ does not mix with any of the tree-level induced operators. These logarithmic contributions can be traced to the non-renormalisabiltiy of the UV theory. For example, the contribution to the Wilson coefficient $C_W$ proportional to $g_{2VW}^3$ arises from the vertex of the form
\begin{align}
    -\frac{g_{2VW}}{2}f^{ABC}V^{\mu B} V^{\nu C} W_{\mu \nu}^A,
    \label{eq: relevant_coupling}
\end{align}
which yields a triangle diagram contributing to $Q_W$, schematically depicted in Fig.\ \ref{Fig: UV_div_matching_illustration} below.
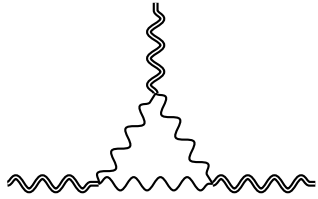
\begin{figure}[h]
\center
\begin{tikzpicture}[scale = 1.5]
\draw[thick, double, snake it] (3.5,2.7) -- (4.3,2.7);
\draw[thick, double, snake it] (5.3,2.7) -- (6.2,2.7);
\draw[thick, double, snake it] (4.8,3.5) -- (4.8,4.3);
\draw[thick, snake it] (4.3,2.7) -- (5.3,2.7);
\draw[thick, snake it] (4.3,2.7) -- (4.8,3.5);
\draw[thick, snake it] (4.8,3.5) -- (5.3,2.7);
\end{tikzpicture}
\caption{Schematic depiction of the $g_{2VW}^3$ contribution to $C^{1\Loop}_W$. The double line corresponds to the $SU(2)_L$ field strength tensor and the single line to the vector triplet.}
\label{Fig: UV_div_matching_illustration}
\end{figure}
The momentum integral appearing in this diagram, at zero external momentum, is given by
\begin{align}
    \int \frac{\rd^d q}{(2 \pi)^d} \frac{1}{(q^2-m_V^2)^3}\left(g_{\mu \nu}-\frac{q_\mu q_\nu}{m_V^2}\right)\left(g_{\rho \sigma}-\frac{q_\rho q_\sigma}{m_V^2}\right)\left(g_{\kappa \lambda}-\frac{q_\kappa q_\lambda}{m_V^2}\right),
\end{align}
and yields divergent contributions due to the parts of the propagator, which are not momentum-suppressed. If these parts were absent, the integral would be finite. Thus, the divergences come exactly from the part of the vector field propagator that renders the theory non-renormalisable in the strict sense. It was checked that in the UOLEA approach, the logarithmic contributions to $C_W$ proportional to $g_{2VW}^3$ arise from operators of the form $X_{VV}^3$, where the relevant part of $X_{VV}$ is precisely due to the coupling of Eq.\ \eqref{eq: relevant_coupling}. These contributions can be interpreted as the diagram in Fig.\ \ref{Fig: UV_div_matching_illustration}, with the field strength tensors constructed from background fields. Clearly, divergent contributions to several correlation functions arise in this way. In order to absorb these divergences one has to add to the Lagrangian of Eq.\ \eqref{eq: redef_lag} a Lagrangian containing all necessary higher dimensional operators. This means that these higher dimensional operators are present in the UV theory at tree level and in principle contribute to the matching, both at tree level and at the one-loop level. This introduces a plethora of input parameters in the UV theory, which have been set to zero in the matching presented in this thesis. In this sense, the matching presented here is incomplete. However, one has to also consider the physical origin of these higher dimensional operators. They arise from the remaining NP sector when it is integrated out. Therefore, if the mass hierarchy between the lightest state in this NP sector, represented by the resonance $V^{\mu A}$, and the next-to-lightest state of the NP sector is sufficiently large the results presented here can be used. Sufficiently large in this case means that the suppression of higher dimensional operators by the mass of the next-to-lightest state is larger than the suppression by the loop factor, since we are including one-loop effects in the matching. Note that if the full one-loop renormalisation up to mass dimension six of the model is performed, one can use that result to check the logarithmic terms appearing in the one-loop Wilson coefficients. 

\clearpage
\lhead{\emph{Conclusions and outlook}}
\chapter{Conclusions and outlook}
\label{ch: conclusions and outlook}
The various shortcomings of the SM combined with exclusion limits on the masses of new particles which often lie in the TeV range suggest the presence of a mass hierarchy between the SM and NP. EFTs are an excellent tool for the study of such models, as they allow for the resummation of large logarithmic corrections as well as the systematic inclusion of NP effects in terms of a perturbative expansion in inverse masses of the new particles. In order to construct a suitable EFT from a given NP model, a matching computation has to be performed. To capture most of the relevant NP effects it is necessary to perform this matching computation at the one-loop level including operators up to mass dimension six in the EFT. To avoid redundant steps associated with such a matching computation the UOLEA was introduced in Refs.\ \cite{Drozd:2015rsp, Ellis:2017jns}. The expressions presented there represent a master formula for the effective action including operators up to mass dimension six and can be applied to scalar theories.

In this thesis we introduced the BSUOLEA, which is an extension of the results of Refs.\ \cite{Drozd:2015rsp, Ellis:2017jns}, that allows for the matching of any renormalisable, Lorentz invariant UV theory, containing scalar fields, vector fields and spin-$1/2$ fermions, to an EFT including operators up to mass dimension six. The specific contributions that were calculated for that purpose are summarised in Table~\ref{table: UOLEA_table_concl}, where they are marked by BS.
\begin{table}[h]
\centering
\begin{tabular}{|c|ccc|c}
\cline{1-4}
 & Heavy-only & heavy-light & derivative couplings  &  \\ \cline{1-4}
Bosonic & \cite{Drozd:2015rsp} & \cite{Ellis:2017jns} & \color{green!65!black}BS &  \\
Fermionic (pure) & \color{green!65!black}BS \color{black}, \cite{Angelescu:2020yzf}, \cite{Ellis:2020ivx} & \color{green!65!black}BS & NR &  \\
Fermionic (mixed) & \color{green!65!black}BS & \color{green!65!black}BS & NR &  \\
Mixed statistics & \color{green!65!black}BS & \color{green!65!black}BS & NR &  \\ \cline{1-4}
\end{tabular}
\caption{Overview of the different contributions to the UOLEA divided into contributions arising from heavy-only and heavy-light loops as well as loops depending on derivative couplings. Fermionic (pure) denotes loops with only one kind of fermion, i.e.\ either Dirac or Majorana. Fermionic (mixed) denotes loops with at least one Dirac and one Majorana fermion. Mixed statistics denotes loops containing at least one boson and one fermion. Contributions that were computed as a part of this thesis are marked by BS. Contributions marked by NR only arise when the UV theory is non-renormalisable.}
\label{table: UOLEA_table_concl}
\end{table}

In order to facilitate the use of the BSUOLEA, it was implemented into the currently private Mathematica package \tofu. Supplemented with the code \route, which was also developed for the purpose of this thesis and allows for the translation of a set of redundant SMEFT operators into the Warsaw basis, this solves the problem of matching a generic, renormalisable, Lorentz invariant UV theory to the dimension six SMEFT at the one-loop level. In addition, \tofu allows for the matching of UV theories to more general low-energy EFTs. It is our ambition to make these matching tools publicly available in the near future. 

Currently it is not possible to apply the BSUOLEA to the matching of a non-renormalisable UV theory to a low-energy EFT. Such scenarios can be of interest when the NP model itself has a limited range of validity as is the case for the model that was considered in Chapter~\ref{ch: resonances}. In fact, Section~\ref{sec: constraints_computation} can be regarded as a first step towards the inclusion of the terms necessary to treat such models using the BSUOLEA. It is desirable to extend the BSUOLEA to fully cover such models as this would also allow for the computation of anomalous dimensions in generic EFTs as discussed in Refs.\ \cite{Henning:2016lyp,Kramer:2019fwz}. Furthermore, the functional methods employed throughout this work can be used to compute the divergent parts of 1PI correlation functions of a given model at the one-loop level. They could therefore be used to study which higher dimensional operators are needed for the renormalisation of the model considered in Chapter~\ref{ch: resonances} at mass dimension six. Combined with the possibility of computing the anomalous dimensions for this model, this would allow for the confirmation of the logarithmic terms of the one-loop Wilson coefficients presented in Section~\ref{sec: resonance_matching_res}.







\addtocontents{toc}{\vspace{2em}} 

\appendix 

\clearpage
\lhead{\emph{Appendix A}}
\chapter{Generating functional of 1LPI correlation functions}
\label{app: 1LPI_gen}
\begin{figure}[h]
\center
\begin{tikzpicture}[scale = 1.5]
\draw[thick] (4.8,1.8) -- (4.4,1.4);
\draw[thick] (4.8,1.8) -- (5.2,1.4);
\draw[thick] (4.8,3) -- (4.8,3.6);
\draw[thick,dashed] (4.8,2.4) -- (4.8,1.8);
\draw[thick, dashed] (4.8,2.7) circle [radius=0.3];;
\node[circle, fill, inner sep=1pt,label=below:x] at (4.4,1.4)%
{};
\node[circle, fill, inner sep=1pt,label=below:y] at (5.2,1.4)%
{};
\node[circle, fill, inner sep=1pt,label=above:z] at (4.8,3.6)%
{};
\node[align=left, above] at (6,2.7)%
{};
\end{tikzpicture}
\caption{Example of a diagram that is 1LPI, but not 1PI. The dashed line represents the heavy field and the solid line represents the light field.}
\label{Fig: counterex}
\end{figure}
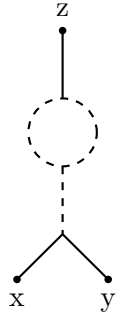

We here briefly discuss the generating functional of 1LPI correlation functions for the case of a theory containing heavy fields $\Phi$ and light fields $\phi$. The results presented here are well-known and can be found in several textbooks such as in Chapters 9 and 11 of Ref.\ \cite{Peskin:1995ev}. In Chapter \ref{chap: fcuntional matching} the generating functional of 1LPI correlation functions was given as
\begin{align}
\Gamma_\text{L}[\bg{\phi}]=-i \log Z[J_\Phi=0,J_\phi]-\int \rd^d x \, J_\phi(x) \bg{\phi}(x),
\label{eq: 1LPI_functional_def}
\end{align}
with $Z$ being the generating functional of all correlation functions given by
\begin{align}
Z[J_\Phi,J_\phi]=\int \measure \Phi \measure \phi \, \exp\left \{i \int \rd^d x \, \big[\LagNoT[\Phi,\phi]+J_{\Phi}(x) \Phi(x)+J_{\phi}(x) \phi(x) \big]\right\},
\end{align}
where $J_\Phi$ and $J_\phi$ are the sources for the heavy and light fields, respectively. The generating functional of 1PI correlation functions, $\Gamma[\bg{\Phi},\bg{\phi}]$, is obtained from $Z[J_\Phi,J_\phi]$ as
\begin{align}
   \Gamma[\bg{\Phi},\bg{\phi}] &= -i \log Z[J_\Phi,J_\phi]- \left(J_\phi\right)_x \left(\bg{\phi}\right)_x- \left(J_\Phi\right)_x \left(\bg{\Phi}\right)_x,
\end{align}
where we remind the reader that $W[J_\Phi,J_\phi]\equiv -i \log Z[J_\Phi,J_\phi]$ is the generating functional of connected correlation functions. Generally, $\bg{\phi}$ and $\bg{\Phi}$ are defined as
\begin{align}
    \bg{\Phi} &= \frac{\delta W}{\delta J_\Phi}, \\
    \bg{\phi} &= \frac{\delta W}{\delta J_\phi},
\end{align}
and depend on both $J_\Phi$ and $J_\phi$. However, setting $J_\Phi = 0$ the light background field appearing in Eq.\ \eqref{eq: 1LPI_functional_def} satisfies
\begin{align}
    \bg{\phi} = \bg{\phi}[J_\Phi = 0, J_\phi] = \left . \frac{\delta W}{\delta J_\phi}\right \vert_{J_\Phi=0}.
    \label{eq: light_bg_field}
\end{align}
Denoting $W[J_\Phi=0,J_\phi]$ by $\tilde{W}[J_\phi]$ we find 
\begin{align}
    \frac{\delta \Gamma_L[\bg{\phi}]}{\delta \left(\bg{\phi}\right)_y} &= \frac{\delta \tilde{W}[J_\phi]}{\delta \left(\bg{\phi}\right)_y}-\frac{\delta \left(J_\phi\right)_x}{\delta \left(\bg{\phi}\right)_y}\left(\bg{\phi}\right)_x-\left(J_\phi\right)_y \nonumber \\
    &= \frac{\delta \tilde{W}[J_\phi]}{\delta \left(J_\phi\right)_x} \frac{\delta \left(J_\phi\right)_x}{\delta \left(\bg{\phi}\right)_y}-\frac{\delta \tilde{W}[J_\phi]}{\delta \left(J_\phi\right)_x} \frac{\delta \left(J_\phi\right)_x}{\delta \left(\bg{\phi}\right)_y}-\left(J_\phi\right)_y \nonumber \\
    &= -\left(J_\phi\right)_y,
\end{align}
where in the second line we used the chain rule and Eq.\ \eqref{eq: light_bg_field}. It follows that 
\begin{align}
    \frac{\delta^2 \Gamma_L[\bg{\phi}]}{\delta \left(J_\phi\right)_y \delta \left(\bg{\phi}\right)_x} = -\delta(x-y).
    \label{eq: delta_eq}
\end{align}
On the other hand, using the chain rule and Eq.\ \eqref{eq: light_bg_field} again we find
\begin{align}
   \frac{\delta^2 \Gamma_L[\bg{\phi}]}{\delta \left(J_\phi\right)_y \delta \left(\bg{\phi}\right)_x} = \frac{\delta^2 \tilde{W}[J_\phi]}{\delta \left(J_\phi\right)_y \delta \left(J_\phi\right)_z} \frac{\delta^2 \Gamma_L[\bg{\phi}]}{\delta \left(\bg{\phi}\right)_z \delta \left(\bg{\phi}\right)_x},
\end{align}
which together with Eq.\ \eqref{eq: delta_eq} implies that 
\begin{align}
    \left(\frac{\delta^2 \Gamma_L[\bg{\phi}]}{\delta \bg{\phi} \delta \bg{\phi}}\right)_{xy} = -\left(\frac{\delta^2 \tilde{W}[J_\phi]}{\delta J_\phi \delta J_\phi}\right)_{xy}^{-1}.
\end{align}
In our conventions
\begin{align}
    \left(\frac{\delta^2 \tilde{W}[J_\phi]}{\delta J_\phi \delta J_\phi}\right)_{xy} = i D_{xy},
\end{align}
is the connected two-point function of the light field and hence $D_{xy}$ is the full propagator of the light field. It follows that  
\begin{align}
    \left(\frac{\delta^2 \Gamma_L[\bg{\phi}]}{\delta \bg{\phi} \delta \bg{\phi}}\right)_{xy} =  i D^{-1}_{xy}.
\end{align}
Now consider the connected three-point function with light external fields given by
\begin{align}
    -\frac{\delta^3 \tilde{W}[J_\phi]}{\delta \left(J_\phi\right)_x \delta \left(J_\phi\right)_y \delta \left(J_\phi\right)_z} &= \frac{\delta}{\delta \left(J_\phi\right)_x} \left(\frac{\delta^2 \Gamma_L[\bg{\phi}]}{\delta \bg{\phi} \delta \bg{\phi}}\right)_{yz}^{-1} \nonumber \\
    &= i D_{xw}\frac{\delta}{\delta \left(\bg{\phi}\right)_w} \left(\frac{\delta^2 \Gamma_L[\bg{\phi}]}{\delta \bg{\phi} \delta \bg{\phi}}\right)_{yz}^{-1} \nonumber \\
    &= iD_{xw} (-1) iD_{yu} \left(\frac{\delta^3 \Gamma_L[\bg{\phi}]}{\delta \bg{\phi} \delta \bg{\phi} \delta \bg{\phi}}\right)_{wuv} iD_{vz} \nonumber \\
    &= i D_{xw} D_{yu} D_{vz} \left(\frac{\delta^3 \Gamma_L[\bg{\phi}]}{\delta \bg{\phi} \delta \bg{\phi} \delta \bg{\phi}}\right)_{wuv}.
    \label{eq: three_point_func}
\end{align}
This is most easily interpreted diagrammatically.
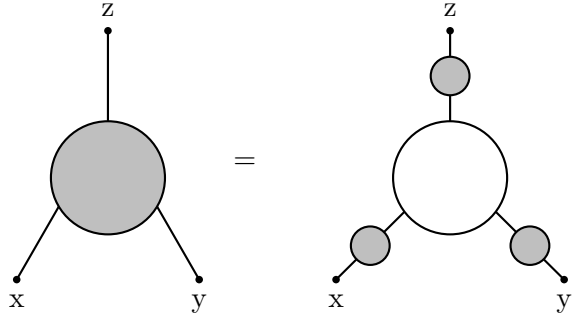
\begin{figure}[h]
\center
\begin{tikzpicture}[scale = 1.5]
\draw[thick] (4,1.8) -- (4.4,2.5);
\draw[thick] (5.2,2.5) -- (5.6,1.8);
\draw[thick] (4.8,3.2) -- (4.8,4);
\draw[thick, fill=lightgray] (4.8,2.7) circle [radius=0.5];;
\node[circle, fill, inner sep=1pt,label=below:x] at (4,1.8)%
{};
\node[circle, fill, inner sep=1pt,label=below:y] at (5.6,1.8)%
{};
\node[circle, fill, inner sep=1pt,label=above:z] at (4.8,4)%
{};
\node[align=left, above] at (6,2.7)%
{$=$};
\draw[thick] (6.8,1.8) -- (7.4,2.4);
\draw[thick] (8.2,2.4) -- (8.8,1.8);
\draw[thick] (7.8,3.2) -- (7.8,4);
\draw[thick] (7.8,2.7) circle [radius=0.5];;
\draw[thick, fill=lightgray] (7.8,3.6) circle [radius=0.17];;
\draw[thick,fill=lightgray] (8.5,2.1) circle [radius=0.17];;
\draw[thick,fill=lightgray] (7.1,2.1) circle [radius=0.17];;
\node[circle, fill, inner sep=1pt,label=below:x] at (6.8,1.8)%
{};
\node[circle, fill, inner sep=1pt,label=below:y] at (8.8,1.8)%
{};
\node[circle, fill, inner sep=1pt,label=above:z] at (7.8,4)%
{};
\end{tikzpicture}
\caption{Diagrammatic representation of Eq.\ \eqref{eq: three_point_func}.}
\label{Fig: app_A_ex}
\end{figure}
In Figure~\ref{Fig: app_A_ex}, the gray circles represent the sum of connected correlation functions, whereas the white circle depicts the third derivative of $i\Gamma_\text{L}[\bg{\phi}]$. We see that this is the connected correlation function of three external light fields, with all three full propagators removed. That means that it is 1PI with respect to light fields. That it is not 1PI with respect to heavy fields can be seen by the fact that the diagram in Figure \ref{Fig: counterex}, where the dashed line represents the heavy field, is part of the third derivative of $i\Gamma_\text{L}[\bg{\phi}]$. Hence, the white circle is the 1LPI three-point function. This generalises to higher derivatives of $\Gamma_\text{L}[\bg{\phi}]$, which therefore, is the generating functional of 1LPI correlation functions. 

\clearpage
\lhead{\emph{Appendix B}}
\chapter{BSUOLEA ingredients}
\label{app: UOLEA_ops}
We here collect the different matrices defined in Section~\ref{sec: Mixed_statistics}, which appear in the BSUOLEA after the factors of $\tilde{\mathds{1}}_B$, $\tilde{\mathds{1}}_F$ and $\cc$ have been absorbed and relate them to the symbols appearing in the Mathematica file found at \cite{code}. We define
\begingroup
\allowdisplaybreaks
\begin{align*}
X[\text{S},\text{F}] &\equiv \mathbf{X}_{\Phi \Xi}= \begin{pmatrix}
X_{\Sigma ^* \Omega} & X_{\Sigma ^* \bar{\Omega}} \cc ^{-1} & X_{\Sigma ^* \Lambda} \\
-X^{\mu}_{W ^* \Omega} & -X^{\mu}_{W ^* \bar{\Omega}} \cc ^{-1} & -X_{W ^* \Lambda} \\
X_{\Sigma \Omega} & X_{\Sigma \bar{\Omega}} \cc ^{-1} & X_{\Sigma \Lambda} \\
-X^{\mu}_{W \Omega} & -X^{\mu}_{W \bar{\Omega}} \cc ^{-1} & -X_{W \Lambda} \\
X_{\Theta \Omega} & X_{\Theta \bar{\Omega}} \cc ^{-1} & X_{\Theta \Lambda} \\
-X^{\mu}_{V \Omega} & -X^{\mu}_{V \bar{\Omega}} \cc ^{-1} & -X_{V \Lambda}
\end{pmatrix}, \nonumber \\
X[\text{s},\text{F}] &\equiv \mathbf{X}_{\phi \Xi}= \begin{pmatrix}
X_{\sigma ^* \Omega} & X_{\sigma ^* \bar{\Omega}} \cc ^{-1} & X_{\sigma ^* \Lambda} \\
-X^\mu_{w ^* \Omega} & -X^\mu_{w ^* \bar{\Omega}} \cc ^{-1} & -X^\mu_{w ^* \Lambda} \\
X_{\sigma \Omega} & X_{\sigma \bar{\Omega}} \cc ^{-1} & X_{\sigma \Lambda} \\
-X^\mu_{w \Omega} & -X^\mu_{w \bar{\Omega}} \cc ^{-1} & -X^\mu_{w \Lambda} \\
X_{\theta \Omega} & X_{\theta \bar{\Omega}} \cc ^{-1} & X_{\theta \Lambda} \\ 
-X^\mu_{v \Omega} & -X^\mu_{v \bar{\Omega}} \cc ^{-1} & -X^\mu_{v \Lambda} \\
\end{pmatrix}, \nonumber \\
X[\text{S},\text{f}] &\equiv \mathbf{X}_{\Phi \xi}=\begin{pmatrix}
X_{\Sigma ^* \omega} & X_{\Sigma ^* \bar{\omega}} \cc ^{-1} & X_{\Sigma ^* \lambda} \\
-X^\mu_{W ^* \omega} & -X^\mu_{W ^* \bar{\omega}} \cc ^{-1} & -X^\mu_{W ^* \lambda} \\
X_{\Sigma \omega} & X_{\Sigma \bar{\omega}} \cc ^{-1} & X_{\Sigma \lambda} \\
-X^\mu_{W \omega} & -X^\mu_{W \bar{\omega}} \cc ^{-1} & -X^\mu_{W \lambda} \\
X_{\Theta \omega} & X_{\Theta \bar{\omega}} \cc ^{-1} & X_{\Theta \lambda}\\
-X^\mu_{V \omega} & -X^\mu_{V \bar{\omega}} \cc ^{-1} & -X^\mu_{V \lambda}
\end{pmatrix}, \nonumber \\
X[\text{s},\text{f}] &\equiv \mathbf{X}_{\phi \xi}= \begin{pmatrix}
X_{\sigma ^* \omega} & X_{\sigma ^* \bar{\omega}} \cc ^{-1} & X_{\sigma ^* \lambda} \\
-X^\mu_{w ^* \omega} & -X^\mu_{w ^* \bar{\omega}} \cc ^{-1} & -X^\mu_{w ^* \lambda} \\
X_{\sigma \omega} & X_{\sigma \bar{\omega}} \cc ^{-1} & X_{\sigma \lambda} \\
-X^\mu_{w \omega} & -X^\mu_{w \bar{\omega}} \cc ^{-1} & -X^\mu_{w \lambda} \\
X_{\theta \omega} & X_{\theta \bar{\omega}} \cc ^{-1} & X_{\theta \lambda}\\
-X^\mu_{v \omega} & -X^\mu_{v \bar{\omega}} \cc ^{-1} & -X^\mu_{v \lambda}
\end{pmatrix}, \nonumber \\
X[\text{F},\text{S}] &\equiv \mathbf{X}_{\Xi \Phi} = \begin{pmatrix}
X_{\bar{\Omega} \Sigma} & X^\mu_{\bar{\Omega} W} &  X_{\bar{\Omega} \Sigma ^*} & X^\mu_{\bar{\Omega} W^*} &  X_{\bar{\Omega} \Theta} & X^\mu_{\bar{\Omega} V} \\ 
\cc^{-1} X_{\Omega \Sigma} & \cc^{-1} X^\mu_{\Omega W} & \cc^{-1} X_{\Omega \Sigma^{*}} & \cc^{-1} X^\mu_{\Omega W^*} & \cc^{-1} X_{\Omega \Theta} & \cc^{-1} X^\mu_{\Omega V} \\ 
\cc^{-1} X_{\Lambda \Sigma} & \cc^{-1} X^\mu_{\Lambda W} & \cc^{-1} X_{\Lambda \Sigma ^*} & \cc^{-1} X^\mu_{\Lambda W^*} & \cc^{-1} X_{\Lambda \Theta} & \cc^{-1} X^\mu_{\Lambda V}
\end{pmatrix}, \nonumber \\
X[\text{f},\text{S}] &\equiv \mathbf{X}_{\xi \Phi} = \begin{pmatrix}
 X_{\bar{\omega} \Sigma} & X^\mu_{\bar{\omega} W} &  X_{\bar{\omega} \Sigma ^*} & X_{\bar{\omega} W^*} &  X_{\bar{\omega} \Theta} & X_{\bar{\omega} V} \\ 
\cc^{-1} X_{\omega \Sigma} & \cc^{-1} X^\mu_{\omega W} & \cc^{-1} X_{\omega \Sigma^{*}} & \cc^{-1} X^\mu_{\omega W ^{*}} &  \cc^{-1} X_{\omega \Theta} & \cc^{-1} X^\mu_{\omega V}\\ 
\cc^{-1} X_{\lambda \Sigma} & \cc^{-1} X^\mu_{\lambda W} & \cc^{-1} X_{\lambda \Sigma ^*} & \cc^{-1} X^\mu_{\lambda W ^*} & \cc^{-1} X_{\lambda \Theta} & \cc^{-1} X^\mu_{\lambda V}
\end{pmatrix}, \nonumber \\
X[\text{F},\text{s}] &\equiv \mathbf{X}_{\Xi \phi} = \begin{pmatrix}
 X_{\bar{\Omega} \sigma} & X^\mu_{\bar{\Omega} w} &  X_{\bar{\Omega} \sigma ^*} & X^\mu_{\bar{\Omega} w^*} &  X_{\bar{\Omega} \theta} & X^\mu_{\bar{\Omega} v} \\ 
\cc^{-1} X_{\Omega \sigma} & \cc^{-1} X^\mu_{\Omega w} & \cc^{-1} X_{\Omega \sigma^{*}} & \cc^{-1} X^\mu_{\Omega w^*} &  \cc^{-1} X_{\Omega \theta} & \cc^{-1} X^\mu_{\Omega v} \\ 
\cc^{-1} X_{\Lambda \sigma} & \cc^{-1} X^\mu_{\Lambda w} & \cc^{-1} X_{\Lambda \sigma ^*} & \cc^{-1} X^\mu_{\Lambda w^*} & \cc^{-1} X_{\Lambda \theta} & \cc^{-1} X^\mu_{\Lambda v}
\end{pmatrix}, \nonumber \\
X[\text{f},\text{s}] &\equiv \mathbf{X}_{\xi \phi} = \begin{pmatrix}
 X_{\bar{\omega} \sigma} & X^\mu_{\bar{\omega} w} &  X_{\bar{\omega} \sigma ^*} & X^\mu_{\bar{\omega} w^*} &  X_{\bar{\omega} \theta} & X^\mu_{\bar{\omega} v} \\ 
\cc^{-1} X_{\omega \sigma} & \cc^{-1} X^\mu_{\omega w} & \cc^{-1} X_{\omega \sigma^{*}} & \cc^{-1} X^\mu_{\omega w^*} & \cc^{-1} X_{\omega \theta} & \cc^{-1} X^\mu_{\omega v} \\ 
\cc^{-1} X_{\lambda \sigma} & \cc^{-1} X^\mu_{\lambda w} & \cc^{-1} X_{\lambda \sigma ^*} & \cc^{-1} X^\mu_{\lambda w^*} & \cc^{-1} X_{\lambda \theta} & \cc^{-1} X^\mu_{\lambda v} 
\end{pmatrix}, \nonumber \\
X[\text{F},\text{F}] &\equiv \mathbf{X}_{\Xi \Xi}=\begin{pmatrix}
X_{\bar{\Omega} \Omega} & X_{\bar{\Omega} \bar{\Omega}} \cc^{-1} &  X_{\bar{\Omega} \Lambda} \\
\cc^{-1} X_{\Omega \Omega} &  \cc ^{-1} X_{\Omega \bar{\Omega}}\cc^{-1} & \cc ^{-1} X_{\Omega \Lambda} \\
\cc ^{-1} X_{\Lambda \Omega} & \cc ^{-1} X_{\Lambda \bar{\Omega}}\cc ^{-1} &  \cc ^{-1} X_{\Lambda \Lambda}
\end{pmatrix}, \nonumber \\
X[\text{f},\text{f}] &\equiv \mathbf{X}_{\xi \xi}=\begin{pmatrix}
X_{\bar{\omega} \omega} & X_{\bar{\omega} \bar{\omega}} \cc ^{-1} &  X_{\bar{\omega} \lambda} \\
\cc ^{-1} X_{\omega \omega} &  \cc ^{-1} X_{\omega \bar{\omega}}\cc^{-1} & \cc ^{-1} X_{\omega \lambda} \\
\cc ^{-1} X_{\lambda \omega} & \cc ^{-1} X_{\lambda \bar{\omega}}\cc ^{-1} &  \cc ^{-1} X_{\lambda \lambda}
\end{pmatrix}, \nonumber \\
X[\text{F},\text{f}] &\equiv \mathbf{X}_{\Xi \xi}=\begin{pmatrix}
X_{\bar{\Omega} \omega} & X_{\bar{\Omega} \bar{\omega}} \cc^{-1} &  X_{\bar{\Omega} \lambda} \\
\cc ^{-1} X_{\Omega \omega} & \cc ^{-1} X_{\Omega \bar{\omega}}\cc ^{-1} & \cc^{-1} X_{\Omega \lambda} \\
\cc ^{-1} X_{\Lambda \omega} & \cc ^{-1} X_{\Lambda \bar{\omega}} \cc ^{-1} &  \cc^{-1} X_{\Lambda \lambda}
\end{pmatrix}, \nonumber \\
X[\text{f},\text{F}] &\equiv \mathbf{X}_{\xi \Xi}=\begin{pmatrix}
X_{\bar{\omega} \Omega} & X_{\bar{\omega} \bar{\Omega}} \cc ^{-1} &  X_{\bar{\omega} \Lambda} \\
\cc ^{-1} X_{\omega \Omega} & \cc ^{-1} X_{\omega \bar{\Omega}}\cc ^{-1} & \cc^{-1} X_{\omega \Lambda} \\
\cc ^{-1} X_{\lambda \Omega} & \cc ^{-1} X_{\lambda \bar{\Omega}} \cc ^{-1} &  \cc^{-1} X_{\lambda \Lambda}
\end{pmatrix}, \nonumber \\
X[\text{S},\text{S}] &\equiv \mathbf{X}_{\Phi \Phi}=\begin{pmatrix} 
X_{\Sigma ^* \Sigma} & X_{\Sigma ^* W}^\nu & X_{\Sigma ^* \Sigma ^*} & X_{\Sigma ^* W^*}^\nu & X_{\Sigma^* \Theta} & X_{\Sigma ^* V}^\nu \\
-X^\mu_{W ^* \Sigma} & -X^{\mu\nu}_{W ^* W} & -X^\mu_{W ^* \Sigma ^*} & -X^{\mu \nu}_{W ^* W^*} & -X^\mu_{W^* \Theta} & -X^{\mu \nu}_{W ^* V} \\
X_{\Sigma \Sigma} & X^\nu_{\Sigma W} & X_{\Sigma \Sigma ^{*}} & X^\nu_{\Sigma W^*} &  X_{\Sigma \Theta} & X^\nu_{\Sigma V} \\
-X^\mu_{W \Sigma} & -X^{\mu\nu}_{W W} & -X^\mu_{W \Sigma ^*} & -X^{\mu \nu}_{W W^*} & -X^\mu_{W \Theta} & -X^{\mu \nu}_{W V} \\
X_{\Theta \Sigma} & X^\nu_{\Theta W} &  X_{\Theta \Sigma ^*} & X^\nu_{\Theta W^*} & X_{\Theta \Theta} & X^\nu_{\Theta V}\\ 
-X^\mu_{V \Sigma} & -X^{\mu\nu}_{V W} & -X^\mu_{V \Sigma ^*} & -X^{\mu \nu}_{V W^*} & -X^\mu_{V \Theta} & -X^{\mu \nu}_{V V}
\end{pmatrix}, \nonumber \\
X[\text{S},\text{s}] &\equiv \mathbf{X}_{\Phi \phi}=\begin{pmatrix} 
X_{\Sigma ^* \sigma} & X_{\Sigma ^* w}^\nu & X_{\Sigma ^* \sigma ^*} & X_{\Sigma ^* w^*}^\nu & X_{\Sigma^* \theta} & X_{\Sigma ^* v}^\nu \\
-X^\mu_{W ^* \sigma} & -X^{\mu\nu}_{W ^* w} & -X^\mu_{W ^* \sigma ^*} & -X^{\mu \nu}_{W ^* w^*} & -X^\mu_{W^* \theta} & -X^{\mu \nu}_{W ^* v} \\
X_{\Sigma \sigma} & X^\nu_{\Sigma w} & X_{\Sigma \sigma ^{*}} & X^\nu_{\Sigma w^*} &  X_{\Sigma \theta} & X^\nu_{\Sigma v} \\
-X^\mu_{W \sigma} & -X^{\mu\nu}_{W w} & -X^\mu_{W \sigma ^*} & -X^{\mu \nu}_{W w^*} & -X^\mu_{W \theta} & -X^{\mu \nu}_{W v} \\
X_{\Theta \sigma} & X^\nu_{\Theta w} &  X_{\Theta \sigma ^*} & X^\nu_{\Theta w^*} & X_{\Theta \theta} & X^\nu_{\Theta v}\\ 
-X^\mu_{V \sigma} & -X^{\mu\nu}_{V w} & -X^\mu_{V \sigma ^*} & -X^{\mu \nu}_{V w^*} & -X^\mu_{V \theta} & -X^{\mu \nu}_{V v}
\end{pmatrix}, \nonumber \\
X[\text{s},\text{S}] &\equiv \mathbf{X}_{\phi \Phi}=\begin{pmatrix} 
X_{\sigma ^* \Sigma} & X_{\sigma ^* W}^\nu & X_{\sigma ^* \Sigma ^*} & X_{\sigma ^* W^*}^\nu & X_{\sigma^* \Theta} & X_{\sigma ^* V}^\nu \\
-X^\mu_{w ^* \Sigma} & -X^{\mu\nu}_{w ^* W} & -X^\mu_{w ^* \Sigma ^*} & -X^{\mu \nu}_{w ^* W^*} & -X^\mu_{w^* \Theta} & -X^{\mu \nu}_{w ^* V} \\
X_{\sigma \Sigma} & X^\nu_{\sigma W} & X_{\sigma \Sigma ^{*}} & X^\nu_{\sigma W^*} &  X_{\sigma \Theta} & X^\nu_{\sigma V} \\
-X^\mu_{w \Sigma} & -X^{\mu\nu}_{w W} & -X^\mu_{w \Sigma ^*} & -X^{\mu \nu}_{w W^*} & -X^\mu_{w \Theta} & -X^{\mu \nu}_{w V} \\
X_{\theta \Sigma} & X^\nu_{\theta W} &  X_{\theta \Sigma ^*} & X^\nu_{\theta W^*} & X_{\theta \Theta} & X^\nu_{\theta V}\\ 
-X^\mu_{v \Sigma} & -X^{\mu\nu}_{v W} & -X^\mu_{v \Sigma ^*} & -X^{\mu \nu}_{v W^*} & -X^\mu_{v \Theta} & -X^{\mu \nu}_{v V}
\end{pmatrix}, \nonumber \\
X[\text{s},\text{s}] &\equiv \mathbf{X}_{\phi \phi}=\begin{pmatrix} 
X_{\sigma ^* \sigma} & X_{\sigma ^* w}^\nu & X_{\sigma ^* \sigma ^*} & X_{\sigma ^* w^*}^\nu & X_{\sigma^* \theta} & X_{\sigma ^* v}^\nu \\
-X^\mu_{w ^* \sigma} & -X^{\mu\nu}_{w ^* w} & -X^\mu_{w ^* \sigma ^*} & -X^{\mu \nu}_{w ^* w^*} & -X^\mu_{w^* \theta} & -X^{\mu \nu}_{w ^* v} \\
X_{\sigma \sigma} & X^\nu_{\sigma w} & X_{\sigma \sigma ^{*}} & X^\nu_{\sigma w^*} &  X_{\sigma \theta} & X^\nu_{\sigma v} \\
-X^\mu_{w \sigma} & -X^{\mu\nu}_{w w} & -X^\mu_{w \sigma ^*} & -X^{\mu \nu}_{w w^*} & -X^\mu_{w \theta} & -X^{\mu \nu}_{w v} \\
X_{\theta \sigma} & X^\nu_{\theta w} &  X_{\theta \sigma ^*} & X^\nu_{\theta w^*} & X_{\theta \theta} & X^\nu_{\theta v}\\ 
-X^\mu_{v \sigma} & -X^{\mu\nu}_{v w} & -X^\mu_{v \sigma ^*} & -X^{\mu \nu}_{v w^*} & -X^\mu_{v \theta} & -X^{\mu \nu}_{v v}
\end{pmatrix}, \nonumber \\
\end{align*}
\endgroup
where 
\begin{align}
\begin{cases}
    \mathbf{X}_{A B} = \mathbf{U}_{A B} + P_\mu \mathbf{Z}_{AB}^\mu+\mathbf{Z}^{\dagger\mu}_{AB} P_\mu, \text{ if } A \wedge B \in \{\phi, \Phi\} \\
    \mathbf{X}_{A B} = \mathbf{U}_{A B} \text{ otherwise}
\end{cases}.
\end{align}
In the operators published at \cite{code} we further defined
\begin{align}
    \mathbf{U}_{A B} &= U[\text{A},\text{B}], \\
    \mathbf{Z}_{AB}^\mu &= Z[\text{A},\text{B}][\text{lor}], \\
    \mathbf{Z}_{AB}^{\dagger \mu} &= ZT[\text{A},\text{B}][\text{lor}].
\end{align}
Finally, covariant derivatives and field strength tensors are denoted by  $P[\text{lor}]$ and  \newline
$G[\text{A},\text{A}][\text{lor1},\text{lor2}]$, respectively. The field type A in the field strength tensor defines which covariant derivative is to be used in its construction. For example, $G[\text{F},\text{F}]$ is obtained from the commutator of two covariant derivatives acting on the heavy fermion multiplet $\Xi$. A commutator is denoted by $c$, for instance $c[P[\text{lor}],U[\text{A},\text{B}]]=[P^\mu,\mathbf{U}_{A B}]$, where the apropriate covariant derivative is the one acting on multiplet $A$.

\clearpage
\lhead{\emph{Appendix C}}
\chapter{Proof of consistency of shifts}
\label{app: shifts}
In this appendix we discuss the consistency of the shifts given in Eqs.\
\eqref{eq:xishift} and \eqref{eq:xishift_T}. The treatment of the shifts given in
Eqs.\ \eqref{eq:Xishift} and \eqref{eq:Xishift_T} is analogous. Since
$\xi$ is a multiplet of Majorana-like component spinors,
for the shifts
\begin{align}
\label{eq:xishift2}
  \delta \xi' &= \delta \xi+\mathbf{\Delta}_\xi^{-1}\left[\tilde{\mathbf{X}}_{\xi \Xi} \delta \Xi-\tilde{\mathbf{X}}_{\xi \Phi} \delta \Phi-\tilde{\mathbf{X}}_{\xi \phi} \delta \phi\right], \\
  \delta \xi'^T &= \delta \xi^T+\left[\delta \Xi^T \tilde{\mathbf{X}}_{\Xi \xi}+\delta \Phi^T \tilde{\mathbf{X}}_{\Phi \xi}+\delta \phi^T \tilde{\mathbf{X}}_{\phi \xi}\right]\overleftarrow{\mathbf{\Delta}}_\xi^{-1},
  \label{eq:xishift2_T}
\end{align}
to be consistent it is necessary and sufficient that
\begin{align}
\left(\mathbf{\Delta}_\xi^{-1}\left[\tilde{\mathbf{X}}_{\xi \Xi} \delta \Xi-\tilde{\mathbf{X}}_{\xi \Phi} \delta \Phi-\tilde{\mathbf{X}}_{\xi \phi} \delta \phi\right]\right)^{T}=\left[\delta \Xi^T \tilde{\mathbf{X}}_{\Xi \xi}+\delta \Phi^T \tilde{\mathbf{X}}_{\Phi \xi}+\delta \phi^T \tilde{\mathbf{X}}_{\phi \xi}\right]\overleftarrow{\mathbf{\Delta}}_\xi^{-1}.
\label{eq:shiftCondition}
\end{align}
In the following we show that \eqref{eq:shiftCondition} holds.
We first construct $\mathbf{\Delta}_\xi ^{-1}$ in position space through
its Neumann series
\begin{align}
\mathbf{\Delta}_\xi ^{-1}(x,y) &= \sum _{n=0} ^{\infty} \left(\prod _{\substack{i=1 \\ n>0}} ^{n} \int \rd^d x_i \; \mathbf{S}(x_{i-1},x_i) \left(-\mathbf{X}_{\xi \xi}(x_i)\right)\right) \mathbf{S}(x_n,y) \tilde{\mathds{1}}_F \cc ^{-1} \nonumber \\
& \equiv \sum _{n=0} ^{\infty} \left(\prod _{\substack{i=1 \\ n>0}} ^{n}  \mathbf{S}_{x_{i-1} x_i} \left(-\mathbf{X}_{\xi \xi x_i}\right)\right) \mathbf{S}_{x_n y} \tilde{\mathds{1}}_F \cc ^{-1}, 
\end{align}
where $x_0\equiv x$ and $\mathbf{S}(x,y)$ is the matrix-valued Green's
function for $(\slashed{P}-M_\xi)$, which itself can be expressed
through a Neumann series. We may write
$(\slashed{P}-M_\xi) = (i\slashed{\partial}-M_\xi-\mathbf{A})$ with
\begin{align}
\mathbf{A}=i\sum_j g_j\slashed{A}_j^a T_j^a,
\end{align}
where we sum over all factors of the gauge group for a direct product
group and $T_j^a$ is a block-diagonal matrix which generates the
reducible representation of $\xi$. Due to the fact that $\xi$ contains
$\omega$, $\ccfield{\omega}$ and $\lambda$ (see Table \ref{table: field_content}),
the generator is of the form
\begin{align}
T^a=\begin{pmatrix}
T^a _{R(\omega)} && 0 && 0 \\
0 && T^a _{\bar{R}(\omega)} && 0 \\
0 && 0 && T^a_{R( \lambda )}
\end{pmatrix},
\end{align}
where $R(\omega)$ is the representation under which $\omega$
transforms, $\bar{R}(\omega)$ its conjugate representation and
$R(\lambda)$ is the representation of $\lambda$, which is necessarily
real. We then have
\begin{align}
\mathbf{S}_{x y}=\sum _{k=0} ^{\infty} \left(\prod _{\substack{i=1 \\ k>0}} ^{k}  \mathbf{S}_{f,x_{i-1} x_i} \mathbf{A}_{x_i}\right) \mathbf{S}_{f, x_k y},
\end{align}
where again $x_0 \equiv x$ and $\mathbf{S}_{f,x y}$ is the matrix
containing the Green's function of the free Dirac equation on its
diagonal. It can be verified by explicit calculation that
\begin{align}
\mathbf{S}_{x y} \left(-i\overleftarrow{\slashed{\partial}_y}-M_\xi-\mathbf{A}_y \right) &= \delta_{x y},
\end{align} 
which means that
\begin{align}
\mathbf{\Delta}_{\xi,x y} ^{-1}\overleftarrow{\mathbf{\Delta}}_{\xi, y}=\delta_{x y}
\end{align}
and therefore
$\overleftarrow{\mathbf{\Delta}}_{\xi,yx}
^{-1}=\mathbf{\Delta}_{\xi,yx}^{-1}$. Hence \eqref{eq:shiftCondition}
reads
\begin{align}
\left(\mathbf{\Delta}_{\xi, x y}^{-1}  \left[\tilde{\mathbf{X}}_{\xi \Xi} \delta \Xi-\tilde{\mathbf{X}}_{\xi \Phi} \delta \Phi-\tilde{\mathbf{X}}_{\xi \phi} \delta \phi\right]_y\right)^{T}=\left[\delta \Xi^T \tilde{\mathbf{X}}_{\Xi \xi}+\delta \Phi^T \tilde{\mathbf{X}}_{\Phi \xi}+\delta \phi^T \tilde{\mathbf{X}}_{\phi \xi}\right]_y\mathbf{\Delta}_{\xi, yx}^{-1}.
\end{align}
It is then useful to calculate
\begingroup
\allowdisplaybreaks
\begin{align}
\label{eq: id_needed}
\cc \tilde{\mathds{1}}_F \mathbf{S}^T_{xy}&=\cc \tilde{\mathds{1}}_F\sum _{k=0} ^{\infty} \mathbf{S}^T_{f,x_k y} \left(\prod _{\substack{i=k \\ k>0}} ^{1}  \mathbf{A}^T_{x_i}\mathbf{S}^T_{f,x_{i-1}x_i} \right) \nonumber  \\
&=  \cc \tilde{\mathds{1}}_F\sum _{k=0} ^{\infty} \cc \mathbf{S}_{f,y x_k} \cc^{-1} \left(\prod _{\substack{i=k \\ k>0}} ^{1}  \mathbf{A}^T _{x_i}\cc \mathbf{S}_{f,x_i x_{i-1}}\cc^{-1} \right) \nonumber \\
&=  -\sum _{k=0} ^{\infty}  \tilde{\mathds{1}}_F \mathbf{S}_{f,y x_k} \tilde{\mathds{1}}_F \tilde{\mathds{1}}_F \left(\prod _{\substack{i=k \\ k>0}} ^{1}   (-\mathbf{A}^t _{x_i})\tilde{\mathds{1}}_F \tilde{\mathds{1}}_F \mathbf{S}_{f,x_i x_{i-1}} \tilde{\mathds{1}}_F \tilde{\mathds{1}}_F \right)\cc^{-1} \nonumber
 \\
&=  -\sum _{k=0} ^{\infty}  \mathbf{S}_{f, y x_k} \left(\prod _{\substack{i=k \\ k>0}} ^{1} (-\tilde{\mathds{1}}_F\mathbf{A}^t_{x_i}\tilde{\mathds{1}}_F)\mathbf{S}_{f,x_i x_{i-1}} \right)\tilde{\mathds{1}}_F\cc^{-1} \nonumber \\
&=  -\sum _{k=0} ^{\infty}  \mathbf{S}_{f,y x_k} \left(\prod _{\substack{i=k \\ k>0}} ^{1}  \mathbf{A}_{x_i} \mathbf{S}_{f,x_i x_{i-1}} \right)\tilde{\mathds{1}}_F\cc^{-1}  \nonumber \\
&= -\mathbf{S}_{yx}\tilde{\mathds{1}}_F\cc ^{-1},
\end{align}
\endgroup
where $\mathbf{A}^t$ means taking the transpose of the gauge group
generators only and we used that
\begin{align}
\tilde{\mathds{1}}_F\begin{pmatrix}
A && 0 && 0 \\
0 && B && 0 \\
0 && 0 && C 
\end{pmatrix}
\tilde{\mathds{1}}_F=\begin{pmatrix}
B && 0 && 0 \\
0 && A && 0 \\
0 && 0 && C 
\end{pmatrix}.
\end{align}
We note in passing that Eq.\ \eqref{eq: id_needed} also holds if the generators contain projectors in spinor space, since the charge conjugation operator commutes with these projectors in the chiral representation, which we have adopted. Thus, this also covers the construction of Section \ref{sec: SSM_to_SM}. We now find
\begin{align}
\left(\mathbf{\Delta}_{\xi, xy} ^{-1}\right)^T &= \cc \tilde{\mathds{1}}_F\sum _{n=0} ^{\infty} \mathbf{S}_{x_n,y}^T \left(\prod _{\substack{i=1 \\ n>0}} ^{n}   \left(-\mathbf{X}_{\xi \xi, x_i}\right)^T \mathbf{S}_{x_{i-1}x_i}^T \right) \nonumber  \\
&= \sum _{n=0} ^{\infty} \mathbf{S}_{y x_n} \tilde{\mathds{1}}_F\cc \left(\prod _{\substack{i=1 \\ n>0}} ^{n}  \left(-\mathbf{X}_{\xi \xi, x_i}\right)^T \mathbf{S}_{x_{i-1} x_i}^T \right) \nonumber  \\
&= \sum _{n=0} ^{\infty} \mathbf{S}_{y x_n} \tilde{\mathds{1}}_F\cc \left(\prod _{\substack{i=1 \\ n>0}} ^{n} \left(-\mathbf{X}_{\xi \xi, x_i}\right)^T \tilde{\mathds{1}}_F\cc ^{-1} \tilde{\mathds{1}}_F\cc  \mathbf{S}_{x_{i-1} x_i}^T \tilde{\mathds{1}}_F\cc ^{-1} \tilde{\mathds{1}}_F\cc \right) \nonumber  \\
&= \sum _{n=0} ^{\infty} \mathbf{S}_{y x_n} \tilde{\mathds{1}}_F\cc \left(\prod _{\substack{i=1 \\ n>0}} ^{n}    \left(-\mathbf{X}_{\xi \xi, x_i}\right)^T \tilde{\mathds{1}}_F\cc ^{-1}   \mathbf{S}_{x_i x_{i-1}}  \tilde{\mathds{1}}_F\cc \right) \nonumber   \\
&= -\sum _{n=0} ^{\infty} \mathbf{S}_{y x_n}  \left(\prod _{\substack{i=1 \\ n>0}} ^{n} \left(-\mathbf{X}_{\xi \xi, x_i}\right)    \mathbf{S}_{x_i x_{i-1}} \right)  \tilde{\mathds{1}}_F\cc ^{-1} \nonumber  \\ 
&= -\mathbf{\Delta}_{\xi yx} ^{-1},
\end{align}
where we used that 
\begin{align}
\cc \tilde{\mathds{1}}_F \mathbf{X}^T_{\xi \xi} \tilde{\mathds{1}} _F\cc^{-1}=\mathbf{X}_{\xi \xi}.
\end{align}
Noting that
\begin{align}
\tilde{\mathbf{X}}^T_{\xi \Xi}&=-\tilde{\mathbf{X}}_{\Xi \xi}, \\
\tilde{\mathbf{X}}^T_{\xi \Phi}&=\tilde{\mathbf{X}}_{\Phi \xi}, \\
\tilde{\mathbf{X}}^T_{\xi \phi}&=\tilde{\mathbf{X}}_{\phi \xi},
\end{align}
the validity of \eqref{eq:shiftCondition} follows immediately. 

\clearpage
\lhead{\emph{Appendix D}}
\chapter{Loop functions}
\label{app:loop_functions}
The integrals $\ZI [q^{2n_c}]^{n_i n_j \dots n_L} _{i j \dots 0}$ are defined as in Ref. \cite{Zhang:2016pja}, that is 
\begin{align}
  \int \frac{\rd^d q}{(2\pi)^d} \frac{q^{\mu_1} q^{\mu_2} \dots q^{\mu_{2n_c}}}{(q^2-M_i^2)^{n_i}(q^2-M_j^2)^{n_j}\dots (q^2)^{n_L}}\equiv \frac{i}{16 \pi ^2} g^{\mu_1 \mu_2 \dots \mu_{2n_c}} \ZI [q^{2n_c}]^{n_i n_j \dots n_L} _{i j \dots 0},
\end{align}
where $g^{\mu_1 \mu_2 \dots \mu_{2n_c}}$ is the completely symmetric
combination of metric tensors with $2n_c$ indices, for instance
$g^{\mu \nu \rho \sigma}=g^{\mu \nu}g^{\rho \sigma}+g^{\mu \rho}g^{\nu
  \sigma}+g^{\mu \sigma}g^{\nu \rho}$.  For $n_c=0$ we define the
shorthand notation
$\ZI[q^0]^{n_i n_j \dots n_L} _{i j \dots 0} \equiv \ZI^{n_i n_j \dots
  n_L} _{i j \dots 0}$.  The integrals can be reduced to basis
integrals using the following reduction relations, which were already stated in Ref.\ \cite{Zhang:2016pja}
\begin{align}
\label{eq: reduction_formula_1}
  \ZI[q^{2n_c}] ^{n_i n_j \dots n_L} _{i j \dots 0} &=
  \frac{1}{\Delta^2 _{ij}}\left(\ZI[q^{2n_c}]^{n_i n_j-1 \dots n_L}-\ZI[q^{2n_c}]^{n_i-1 \, n_j \dots n_L}\right), \\
  \label{eq: reduction_formula_2}
  \ZI[q^{2n_c}] ^{n_i n_j \dots n_L} _{i j \dots 0} &=
  \frac{1}{M_i^2}\left(\ZI[q^{2n_c}]^{n_i n_j \dots n_L-1}-\ZI[q^{2n_c}]^{n_i-1 \, n_j \dots n_L}\right),
\end{align}
where $\Delta^2 _{ij}=M_i^2-M_j^2$. The basis integrals can be computed using the formula 
\begin{align}
    \ZI[q^{2n_c}]^{n_i n_L}_{i0} ={}& \left(-M_i^2\right)^{2+n_c-n_i-n_L} \frac{1}{2^{n_c}(n_i-1)!}\frac{\Gamma(\frac{\epsilon}{2}-2-n_c+n_i+n_L)}{\Gamma(\frac{\epsilon}{2})}\nonumber \\ &\times \frac{\Gamma(-\frac{\epsilon}{2}+2+n_c-n_L)}{\Gamma(-\frac{\epsilon}{2}+2+n_c)}\left(\frac{2}{\bar{\epsilon}}-\log \frac{M_i^2}{\mu^2}\right),
    \label{eq: basis_integrals}
\end{align}
where $\epsilon = 4-d$ and $\frac{2}{\bar{\epsilon}}=\frac{2}{\epsilon}-\gamma+\log 4\pi$.
In the lists of BSUOLEA operators found at \cite{code} the integrals are denoted by
\begin{align}
  J[n_c, \{\{M_i,n_i\}, \{M_j,n_j\}, \ldots\}, n_L] \equiv \ZI [q^{2n_c}]^{n_i n_j \dots n_L} _{i j \dots 0},
\end{align}
and in the code \tofu the insertion of the correct integrals is performed automatically. 

\clearpage
\lhead{\emph{Appendix E}}
\chapter{The Berezin algebra}
\label{App: berezin_algebra}
\section{Generalities}
In this appendix we introduce some properties of the Berezin algebra which are used throughout this thesis. For more details we refer the reader to Appendix D of Ref.\ \cite{guitman1990quantization} and to Ref.\ \cite{berezin1987introduction}. We will define the Berezin algebra $B$ as the algebra over the field $\mathds{C}$ generated by the elements $z^a = (\theta^i,\xi^\alpha)$, where the generating elements $\xi^\alpha$ satisfy 
\begin{align*}
    \xi^\alpha \xi^\beta + \xi^\beta \xi^\alpha = 0,
\end{align*}
and any $b \in B$ may be written as 
\begin{align}
    b = f_0(\theta)+f_\alpha(\theta)\xi^\alpha+\ldots+f_{\alpha_1 \ldots \alpha_n}(\theta)\xi^{\alpha_1} \cdots \xi^{\alpha_n},
    \label{eq: general_b_elem}
\end{align}
where the functions $f_{{\alpha_1 \ldots \alpha_n}}$ are anti-symmetric in their indices. Any $b \in B$ may be written as an expansion in $z^a$ by Taylor expanding these functions. We say that an element of $B$ is Grassmann odd if its representation of the form given in Eq.\ \eqref{eq: general_b_elem} only contains odd powers of $\xi$. Similarly, any element of $B$ whose representation only contains even powers of $\xi$ will be referred to as Grassmann even. For an element $b \in B$ we introduce the parity $P_b$ as 
\begin{align}
    \begin{cases}
    P_b = 1, \text{ if $b$ is Grassmann odd}\\
    P_b = 0, \text{ if $b$ is Grassmann even}
    \end{cases},
\end{align}
whenever this is well-defined. The only function of the parity $P$ relevant for this thesis is $(-1)^P$ and hence the parity of $b_3=b_1b_2$ can be taken to be $P_{b_3}=P_{b_1}+P_{b_2}$, whenever $b_1$ and $b_2$ have definite parities. We also introduce the parity of the index $a$, $P_{(a)}$, corresponding to the element $z^a$, as
\begin{align}
    P_{(a)}=P_{z^a}.
\end{align}
This definition allows for the introduction of supermatrices as follows. A supermatrix $M$ is a set of matrix elements $M_{ab}\in B$, with definite parity, together with a set of parities of its indices $\left(P_{(a)},P_{(b)}\right)$ such that
\begin{align}
    P_{M_{ab}} = P_{(a)}+P_{(b)}.
\end{align}
For any supermatrix $M$ we introduce the normal form $M^{(N)}$, which is obtained from $M$ by simultaneous permutation of rows and columns so that $M^{(N)}$ has a definite order of indices: first come all Grassmann even and then all Grassmann odd indices. This matrix can be represented as
\begin{align}
    M^{(N)} = \begin{pmatrix}
    \left(M_1\right)_{ij} & \left(M_2\right)_{i \beta} \\
    \left(M_3\right)_{\alpha j} & \left(M_4\right)_{\alpha \beta}
    \end{pmatrix},
\end{align}
with $P_{(i)}=P_{(j)}=0$ and $P_{(\alpha)}=P_{(\beta)}=1$. Given any supermatrix $M$ we may define its supertrace $\Str \, M$ and its superdeterminant $\Sdet \, M$ as
\begin{align}
    \Str \, M &= \sum_a (-1)^{P_{(a)}} M_{aa},\\
    \Sdet \, M &= \exp \Str\log \, M.
\end{align}
Several properties of these quantities can be found in Appendix D of Ref.\ \cite{guitman1990quantization}. The relevant ones for this thesis are 
\begin{align}
    \Sdet \, M = \Sdet \, M^{(N)} &= \det M_1 \det(M_4-M_3 M_1^{-1}M_2)^{-1} \\
    &= \det(M_1-M_2 M_4^{-1}M_3) \det M_4^{-1}.
\end{align}

We finally define the left and right derivatives w.r.t.\ generating elements. It is sufficient to define these on a product of generating elements. The left derivative is thus defined through 
\begin{align}
    \frac{\partial _l}{\partial z^a}z^{a_1}z^{a_2}\ldots z^{a_n}=\sum_{i=1}^n (-1)^{\sum_{j=1}^{i-1}P_{(a)}P_{(a_j)}}\delta^{a_i}_a z^{a_1}\ldots z^{a_{i-1}}z^{a_{i+1}}\ldots z^{a_n},
\end{align}
whereas the right derivative is defined through
\begin{align}
    \frac{\partial _r}{\partial z^a}z^{a_1}z^{a_2}\ldots z^{a_n}=\sum_{i=1}^n (-1)^{\sum_{j=n}^{i+1}P_{(a)}P_{(a_j)}}\delta^{a_i}_a z^{a_1}\ldots z^{a_{i-1}}z^{a_{i+1}}\ldots z^{a_n}.
\end{align}
\section{The Poisson bracket in the Berezin algebra}
In order to formulate a classical theory in which some of the degrees of freedom, collectively denoted by $q^i$, are Grassmann even and some Grassmann odd one has to generalise the Poisson bracket. Taking $q^i$ and the associated conjugate momenta $p_i$ as the generators of the Berezin algebra, any function $\mathcal{F}$ of these variables is itself an element of the generated Berezin algebra. Then, considering the functions $\mathcal{F}$ and $\mathcal{G}$ of definite parity, we define their Poisson bracket to be
\begin{align}
    \{\mathcal{F},\mathcal{G}\}= \frac{\partial _r \mathcal{F}}{\partial q^i} \frac{\partial _l \mathcal{G}}{\partial p_i}-(-1)^{P_\mathcal{F} P_\mathcal{G}}\frac{\partial _r \mathcal{G}}{\partial q^i} \frac{\partial _l \mathcal{F}}{\partial p_i}.
    \label{eq: PB_def}
\end{align}
In Eq.\ \eqref{eq: PB_def} a summation over the index $i$ is implied. In the case of a field theory, the partial derivatives are to be interpreted as functional derivatives taken at a fixed time and the summation over $i$ includes an integration over space. Two useful properties of this generalised Poisson bracket are 
\begin{align}
\label{eq: PB_prop1}
    \{\mathcal{F},\mathcal{G}\} &= -(-1)^{P_\mathcal{F} P_\mathcal{G}} \{\mathcal{G},\mathcal{F}\}, \\
\label{eq: PB_prop2}    
\{\mathcal{F},\mathcal{G}\mathcal{K}\} &= \{\mathcal{F},\mathcal{G}\}\mathcal{K}+(-1)^{P_\mathcal{F} P_\mathcal{G}} \mathcal{G} \{\mathcal{F},\mathcal{K}\}.
\end{align} 

\clearpage
\lhead{\emph{Appendix F}}
\chapter{Some details of the constraint analysis}
\label{app: constraint_comp}
In this section we show the validity of Eq.\ \eqref{eq: conservation_secondary_gauge_constraint}, that is, we prove that
\begin{align}
\{\Phi_{1(2)}^A,H_T\} \approx -g f^{ABC} \left(W^{0B} \Phi^C_{1(2)}+V^{0B} \Phi^C_{2(2)}\right).    
\end{align}
For convenience we repeat the two relevant constraints,
\begin{align}
\label{eq: secondary_gauge_constraint_convenience}
    \Phi_{1(2)}^A ={}&  D^i \pi(W)_i^A + g f^{DAC} \pi(V)^D_\mu V^{\mu C} +\frac{i}{2}g\Hdagg \tau^A \pi(\Hdagg) -\frac{i}{2}g\pi(H) \tau^A H \nonumber \\ & -\frac{i}{2} g \bar{\pi}(f) \tau^A f - \frac{i}{2}g \bar{f} \tau^A \pi(\bar{f}), \\
     \label{eq: secondary_resonance_constraint_convenience}
    \Phi_{2(2)}^A  ={}& D^i \pi(V)_i^A+\frac{i}{2}g_H \Hdagg \tau^A \pi(\Hdagg)-\frac{i}{2}g_H \pi(H) \tau^A H \nonumber \\
    &+ V^{0A} |H|^2 \left(g_{VH}-\frac{g_H^2}{2}\right)+ g_f J_0^{fA} + m_V^2 V^{0A}.
\end{align}
Using the linearity of the PB we may write
\begin{align}
    \{\Phi_{1(2)}^A,H_T\} ={}& \{D^i \pi(W)_i^A,H_T\}+gf^{DAC} \{\pi(V)^D_\mu V^{\mu C},H_T\} \nonumber \\
    &+\frac{i}{2}g \{\Hdagg \tau^{A} \pi(\Hdagg),H_T\}-\frac{i}{2}g \{\pi(H) \tau^{A} H,H_T\} \nonumber \\
    &-\frac{i}{2}g\{\bar{\pi}(f) \tau^{A} f,H_T\}-\frac{i}{2}g\{\bar{f} \tau^{A} \pi(\bar{f}),H_T\}.
    \label{eq: full_PB}
\end{align}
One may then compute the different contributions independently using Eq.\ \eqref{eq: PB_prop2} and the definition as given in Eq.\ \eqref{eq: PB_def}. It should be noted that the PBs have to be interpreted as functionals and the equalities below hold when acting upon appropriate test functions. Consider the first PB of Eq.\ \eqref{eq: full_PB}, which can be written as
\begin{align}
    \{D^i \pi(W)^A_i(x),H_T\} ={}& \{\partial^i \pi(W)^A_i(x),H_T\} + g f^{ABC} W^{iB}(x)\{\pi(W)_i^C(x),H_T\} \nonumber \\
    &+ g f^{ABC} \pi(W)_i^C(x) \{W^{iB}(x),H_T\},
    \label{eq: full_PB_CD_part}
\end{align}
where we have written out the dependence of the field on the spacetime point explicitly. The first term in the first line is given by
\begin{align}
   \{\partial^i \pi(W)^A_i(x),H_T\} = - \int \rd^3 y \int \rd^3 z \left(\partial^i_x \delta_{xy}\right) \frac{\delta_t\Ham (z)}{\delta W^{iA}(y)},
\end{align}
which acting on a test-function $\varphi^A(x)$ yields 
\begin{align}
    -\int \rd^3 x \int \rd^3 y \int \rd^3 z \, \varphi^A(x) \left(\partial^i_x \delta_{xy}\right) \frac{\delta_t\Ham (z)}{\delta W^{iA}(y)} = -\int \rd^3 y \int \rd^3 z \, \varphi^A(y) \partial^i_y \frac{\delta_t\Ham (z)}{\delta W^{iA}(y)}.
    \label{eq: full_PB_partial_part}
\end{align}
To arrive at Eq.\ \eqref{eq: full_PB_partial_part} we first used the delta function, $\delta_{xy}\equiv \delta(\mathbf{x}-\mathbf{y})$, to perform the $x$-integral and then integrated by parts in the $y$-integral. It follows from this that 
\begin{align}
    \{\partial^i \pi(W)^A_i(x),H_T\} = - \int \rd^3 z \, \partial^i_x \frac{\delta_t\Ham (z)}{\delta W^{iA}(x)},
\end{align}
and in a similar fashion one can show that 
\begin{align}
    g f^{ABC} W^{iB}(x) \{\pi(W)_i^C(x),H_T\} = - g f^{ABC} \int \rd^3 z \, W^{iB}(x) \frac{\delta_t\Ham (z)}{\delta W^{iA}(x)},
\end{align}
so that the first line of Eq.\ \eqref{eq: full_PB_CD_part} is given by 
\begin{align}
 \{\partial^i \pi(W)^A_i(x),H_T\} + g f^{ABC} W^{iB}(x)\{\pi(W)_i^C(x),H_T\} = - \int \rd^3 z \, D^i_x \frac{\delta_t\Ham (z)}{\delta W^{iA}(x)}.
\label{eq: first_line}
\end{align}
The relevant functional derivative of the Hamiltonian density can be computed from Eq.\ \eqref{eq: modified_Hamiltonian_density} and reads
\begin{align}
    -\frac{\delta_t\Ham (z)}{\delta W^{iA}(x)} ={}& \Big(-g f^{DAF} \pi(W)_i^D(z) W^{0F}(z)-g f^{DAF} \pi(V)_i^D(z) V^{0F}(z)\nonumber \\
    &+\frac{g_H g}{4} V_i^B(z) \Hdagg (z) [\tau^B,\tau^A]_{+} H(z) + \frac{g}{2}\bar{f}(z)\tau^A \gamma_i f(z) \nonumber \\ 
    &+ \frac{ig}{2}\Hdagg(z) \tau^A \overleftrightarrow{D}_i H(z) + g f^{ABC} V^B_{ij}(z)V^{jC}(z) + g g_M f^{ABC} W_{ij}^{B}(z) V^{jC}(z) \nonumber \\
    & +W^B_{ik}(z)D^{kBA}_z+g_M V^B_{ik}(z)D^{kBA}_z\Big) \delta_{zx},
    \label{eq: full_PB_CD_kernel}
\end{align}
where 
\begin{align}
    D^{kBA}_z = \partial_z^k \delta^{AB} + g f^{BCA} W^{kC}(z). 
\end{align}
For the first three lines of Eq.\ \eqref{eq: full_PB_CD_kernel} the delta function ensures that we obtain a total covariant derivative of each individual term upon insertion into Eq.\ \eqref{eq: first_line}. For the last line of Eq.\ \eqref{eq: full_PB_CD_kernel} the delta function allows us to move the covariant derivative to the field strength tensors at the cost of a minus sign. We may therefore write Eq.\ \eqref{eq: first_line} as
\begin{align}
  \{\partial^i \pi(W)^A_i(x),H_T\} + g f^{ABC} W^{iB}(x)\{\pi(W)_i^C(x),H_T\} = D^i \mathcal{R}_i^A -\frac{g g_M}{2}f^{ABC} W^{ikB}V_{ik}^C  ,
  \label{eq: full_PB_CD_kernel_final}
\end{align}
where the r.h.s.\ is evaluated at the spacetime point $x$. We here defined 
\begin{align}
    \mathcal{R}_i^A ={}& -g f^{DAF} \pi(W)_i^D W^{0F}-g f^{DAF} \pi(V)_i^D V^{0F}\nonumber \\
    &+\frac{g_H g}{4} V_i^B \Hdagg  [\tau^B,\tau^A]_{+} H + \frac{g}{2}\bar{f}\tau^A \gamma_i f \nonumber \\ 
    &+ \frac{ig}{2}\Hdagg \tau^A \overleftrightarrow{D}_i H + g g_M f^{ABC} W_{ij}^B V^{jC} +  g f^{ABC} V^B_{ij}V^{jC},
\end{align}
and used that 
\begin{align}
    -D^i D^k V_{ik}^B = -\frac{g}{2} f^{BCD}W^{ikC} V_{ikB}.
\end{align}
A similar term arises from $-D^i D^k W_{ik}^B$ and vanishes due the antisymmetry of the structure constants. To Eq.\ \eqref{eq: full_PB_CD_kernel_final} we have to add
\begin{align}
  g f^{ABC} \pi(W)_i^C \{W^{iB},H_T\} ={}& -\frac{g_M}{g_M^2-1}g f^{ABC} \pi(W)_i^C \pi(V)^{iB} \nonumber \\
  &+ g f^{ABC} \pi(W)_i^C(x) D^i W^{0B},  
  \label{eq: full_PB_CD_part_gauge_cont}
\end{align}
to obtain a full expression for the first term contributing to Eq.\ \eqref{eq: full_PB}. It should be noted that combining the first term of $D^i \mathcal{R}^A_i$ with the last line of Eq.\ \eqref{eq: full_PB_CD_part_gauge_cont} one obtains 
\begin{align}
    -g f^{ABC} W^{0B}D^i \pi(W)^C_i \subset -g f^{ABC} W^{0B} \Phi_{1(2)}^C.
    \label{eq: Phi_12_1}
\end{align}
Having treated the first term contributing to Eq.\ \eqref{eq: full_PB} in detail we simply state the results for the remaining ones. We find
\begin{align}
    gf^{DAC} \{\pi(V)^D_\mu V^{\mu C},H_T\} \approx {}& gf^{DAC}\Big(\frac{-g_M}{g_M^2-1}\pi(V)_i^D \pi(W)^{iC}+\pi(V)_i^D D^i V^{0C} \nonumber \\
    &~~~~~~~~~~- g f^{CFE} \pi(V)_i^D W^{0F}V^{iE}+g f^{EBD} \pi(V)_i^E V^{iC} W^{0B}\nonumber \\ 
    &~~~~~~~~~~+g_H V^{iC}J^{HD}_i+g_f V^{iC}J^{fD}_i \nonumber \\
    &~~~~~~~~~~+V^{iC}D^{jDB}V_{ji}^B+g_MV^{iC}D^{jDB}W_{ji}^B\Big), 
    \label{eq: full_PB_resonances_contr}
\end{align}
where the first term in Eq.\ \eqref{eq: full_PB_resonances_contr} cancels the first term in Eq.\ \eqref{eq: full_PB_CD_part_gauge_cont} and the second term combines with the second term of $D^i \mathcal{R}^A_i$ to yield
\begin{align}
    -g f^{ABC} V^{0B}D^i \pi(V)^C_i \subset -g f^{ABC} V^{0B} \Phi_{2(2)}^C.
    \label{eq: Phi_2_2_1}
\end{align}
Furthermore, the last term in the last line in the definition of $\mathcal{R}^A_i$, when acted on with a covariant derivative, combines with the first term in the last line of Eq.\ \eqref{eq: full_PB_resonances_contr} to yield a contribution that is proportional to $f^{ABC}V_{ij}^{B}V^{ijC}=0$. Similarly, the second term in the last line of Eq.\ \eqref{eq: full_PB_resonances_contr} combines with the second to last term of $D^i \mathcal{R}_i^A$ to yield $\frac{gg_M}{2}f^{ABC}V^{jiC}W_{ji}^B$, which then cancels the second term on the r.h.s.\ of Eq.\ \eqref{eq: full_PB_CD_kernel_final}. Using the Jacobi identity, the second line of Eq.\ \eqref{eq: full_PB_resonances_contr} can be brought into the form
\begin{align}
    -g f^{AFC}W^{0F} \left(gf^{DCE} \pi(V)^D_iV^{iE}\right) \subset -g f^{AFC} W^{0F} \Phi^C_{1(2)}.
    \label{eq: Phi_12_2}
\end{align}
This eliminates all contributions to Eqs.\ \eqref{eq: full_PB_CD_kernel_final} and \eqref{eq: full_PB_resonances_contr} that do not depend on at least one fermionic or one scalar field. The second line of Eq.\ \eqref{eq: full_PB} is given by
\begin{align}
    \frac{i}{2}g \{\Hdagg \tau^{A} \pi(\Hdagg),H_T\}-\frac{i}{2}g \{\pi(H) \tau^{A} H,H_T\} \approx{}& -\frac{ig}{2}f^{ABC} \Hdagg \tau^C \pi(\Hdagg) (gW_0^B+g_H V_0^B) \nonumber \\ &+\frac{ig}{2}f^{ABC} \pi(H) \tau^C H (gW_0^B+g_H V_0^B) \nonumber \\
    &-\frac{g g_H}{4} \Hdagg \tau^A \tau^B \left(2V^{iB}D_iH+HD_iV^{iB}\right) \nonumber \\
    &-\frac{g g_H}{4} \left(2V^{iB}D_i\Hdagg+\Hdagg D_iV^{iB}\right) \tau^B \tau^A H \nonumber \\
    & -\frac{ig}{2} \Hdagg \tau^A D_iD^i H +\frac{ig}{2} \left(D_iD^i\Hdagg\right) \tau^A H \nonumber \\
    & +\frac{ig}{2} \Hdagg \tau^A \frac{\partial \Lag{rest}}{\partial \Hdagg}-\frac{ig}{2} \frac{\partial \Lag{rest}}{\partial H} \tau^A H, 
    \label{eq: full_PB_scalar_contr}
\end{align}
the first two lines of which are recognised as
\begin{align}
    -gf^{ABC}W_0^B \left(\frac{ig}{2} \Hdagg \tau^C \pi(\Hdagg) - \frac{ig}{2} \pi(H) \tau^C H \right) \subset  -gf^{ABC}W_0^B \Phi^C_{1(2)},
    \label{eq: Phi_12_3}
\end{align}
and
\begin{align}
    -gf^{ABC}V_0^B \left(\frac{ig_H}{2} \Hdagg \tau^C \pi(\Hdagg) - \frac{ig_H}{2} \pi(H) \tau^C H \right) \subset  -gf^{ABC}V_0^B \Phi^C_{2(2)},
    \label{eq: Phi_22_2}
\end{align}
completing the first line of the constraints of Eqs.\ \eqref{eq: secondary_gauge_constraint_convenience} and \eqref{eq: secondary_resonance_constraint_convenience}. Noting that 
\begin{align}
    D^i\left(\frac{ig}{2}\Hdagg \tau^A \overleftrightarrow{D}_i H\right) = \frac{ig}{2}\left(-\left(D_iD^i \Hdagg\right)\tau^A H + \Hdagg \tau^A D^iD_i H \right),
\end{align}
we find that this contribution to $D^i \mathcal{R}_i^A$ cancels the second to last line of Eq.\ \eqref{eq: full_PB_scalar_contr}. In order to treat the third line of Eq.\ \eqref{eq: full_PB_scalar_contr} we act on a test function $\varphi^A$ and integrate one of the two terms where the covariant derivative acts on the scalar doublet by parts
\begin{align}
   &-\frac{g g_H}{4} \int \rd^3 x\, \Hdagg \tau^A \tau^B \left(2V^{iB}D_iH+HD_iV^{iB}\right) \varphi^A \nonumber \\
   &= \frac{g g_H}{4} \int \rd^3 x \, V^{iB} \left[\left(D_i\Hdagg\right) \tau^A \tau^BH\varphi^A+\Hdagg \tau^A \tau^BHD_i\varphi^A-\Hdagg \tau^A \tau^B\left(D_iH\right)\varphi^A\right]. 
\end{align}
Performing the corresponding operation on the fourth line of Eq.\ \eqref{eq: full_PB_scalar_contr} and combining the two results we find
\begin{align}
    &-\frac{g g_H}{4} \! \int \rd^3x \! \left[\Hdagg \tau^A \tau^B \left(2V^{iB}D_iH+HD_iV^{iB}\right)+\left(2V^{iB}D_i\Hdagg+\Hdagg D_iV^{iB}\right) \tau^B \tau^A H\right] \varphi^A \nonumber \\
    &= -\frac{g g_H}{4} \int \rd^3 x \left[V^{iB}\Hdagg [\tau^A,\tau^B]\left(D_iH\right)-V^{iB}\left(D_i\Hdagg\right) [\tau^A,\tau^B]H\right]\varphi^A \nonumber \\
    &~~~~ -\frac{g g_H}{4} \int \rd^3 x\, D_i \left(V^{iB} \Hdagg [\tau^A,\tau^B]_{+}H\right)\varphi^A,
    \label{eq: full_PB_scalar_contr_intermediate_step}
\end{align}
where $[\tau^A,\tau^B]$ is the commutator of $\tau^A$ and $\tau^B$. Removing the test function we note that the last line of Eq.\ \eqref{eq: full_PB_scalar_contr_intermediate_step} cancels with the first term of the second line of $D^i \mathcal{R}_i^A$. The first line of Eq.\ \eqref{eq: full_PB_scalar_contr_intermediate_step} yields 
\begin{align}
    -\frac{g g_H}{4} \left[V^{iB}\Hdagg [\tau^A,\tau^B]\left(D_iH\right)-V^{iB}\left(D_i\Hdagg\right) [\tau^A,\tau^B]H\right] = -g g_H f^{CAB}V^{iB}J^{HC}_i,
\end{align}
which cancels the first term of the third line of Eq.\ \eqref{eq: full_PB_resonances_contr}. This only leaves the last line of Eq.\ \eqref{eq: full_PB_scalar_contr}, which we will comment on shortly. First, we address the last line of Eq.\ \eqref{eq: full_PB}, which is given by
\begin{align}
-\frac{i}{2}g\{\bar{\pi}(f) \tau^{A} f,H_T\}-\frac{i}{2}g\{\bar{f} \tau^{A} \pi(\bar{f}),H_T\} \approx{}& -g g_F f^{ABC} V^{\mu B} J_\mu^{fC}-g^2 f^{ABC} W^{\mu B} J_\mu^{fC} \nonumber \\
&-ig \frac{\partial_r \Lag{rest}}{\partial f}\frac{\tau^A}{2}f-ig \frac{\partial _r \Lag{rest}}{\partial \bar{f}} \frac{\tau^A}{2} \bar{f} \nonumber \\
&-\frac{g}{2}\partial_i\left(\bar{f}\tau^A\gamma^i f\right).
\label{eq: full_PB_fermionic_contr}
\end{align}
The spatial part of the first term of Eq.\ \eqref{eq: full_PB_fermionic_contr} cancels the last remaining term of Eq.\ \eqref{eq: full_PB_resonances_contr} and the time component yields
\begin{align}
    -g g_F f^{ABC} V^{0 B} J_0^{fC} \subset -g f^{ABC} V^{0B} \Phi^C_{2(2)}.
    \label{eq: Phi_22_3}
\end{align}
The time component of the second term of Eq.\ \eqref{eq: full_PB_fermionic_contr} is given by 
\begin{align}
    -g^2 f^{ABC}W^{0B} \bar{f}\gamma^0 \frac{\tau^A}{2}f \approx \frac{i}{2} g^2 f^{ABC} W^{0B} \left( \bar{\pi}(f) \tau^A f + \bar{f} \tau^A \pi(\bar{f})\right) \subset -g f^{ABC}W^{0B}\Phi^C_{1(2)},
    \label{eq: Phi_12_4}
\end{align}
where we used the fermionic constraints. The spatial part of this term combines with the last line of Eq.\ \eqref{eq: full_PB_fermionic_contr} to yield 
\begin{align}
    -\frac{g}{2}D_i\left(\bar{f}\tau^A\gamma^if\right),
\end{align}
which cancels the last remaining term of $D^i \mathcal{R}_i^A$. This only leaves the terms 
\begin{align}
    \frac{ig}{2} \Hdagg \tau^A \frac{\partial \Lag{rest}}{\partial \Hdagg}-\frac{ig}{2} \frac{\partial \Lag{rest}}{\partial H} \tau^A H-ig \frac{\partial_r \Lag{rest}}{\partial f}\frac{\tau^A}{2}f-ig \frac{\partial _r \Lag{rest}}{\partial \bar{f}} \frac{\tau^A}{2} \bar{f},
\end{align}
which vanish as long as $\Lag{rest}$ is gauge invariant. For a proof see Ref.\ \cite{guitman1990quantization}. Collecting Eqs.\ \eqref{eq: Phi_12_1}, \eqref{eq: Phi_12_2}, \eqref{eq: Phi_12_3} and \eqref{eq: Phi_12_4} we find precisely $-gf^{ABC}W^{0B}\Phi_{1(2)}^C$. Similarly, collecting Eqs.\ \eqref{eq: Phi_2_2_1}, \eqref{eq: Phi_22_2} and \eqref{eq: Phi_22_3} one finds $-gf^{ABC}V^{0B}\Phi_{2(2)}^C$ since $f^{ABC}V^{0B}V^{0C}=0$.

\addtocontents{toc}{\vspace{2em}}  
\backmatter

\printbibliography
\end{refsection}

\setstretch{1.3}  

\cleardoublepage

\acknowledgements{
\addtocontents{toc}{\vspace{1em}}  

First of all I would like to thank Prof.\ Michael Kr\"amer for making this thesis possible by accepting me as a PhD student in his group. I appreciate that I was given the opportunity to follow my interests while also being provided with guidance when needed. 

I also thank Prof.\ Robert Harlander for agreeing to be my second supervisor.

For encouraging me to work on this project and for countless hours of useful discussions, and useful comments on earlier versions of this thesis, I would like to thank Dr.\ Alexander Voigt. I also thank Jonas Klappert and Fabian Lange for useful discussions, and Jonas Klappert in particular for comments on earlier versions of this thesis. Furthermore, I thank Jonas Klappert and Fabian Lange for making my time in the office a very pleasant experience.

A special thank you goes to everybody who supported me with encouragement throughout the completion of this work. This in particular includes my parents Roland and Sabine, without whom I would not be the person I am today, as well as my sister Alice. Furthermore I thank Hesam and Moncef, not only for encouragement, support and friendship throughout the completion of this thesis, but also the five years of studies that led up to it. I thank Jonas for his friendship and support and for the many rounds of Swedish basketball that we played together.

Last but not least I thank Physics itself for being such a rich subject that never seizes to surprise me. Thank you for the smile created by the feeling that a new layer of reality has revealed itself.

}
\cleardoublepage  
\Eid{
Ich, Benjamin Summ erkläre hiermit, dass diese Dissertation und die darin dargelegten Inhalte die eigenen sind und selbstständig, als Ergebnis der eigenen originären Forschung, generiert wurden.
Hiermit erkläre ich an Eides statt
\begin{enumerate}
    \item Diese Arbeit wurde vollst\"andig oder gr\"o{\ss}tenteils in der Phase als Doktorand dieser Fakult\"at und Universit\"at angefertigt;
    \item Sofern irgendein Bestandteil dieser Dissertation zuvor f\"ur einen akademischen Abschluss oder eine andere Qualifikation an dieser oder einer anderen Institution verwendet wurde, wurde dies klar angezeigt;
    \item Wenn immer andere eigene- oder Ver\"offentlichungen Dritter herangezogen wurden, wurden diese klar benannt;
    \item Wenn aus anderen eigenen- oder Ver\"offentlichungen Dritter zitiert wurde, wurde stets die Quelle hierf\"ur angegeben. Diese Dissertation ist vollst\"andig meine eigene Arbeit, mit der Ausnahme solcher Zitate; 
    \item Alle wesentlichen Quellen von Unterst\"utzung wurden benannt;
    \item Wenn immer ein Teil dieser Dissertation auf der Zusammenarbeit mit anderen basiert, wurde von mir klar gekennzeichnet, was von anderen und was von mir selbst erarbeitet wurde;
    \item Ein Teil oder Teile dieser Arbeit wurden zuvor ver\"offentlicht und zwar in:
    \begin{itemize}
    \item[{[BS1]}] \fullcite{Summ:2018oko}.
    \item[{[BS2]}] \fullcite{Kramer:2019fwz}. 
\end{itemize}
\end{enumerate}
}

\end{document}